# Data-driven relationship of atomic structure and physical properties as the holistic view on the materials science fundamentals

Pierre Villars,
Material Phases Data System,
Unterschwanden 6, 6354, Vitznau, Switzerland
E-mail: villars.mpds@bluewin.ch

Evgeny Blokhin,
Materials Platform for Data Science,
Sepapaja 6, 15551, Tallinn, Estonia, and
Tilde Materials Informatics,
Straßmannstraße 25, 10249, Berlin, Germany
E-mail: eb@tilde.pro

Shuichi Iwata,
The University of Tokyo,
7-3-1, Hongo, Bunkyo-ku, 113-8654, Tokyo, Japan
E-mail: iwatacodata@mac.com

## Abstract

The fundamental relationship of the atomic structure (represented by its atomic property parameters, APPs) and its physical properties of a specific inorganic substance can be realized in the bottom-up data-centric and the top-down knowledge physics-centric ways. Nowadays these two approaches compete and enhance one another qualitatively and quantitatively. We present our own holistic method and implementation, based on the PAULING FILE peer-reviewed inorganic substances database, the world largest materials database containing under one shelter crystallographic structures, phase diagrams and large variety of physical properties of single-phase inorganic substances. In addition we present generated machine-learning data, as well as simulated DFT physics-centered data, which are in close connection and comparison with the PAULING FILE peer-reviewed reference data.

## Content

## 1. Introduction

This review is largely based on our previous publications [1-9]. Using the most recent release of the online platform called Materials Platform for Data Science, we have updated all the relevant numbers, figures, and tables with the quantitative information [17a,b]. The review's content is prepared in a didactic 'build-up' way, which demonstrates a path dependency of building materials databases and shows future directions of the data-centric materials research. We give below a brief overview on the inspiring studies and their framing prior to the detailed description of our efforts.

Getting the holistic views on the inorganic substances is the ultimate objective not only for materials scientists, but for engineers, technology adopters, and other advanced users as well. Deep insight into the inorganic substances has been developed via ongoing interaction between observation, experiments, data compilation, theoretical modeling, and computation, from *ab initio* to quantitative reasoning. Most examples of prominent scientists such as Mendeleyev, Meyer, and Pauling illustrated it through the history. Today, entering the 'Data Era', such interactions can be carried out on a huge scale and relatively easily by means of the programming logic and data. For instance, introducing the topology into the data sets allows getting a variety of relations freely by tracing the knowledge that has been compiled inductively, deductively, or abductively, as working hypotheses. About 30 years ago, two of us (P.V. and S.I.) made an attempt to do so, following the spirit of Linus Pauling, dealing with the numerous challenges and learning from mistakes in developing inorganic substance data systems since the 1960s. This became the origin of the PAULING FILE project, which was launched in 1992 as a joint undertaking of Japan Science and Technology Corporation, Tokyo, Japan (JST), Material Phases Data system, Vitznau, Switzerland (MPDS), and the University of Tokyo, RACE, Japan [9].

There are two fundamental approaches to the 'holistic view' on the inorganic substances. The first one is the bottom-up data-centric approach, relied on inorganic substance *data*. This data-driven concept is the core of the PAULING FILE project. The second approach is known as the top-down knowledge physics-centric approach. According to it, the guidelines and logics are taken from the outside, for example from the models elaborated in other scientific fields (mathematics, physics, chemistry, and engineering) and surrounding monitored environments (nature and artifacts). The top-down approach may serve as a combination of potent sets of scientific models integrated strategically through the eyes of discipline-based scientific rationality and (or) ad hoc constitutive engineering models within the tolerance range, driven by the market demands. Such integrations have been oscillating between facts and logics similarly to a pendulum in human history, converging into the united agenda of the materials design, as stated in 1963 in a volume by Arthur R. von Hippel [102]: "Nature designs everything from atoms; hence, we should be able to create any feasible kind of material and device with foresight, if we understand the periodic system in all its implications. Yet – like the weather forecast – we find ourselves still members of the somewhat gambling profession".

It was the liftoff of the integration of science and engineering on the materials: practical applications of prediction began to appear [103] and theoretical basis [104] for *ab initio* calculations was created. Such pioneering studies inspired networking of multi-facet knowledge models into an uninterrupted process of materials development by filling gaps and harmonizing available data with associated models, i.e. a Computer-Assisted Design (CAD) system by E. J. Corey et al. [105]. An explicit representation on molecular structures of organic materials enables to reuse structure-property correlations as explicitly digitized scientific knowledge (not tacit knowledge usually mentioned in engineering applications). As a result, materials data can be used not only as inputs to simulation programs estimating properties with embedded algorithms, but also as inputs to causality, predefined rules applied by Artificial Intelligence (AI) systems. The following chapters demonstrate the introduction of semi-explicit representations on structures of inorganic materials by its atomic environment type (AET) introduced by Daams, van Vucht and Villars [26]. We have presented the description of spatial relations among structures and states of materials in organized ways via different diagrams. Thus, there are phase diagrams which were theoretically formulated by Gibbs in the $19^{th}$ century and then compiled as handbooks by successors motivated by industrial needs, Hansen among others. There are also structure maps by Pettifor [101] and others, CCT and TTT diagrams considering cooling rate and time, and other state diagrams in regard to magnetic field, irradiation conditions, etc. Interactions of materials' structural primitives have been brought together in so-called mechanism maps, deformation mechanism maps. Integrating models of deviant behaviors and crucial data of materials illustrate general mechanisms of deformation as called Ashby maps [106]. Pourbaix diagrams show electrochemical stability for variable redox states of an element as a function of pH, and so on.
The data-driven approach to obtain a holistic view needs a digitized system, whose prototype was elaborated as a Computer-Aided Design (CAD) system of databases, simulation and AI in the mid-1970s. In the mid-1980s, almost all ideas (i.e. meta-data, meta-knowledge, distributed data systems, intelligent systems of data classification and data mining, knowledge management, learning, reasoning, and heuristics as deduction, induction, and abduction on the inference logics), had been re-accessed on the threshold of the new networked information environment and the augmenting accessibility of high-performance computers.

Holistic views are got through converging interplay between the bottom-up and top-down approaches. For the first time it was successfully demonstrated by Pierre Villars 'manually' in 1985 [10], and can strategically be transcribed into computational algorithms. The PAULING FILE project suggested the possibility of such a converging interplay through a digital platform. Ideas and prototypes for digital systems supporting similar interplay had also been proposed by E.J. Corey in 1969, and then developed by S. Iwata et al. [116] in 1975, but the key difference of the PAULING FILE project is to discover knowledge directly from data, rather than to re-use a range of pre-defined knowledge. In fact, the new knowledge is emerging from the data, even if we reuse a set of predefined knowledge patterns for convenience. However, making it happen requires some information infrastructure. Triggered by a workshop that took place in Como in 1992 [11], huge contributions towards 'holistic views' on inorganic substances were envisaged in the shape of networked intelligent systems of key data and models with powerful PCs and high-performance computers. A few projects to build inorganic substance data systems, targeting practical needs for materials users in industry, were started in the mid-1990s. The Virtual Experiment for Materials Design (VEMD) project [107] and the PAULING FILE project, concentrating on the top-down and bottom-up approaches, respectively, were typical projects kicking off at that time. But, as proved by this review, the implementation of a digital system of practical impact is not fast. The editorial team (24 full-time researchers) of the PAULING FILE has devoted their time for almost three decades after its launch to develop the PAULING FILE to practical competitiveness. Without a comprehensive compilation of the highly quality-controlled inorganic substance data, as written in the following chapters, even if followed by strategic mixing of inductive and deductive inferences, holistic views have NOT yet 'emerged' digitally.

The top-down approach for engineering applications aimed by the VEMD project requires extensions of physics-centered way for substance such as piling up of comparative studies by allocating logics for each fact, enhancing logical coherence of the accumulated facts and logics along hierarchical structures from spin level to engineering product, balancing combined uncertainties to meet requirements with enough accuracy for potential users. Though we are still in an incubation period for such engineering applications, lessons obtained through the PAULING FILE project experience at the first phase need careful applying. Such lessons are, for instance, importance of the total quality control with explicit and transparent linkage of associated data like phase diagrams, crystallography, intrinsic and fundamental properties of materials, data-centric inverse approaches on models, strategic organizations of the best teams for the second phase development market-out and market-in as well as the first phase development product-out and market-in. The first milestone of this assumed the top-down approach can be fulfilled by carrying out following agenda:

• Appliance of Internet of Things (IoT) (integration of huge monitoring data, data mining, data assimilation and adaptive data utilization technologies, etc.) for engineering products as proposed as 'digital twins' for maintenance services of recent General Electric jet engines. It is an implementation of Jack. H. Westbrook's ideas [108] in the 1970s and 1980s integrating all indispensable data and knowledge for one engineering product covering from its scientific basics to commercial service data including associated claims and troubles.

• Compiling the above challenges into common platforms, flexible enough activating weak ties of specialized experts in different domains, tipping points to converge into a solution and other creative dynamics as discussed by a sociologist M. Granovetter [117]. This evolving ecology has been becoming a reality through the basic feature of the PAULING FILE, namely, interoperability not only as a digital system but also as the scientific semantics of systematically and hopefully perfectly organized structural data as described precisely in the following chapters.

The below passage has been written in 2002 in the manual of the first output of the PAULING FILE Binaries Edition [8] published by ASM International:

One of the most challenging tasks in material science is the design of novel inorganic substances with predefined physical properties. In general, two different approaches are explored:

- Top-down, considering the atomic motion in the inorganic substance as well as their electronic interactions as close to reality as possible by using simulations on the quantum-mechanical level.

- Bottom-up, which seems to be less challenging at first glance, since it does not ask for the really fundamental principles but remains on a pragmatic level. Nevertheless most of our current knowledge in chemistry and material science has been collected *empirically*, just by searching for patterns and rules within experimental results published in the world literature.

Beside the classical use of these databases for analysis, phase identification and teaching purposes, *it should be worthwhile trying to use modern computer technology to search for additional rules and correlations implied in these data and use them in inorganic substances design.*

The shortcomings of the empirical bottom-up approach provided the initial motivation for the launch of the PAULING FILE project in 1992. The project envisaged three steps:

- The initial objective is to invent and maintain a comprehensive database for all solid, mainly single-phase, inorganic substances / systems (no C-H bonds), covering crystallographic data, diffraction patterns, phase diagrams, and a vast variety of physical properties. An important aspect along with completeness is the data quality. The information needs screening with extreme care, for unrecognized errors will confuse the correlation tools or even result in deducing wrong rules.

- Simultaneously to the database creation, appropriate retrieval software is elaborated. This makes the four different groups of inorganic substances data mentioned above accessible by a single user interface.

- In longer term, novel inorganic substance design tools will be invented, which will more or less automatically search the database for correlations that would help with the purposeful and rapid design of novel inorganic substances with distinct predefined physical properties.

Our first in-house developed inorganic substances design instrument was the DISCOVERY program [12]. As soon as one has extracted some interesting dataset from the PAULING FILE database, one probably would like to search for some hidden links between the inorganic substances and atomic property parameters (APPs) of its constituent chemical elements. Thus the DISCOVERY is a tool for searching the links between the specific inorganic substance property (e.g. structure prototype or enthalpy of formation) and APPs of its constituent chemical elements (e.g. atomic number AN or electronegativity EN values). For the binary case, the APPs of the two constituent elements are combined by some mathematical expression (e.g. the difference of its atomic numbers) to an atomic property parameter expression (APPE). The varieties of APPs as well as the range of mathematical operations that may be applied to them are pre-selected by the user before the computation. The DISCOVERY explores all possible ‘features sets’ of two or three of these combinations; the goal of this exploration is to find features sets, which causes a maximum separation between the different inorganic substance property groups (like e.g. structure prototypes). Let’s work through an example: A large variety of binary inorganic substances crystallize either in the NaCl,*cF8*,225 or CsCl,*cP2*,221 structure prototypes. The task is to establish a correlation between the structure prototype of these inorganic substances and some mathematical expressions based on APPs of their constituent chemical elements. Since several inorganic substances crystallize in both structure prototypes depending on certain conditions (e.g. high pressure), one concentrates on environmental conditions for stable inorganic substances.

A graphical representation of the discovery space is presented in Fig. 1. There is an excellent distinction between the blue dots (representing the inorganic substances, which crystallize in NaCl,*cF8*,225 structure prototype) and the yellow dots (representing the ones with CsCl,*cP2*,221 structure prototype). This 'best' separation was obtained by using as APPE: periodic number $PN_{MD}$ (on the basis on the periodic system from Mendeleyev) difference and maximum values (for details see chapter 2.).

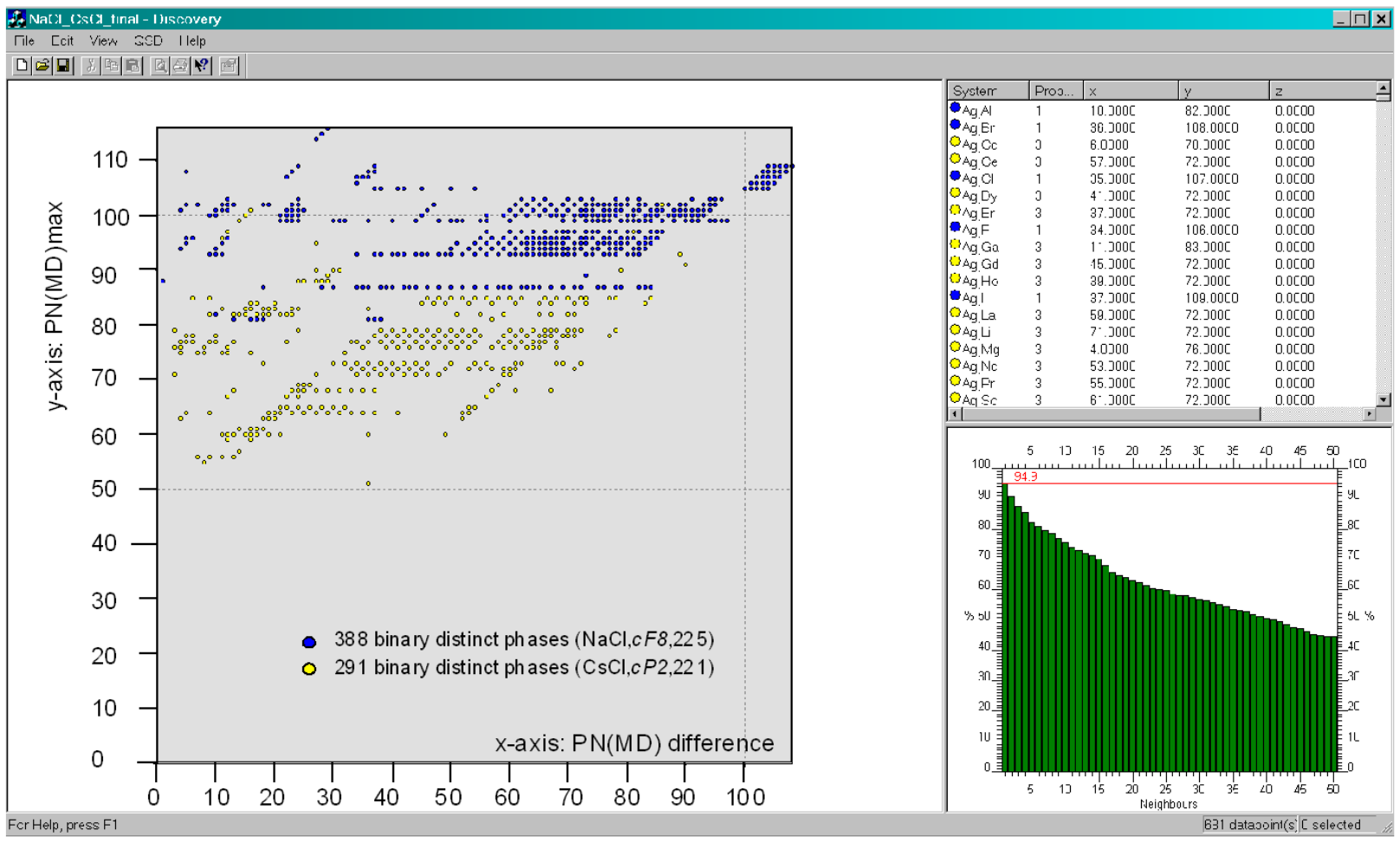

Fig. 1. "Discovery Space" view of the best result.

## 2. Description of the atomic structures shown through their atomic property parameters

A vast majority of the chemical elements are well characterized by different experimental and theoretical atomic property parameters (APPs), for instance, the atomic number AN, experimental and theoretical radii, ionization potential, bulk modulus, melting temperature, etc. Plotting these APPs versus AN respectively PN, most of them may be divided into five evidently different pattern groups which are shown in Figs. 2a+b, 3a+b, 4.

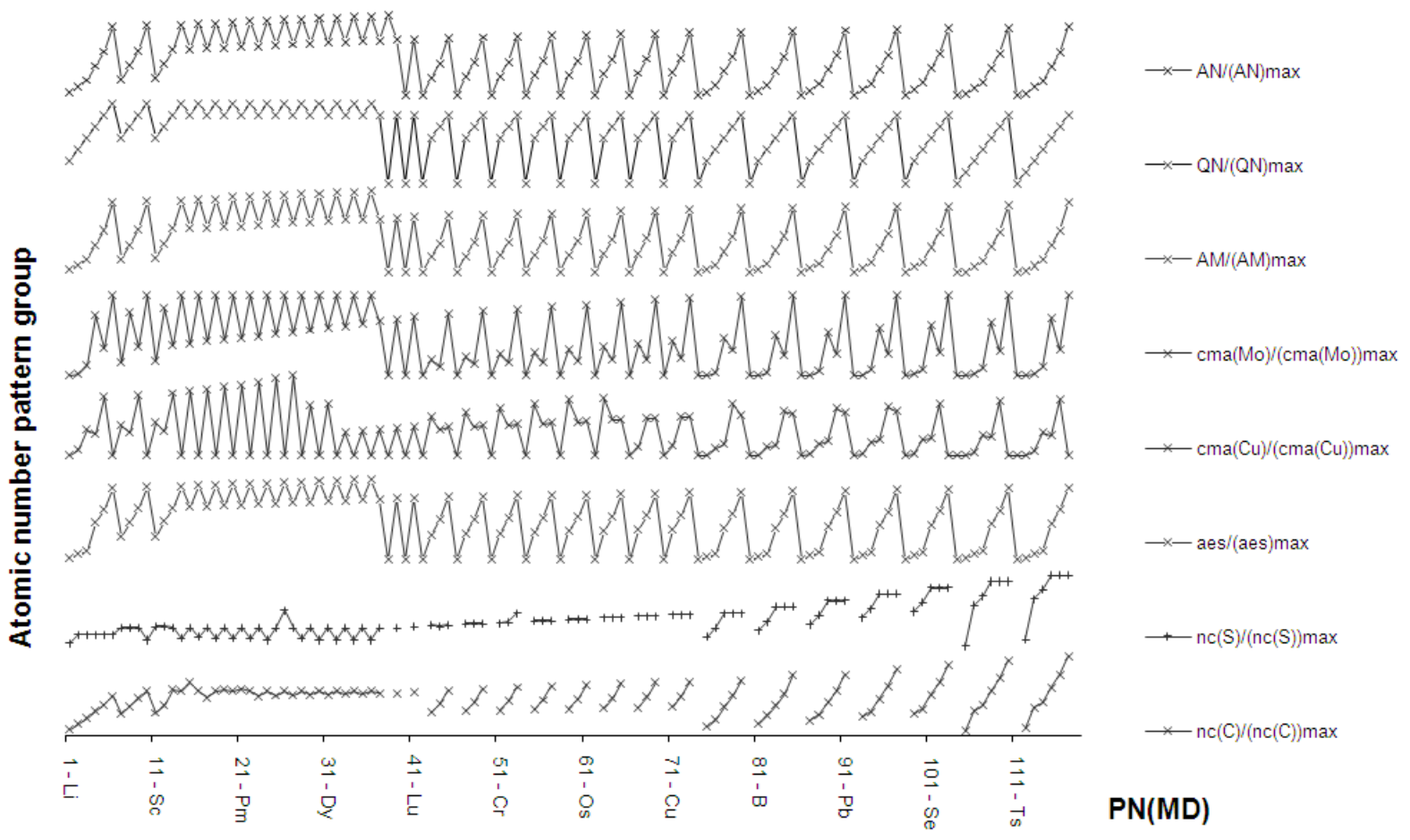

Fig. 2a. The 8 atomic property parameters (normalized to their maximum values) belonging to the atomic number pattern group versus $PN_{MD}$, where $PN_{MD}$ is periodic number using Mendeleyev periodic system.

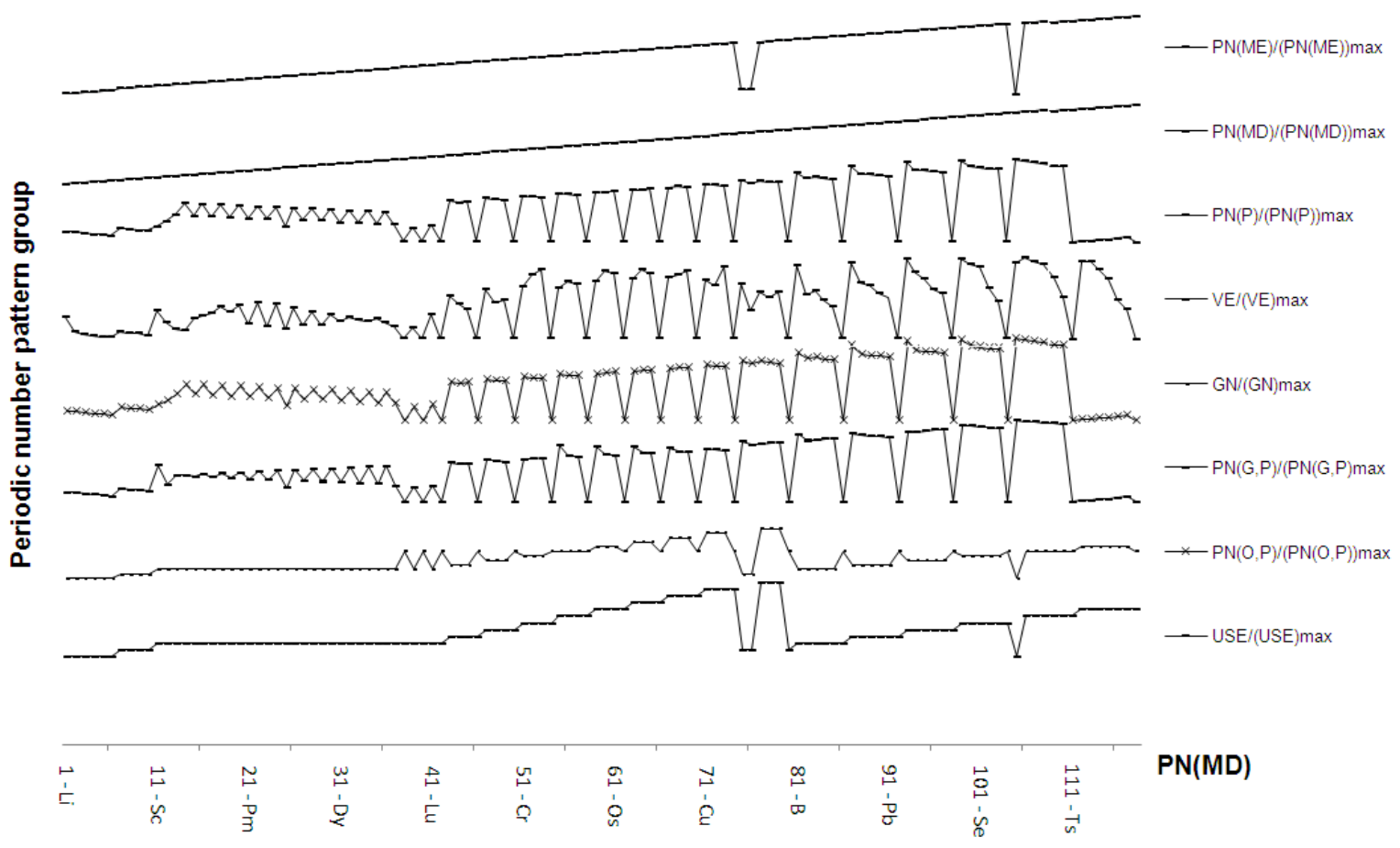

Fig. 2b. The 8 atomic property parameters (normalized to their maximum values) belonging to the periodic number pattern group versus $PN_{MD}$, where $PN_{MD}$ is periodic number using Mendeleyev periodic system.

---

I) **Atomic number pattern group** (see Fig. 2a)
AN: Atomic number [/] [14]
QN: Main quantum number [/] [14]
AM: Atomic mass [$10^{-3}$ kg] [13]
cma(Mo): Coefficient of mass attenuation for Mo Ka [$cm^2$ $g^{-1}$] [91]
cma(Cu): Coefficient of mass attenuation for Cu Ka [$cm^2$ $g^{-1}$] [91]
aes: Atomic electron scattering factor [/] [91]
nc(S): Nuclear effective charge according to Slater [/] [92]
nc(C): Nuclear effective charge according to Clementi [/] [91]

II) **Periodic number pattern group** (see Fig. 2b)
PN(ME): Periodic number according to Meyer [/] [14]
PN(MD): Periodic number according to Mendeleyev [/] [5]

PN(P): Periodic number according to Pettifor's sequence [/] [93]
VE: Valence electron number [/] [14]
GN: Group number [/] [14]
PN(G,P) Glawe's modified periodic number according to Pettifor [/] [58]
PN(O,P) Oganov's modified periodic number according to Pettifor [/] [58]
USE Universal sequence of elements according to Oganov [/] [58]

III) **Atomic size pattern group** (see Fig. 3a)
R(Z): Pseudo-potential radii according to Zunger [a.u.] [52]
Ri(Y): Ionic radii according to Yagoda [Å] [91]
Rc(P): Covalent radii according to Pauling [pm] [91]
Rm(WG): Metal radii according to Waber and Gschneidner [Å] [94]
Rve(S): Valence electron distance according to Schubert [Å] [95]
Rce(S): Core electron distance according to Schubert [Å] [95]
R(M): Radii according to Miedema (derived from his $V^{2/3}$ compilation) [$cm^2$] [96]
Va: Atomic volume according to Busch [$cm^3$ g-atom$^{-1}$] [13]

IV) **Atomic reactivity pattern group** (see Fig. 3b)
EN(MB): Electronegativity according to Martynov and Batsanov [/] [98]
EN(P): Electronegativity according to Pauling [/] [13]
EN(AR): Electronegativity according to Allred and Rochow [/] [91]
EN(abs): Absolute electronegativity [/] [95]
IE(first): First ionization energy [kJ mole$^{-1}$] [13]
CP(M): Chemical potential according to Miedema [V] [96]
wf: Work function [eV] [13]
n(WS): n(Wigner and Seitz) according to Miedema [a.u.$^{-1/3}$] [96]

V) **Atomic affinity pattern group** (see Fig. 4)
T(fus): Temperature of melting [K] [13]
T(vap): Temperature of boiling [K] [13]
H(vap): Enthalpy of vaporization [kJ mole$^{-1}$] [13]
H(fus): Enthalpy of melting [kJ mole$^{-1}$] [13]
H(atom): Enthalpy of atomization [kJ mole$^{-1}$] [13]
Hsurf(M): Surface energy according to Miedema [kJ mole$^{-1}$] [96]
CE(B): Cohesion energy according to Brewer [J mole$^{-1}$] [99]
VC: Isothermal volume compressibility [GPa] [99]

---

Table 1. The 40 atomic property parameters (APPs) considered are grouped into the five fundamental groups, the atomic number AN pattern group and the periodic number PN pattern group, the atomic size pattern group, the atomic reactivity pattern group, and the atomic affinity pattern group.

The APPs belonging to the same pattern group are qualitatively equivalent despite their quantitative difference. The 40 APPs considered are shown in Table 1. The remaining 60 APPs not considered in the absence of completeness (e.g. values only known for a limited number of chemical elements) and (or) considered APP is too sensitive to impurities and other defects (e.g. electrical conductivity and APP derived from a combination of two pattern groups like e.g. mass density $D_X$). Looking closer at the data sets plotted in Table 1 we find out that the integer parameter values are present only in the first two pattern groups. The only comprehensive ones are AN and PN. The atomic number AN and the periodic number PN determine, in a different way, the position of each chemical element in the periodic system (also called periodic table). The AN is a simple count of the protons of the nucleus, which is equal to the total number of electrons of this chemical element. The periodic number PN is the result of a slightly other enumeration of the chemical elements, since it accentuates the role of the valence electrons, in particular, it implicitly reflects the construction principle of the periodic system, which is less obvious from the AN. For the periodic number PN the enumeration of the elements is performed in first priority within the same group number GN with increasing main quantum number QN, and in second priority with increasing group number GN.

The familiar group names of the periodic system of Meyer (groups IA, IIA, IIIB, IVB, VB, VIB, VIIB, VIIIB (for the Fe, Co and Ni groups), IB, IIB, IIIA, IVA, VA, VIA, VIIA, VIIIA) are replaced by GN 1, 2, … 18
(18 = maximum number of s, p, and d electrons, 2 + 6 + 10). The details, on the other hand, depend upon the arrangement of the underlying periodic system. In the Sanderson representation of the periodic system [13], e.g. the transition elements (including Cu, Ag and Au) are rated as insert below Be, Mg and between Ca, Sr, Ba, Ra and Zn, Cd, Hg. In the same way as the rare-earth elements Ce…Yb and the actinide elements Th…No are inserted below Sc, Y and between La, Ac and Lu, Lr. We refer to this kind of periodic systems since they visualize the uncertainty in placing the elements Be, Mg, which are found above Ca, Sr, Ba, Ra in the Meyer periodic system (here designated as $PN_{ME}$) (see Table 2) [14], but above Zn, Cd, Hg in the Mendeleyev periodic system (here designated as $PN_{MD}$) (see Table 3) [15].

AN $PN_{ME}$
$SZ_{aME}/(SZ_{aME})_{max}$
$RE_{aME}/(RE_{aME})_{max}$

| | IA | IIA | IIIB | IVB | VB | VIB | VIIB | VIIIB | VIIIB | VIIIB | IB | IIB | IIIA | IVA | VA | VIA | VIIA | VIIIA |
|---|---|---|---|---|---|---|---|---|---|---|---|---|---|---|---|---|---|---|
| GN | 1 | 2 | 3 | 4 | 5 | 6 | 7 | 8 | 9 | 10 | 11 | 12 | 13 | 14 | 15 | 16 | 17 | 18 |
| 1 | 1 H 1<br>0.16383<br>0.73024 | | | | | | | | | | | | | | | | | 2 He 112<br>0.05596<br>2.13768 |
| 2 | 3 Li 2<br>0.32684<br>0.36603 | 4 Be 8<br>0.35485<br>0.33713 | | | | | | | | | | | 5 B 82<br>0.15287<br>0.78257 | 6 C 88<br>0.15167<br>0.78879 | 7 N 94<br>0.14730<br>0.81216 | 8 O 100<br>0.14058<br>0.85098 | 9 F 106<br>0.13207<br>0.90585 | 10 Ne 113<br>0.11963<br>1.0000 |
| 3 | 11 Na 3<br>0.58150<br>0.20573 | 12 Mg 9<br>0.55821<br>0.21431 | | | | | | | | | | | 13 Al 83<br>0.22186<br>0.53922 | 14 Si 89<br>0.20781<br>0.57567 | 15 P 95<br>0.19319<br>0.61926 | 16 S 101<br>0.17810<br>0.67171 | 17 Cl 107<br>0.16265<br>0.73553 | 18 Ar 114<br>0.14382<br>0.83180 |
| 4 | 19 K 4<br>0.69396<br>0.17239 | 20 Ca 10<br>0.65398<br>0.18293 | 21 Sc 14<br>0.63030<br>0.18980 | 22 Ti 46<br>0.43051<br>0.27789 | 23 V 50<br>0.41533<br>0.28804 | 24 Cr 54<br>0.40019<br>0.29894 | 25 Mn 58<br>0.38510<br>0.31065 | 26 Fe 62<br>0.37006<br>0.32328 | 27 Co 66<br>0.35507<br>0.33693 | 28 Ni 70<br>0.34013<br>0.35172 | 29 Cu 74<br>0.32526<br>0.36781 | 30 Zn 78<br>0.31044<br>0.38536 | 31 Ga 84<br>0.28706<br>0.41675 | 32 Ge 90<br>0.26414<br>0.45291 | 33 As 96<br>0.24164<br>0.49508 | 34 Se 102<br>0.21954<br>0.54493 | 35 Br 108<br>0.19779<br>0.60484 | 36 Kr 115<br>0.17263<br>0.69301 |
| 5 | 37 Rb 5<br>0.83298<br>0.14362 | 38 Sr 11<br>0.77674<br>0.15402 | 39 Y 15<br>0.74258<br>0.16110 | 40 Zr 47<br>0.50364<br>0.23753 | 41 Nb 51<br>0.48244<br>0.24797 | 42 Mo 55<br>0.46178<br>0.25907 | 43 Tc 59<br>0.44162<br>0.27090 | 44 Ru 63<br>0.42190<br>0.28356 | 45 Rh 67<br>0.40260<br>0.29715 | 46 Pd 71<br>0.38367<br>0.31181 | 47 Ag 75<br>0.36509<br>0.32768 | 48 Cd 79<br>0.34684<br>0.34493 | 49 In 85<br>0.31920<br>0.37479 | 50 Sn 91<br>0.29236<br>0.40920 | 51 Sb 97<br>0.26623<br>0.44935 | 52 Te 103<br>0.24076<br>0.49689 | 53 I 109<br>0.21589<br>0.55412 | 54 Xe 116<br>0.18743<br>0.63826 |
| 6 | 55 Cs 6<br>0.91053<br>0.13139 | 56 Ba 12<br>0.84609<br>0.14139 | 6' | 72 Hf 48<br>0.57475<br>0.20815 | 73 Ta 52<br>0.54867<br>0.21804 | 74 W 56<br>0.52343<br>0.22855 | 75 Re 60<br>0.49896<br>0.23976 | 76 Os 64<br>0.47519<br>0.25176 | 77 Ir 68<br>0.45206<br>0.26464 | 78 Pt 72<br>0.42952<br>0.27853 | 79 Au 76<br>0.40752<br>0.29356 | 80 Hg 80<br>0.38602<br>0.30991 | 81 Tl 86<br>0.35416<br>0.33779 | 82 Pb 92<br>0.32335<br>0.36997 | 83 Bi 98<br>0.29350<br>0.40761 | 84 Po 104<br>0.26452<br>0.45227 | 85 At 110<br>0.23633<br>0.50621 | 86 Rn 117<br>0.20429<br>0.58560 |
| 7 | 87 Fr 7<br>1.00000<br>0.11963 | 88 Ra 13<br>0.92721<br>0.12902 | 7' | 104 Rf 49<br>0.61579<br>0.19427 | 105 Db 53<br>0.58710<br>0.20377 | 106 Sg 57<br>0.55939<br>0.21386 | 107 Bh 61<br>0.53255<br>0.22464 | 108 Hs 65<br>0.50653<br>0.23618 | 109 Mt 69<br>0.48124<br>0.24859 | 110 Ds 73<br>0.45664<br>0.26199 | 111 Rg 77<br>0.43267<br>0.27650 | 112 Cn 81<br>0.40928<br>0.29230 | 113 Nh 87<br>0.37488<br>0.31913 | 114 Fl 93<br>0.34165<br>0.35016 | 115 Mc 99<br>0.30950<br>0.38654 | 116 Lv 105<br>0.27832<br>0.42983 | 117 Ts 111<br>0.24806<br>0.48228 | 118 Og 118<br>0.21372<br>0.55975 |
| QN | | | | | | | | | | | | | | | | | | |
| | | | 6' | 57 La 16<br>0.80695<br>0.14825 | 58 Ce 18<br>0.78996<br>0.15144 | 59 Pr 20<br>0.77343<br>0.15468 | 60 Nd 22<br>0.75734<br>0.15796 | 61 Pm 24<br>0.74165<br>0.16131 | 62 Sm 26<br>0.72634<br>0.16471 | 63 Eu 28<br>0.71138<br>0.16817 | 64 Gd 30<br>0.69675<br>0.17170 | 65 Tb 32<br>0.68242<br>0.17531 | 66 Dy 34<br>0.66837<br>0.17899 | 67 Ho 36<br>0.65459<br>0.18276 | 68 Er 38<br>0.64105<br>0.18662 | 69 Tm 40<br>0.62774<br>0.19058 | 70 Yb 42<br>0.61466<br>0.19463 | 71 Lu 44<br>0.60177<br>0.19880 |
| | | | 7' | 89 Ac 17<br>0.88292<br>0.13550 | 90 Th 19<br>0.86292<br>0.13864 | 91 Pa 21<br>0.84352<br>0.14183 | 92 U 23<br>0.82469<br>0.14506 | 93 Np 25<br>0.80639<br>0.14836 | 94 Pu 27<br>0.78857<br>0.15171 | 95 Am 29<br>0.77122<br>0.15512 | 96 Cm 31<br>0.75429<br>0.15860 | 97 Bk 33<br>0.73775<br>0.16216 | 98 Cf 35<br>0.72159<br>0.16579 | 99 Es 37<br>0.70577<br>0.16951 | 100 Fm 39<br>0.69028<br>0.17331 | 101 Md 41<br>0.67509<br>0.17721 | 102 No 43<br>0.66019<br>0.18121 | 103 Lr 45<br>0.64556<br>0.18532 |

Table 2

Table 2. Atomic number AN, periodic number $PN_{ME}$, atomic size $SZ_{aME}$, and atomic reactivity $RE_{aME}$ using the Meyer periodic system.

Legend: AN PN$_{MD}$ / $SZ_{aMD}/(SZ_{aMD})_{max}$ / $RE_{aMD}/(RE_{aMD})_{max}$

| | IA | IIA | IIIB | IVB | VB | VIB | VIIB | VIIIB | VIIIB | VIIIB | IB | IIB | IIIA | IVA | VA | VIA | VIIA | VIIIA |
|---|---|---|---|---|---|---|---|---|---|---|---|---|---|---|---|---|---|---|
| GN | 1 | 2 | 3 | 4 | 5 | 6 | 7 | 8 | 9 | 10 | 11 | 12 | 13 | 14 | 15 | 16 | 17 | 18 |
| 1 | | | | | | | | | | | | | | | | | 1 H 105<br>0.04000<br>2.95309 | 2 He 112<br>0.05526<br>2.13768 |
| 2 | 3 Li 1<br>0.32352<br>0.36512 | | | | | | | | | | | 4 Be 75<br>0.14987<br>0.78817 | 5 B 81<br>0.15317<br>0.77120 | 6 C 87<br>0.15208<br>0.77673 | 7 N 93<br>0.14784<br>0.79901 | 8 O 99<br>0.14125<br>0.83626 | 9 F 106<br>0.13040<br>0.90585 | 10 Ne 113<br>0.11812<br>1.0000 |
| 3 | 11 Na 2<br>0.57846<br>0.20420 | | | | | | | | | | | 12 Mg 76<br>0.23553<br>0.50153 | 13 Al 82<br>0.22232<br>0.53132 | 14 Si 88<br>0.20841<br>0.56679 | 15 P 94<br>0.19392<br>0.60912 | 16 S 100<br>0.17899<br>0.65995 | 17 Cl 107<br>0.16060<br>0.73553 | 18 Ar 114<br>0.14201<br>0.83180 |
| 4 | 19 K 3<br>0.69220<br>0.17065 | 20 Ca 7<br>0.67141<br>0.17593 | 21 Sc 11<br>0.64709<br>0.18255 | 22 Ti 43<br>0.44100<br>0.26786 | 23 V 47<br>0.42558<br>0.27756 | 24 Cr 51<br>0.41024<br>0.28794 | 25 Mn 55<br>0.39497<br>0.29907 | 26 Fe 59<br>0.37977<br>0.31104 | 27 Co 63<br>0.36466<br>0.32393 | 28 Ni 67<br>0.34962<br>0.33787 | 29 Cu 71<br>0.33466<br>0.35297 | 30 Zn 77<br>0.31091<br>0.37993 | 31 Ga 83<br>0.28769<br>0.41060 | 32 Ge 89<br>0.26493<br>0.44586 | 33 As 95<br>0.24261<br>0.48689 | 34 Se 101<br>0.22068<br>0.53528 | 35 Br 108<br>0.19530<br>0.60484 | 36 Kr 115<br>0.17045<br>0.69301 |
| 5 | 37 Rb 4<br>0.83202<br>0.14197 | 38 Sr 8<br>0.79756<br>0.14811 | 39 Y 12<br>0.76224<br>0.15497 | 40 Zr 44<br>0.51595<br>0.22894 | 41 Nb 48<br>0.49439<br>0.23893 | 42 Mo 52<br>0.47342<br>0.24951 | 43 Tc 56<br>0.45299<br>0.26076 | 44 Ru 60<br>0.43305<br>0.27277 | 45 Rh 64<br>0.41355<br>0.28563 | 46 Pd 68<br>0.39446<br>0.29946 | 47 Ag 72<br>0.37574<br>0.31438 | 48 Cd 78<br>0.34740<br>0.34003 | 49 In 84<br>0.31994<br>0.36921 | 50 Sn 90<br>0.29328<br>0.40277 | 51 Sb 96<br>0.26735<br>0.44184 | 52 Te 102<br>0.24207<br>0.48798 | 53 I 109<br>0.21317<br>0.55412 | 54 Xe 116<br>0.18507<br>0.63826 |
| 6 | 55 Cs 5<br>0.91016<br>0.12978 | 56 Ba 9<br>0.86879<br>0.13596 | 6' | 72 Hf 45<br>0.58884<br>0.20060 | 73 Ta 49<br>0.56231<br>0.21007 | 74 W 53<br>0.53670<br>0.22009 | 75 Re 57<br>0.51190<br>0.23076 | 76 Os 61<br>0.48784<br>0.24214 | 77 Ir 65<br>0.46446<br>0.25432 | 78 Pt 69<br>0.44171<br>0.26743 | 79 Au 73<br>0.41953<br>0.28157 | 80 Hg 79<br>0.38669<br>0.30547 | 81 Tl 85<br>0.35503<br>0.33272 | 82 Pb 91<br>0.32443<br>0.36410 | 83 Bi 97<br>0.29478<br>0.40072 | 84 Po 103<br>0.26601<br>0.44406 | 85 At 110<br>0.23335<br>0.50621 | 86 Rn 117<br>0.20171<br>0.58560 |
| 7 | 87 Fr 6<br>1.00000<br>0.11812 | 88 Ra 10<br>0.95203<br>0.12408 | 7' | 104 Rf 46<br>0.63093<br>0.18722 | 105 Db 50<br>0.60177<br>0.19630 | 106 Sg 54<br>0.57364<br>0.20592 | 107 Bh 58<br>0.54644<br>0.21617 | 108 Hs 62<br>0.52011<br>0.22712 | 109 Mt 66<br>0.49455<br>0.23885 | 110 Ds 70<br>0.46971<br>0.25148 | 111 Rg 74<br>0.44554<br>0.26513 | 112 Cn 80<br>0.41004<br>0.28808 | 113 Nh 86<br>0.37584<br>0.31429 | 114 Fl 92<br>0.34284<br>0.34455 | 115 Mc 98<br>0.31091<br>0.37993 | 116 Lv 104<br>0.27996<br>0.42192 | 117 Ts 111<br>0.24493<br>0.48228 | 118 Og 118<br>0.21103<br>0.55975 |
| QN | | | | | | | | | | | | | | | | | | |
| | | | 6' | 57 La 13<br>0.82819<br>0.14263 | 58 Ce 15<br>0.81046<br>0.14575 | 59 Pr 17<br>0.79323<br>0.14891 | 60 Nd 19<br>0.77648<br>0.15213 | 61 Pm 21<br>0.76019<br>0.15539 | 62 Sm 23<br>0.74432<br>0.15870 | 63 Eu 25<br>0.72885<br>0.16207 | 64 Gd 27<br>0.71375<br>0.16550 | 65 Tb 29<br>0.69898<br>0.16899 | 66 Dy 31<br>0.68454<br>0.17256 | 67 Ho 33<br>0.67039<br>0.17620 | 68 Er 35<br>0.65651<br>0.17993 | 69 Tm 37<br>0.64290<br>0.18374 | 70 Yb 39<br>0.62952<br>0.18764 | 71 Lu 41<br>0.61638<br>0.19164 |
| | | | 7' | 89 Ac 14<br>0.90600<br>0.13038 | 90 Th 16<br>0.88516<br>0.13345 | 91 Pa 18<br>0.86498<br>0.13656 | 92 U 20<br>0.84542<br>0.13972 | 93 Np 22<br>0.82645<br>0.14293 | 94 Pu 24<br>0.80802<br>0.14619 | 95 Am 26<br>0.79009<br>0.14951 | 96 Cm 28<br>0.77264<br>0.15288 | 97 Bk 30<br>0.75563<br>0.15633 | 98 Cf 32<br>0.73902<br>0.15984 | 99 Es 34<br>0.72280<br>0.16343 | 100 Fm 36<br>0.70693<br>0.16709 | 101 Md 38<br>0.69140<br>0.17085 | 102 No 40<br>0.67618<br>0.17469 | 103 Lr 42<br>0.66125<br>0.17864 |

Table 3

Table 3. Atomic number AN, periodic number $PN_{MD}$, atomic size $SZ_{aMD}$, and atomic reactivity $RE_{aMD}$ using the Mendeleyev periodic system.

A sort of phenomenological PN (periodic number-like) scale was initially proposed by Villars in 1981 [16] and two years later by Pettifor [100,101]. In spite of some differences, in both studies it was employed as APP to correlate them with crystal structures of binary inorganic substances via the structure prototype classification.

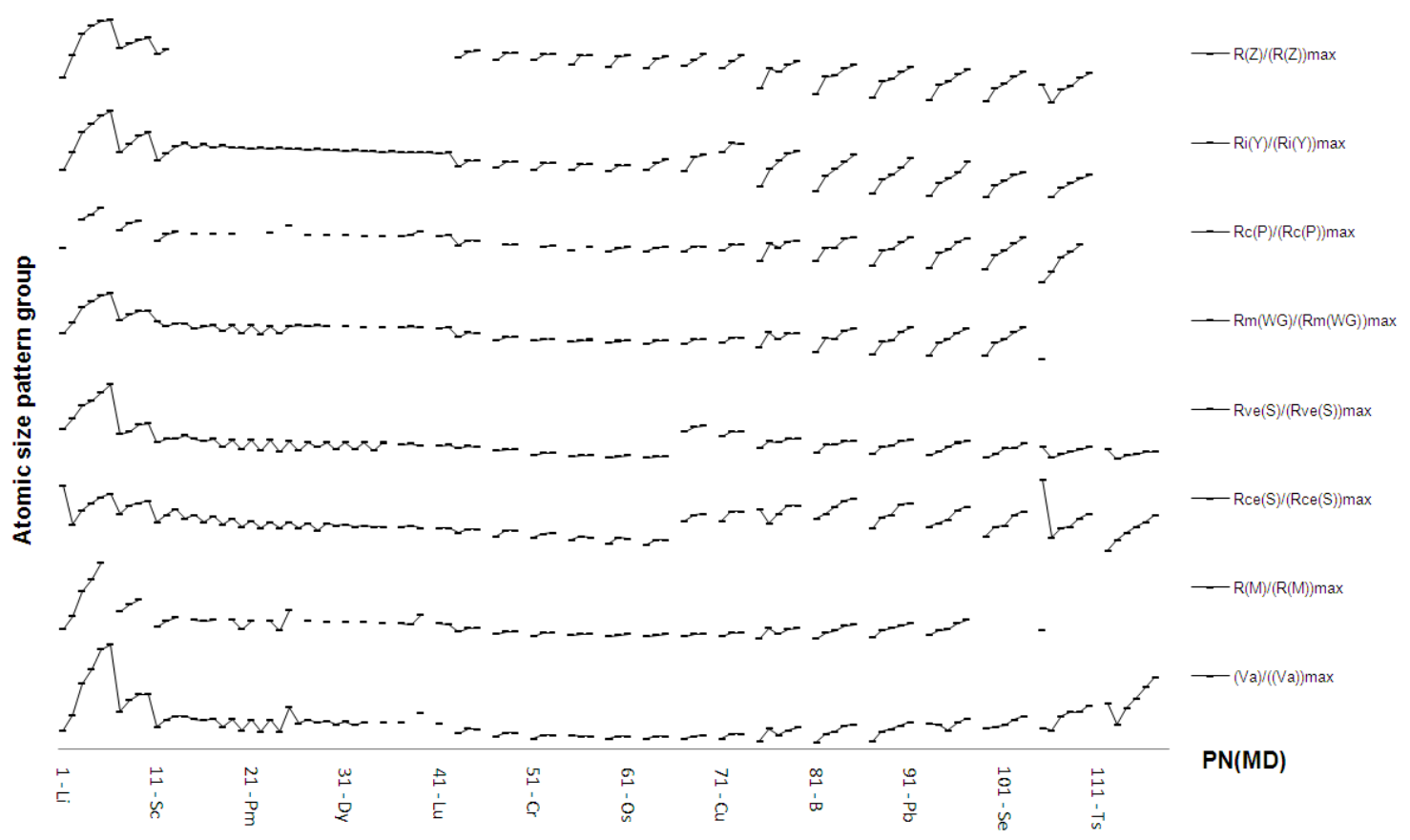

Fig. 3a. The 8 atomic property parameters (normalized to their maximum values) belonging to the atomic size pattern group versus $PN_{MD}$, where $PN_{MD}$ is periodic number using Mendeleyev periodic system.

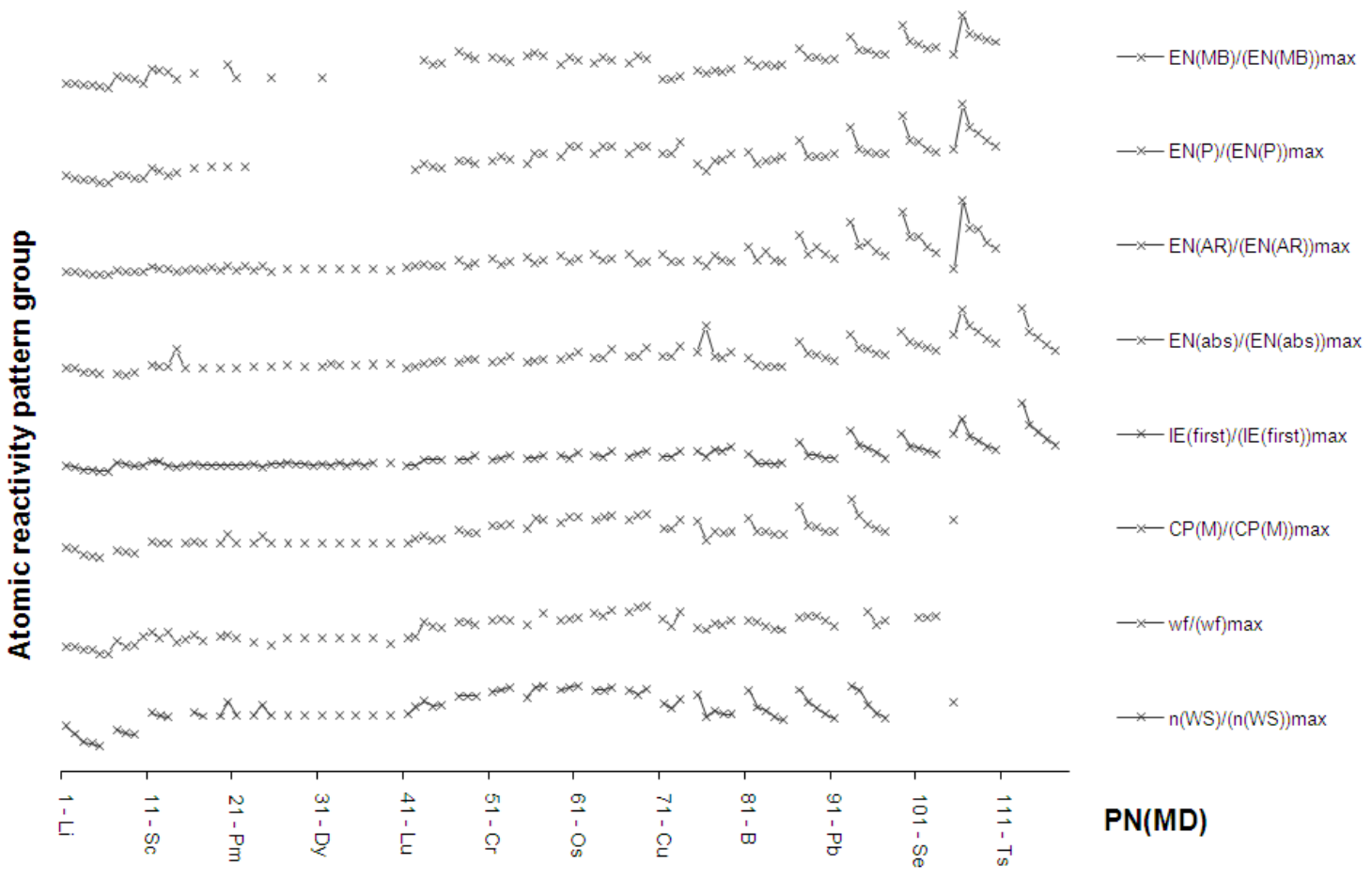

Fig. 3b. The 8 atomic property parameters (normalized to their maximum values) belonging to the atomic reactivity pattern group versus $PN_{MD}$, where $PN_{MD}$ is periodic number using Mendeleyev periodic system.

In Figs. 5a+b we illustrate the step-like functions $PN_{ME}/(PN_{ME})_{max}$ (respectively $PN_{MD}/(PN_{MD})_{max}$) versus AN, and the reversed way in Fig. 6 $AN/(AN)_{max}$ and $QN/(QN)_{max}$ versus $PN_{ME}$, where $(AN)_{max} = (PN_{ME})_{max} = 118$, $(GN)_{max} = 18$, $(QN)_{max} = 7$ for the chemical elements known up to now together with some predicted ones.

In Figs. 2a+b, 3a+b, 4 eight different data sets per atomic property parameter pattern groups are plotted as functions of $PN_{MD}$. Within each pattern group the eight data sets prove rather similar functional behavior (patterns) although they vary quantitatively from each other. Each pattern group demonstrates well-defined functional behavior within each group number GN, as well as within each main quantum number QN. Inspired by this fact and having observed by chance that in first approximation $R_{ionic}$(Yagoda) ≈ (PN-GN), we started studying functions like $(PN/(PN)_{max})^x$, $(\log PN/\log(PN)_{max})^x$, $\{[k-PN]/[k-(PN)_{max}]\}^x$, $\{[k-(\log PN)]/[k-(\log(PN)_{max})]\}^x$,…versus AN and the analogous functions of AN versus PN, x= ±1, ±2,… and k= $((PN)_{max})^x$ or $((AN)_{max})^x$, respectively $(\log(PN)_{max})^x$ or $(\log(AN)_{max})^x$, with the aim to reproduce APPs by a combination of such simple functions of AN and PN. A systematic trial and error approach led us finally to the following approximations for the atomic size $SZ_a$(AN,PN), the atomic reactivity $RE_a$(AN,PN), and in context with this review in the first approximation the atomic affinity $AF_a$(AN,PN):

$$SZ_a = k_{SZ}[\log(AN+1)][k_{PN}-(\log PN)^3], \quad (1)$$
$$RE_a = k_{RE}\{[\log(AN+1)][k_{PN}-(\log PN)^3]\}^{-1} = k_{SZ}k_{RE}(SZ_a)^{-1}, \quad (2)$$
$$AF_a = k_{AF}AN/k_{SZ}(SZ_a)^3 \approx D_X(\text{X-ray density}) \quad \textit{(first approximation)} \quad (3)$$

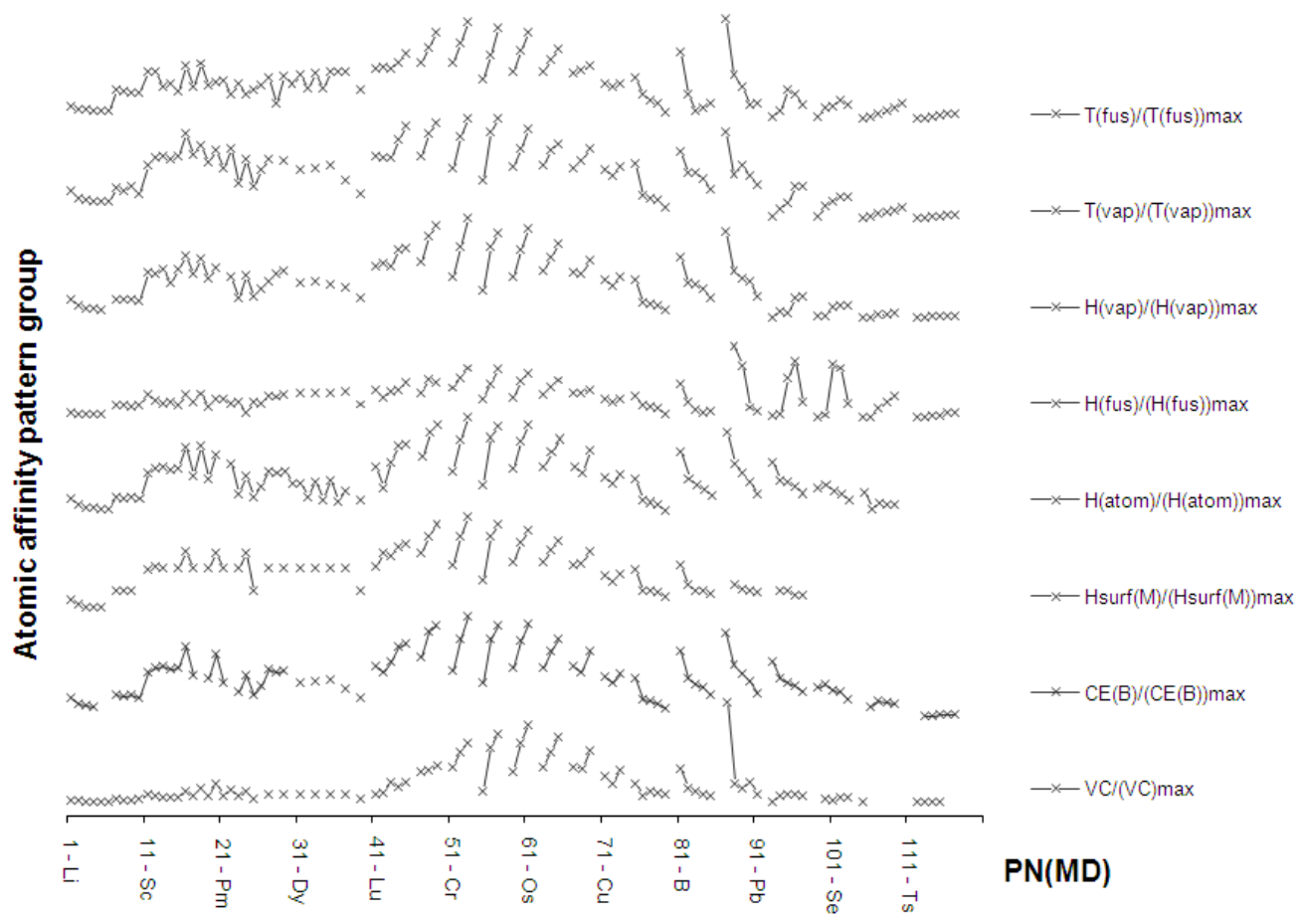

Fig. 4. The 8 atomic property parameters (normalized to their maximum values) belonging to the atomic affinity pattern group versus $PN_{MD}$, where $PN_{MD}$ is periodic number using Mendeleyev periodic system.

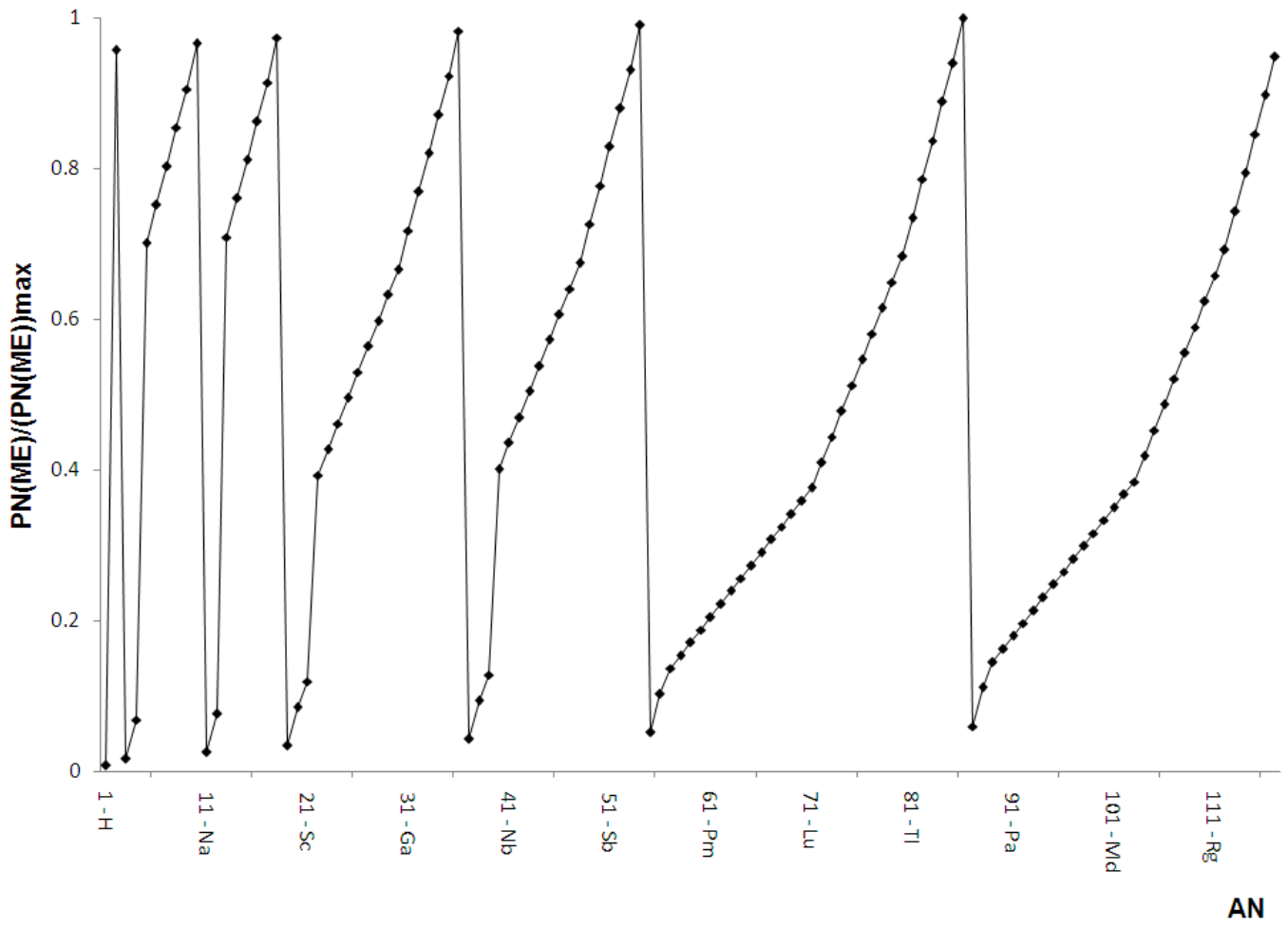

Fig. 5a. $PN_{ME}/(PN_{ME})_{max}$ versus AN, where AN is the atomic number, $PN_{ME}$ is the periodic number using the Meyer periodic system.

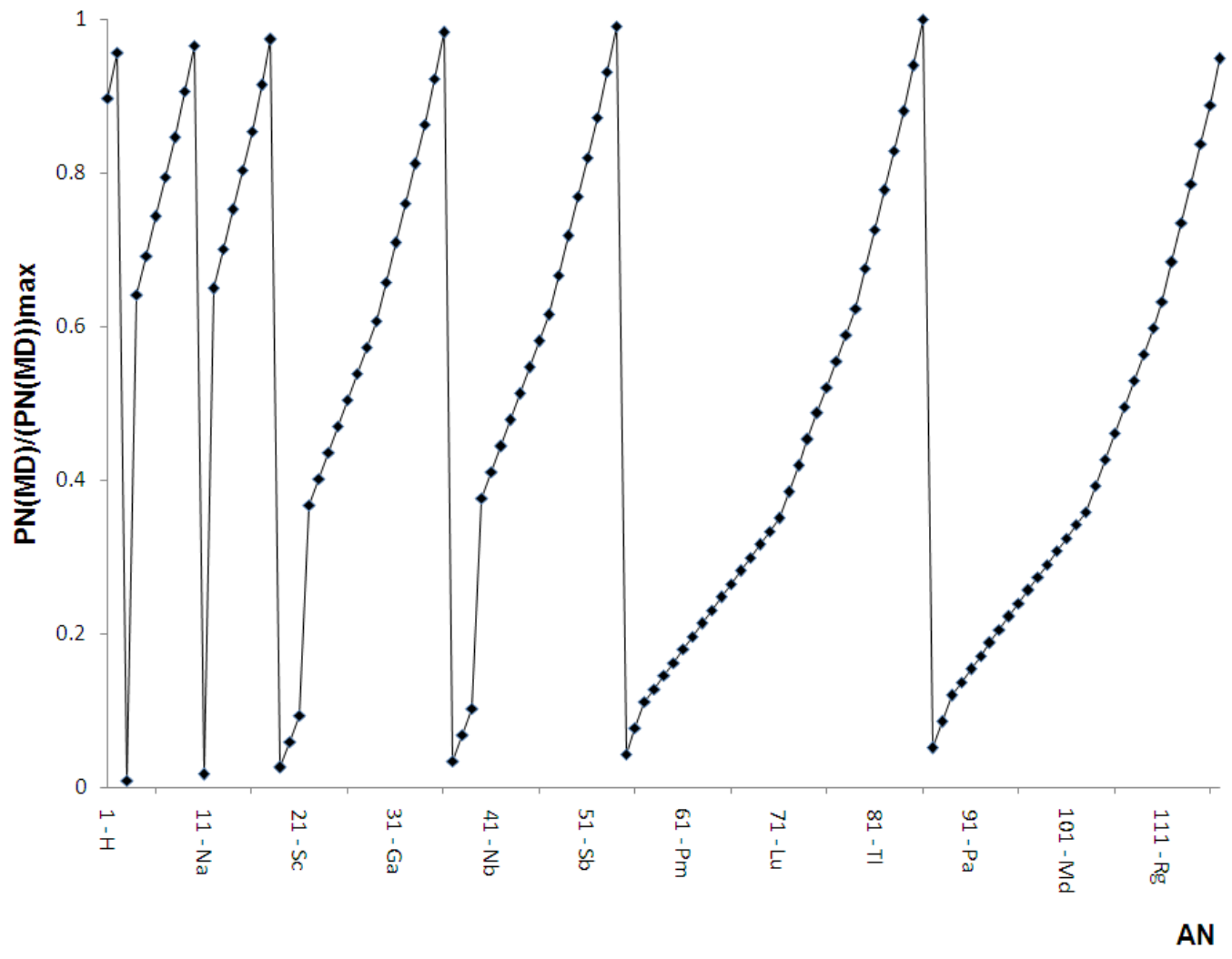

Fig. 5b. $PN_{MD}/(PN_{MD})_{max}$ versus AN, where AN is the atomic number, $PN_{MD}$ is the periodic number using the Mendeleyev periodic system.

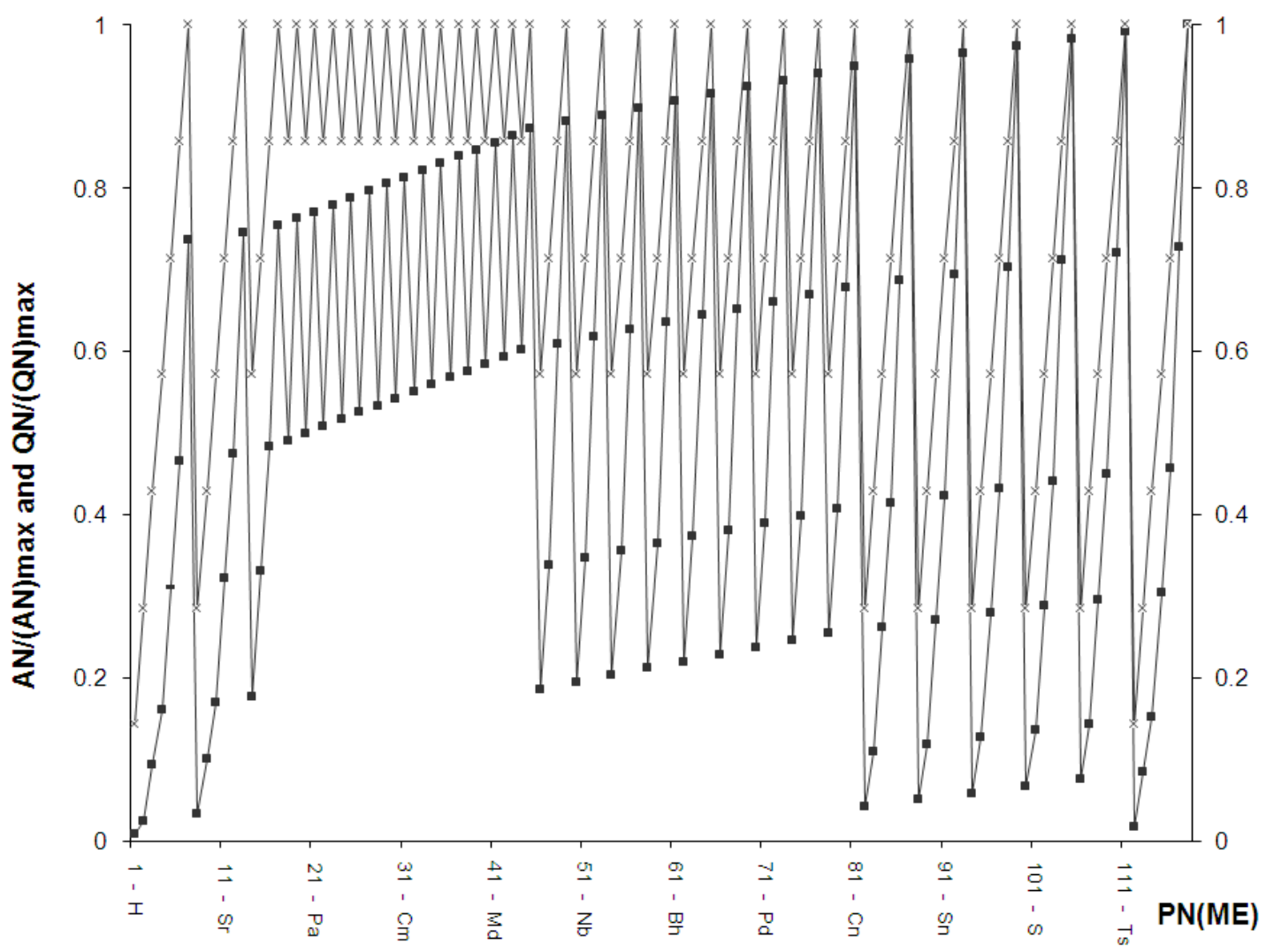

Fig. 6. $(AN/AN)_{max}$ and $QN/(QN)_{max}$ versus $PN_{ME}$, where AN is the atomic number, QN is the main quantum number, and $PN_{ME}$ is the periodic number using the Meyer periodic system; ■ $AN/(AN)_{max}$, x $(QN)/(QN)_{max}$.

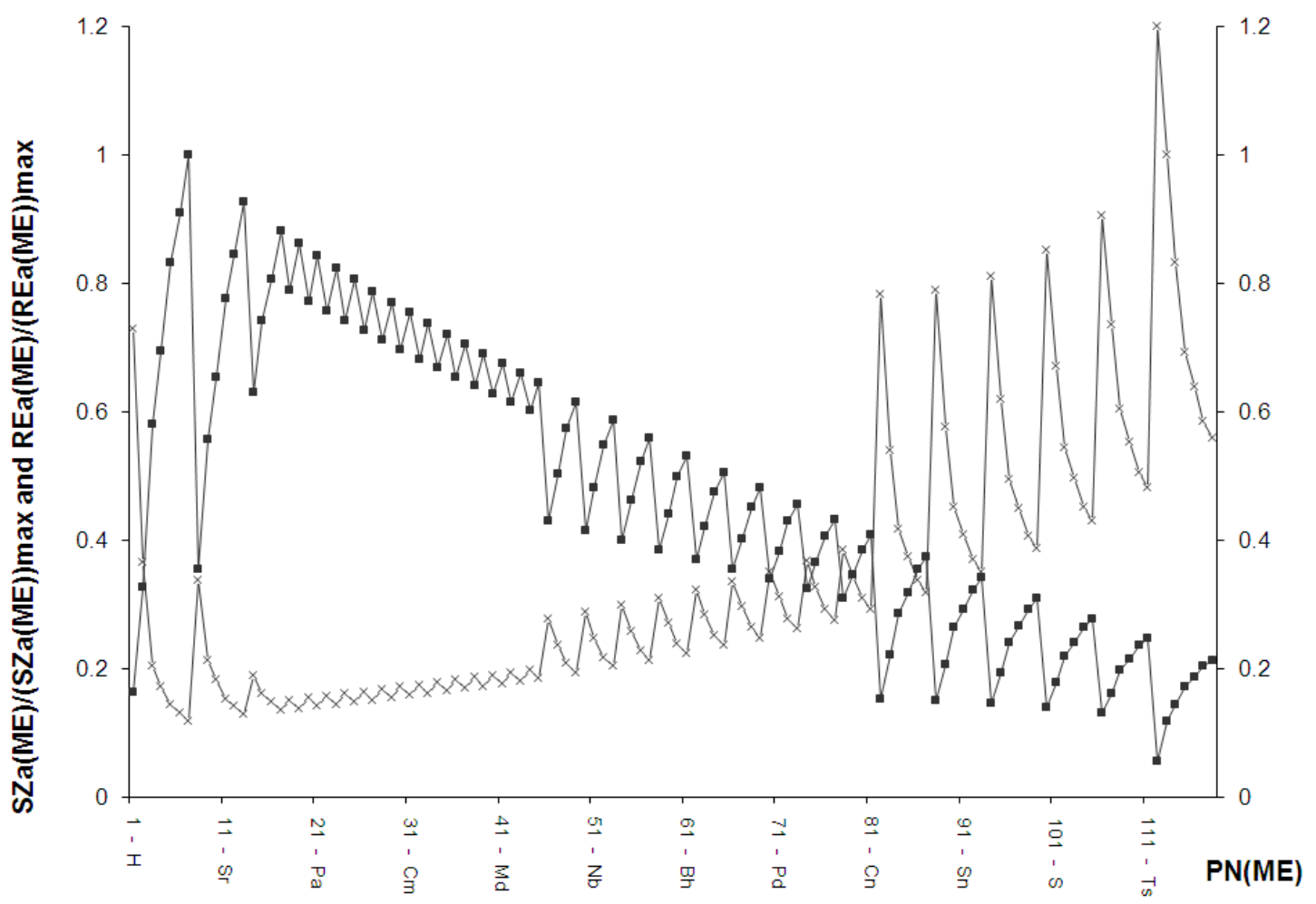

Fig. 7. $SZ_{aME}/(SZ_{aME})_{max}$ and $RE_{aME}/(RE_{aME})_{max}$ versus $PN_{ME}$, where $SZ_{aME}$ is the atomic size after Meyer, $RE_{aME}$ is the atomic reactivity after Meyer, and $PN_{ME}$ is the periodic number using Meyer periodic system; ■ $SZ_{aME}/(SZ_{aME})_{max}$, x $Re_{aME}/Re_{aME}(Ne)$.

The scaling number $k_{PN}$ ended up defined by $k_{PN}= \log(AN+1)_{max} + (\log(PN)_{max})^3 = 10.9695$, for $PN_{ME}$ and $PN_{MD}$. The factors $k_{SZ}$ and $k_{RE}$ are fitting parameters for an adaptation to experimental or theoretical data sets. In Tables 2 and 3 we replaced this fitting by normalized parameters: $SZ_a/(SZ_a)_{max} = SZ_a/SZ_a(Fr)$, here $SZ_{aME}(Fr)= k_{SZME}(\log 88)[k_{PN}-(\log 7)^3]= 20.1564\ k_{SZME}$ for the Meyer periodic system, and $SZ_{aMD}(Fr)= k_{SZMD}(\log 88)[k_{PN}-(\log 6)^3]= 20.4138\ k_{SZMD}$ for the Mendeleyev periodic system. For the normalization of $RE_{aME}$ and $RE_{aMD}$ we did not select $RE_{aME}(He)$ and $RE_{aMD}(He)$ respectively, which revealed to be too high, but $RE_{aME}(Ne)= RE_{aMD}(Ne)$; $RE_a/RE_a(Ne)= SZ_a(Ne)/SZ_a= (\log 11)[k_{PN}-(\log 113)^3]/[\log(AN+1)][k_{PN}-(\log PN)^3]$ or $RE_{aME}/RE_{aME}(Ne)= 2.4114\ \{[\log(AN+1)][k_{PN}-(\log PN_{ME})^3]\}^{-1}$ for the Meyer periodic system. The analogous equation based on Mendeleyev periodic system differs only in the logPN term: $RE_{aMD}/RE_{aMD}(Ne)= 2.4114\{[\log(AN+1)][k_{PN}-(\log PN_{MD})^3]\}^{-1}$. As stated above, while writing this review we were able to derive the first approximation of equation (3) for $AF_a$. As we are about to refine it, its values for $AF_a/(AF_a)_{max}$ have been omitted in Tables 2 and 3 on purpose.

Table 2 represents AN, $PN_{ME}$, $SZ_{aME}/(SZ_{aME})_{max}= SZ_{aME}/SZ_{aME}(Fr)$, and $RE_{aME}/RE_{aME}(Ne)$ in Meyer periodic system, whereas in Table 3 the corresponding values AN, $PN_{MD}$, $SZ_{aMD}/(SZ_{aMD})_{max}= SZ_{aMD}/SZ_{aMD}(Fr)$ and $RE_{aMD}/RE_{aMD}(Ne)$ are juxtaposed in conformity with Mendeleyev periodic system. Figs. 7, 8 show $SZ_{aME}/(SZ_{aME})_{max}$ and $RE_{aME}/RE_{aME}(Ne)$ as function of $PN_{ME}$ respectively AN. Owing to our normalization with (Ne) instead of (He) the values for $RE_{aME}(Ne)/RE_{aME}(He)$, $(RE_{aMD}(Ne)/RE_{aMD}(He)) = 2.13768$. In the Figs. 7 and 8, where this value is used, we set it to 1.2 for illustrative reasons. We have not included $AF_{aME}/(AF_{aME})_{max}$ as function of $PN_{ME}$ on purpose.

As examples and in order to demonstrate the 'linear correlation' between $SZ_{aME}$, and R(Z) in Fig. 9 we plot $SZ_{aME}/(SZ_{aME})_{max}$ versus $R(Z)/R(Z)_{max}$, where R(Z) is pseudo-potential radii according to Alex Zunger. The slope of its linear regression line represents the scaling factor $k_{SZ}$ in our equation (1) for the case $SZ_{aME}$ and R(Z) and its linear regression factor is $R^2= 0.886$. Equations (1) and (2) correlate quantitatively both the size of an atom and its ability to attract different atoms with two different ways of counting the chemical elements in the periodic system.

In context with $AF_{aME}$ it is relevant that the atomic mass (AM) is linearly correlated with the AN ($R^2= 0.998$), as well as the cohesion energy according to L. Brewer (CE(B)) is linearly correlated to the enthalpy of atomization (H(atom)) ($R^2= 0.998$). The X-ray density (Dx) is derived from a combination belonging to the following two pattern groups: the atomic number pattern group and the atomic size pattern group. So far the best matching with Dx has been achieved using from the atomic affinity pattern group the enthalpy of atomization (H(atom)) shown in Fig. 10a as a function of $PN_{MD}$. The linear correlation factor between Dx and H(atom) is $R^2 = 0.831$ (see Fig. 10b). Figs. 11 and 12 show in as overviews the five different APPs as a function of AN respectively $PN_{MD}$. For the atomic affinity pattern group we chose the enthalpy of atomization H(atom).

Thus, chapter 2 clearly points the following facts:

1) An appropriate description of the derived APPs requires the introduction of the periodic number PN in addition to the well-established atomic number AN. AN and PN represent crucial APPs which are independent from each other.

2) The derived APPs atomic size $SZ_a$, and its reciprocal, the atomic reactivity $RE_a$, as well as the newly in first approximation described atomic affinity $AF_a$ can be presented as functions of AN and PN. Other APPs, like the mass density Dx, can be expressed by a combination of two APPs belonging to two different pattern groups (for the mass density it is a combination of atomic weight belong to the atomic number pattern group and e.g. R(Z) belonging to the atomic size pattern group).

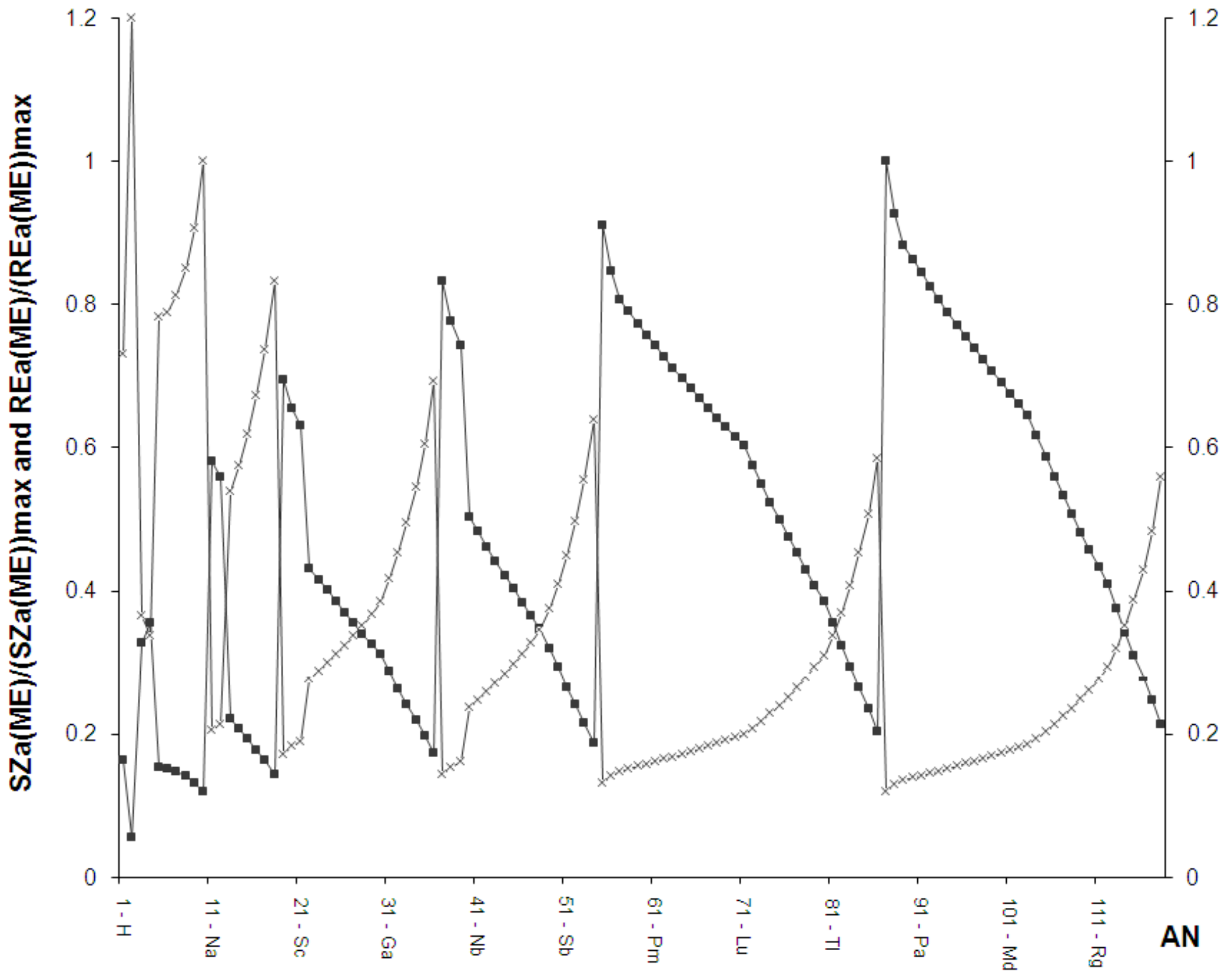

Fig. 8. $SZ_{aME}/(SZ_{aME})_{max}$ and $RE_{aME}/(RE_{aME})_{max}$ versus AN, where $SZ_{aME}$ is the atomic size after Meyer, $RE_{aME}$ is the atomic reactivity after Meyer, and AN is the atomic number; ■ $SZ_{aME}/(SZ_{aME})_{max}$, x $RE_{aME}/RE_{aME}(Ne)$.

3) The result of equation (2) $RE_a = k_{SZ}k_{RE}(SZ_a)^{-1}$, is most remarkable. This means that the atomic reactivity (different electronegativity scales belong to this pattern group) of a chemical element is the reciprocal value of its atomic size. This may explain why the Linus Pauling 'electronegativity concept' was never really accepted in physics. At the same time in chemistry its ability to reflect the reactivity of the chemical elements with each other is accepted and commonly employed. *As far as we know, the result that the atomic reactivity is simply the reciprocal of the corresponding atomic size is not found in the literature*.

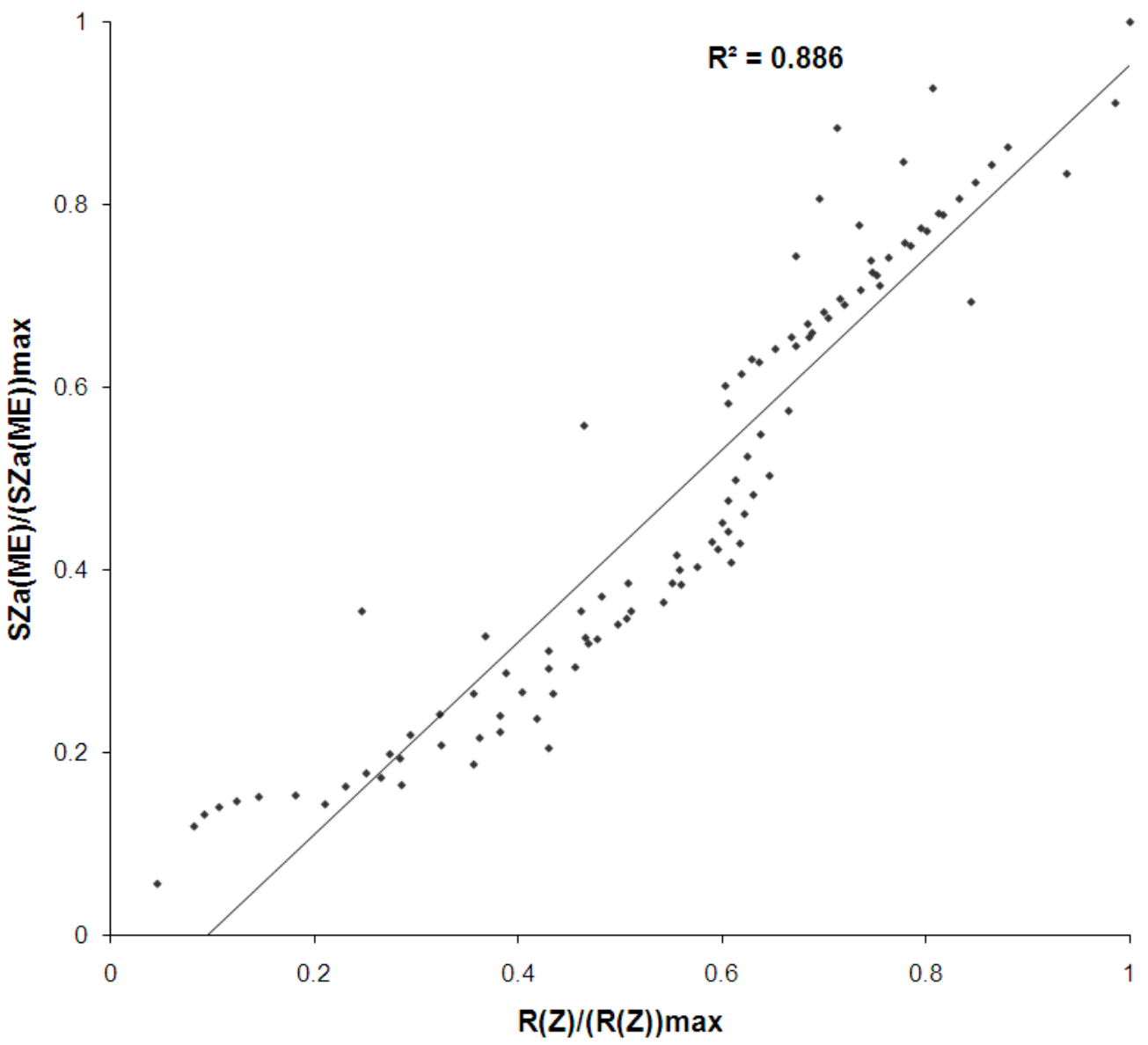

Fig. 9. Atomic size $SZ_{aME}$ (normalized to $(SZ_{aME})_{max}$) versus pseudo-potential radii according to Zunger R(Z) (normalized to $R(Z)_{max}$) with linear regression line ($R^2$= 0.886). In the case of R(Z), the pseudo-potential radii according to Zunger (where no published data exist for the f-elements) we estimated the R(Z) values based on a linear dependence for QN 6 between R(Z) = 3.55 for Ce and R(Z) = 2.64 for Lu, as well as for QN 7 between R(Z) = 3.85 for Th and R(Z) = 2.94 for Lr.

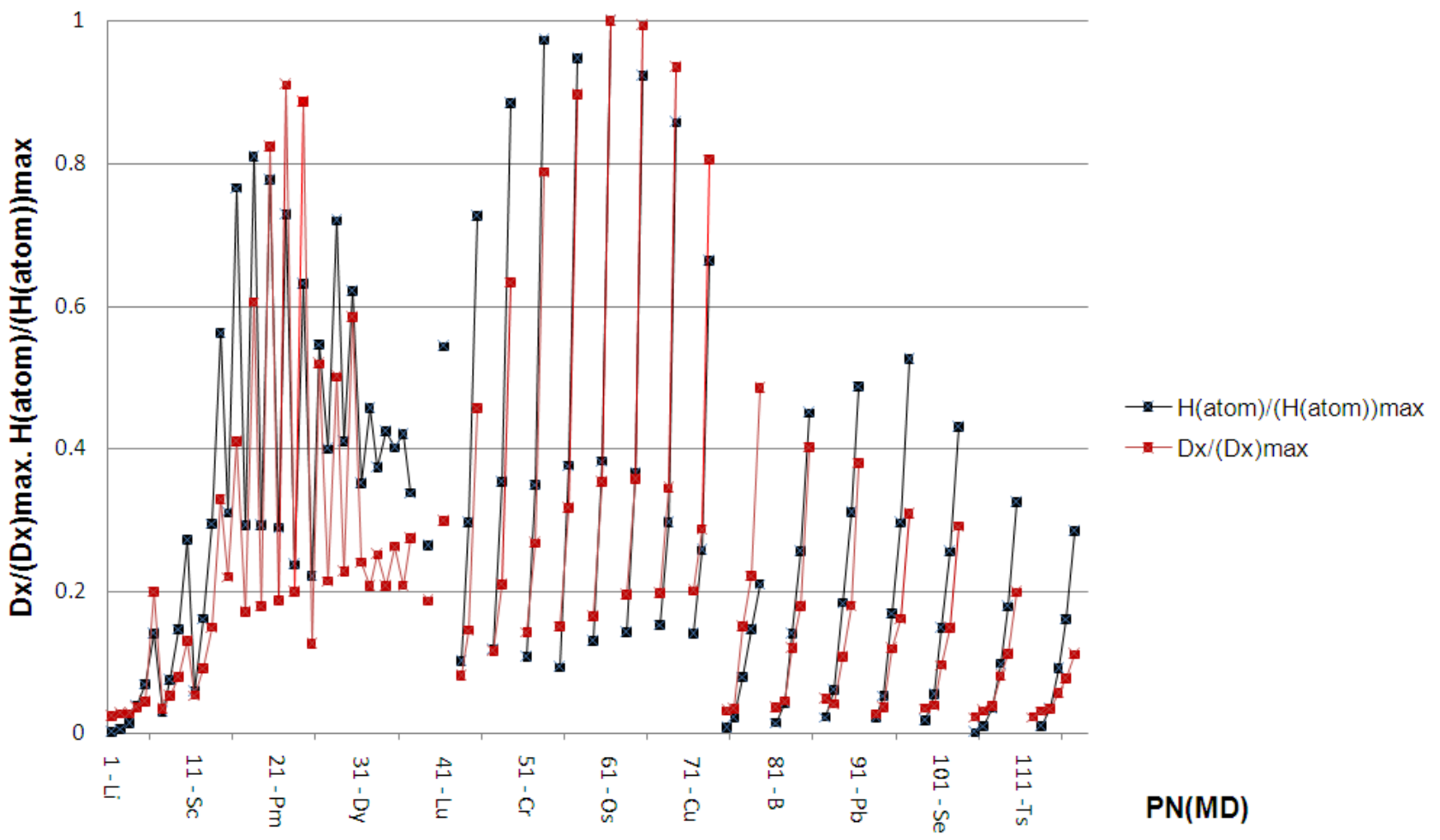

Fig. 10a. $Dx/(Dx)_{max}$ and $H(atom)/(H(atom))_{max}$ versus $PN_{MD}$, where $PN_{MD}$ is the periodic number using Mendeleyev periodic system, Dx is the X-ray density, H(atom) is the enthalpy of atomization.

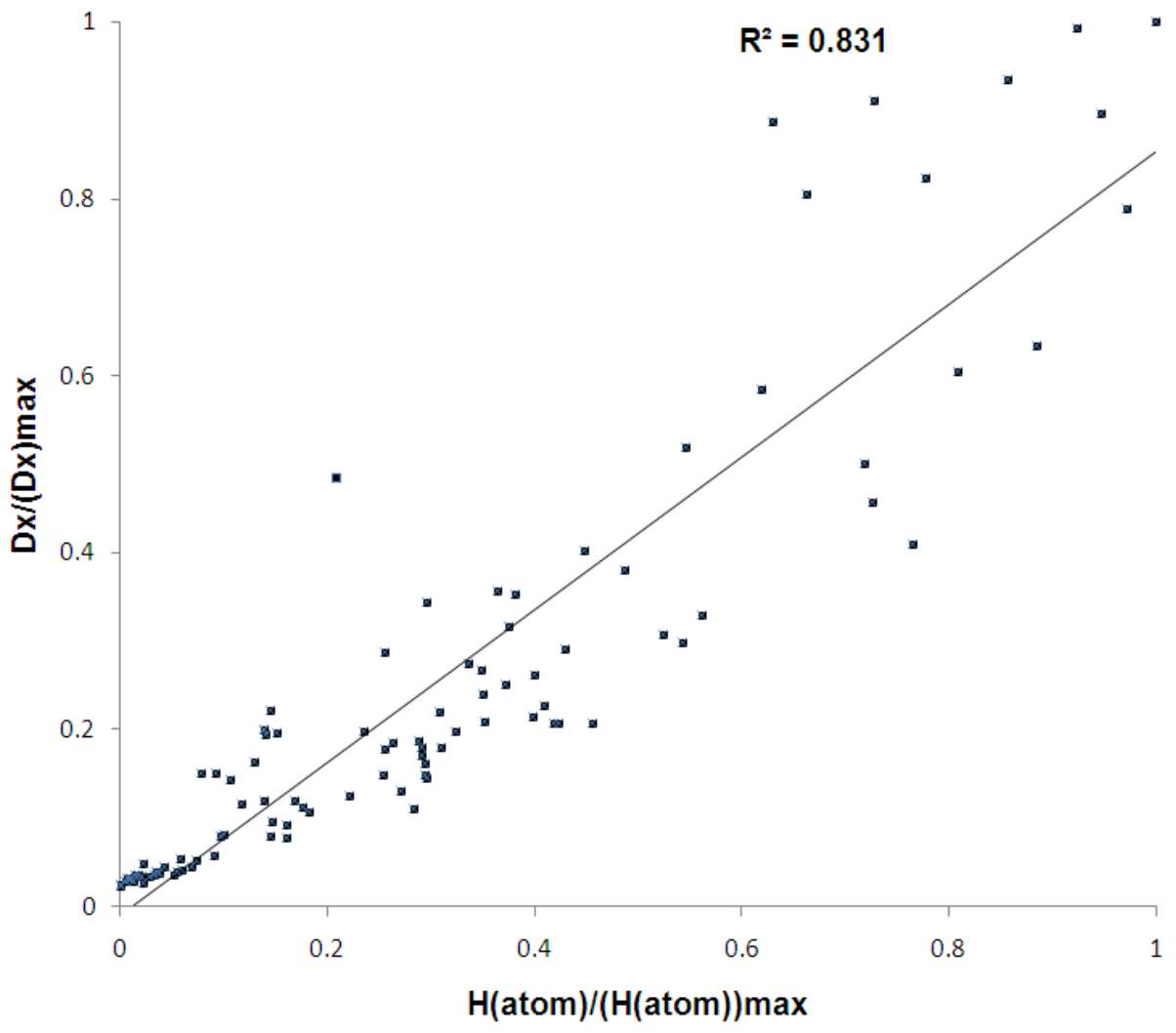

Fig. 10b. $Dx/(Dx)_{max}$ versus $H(atom)/(H(atom))_{max}$ with linear regression line ($R^2$= 0.831) **,** where Dx is the X-ray density, H(atom) is the enthalpy of atomization.

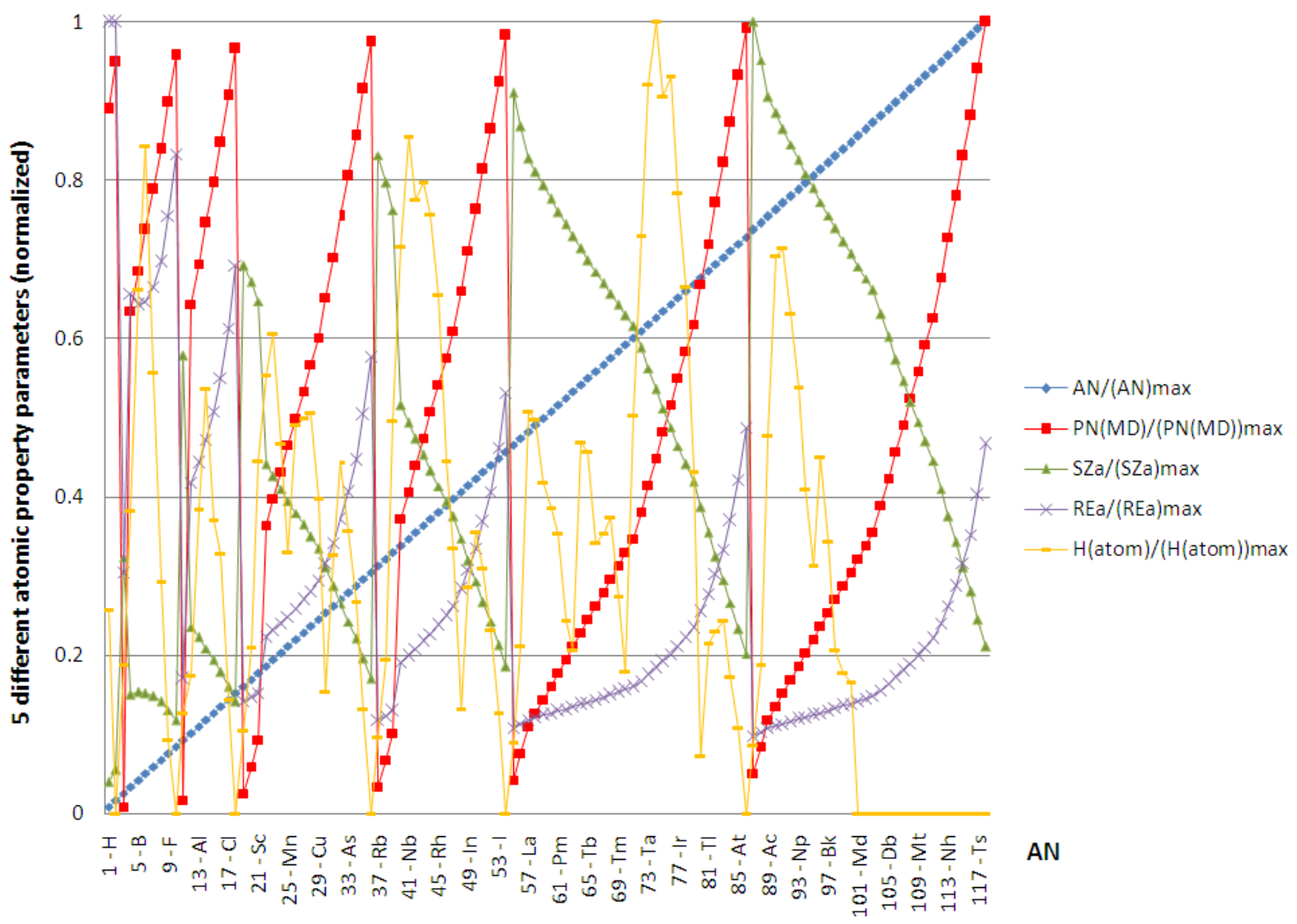

Fig. 11. Atomic property parameters (AN, $PN_{MD}$, $SZ_{aMD}$, $RE_{aMD}$, H(atom)) versus AN (for abbreviations see Table 1).

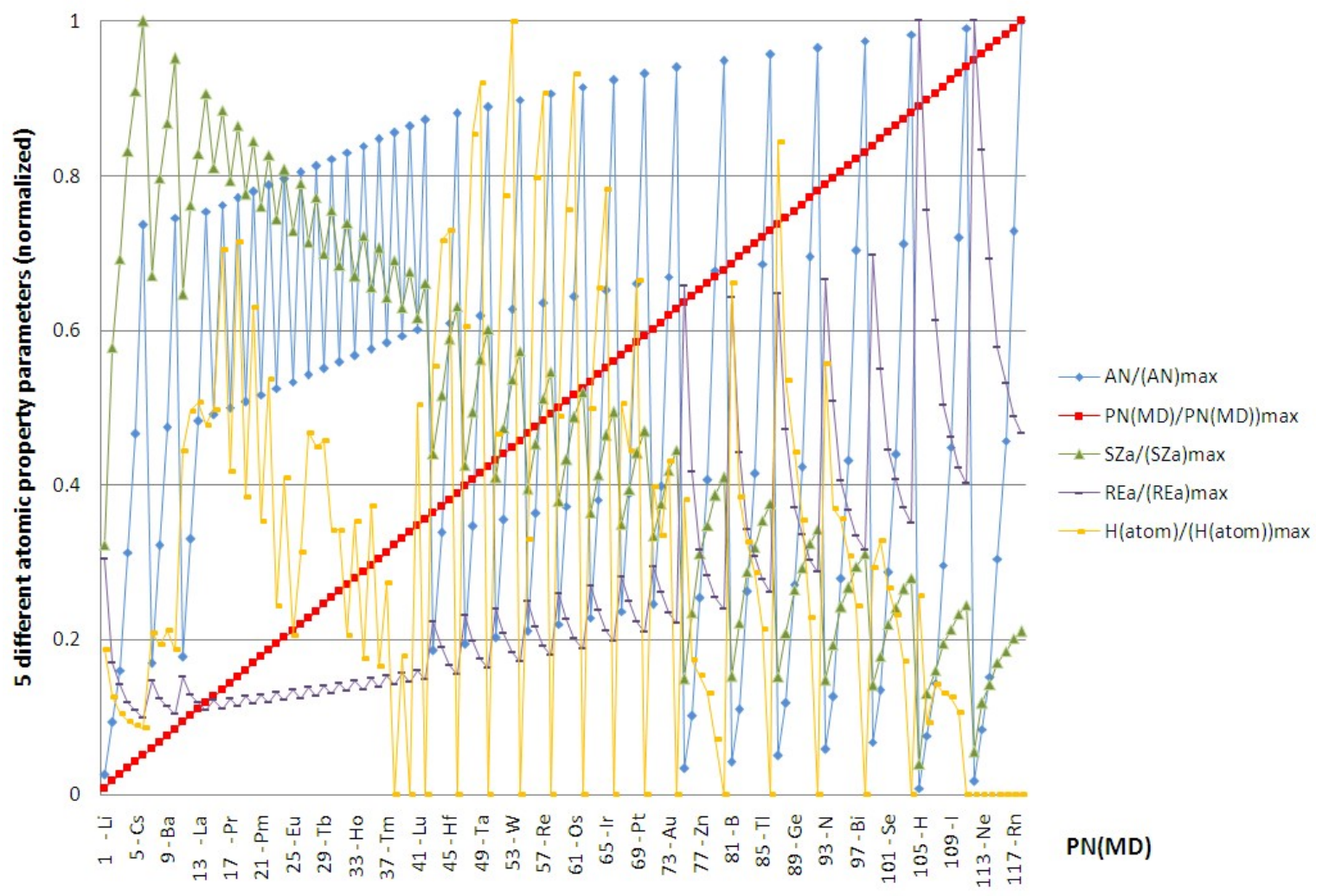

Fig. 12. Atomic property parameters (AN, $PN_{MD}$, $SZ_{aMD}$, $RE_{aMD}$, H(atom)) versus $PN_{MD}$ (for abbreviations see Table 1).

## 3. Inorganic substance formers vs. non-formers: binary, ternary, and quaternary systems

This chapter shows that there is the quantitative link between the inorganic substance properties and the APPs AN and PN (or simple functions of both) of the constituent elements, introducing the "former vs. non-former" terminology. Our work **[8]** on this specific subject together with the atomic environment type (AET) prediction for equiatomic binary inorganic substances **[1]**, do confirm the above statement. From **[17a,b]** we obtained the following figures on the inorganic substance former systems alongside the inorganic substance non-former systems: 2,137 formers and 686 non-formers, as well as 73 ordered phases (crystallizing from extended solid solutions), which gives in total 2,896 binary systems. For ternary and quaternary systems, we obtain the non-formers as follows: Below we bring the two conditions that permitted us to define whether a ternary or quaternary system belongs to a former or non-former system. These two criteria (here explicitly pointed out for the ternary case) for distinguishing formers from non-formers are founded on the following postulates:

(i) Description of crystallographic structure within the structure prototype classification built on the conception of space group theory,
(ii) Gibbs' phase rule.

Definition 1: An inorganic substance former system possesses at least one ternary inorganic substance separated by three two-phase regions including three adjacent chemical element(s) and (or) binary inorganic substance(s) and (or) ternary inorganic substance(s) in at least one complete isothermal section. In those cases where no phase diagram for an A–B–C system is known, but a ternary inorganic substance with a ternary *basic structure prototype* (see also *7.3.2.4.*) is reported, then this system is undoubtedly accepted as an inorganic substance former system.

Definition 2: A ternary system is a non-former when no three-phase region or complete solid solution respectively is shown in at least one complete isothermal section. *In those cases where no complete isothermal section for an A–B–C system is known, a ternary system will be a non-former if all its three binary boundary systems are non-former.* Thanks to this criterion we gained 3,530 ternary and 12,125 quaternary non-former systems from the 759 published (established) binary non-former phase diagrams. This criterion has been proved up to now by all published ternary isothermal sections.

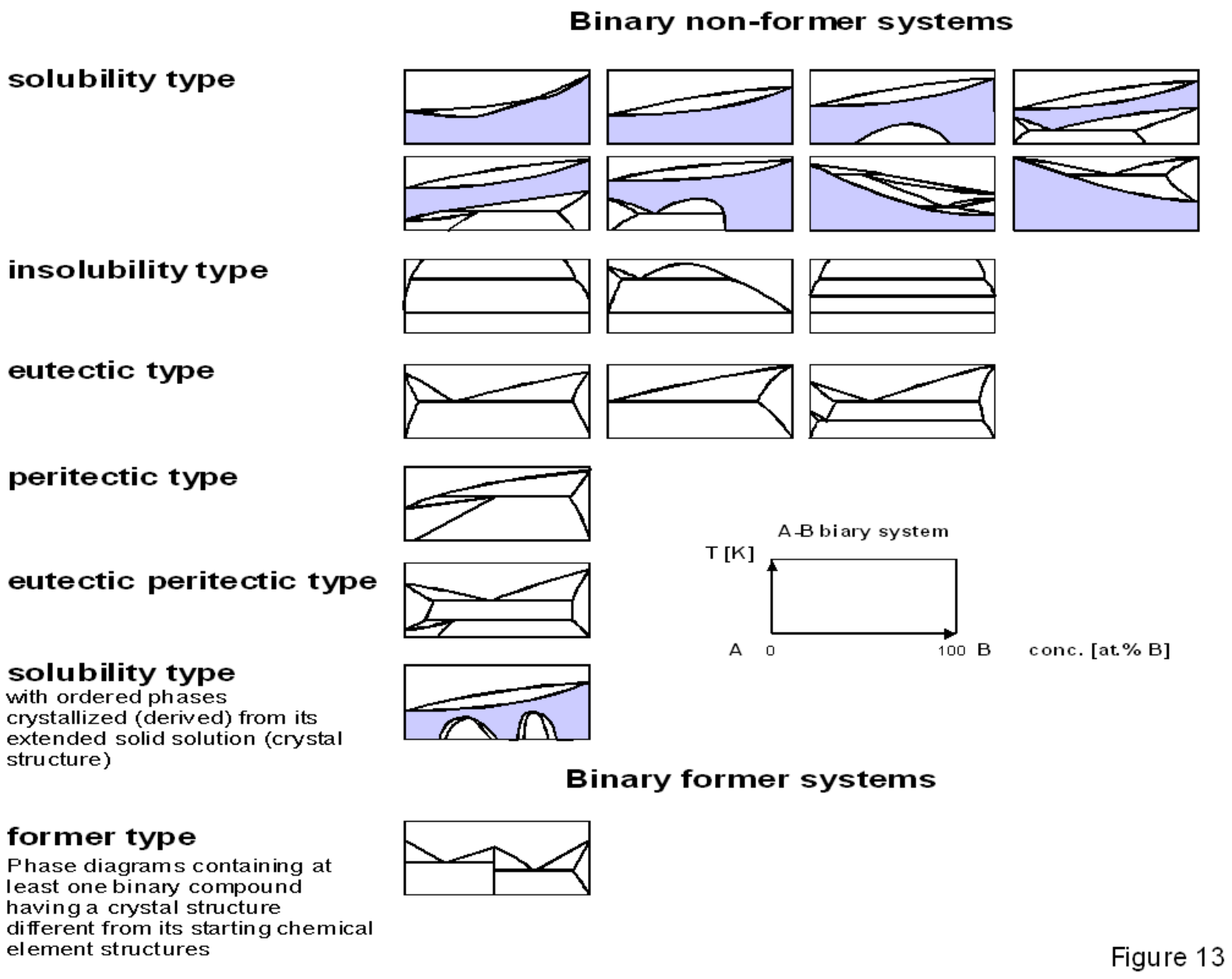

Fig. 13. 6 different types of inorganic substance non-forming systems (solubility without ordered phases(s), insolubility, eutectic, peritectic, eutectic-peritectic, and solubility with ordered phases(s)) in contrast to the inorganic substance forming system.

In parallel we have learned that non-former systems based on experimentally determined ternary isothermal sections are not reliable in this regard, as in about 1/3 of the cases, after about one decade, ternary inorganic substance(s) have been experimentally proven to exist (through crystallographic structure determination). In contrast to binary systems, as for the far majority of the binary systems several independent phase diagram studies have been published. Thus, the binary non-former systems may be considered as established. Fig. 13 displays the different cases of non-former binary systems: 'solubility'-, 'insolubility'-, 'eutectic'-, 'peritectic'-, 'eutectic-peritectic'- types. The 'solubility'-type with ordered phases crystallizing from its extended solid solutions are also considered as non-former systems (so far there exists 73 such cases).

For the 'former'-type we found in our present investigation on the existence of inorganic substances with ternary respectively quaternary basic structure prototypes. Based on the condition that a ternary system is a former when at least one ternary daltonide inorganic substance with a ternary basic structure prototype is published (see 7.3.1), we detected 14,384 ternary inorganic substance former systems and with analog criteria 13,874 quaternary inorganic substance former systems. It is noteworthy that this simple condition eliminates pseudo-ternary (pseudo-quaternary) inorganic substances, which are solid solutions of binary respectively ternary inorganic substances.

In order to monitor the 'parameterization power' of the APPs AN, PN, $SZ_a$(AN,PN), $RE_a$(AN,PN), and $AF_a$(AN,PN) we used the program DISCOVERY [12] on the above-mentioned data sets. The objective of this procedure is to search systematically for 2D- and 3D-features sets, to visualize them and to correlate quantitatively the materials properties (in this case: formers vs. non-formers) with the APP or APPE of their constituent chemical elements. As input data for the ternary systems were taken: 14,384 former + 3,530 non-former systems (quaternary: 13,874 former + 12,125 non-former).

This is done in a three-step mechanism:

1) Pre-selection of published (experimental and theoretical) APPs (see Table 1).

2) Application of an automatic generator for 2D- and 3D-features sets resulting from combinations of pre-selected APPs and mathematical operators. We introduced the simple operators so as not to miss the main essential point as inverse problem solving, namely, +, –, *, /, max. and min. to link the APPs of its constituent chemical elements A, B, C for ternary systems, to form global atomic property parameter expressions $APPE_{1(tot)} = APP_{1(A)}$ op $APP_{1(B)}$ op $APP_{1(C)}$ for e.g. the ternary inorganic substance $(A_xB_yC_z)_1$. For the ternary case e.g.:

$$(T_{fus}/T_{fus}) = 1/3\ (T_{fusC}/T_{fusA} + T_{fusB}/T_{fusA} + T_{fusC}/T_{fusB}),\ \text{with}\ T_{fusA} > T_{fusB} > T_{fusC} \quad (4)$$
$$\Delta R = 1/3\ (|R(Z)_A - R(Z)_B| + R(Z)_A - R(Z)_C| + |R(Z)_B - R(Z)_C|), \quad (5)$$
$$\Delta VE = 1/3\ (|VE_A - VE_B| + |VE_A - VE_C| + |VE_B - VE_C|) \quad (6)$$

3) Automatic high-quality separation detection and its visualization. The general concept is very simple: Assuming we consider 2,000 data points (1,000 ternary formers and 1,000 ternary non-formers) in a graph of a selected 3D-features set. We check now for each point whether their $n^{th}$ nearest neighbors in the graph are of the same category or not (here formers or non-formers) and make a statistical analysis for all 2,000 data points (n = 1, …, 50, depending on the standard assumed).

| 5 fundamental atomic property parameters (APPs) pattern groups | Atomic property parameters (APPs) | Separation result of the best 3D-feature set |
|---|---|---|
| I) Atomic number pattern group | AN, QN, AM, cma(Mo), cma(Cu), aes, nc(C), nc(S) | 98.0% (n=1) → 80.0% (n= 50) |
| II) Periodic number pattern group | PN(ME), PN(MD), PN(P), VE, GN, PN(G,P), PN(O,P), USE | **99.7% (n=1) → 92.6% (n= 50)** |
| III) Atomic size pattern group | R(Z), Ri(Y), Rc(P), Rm(WG), Rve(S), Rce(S), R(M), Ra | 98.5% (n=1) → 76.8% (n= 50) |
| IV) Atomic reactivity pattern group | EN(MB), EN(P), EN(AR), EN(abs), IE(first), CP(M), wf, n(WS) | 98.5% (n=1) → 81.4% (n= 50) |
| V) Atomic affinity pattern group | T(fus), T(vap), H(vap), H(fus), H(atom), Hsurf(M), CE(B), VC | 97.2% (n=1) → 63.8% (n= 50) |

Table 4. Result of the best 3D-features sets using the published atomic property data sets (Table 1). The hit rate per ternary system decreases from e.g. 98.5% with n = 1 correct neighbor to 81.4% with n = 50 correct nearest neighbors for the 'best' case of the reactivity pattern group.

| 5 fundamental atomic property parameters (APPs) pattern groups | Derived atomic property parameters (APPs) | Separation result of the best 2D-feature set |
|---|---|---|
| I) Atomic number pattern group | AN | 88.1% (n=1) → 33.1% (n= 50) |
| II) Periodic number pattern group | $PN_{ME}$, $PN_{MD}$ | **99.4% (n=1) → 91.4% (n= 50)** |
| III) Atomic size pattern group | $SZ_{aME}$, $SZ_{aMD}$ | 97.8% (n=1) → 76.6% (n= 50) |
| IV) Atomic reactivity pattern group | $RE_{aME}$, $RE_{aMD}$ | 98.0% (n=1) → 80.3% (n= 50) |
| V) Atomic affinity pattern group | $AF_{aME}$, $AF_{aMD}$ (first approximation) | 95.0% (n=1) → 60.3% (n= 50) |

Table 5. Result of the best 2D-features sets using the calculated atomic property data sets (Tables 2 + 3). The hit rate per ternary system decreases from e.g. 98.0% with n = 1 correct neighbor to 80.3% with n = 50 correct nearest neighbors for the case of $RE_{aMD}$.

We ran the program DISCOVERY on all APPs belonging to each pattern group indicated in Table 1. With the help of all mathematical operators mentioned above, we give the best 3D-features set results for the ternary systems mentioned in Table 4. The success of the separation is visible by the hit rate, e.g. 99.7% (n = 1) – 92.6% (n = 50). This means that 99.7% of all data points have a nearest neighbor of the same kind and 92.6% of all data points have 50 nearest neighbors of the same kind taking the reference data point as the center of a spherical volume. Analogically, Table 5 illustrates the results of the best 2D-features sets using APPs that belong to the following pattern groups: atomic number, periodic number, atomic size, atomic reactivity, and atomic affinity. Again for ternaries Table 6 shows the best 2D-features sets using the APPs belonging to the atomic size pattern group. It can be seen that with equation (1) derived $SZ_{aMD}$ (AN, $PN_{MD}$) demonstrates the 'best' separation.

| APPs belonging to III) Atomic size pattern group | APP | Separation result of the best 2d-feature set |
|---|---|---|
| Atomic radii according to Busch | Ra | 89.0% (n= 1) → 15.7% (n= 50) |
| Valence electron distance according to Schubert | Rve(S) | 85.8% (n= 1) → 24.7% (n= 50) |
| Core electron distance according to Schubert | Rce(S) | 83.0% (n= 1) → 30.8% (n= 50) |
| Metal radii according to Waber and Gschneidner | Rm(WG) | 90.4% (n= 1) → 38.7% (n= 50) |
| Radii according to Miedema | R(M) | 92.4% (n= 1) → 46.0% (n= 50) |
| Covalent according to Pauling | Rc(P) | 89.0% (n= 1) → 50.4% (n= 50) |
| Ionic radii according to Yagoda | Ri(Y) | 88.0% (n= 1) → 51.7% (n= 50) |
| Pseudo-potential radii according to Zunger | R(Z) | 97.8% (n= 1) → 67.2% (n= 50) |
| Atomic Size using Meier periodic system | $SZ_{aME}$ | 97.8% (n= 1) → 73.8% (n= 50) |
| Atomic Size using Mendeleyev periodic system | $SZ_{aMD}$ | **97.8% (n= 1) → 76.1% (n= 50)** |

Table 6. Results of the best 2D-features sets using only atomic property parameters belonging to the atomic size pattern. The hit rate per ternary system decreases from e.g. 97.8% with n = 1 correct neighbor to 76.1% with 50 correct nearest neighbors for the case of $SZ_{aMD}$.

Detailed comparisons for the 'parameterization power' of the various APPs: belonging to atomic size, atomic reactivity, and atomic affinity patterns groups are given in Tables 4, 5 and 6, respectively showing the 'separation efficiency' of former from non-former systems. It is obvious that using only $PN_{ME}$ (or $PN_{MD}$) instead of the derived $SZ_{aME}$ (AN,$PN_{ME}$), $RE_{aME}$ (AN,$PN_{ME}$), as well as $AF_{aME}$ (AN,$PN_{ME}$) generates comparable separation between inorganic substance former and non-former systems. The separation effect using only AN is definitely less good. Analogous results have been obtained for ternary and quaternary data sets.

The overall simplest and most efficient APP proved to be either $PN_{ME}$ and (or) $PN_{MD}$. Figs. 14-17 show the overall best 2D-features sets for binary systems, once using the Meyer periodic system and once using the Mendeleyev periodic system. In both cases the amazing 'parameterization power' of PN showing the best separation is demonstrated by comparing Figs. 14 and 16 (without inclusion of the ordered phases), and Figs. 15 and 17 (including the 73 ordered phases). All 73 ordered phases containing chemical systems are nicely located in the non-former domains, as it has to be. Figs. 18, 19b and 20b give the 'best' 3D-features sets for the binary, ternary and quaternary chemical systems using as APPE's: $PN_{MD}$ ratio (x-axis), $PN_{MD}$ product (y-axis), and $(PN_{MD})_{max}$ (z-axis).

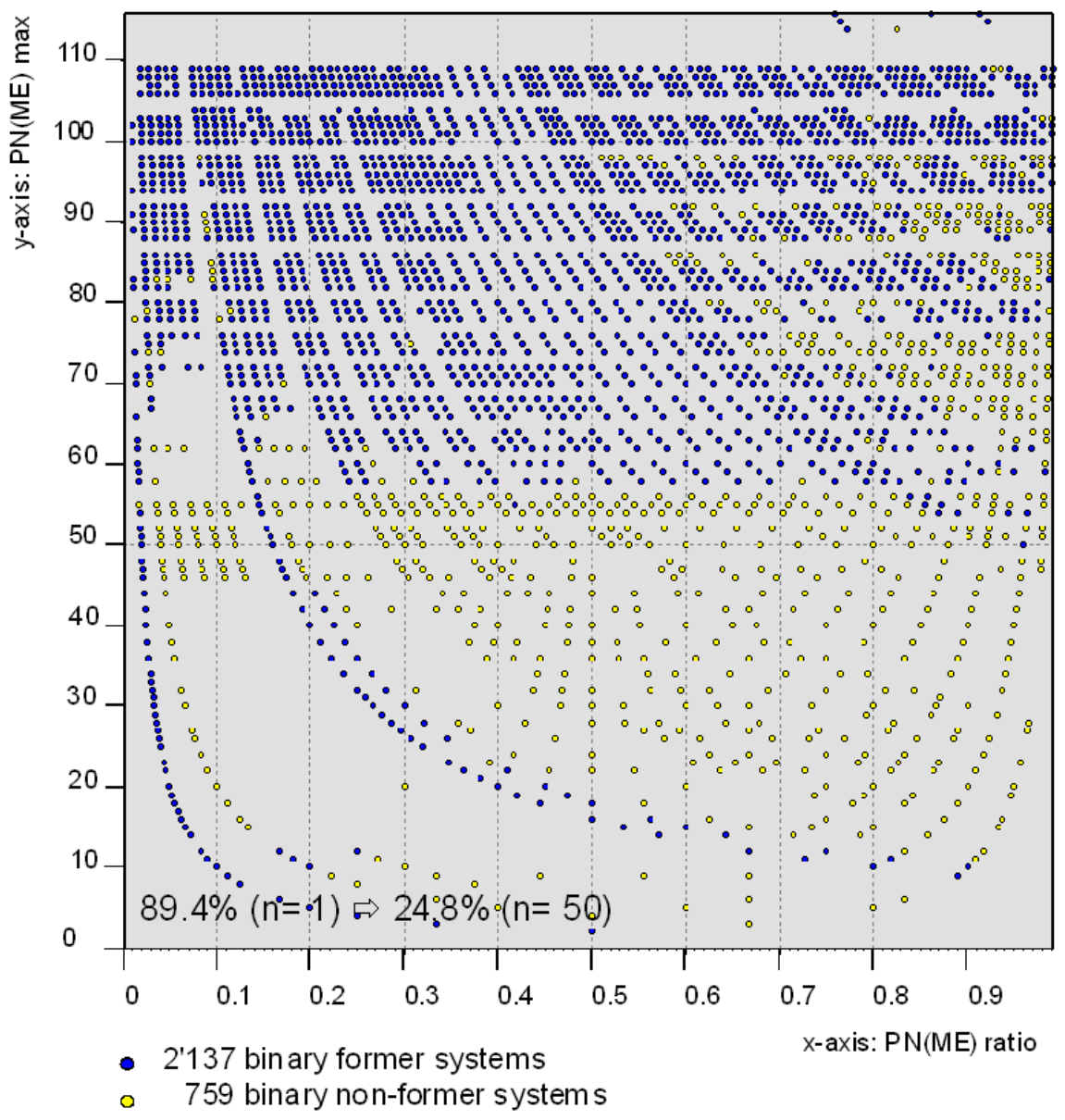

Fig. 14. Separation of 2,896 binary systems into inorganic substance formers (blue circle) and non-formers (yellow circle) based on a plot of $[(PN_{ME})_A,(PN_{ME})_B]_{max}$ (y-axis) versus ratio $(PN_{ME})_A/(PN_{ME})_B$ (x-axis), using Meyer periodic system.

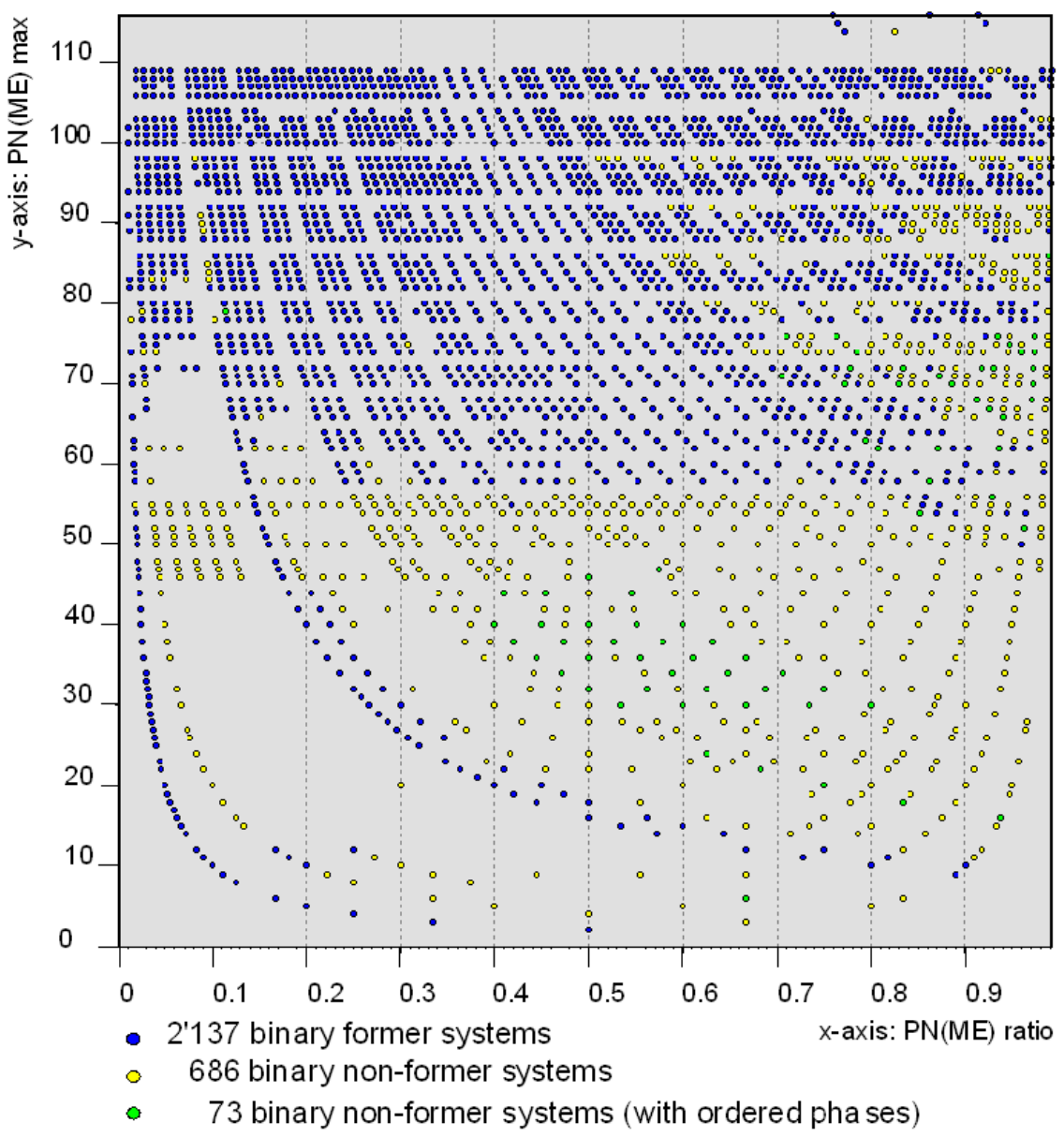

Fig. 15. Separation of 2,969 binary systems into inorganic substance formers (blue circle), non-formers (yellow circle), and non-formers with ordered phases (green circle) based on a plot of $[(PN_{ME})_A,(PN_{ME})_B]_{max}$ (y-axis) versus ratio $(PN_{ME})_A/(PN_{ME})_B$ (x-axis), using Meyer periodic system. The 73 ordered phases belong to the solubility type with ordered phases added to the non-formers.

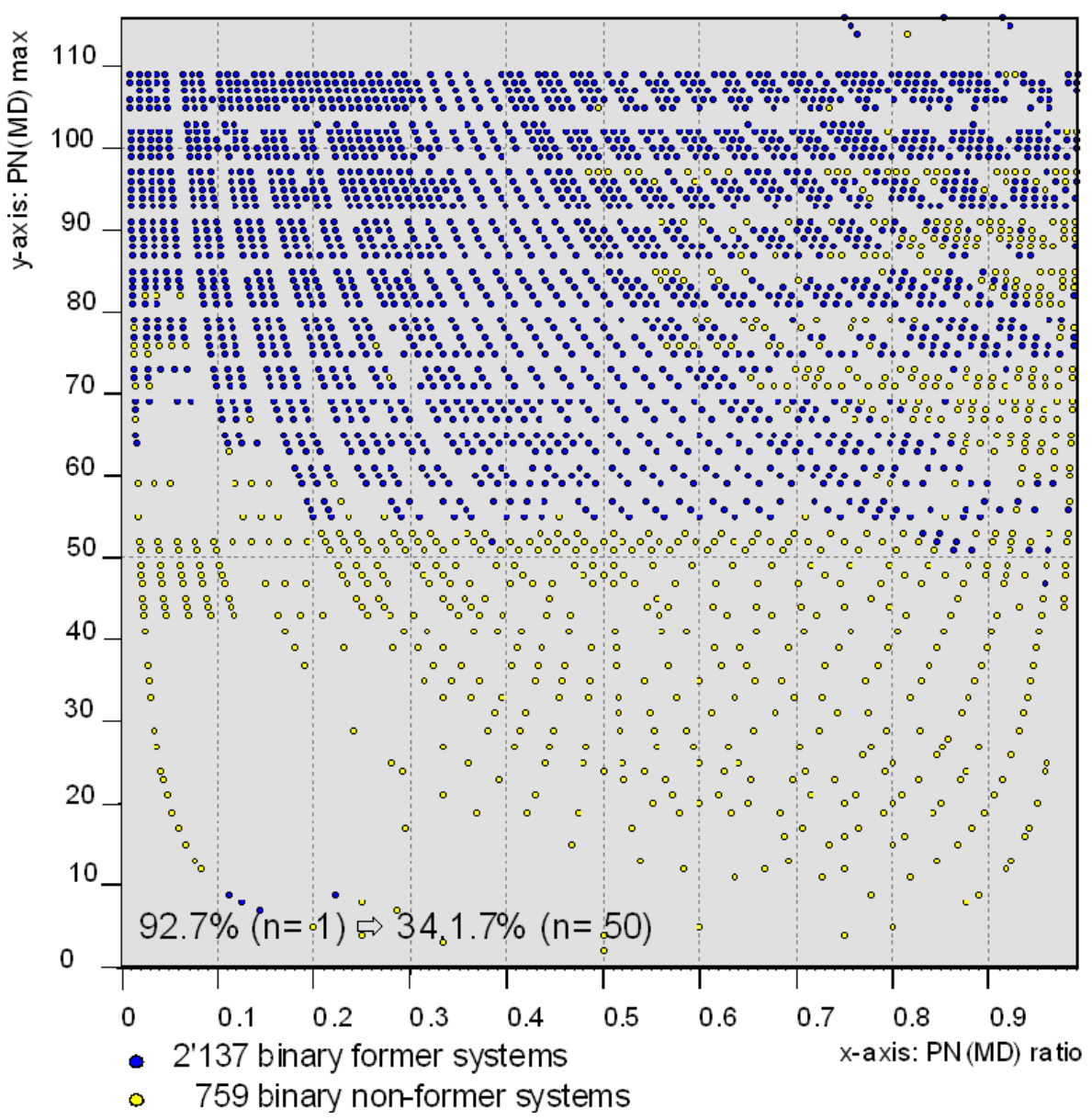

Fig. 16. Separation of 2,896 binary systems into inorganic substance formers (blue circle) and non-formers (yellow circle) based on a plot of $[(PN_{MD})_A,(PN_{MD})_B]_{max}$ (y-axis) versus ratio $(PN_{MD})_A/(PN_{MD})_B$ (x-axis), using Mendeleyev periodic system.

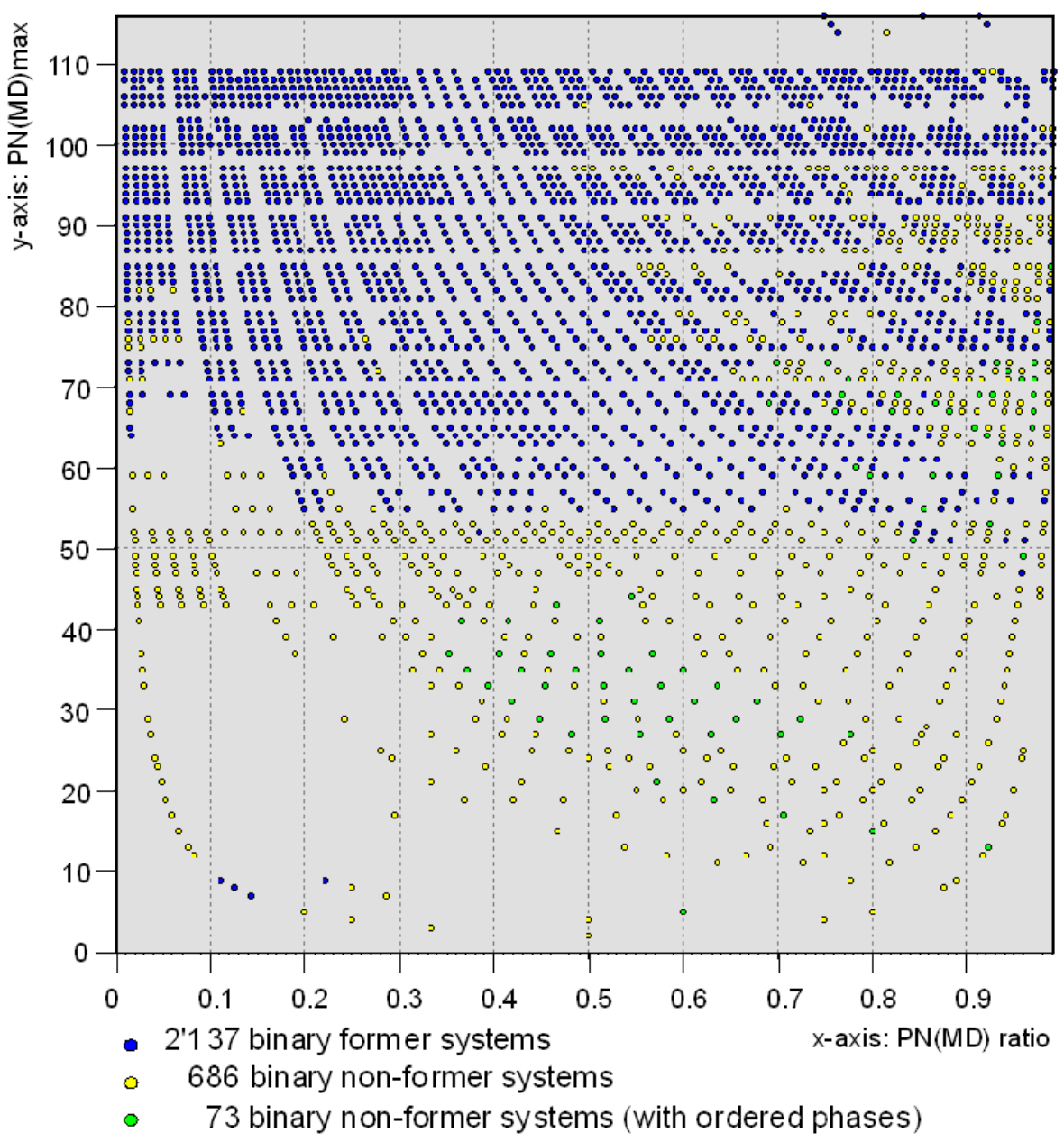

Fig. 17. Separation of 2,969 binary systems into inorganic substance formers (blue circle) and non-formers (yellow circle), and non-formers with ordered phases (green circle) based on a plot of $[(PN_{MD})_A,(PN_{MD})_B]_{max}$ (y-axis) versus ratio $(PN_{MD})_A/(PN_{MD})_B$ (x-axis), using Mendeleyev periodic system. The 73 ordered phases belong to the solubility type with ordered phases added to the non-formers.

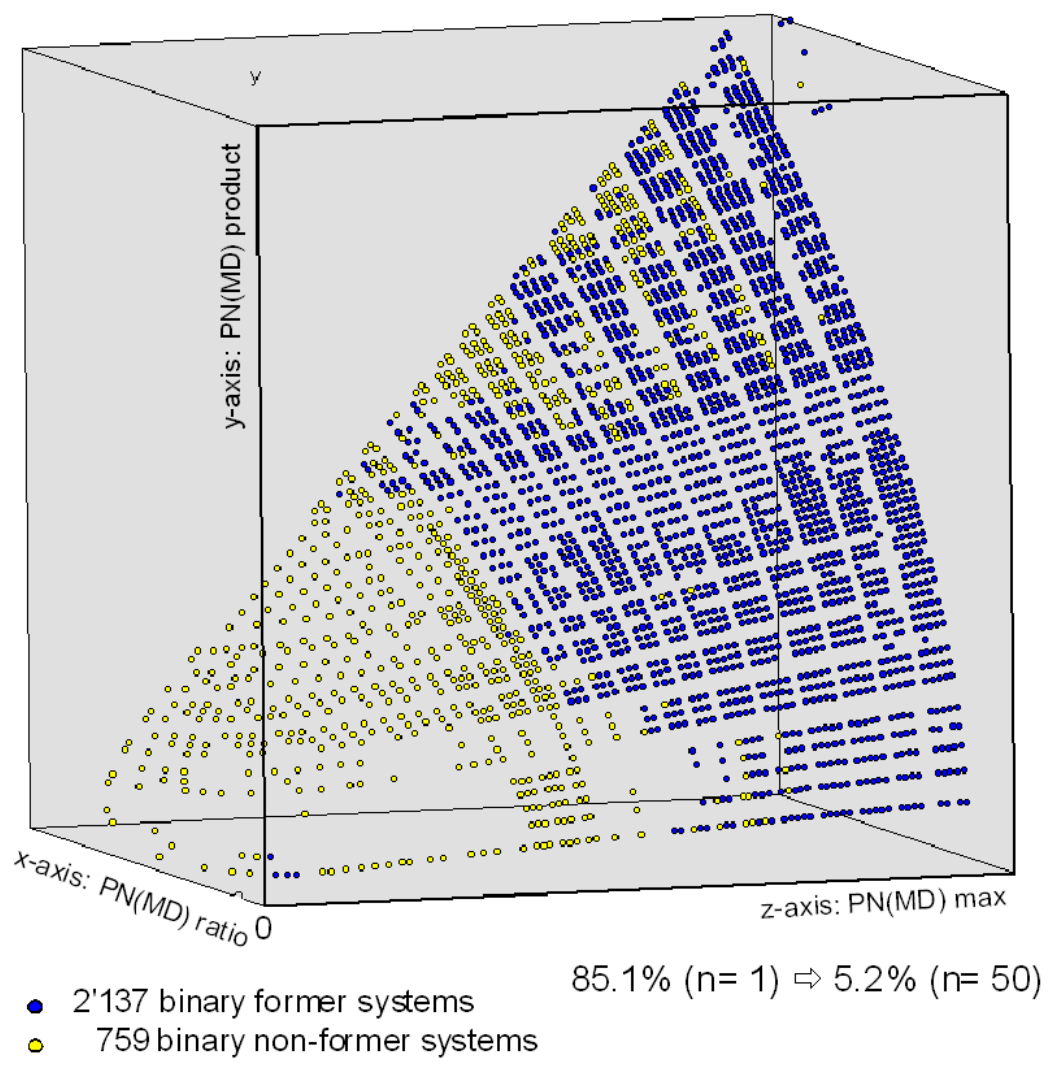

Fig. 18. Separation of 2,896 binary systems into inorganic substance formers (blue circle) and non-formers (yellow circle) based on a 3D-plot of product $(PN_{MD})_A/(PN_{MD})_B$ (x-axis) (x-axis) versus $(PN_{MD})_A*(PN_{MD})_B$ (y-axis) versus, $[(PN_{MD})_A,(PN_{MD})_A]_{max}$ (z-axis) using Mendeleyev periodic system.

With the use of a 2D-features set instead of a 3D-features set, we are losing some separation efficiency. Despite this, using the appropriate 2D-projection instead of the 3D-features provides us the advantage that the given figures become easier to interpret. Curiously, by using $(PN_{MD})_{max}$ as one axis for binaries all, except 4 binary systems with $(PN_{MD})_{max} < 51$ are non-formers. For ternaries and quaternaries this is even almost in all cases true with $(PN_{MD})_{max} < 54$ respectively $(PN_{MD})_{max} < 78$ (see Figs. 19a and 20a). It should be high-lightened that with just three simple $(PN_{MD})_{max}$ vs. $PN_{MD}$ ratio graphs one can in most areas predict precisely the existence respectively non-existence of potential binary, ternary and quaternary inorganic substances.

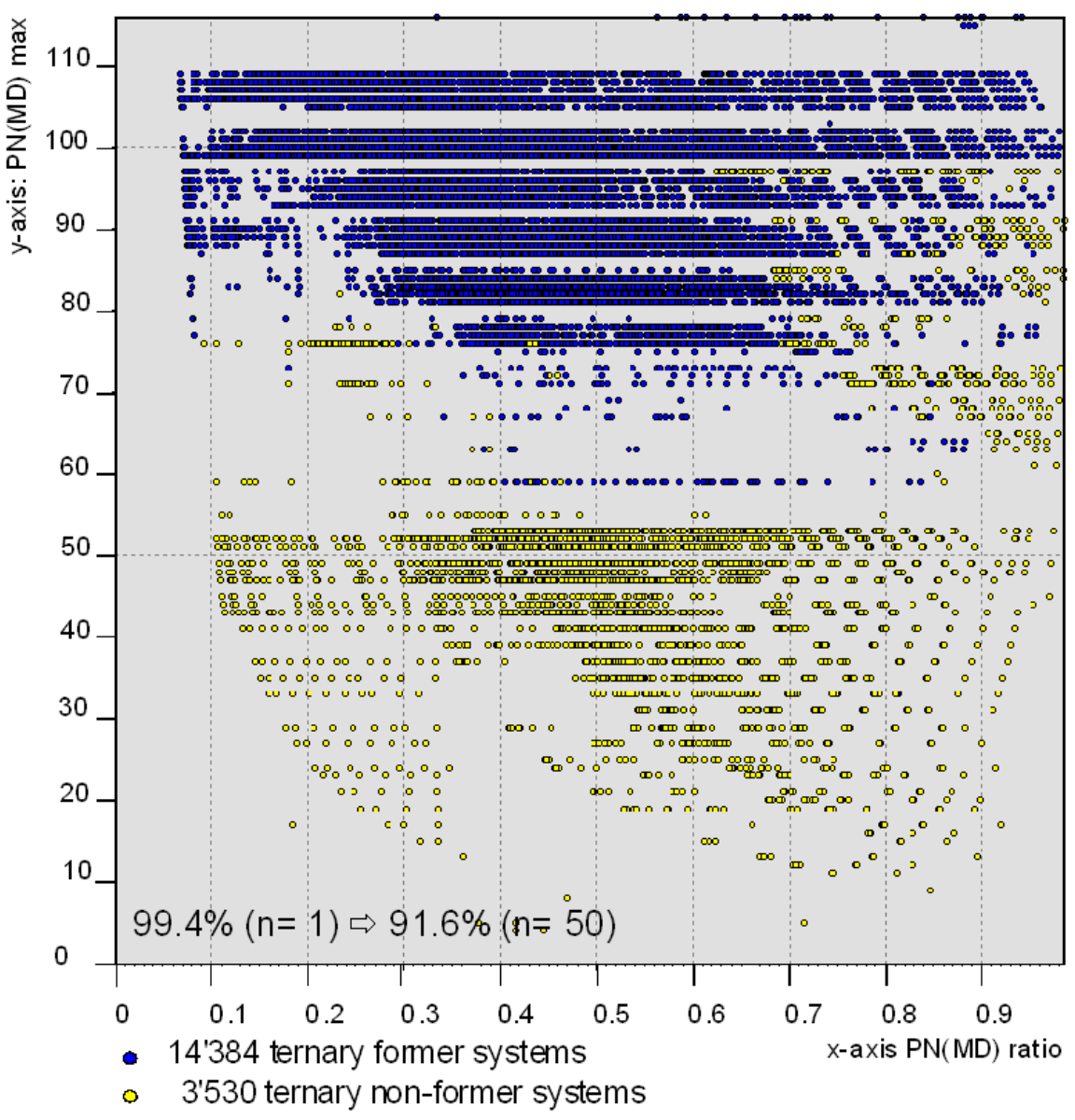

Fig. 19a. Separation of 17,914 ternary systems into inorganic substance formers (blue circle) and non-formers (yellow circle) based on a 2D-plot of $[(PN_{MD})_A,(PN_{MD})_B]_{max}$ (y-axis) versus $(PN_{MD})_A/(PN_{MD})_B$ (x-axis) using Mendeleyev periodic system.

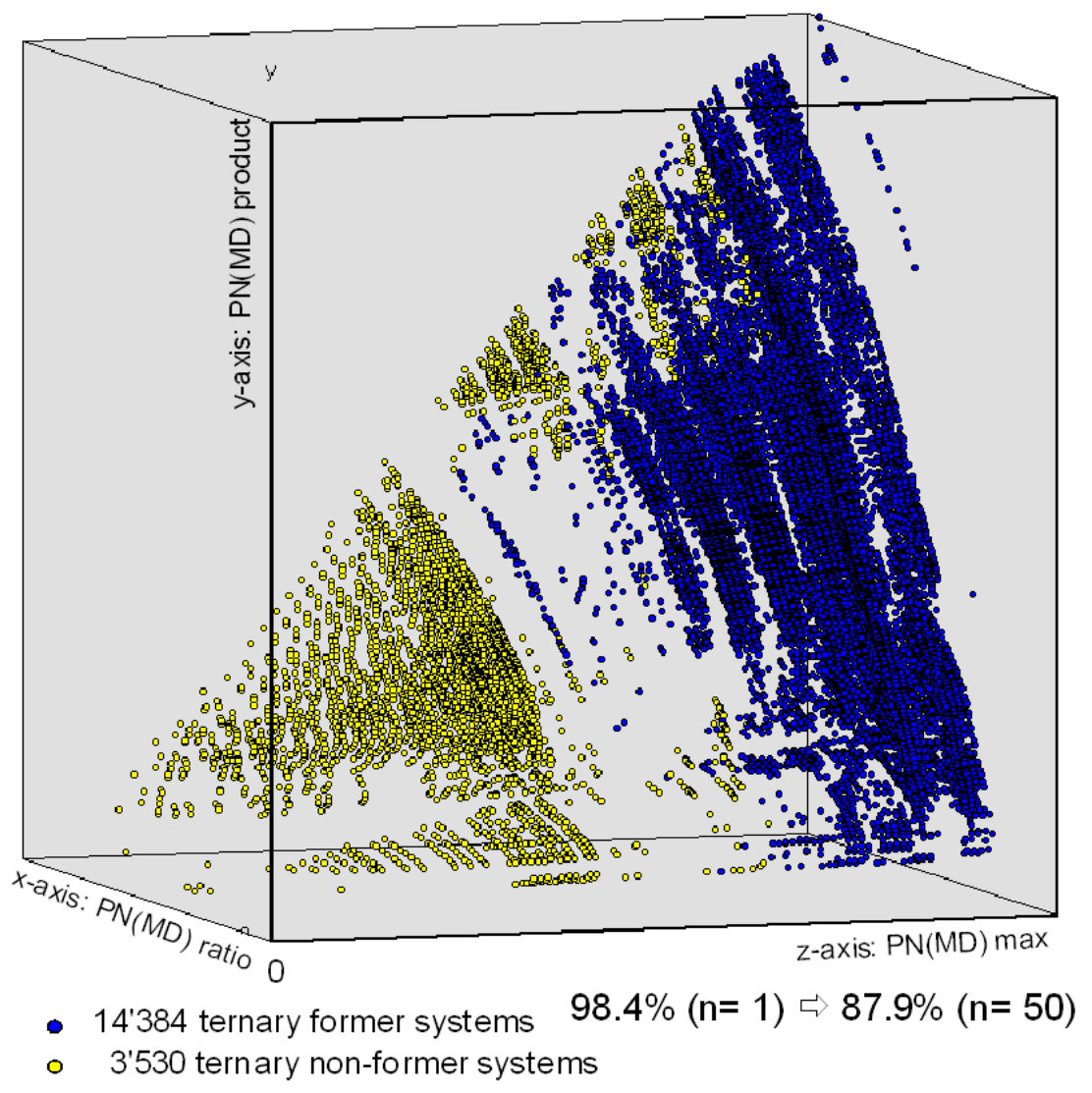

Fig. 19b. Separation of 17,914 ternary systems into inorganic substance formers (blue circle) and non-formers (yellow circle) based on a 3D-plot of product $(PN_{MD})_A/(PN_{MD})_B$ (x-axis) versus $(PN_{MD})_A*(PN_{MD})_B$ (y-axis) versus $[(PN_{MD})_A,(PN_{MD})_B]_{max}$ (z-axis) using Mendeleyev periodic system.

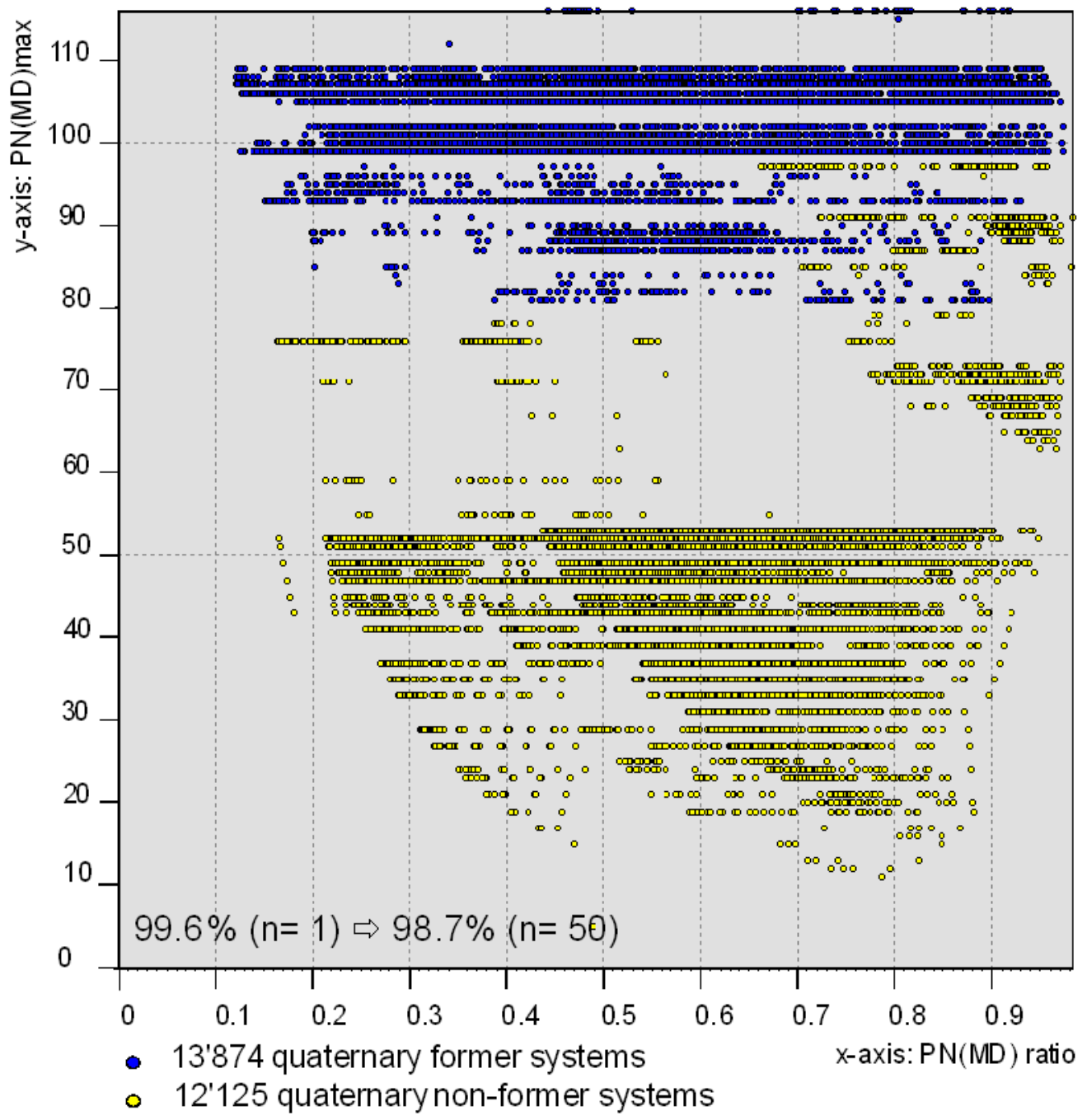

Fig. 20a. Separation of 25,999 quaternary systems into inorganic substance formers (blue circle) and non-formers (yellow circle) based on a 2D-plot of $[(PN_{MD})_A,(PN_{MD})_B]_{max}$ (y-axis) versus $(PN_{MD})_A/(PN_{MD})_B$ (x-axis) using Mendeleyev periodic system.

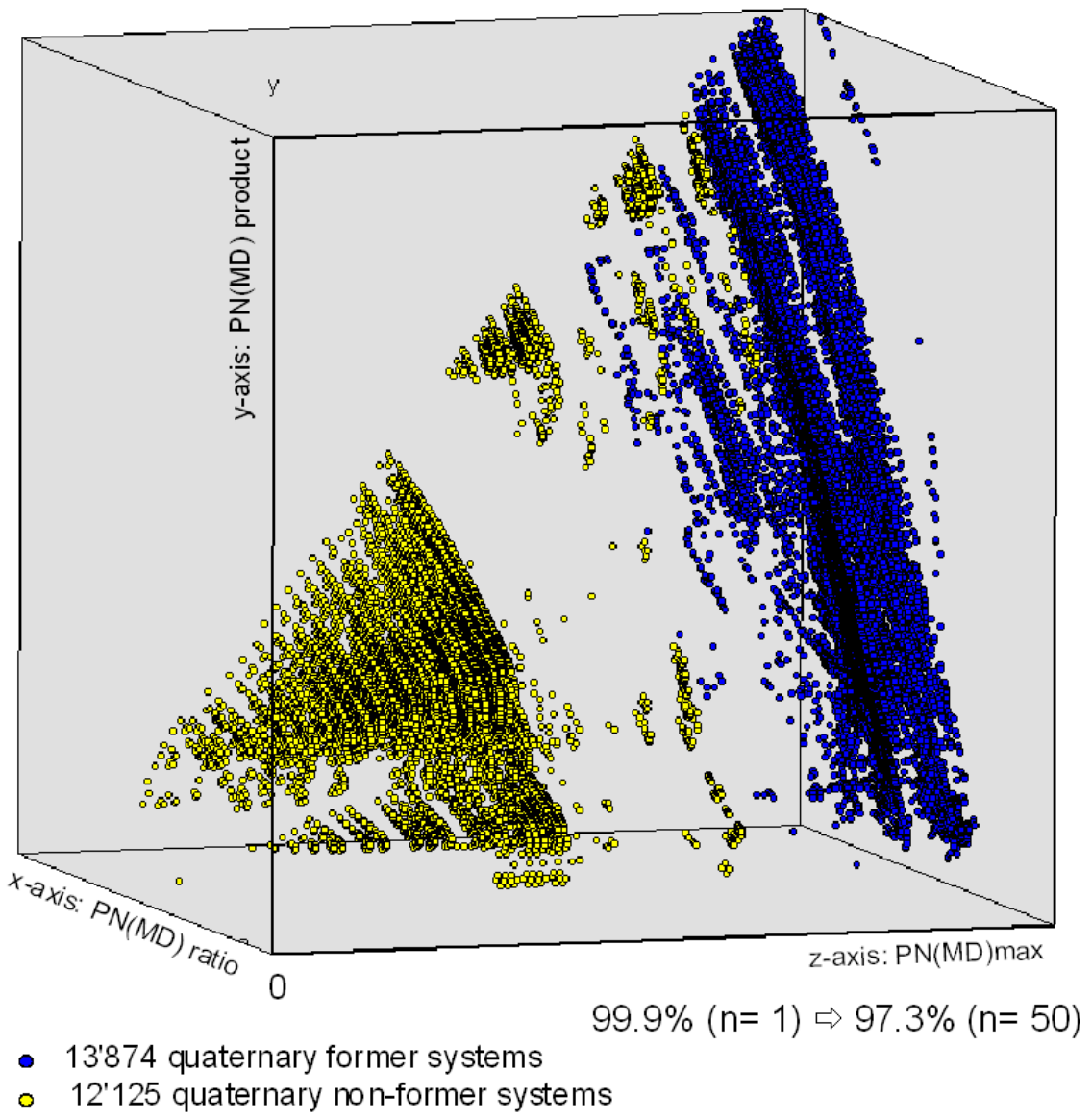

Fig. 20b. Separation of 25,999 quaternary systems into inorganic substance formers (blue circle) and non-formers (yellow circle) based on a 3D-plot of product $(PN_{MD})_A/(PN_{MD})_B$ (x-axis) versus $(PN_{MD})_A*(PN_{MD})_B$ (y-axis) versus, $[(PN_{MD})_A,(PN_{MD})_B]_{max}$ (z-axis) using Mendeleyev periodic system.

To sum up, chapter 3 highlights the following facts:

1) On the example of the problem of separating former from non-former in binary, ternary, and quaternary systems, we check and re-formulate our statement [8]: 'Structure prototype-sensitive properties of inorganic systems are quantitatively described by APPs or APPEs of their constituent chemical elements'. The re-formulation of the statement results in: 'Structure prototype-sensitive material properties are quantitatively described by the APPs AN and PN (or simple mathematical functions of them) of their constituent chemical elements'. The significant implication of this example is, the data-driven approach implemented as a tool DISCOVERY shows a new way of inverse problem solving.

2) We should note some observations in context with the noble gases. It is known that noble gases create inorganic substances only in very exceptional cases. Figs. 14-18 illustrate that the chemical systems containing noble gases are all placed in the inorganic substance former area. The eight mentioned APPs belonging to the atomic reactivity pattern group show only values for the absolute electronegativity EN(abs), and the first ionization energy IE(first). They follow the general trend within the GN 18, like our calculated atomic reactivity $RE_a$. Similar observations are made for the APPs Rve(S), and Rce(S) belonging to the atomic size pattern group, which are also in harmony with our calculated atomic size $SZ_a$. Looking at the binaries, we see only 10 systems have been examined from the about 500 noble-gas-containing systems.

## 4. Experimentally determined inorganic substances data (peer-reviewed) from world literature since 1891 (PAULING FILE project)

### 4.1. The PAULING FILE

The PAULING FILE is a relational database for materials scientists, gathering crystallographic data, phase diagrams, and large set of physical properties of inorganic substances (focusing mainly on single-phase) under the same frame. Focus is on experimental observations and the data are processed from the original publications, encompassing world literature from 1891 to current date. Each individual crystallographic structure, phase diagram, or physical property database entry contains data from a particular publication, but the linkage of the different data sets is achieved via the *Distinct Phases Concept* (see 4.7.), taking into account the chemical system and the crystallographic structure, and (or) domain of existence, of different *distinct phases.*

The goal of the PAULING FILE is to ensure easy access to vast amounts of all sorts of critically evaluated experimental data. This also gives a general overview on inorganic substances and offers opportunities to reveal yet undiscovered patterns among data, at the same time facilitating a sensible and efficient search for novel inorganic substances with tailored predefined physical parameters. Alongside different data mining and knowledge discovery techniques, the PAULING FILE provides patterns of holistic views on inorganic substances, *not* lacking in the 'Gestalt decomposition' into the unrelated datasets and the fragmented models, *but* proving that 'The whole is greater than the sum of its parts'.

### 4.2. Origin and evolution of the PAULING FILE

The shortcomings of the empirical bottom-up approach provided the initial motivation for the launch of the PAULING FILE project in 1992. The PAULING FILE project **[8,9]** envisaged three steps. As already stressed in Chapter 1, the first objective was to invent and maintain a comprehensive database for inorganic crystalline substances, encompassing crystallographic data, diffraction patterns, a large set of physical properties and phase diagrams. The data should be reviewed with extreme care, and the term crystallized (solid) 'inorganic substances' was defined as inorganic substances including no C-H bonds. Simultaneously with the database creation, appropriate retrieval and visualization software must be developed to make the different groups of data mentioned above available by means of a single user and user-friendly interface. In longer term, new instruments for inorganic substance design must be invented, which would more or less automatically search the database for correlations and causalities for intelligent design of novel inorganic substances with the predetermined physical parameters.

To test the approach, the prototype PAULING FILE – Binaries Edition appeared in 2002 **[8,18]**. Since about the same time, the PAULING FILE is under the sole leadership of MPDS, Vitznau, Switzerland; from 2017 onwards it is again Japanese / Swiss collaboration between NIMS (National Institute for Materials Science, Tsukuba, Japan) and MPDS. Selected PAULING FILE data are included in several printed, off-line and on-line products, most of them updated on a yearly basis, and three multinary on-line editions of the PAULING FILE are available since 2017 **[17,19,20]**.

### 4.3. PAULING FILE – Crystalline Structures

The minimal requirement for a crystallographic structure database entry in the PAULING FILE is a set of published cell parameters, assigned to a crystalline inorganic substance of well-defined composition. Whenever the published data are accessible, the crystallographic data contain as well the atom coordinates, (an)isotropic displacement parameters and experimental diffraction lines, and are accompanied by the information concerning preparation, experimental conditions, characteristics of the sample, phase transitions, dependence of the cell parameters on temperature, pressure, and composition. The crystallographic data are collected as published, but have also been standardized on the method by Parthé and Gelato **[21,22,23,24]**. Derived data include Atomic Environment Type (AET) of individual atom sites, based on the maximum gap method **[25,26,27]**, and the reduced Niggli cell. The database entries are checked for inconsistencies within the database entry (e.g. chemical elements, charge balance, interatomic distances, space group, symmetry constraints), and by comparing different database entries (e.g. cell parameters and atom coordinates of isotypic inorganic substances) with a program package including more than 50 modules **[28]**. For 5% of the database entries, one or more misprints in the published crystallographic data are detected and corrected. Warnings concerning remaining short interatomic distances, deviations from the nominal composition, etc., are added in remarks. SI units are used everywhere and the crystallographic terms follow the recommendations by the International Union of Crystallography **[29]**.

*4.3.1. Data selection*

The data are extracted from the primary literature. The thesis works are not processed, the conference abstracts are considered only in exceptional cases. When available, supplementary material deposited as CIF files or in other formats is used as a data source.

*4.3.2. Categories of crystallographic structure entries*

As mentioned above, the minimal requirement for a database entry in the crystallographic structure part of the PAULING FILE is a total set of published cell parameters and composition. The database entries are subdivided into different groups, depending upon the level of research. The most common are:
• Complete structure defined,
• Coordinates of non-H atoms defined,
• Cell parameters defined and prototype with fixed coordinates assigned,
• Cell parameters defined and prototype assigned,
• Cell parameters defined.

*4.3.3. Database fields*

Besides the crystallographic data, large amounts of data regarding the sample preparation and experimental investigation are present in the PAULING FILE. Basic information is accumulated as published (for rapid comparison with the original paper) and standardized (for efficient data checking and retrieval and for a homogeneous presentation). The following database areas can be found in a crystal structure database entry:
• *Classification*
• *Bibliographic data*
• *Published crystallographic data*
• *Standardized crystallographic data*
• *Niggli-reduced cell*
• *Displacement parameters*
• *Published diffraction lines*
• *Preparation*
• *Mineral name*
• *Inorganic substance description*
• *Determination of cell parameters*
• *Structure determination*
• *Editor and error remarks*
• *Figure descriptions*
The data extracted and stored for $Be_{17}Nb_2$ ht (fictitious example) are available in Table 7 (external file).

*4.3.4. Structure prototypes*

The structure prototype is a famous notion in inorganic chemistry, where a large number of inorganic substances often crystallize with very similar atom arrangements. The compilation Strukturbericht **[30]** had started since the early 20$^{th}$ century to classify crystallographic structures into structure prototypes, named by codes such as A1, B1 or A15. Although these notations are still used, today structure prototypes are generally referred to by the name of the inorganic substance for which this particular atom arrangement was first identified, as for the structure prototypes enumerated above: Cu, NaCl, $Cr_3Si$. The PAULING FILE uses a longer notation, which contains also the Pearson symbol (a lower-case letter for the crystal system, an upper-case letter for the Bravais lattice, sum of multiplicities of all, fully or partially occupied atom sites) **[30]** and the number of the space group in the International Tables for Crystallography **[29]**: Cu,*cF4*,225; NaCl,*cF8*,225; $Cr_3Si$,*cP8*,223. All data sets with published atom coordinates are in the PAULING FILE grouped into structure prototypes, according to the criteria defined in TYPIX **[31]**. On the basis of this definition, isotypic inorganic substances must crystallize in the same space group, have similar cell parameter ratios, same Wyckoff positions, should be occupied in the standardized description (see below), with same or similar values of the atom coordinates. If all these criteria are accomplished, the AETs are expected to be similar. When possible, a structure prototype has been assigned also to data sets without atom coordinates.

*4.3.5. Standardized crystallographic data*

There exist multitudes of ways to select the crystallographic data (cell parameters, space group setting, representative atom coordinate triplets) that determine a crystallographic structure. The quantity remains high even when the basic rules recommended by the International Tables for Crystallography **[29]** are respected, owing to space group permitted operations such as permutations, origin shifts, etc. As a result, even identical or very similar atom arrangements may not be recognized as so (see Fig. 21). The grouping of crystallographic structures into structure prototypes is considerably facilitated by the use of standardized crystallographic data (several examples are given in **[32]**). The crystallographic structure data in the PAULING FILE are accumulated and saved as published, but also standardized. This second representation of the same data is such that inorganic substances crystallizing with the same structure prototype (also called isotypic inorganic substances) can be directly compared.

It is prepared in a 3-step procedure:
(1) The published data are checked for the presence of overlooked symmetry elements **[33]** and, if adequate, reorganized into a space group of higher symmetry.
(2) The processed data are standardized with the program STRUCTURE TIDY **[23]**.
(3) The processed data are compared with the standardized data of the type-defining database entry (STP) with the program COMPARE **[24]**.

*(1) Processing the symmetry*
A crystallographic structure can always be refined and described in a subgroup of the actual space group. To an extreme, any crystallographic structure can be described in the triclinic space group P1, having no other symmetry elements than identity and translation. To know the correct space group is important not only for the recognition of isotypic inorganic substances, but also in connection with the physical properties. Particular properties are effectively restrained to certain symmetries, e.g. ferroelectricity can only be observed for polar space groups, whereas pyroelectricity is excluded for crystallographic structures possessing an inversion center. Therefore, the crystallographic data in the PAULING FILE are checked for the presence of overlooked symmetry elements **[33]**. Whenever it is possible to describe the crystallographic structure in a space group of higher symmetry, or with a smaller unit cell, without any approximations, this is done. Fig. 21 shows how the structure of $WAl_5$, reported in space group $P6_3$ (173), can be described in space group $P6_322$ (182), after applying an origin shift of 0 0 3/4 to the published data **[34]**.

*Published crystallographic data*

| | |
|---|---|
| Space group | *P*6$_3$ (173) |
| Cell parameters | *a*= 0.49020(3) nm, *c*= 0.88570(5) nm, *V*= 0.18432 nm$^3$, *c/a*= 1.807 |

Atom coordinates

| Site | Elements | Wyck. | Sym. | *x* | *y* | *z* |
|---|---|---|---|---|---|---|
| W | W | 2*b* | 3.. | 1/3 | 2/3 | 0.5 |
| Al1 | Al | 2*b* | 3.. | 1/3 | 2/3 | 0.0 |
| Al2 | Al | 2*a* | 3.. | 0 | 0 | 0.0 |
| Al3 | Al | 6*c* | 1 | 1/3 | 1/3 | 0.25 |

*Standardized crystallographic data*

| | |
|---|---|
| Space group | *P*6$_3$22 (182) |
| Cell parameters | *a*= 0.49020 nm, *c*= 0.88570 nm, *V*= 0.18432 nm$^3$, *c/a*= 1.807 |

Atom coordinates

| Site | Elements | Wyck. | Sym. | *x* | *y* | *z* |
|---|---|---|---|---|---|---|
| Al3 | Al | 6*g* | .2. | 0.33333 | 0 | 0 |
| W | W | 2*d* | 3.2 | 1/3 | 2/3 | 3/4 |
| Al1 | Al | 2*c* | 3.2 | 1/3 | 2/3 | 1/4 |
| Al2 | Al | 2*b* | 3.2 | 0 | 0 | 1/4 |

Transformation: origin shift 0 0 3/4

Fig. 21. The structure of $WAl_5$, reported in space group $P6_3$ (173), can be described in space group $P6_322$ (182), after applying an origin shift of 0 0 ¾ to the published data. Data set from the PAULING FILE – Binaries Edition [8].

*(2) Standardization*

At the next stage, the crystallographic data are standardized on the method by Parthé and Gelato [21,22,23]. The standardization mechanism applies criteria to select the space group setting, the cell parameters, and the origin of the coordinate system, the representative atom coordinates, and the order of the atom sites. See an example in Fig. 22.

**RbO**

*Published crystallographic data*

| | |
|---|---|
| Space group | *Immm* (71) |
| Cell parameters | *a*= 0.4201(5) nm, *b*= 0.7075(5) nm, *c*= 0.5983(5) nm, *V*= 0.17783 nm$^3$, *a/b*= 0.594, *b/c*= 1.183, *c/a*= 1.424 |

Atom coordinates

| Site | Elements | Wyck. | Sym. | *x* | *y* | *z* |
|---|---|---|---|---|---|---|
| Rb | Rb | 4*g* | *m*2*m* | 0 | 0.25 | 0 |
| O | O | 4*i* | *mm*2 | 0 | 0 | 0.374 |

*Standardized crystallographic data*

| | |
|---|---|
| Space group | *Immm* (71) |
| Cell parameters | *a*= 0.4201 nm, *b*= 0.5983 nm, *c*= 0.7075 nm, *V*= 0.17783 nm$^3$, *a/b*= 0.702, *b/c*= 0.846, *c/a*= 1.684 |

Atom coordinates

| Site | Elements | Wyck. | Sym. | *x* | *y* | *z* |
|---|---|---|---|---|---|---|
| Rb | Rb | 4*i* | *mm*2 | 0 | 0 | 0.25 |
| O | O | 4*g* | *m*2*m* | 0 | 0.374 | 0 |

Transformation: new axes a,-c,b

**CsO**

*Published crystallographic data*

| | |
|---|---|
| Space group | *Immm* (71) |
| Cell parameters | *a*= 0.4322(10) nm, *b*= 0.7517(10) nm, *c*= 0.6430(10) nm, *V*= 0.20890 nm$^3$, *a/b*= 0.575, *b/c*= 1.169, *c/a*= 1.488 |

Atom coordinates

| Site | Elements | Wyck. | Sym. | *x* | *y* | *z* |
|---|---|---|---|---|---|---|
| Cs | Cs | 4*g* | *m*2*m* | 0 | 0.25 | 0 |
| O | O | 4*i* | *mm*2 | 0 | 0 | 0.383 |

*Standardized crystallographic data*

| | |
|---|---|
| Space group | *Immm* (71) |
| Cell parameters | *a*= 0.4322 nm, *b*= 0.643 nm, *c*= 0.7517 nm, *V*= 0.208901 nm$^3$, *a/b*= 0.672, *b/c*= 0.855, *c/a*= 1.739 |

Atom coordinates

| Site | Elements | Wyck. | Sym. | *x* | *y* | *z* |
|---|---|---|---|---|---|---|
| Cs | Cs | 4*i* | *mm*2 | 0 | 0 | 0.25 |
| O | O | 4*g* | *m*2*m* | 0 | 0.383 | 0 |

Transformation: new axes a,-c,b

Fig. 22. Data sets for RbO and CsO, as published and after standardization, revealing their isotypism. Data sets from the PAULING FILE – Binaries Edition [8].

*(3) Comparing with the type-defining data set (Structure type pool (STP))*

In general case, comparable data sets for isotypic inorganic substances are obtained directly out of the standardization procedure. This is, meanwhile, not always correct, since particular situations may occur. In order to tackle with such problems, each standardized data set is compared with the standardized database entry that determines the structure prototype in the PAULING FILE. The program COMPARE [24] creates the different space group permitted crystallographic representations. Each representation is compared with the standardized description of the type-defining entry from the STP.

### *4.3.6. Assigned atom coordinates*

With the purpose to give an approximate notion of the actual crystallographic structure, a total range of atom coordinates and site occupancies is provided for database entries where a structure prototype could be assigned (by the authors or editors), but atom coordinates were not identified. Two different cases are distinguished:

*(1) A structure prototype where all atom coordinates are established by symmetry is assigned.*
The editor, on the grounds of the chemical formula, will assign also a probable atom distribution.
*(2) A structure prototype with determinable atom coordinates is assigned.*
The atom coordinates of the type-defining entry are suggested as the first approximation. The atom distribution is inserted by a program that compares the chemical formula of the type-defining entry with a chemical formula transformed by the editor so that the substitution element by element is outlined [28].

### *4.3.7. Atomic Environment Types (AETs) producing the coordination types classification*

For the approach used here [26,27], the AETs, also known as coordination polyhedron, are considered applying the method of Brunner and Schwarzenbach [25]. This method means that the interatomic distances between an atom and its neighbors are put in a next-neighbor histogram, as we can see on the left hand-side of Fig. 23 for the Li atom in $BaLiF_3$ ($CaTiO_3$,*cP*5,221), the right side illustrates its AET a octahedron. A clear maximum gap is often revealed and the atoms placed at distances to the left of the maximum gap are supposed to belong to the AE of the central atom. This principle is named the ‘maximum gap rule’ and the coordination polyhedron, the AET, is built with the atoms to the left of the maximum gap. The polyhedron around the Li atom in Fig. 23 is an octahedron.

When the maximum gap rule leads to AETs with not only the selected central atoms but also supplementary atoms enclosed in the polyhedron, or to AETs with atoms situated on one or more of the faces or edges of the coordination polyhedron, the so-called ‘maximum-convex-volume rule’ is applicable. This principle is defined as the maximum volume around the central atom delimited by convex faces, with all the atoms of the AE lying at the intersection of three faces at a minimum. This rule is also applied to those cases where no clear maximum gap is determined.

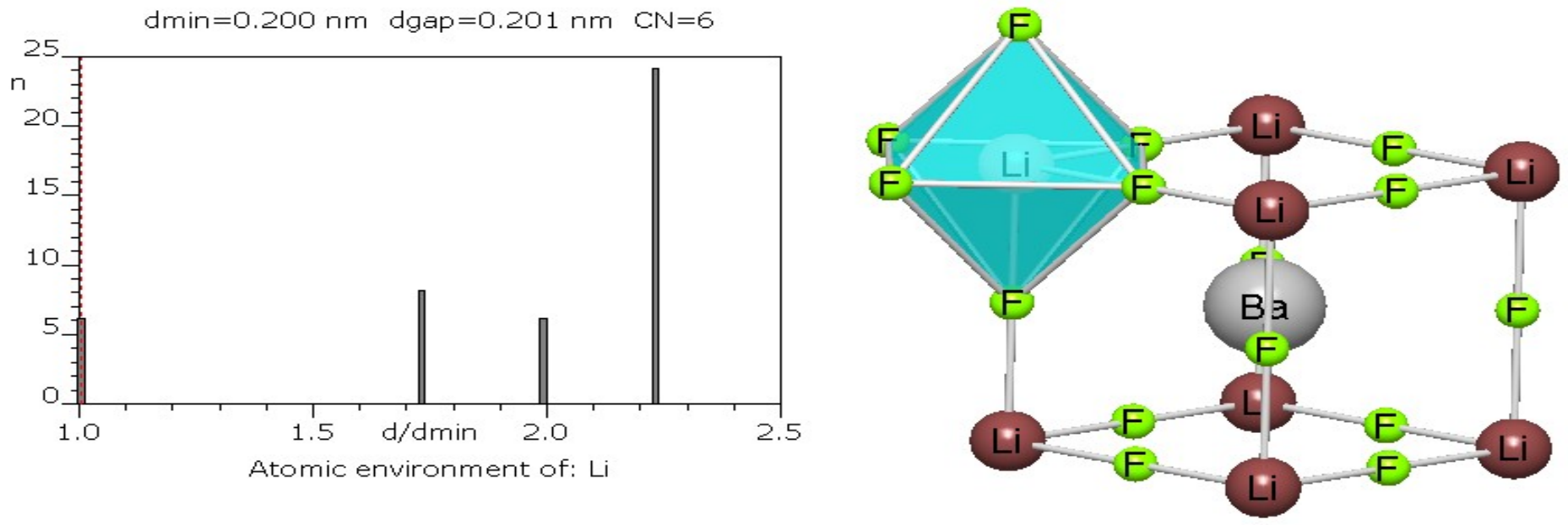

Fig. 23. Next-neighbor histogram (NNH) (top left) and the corresponding coordination polyhedron type (AET) for an entry for $BaLiF_3$ in Pearson’s Crystal Data [62].

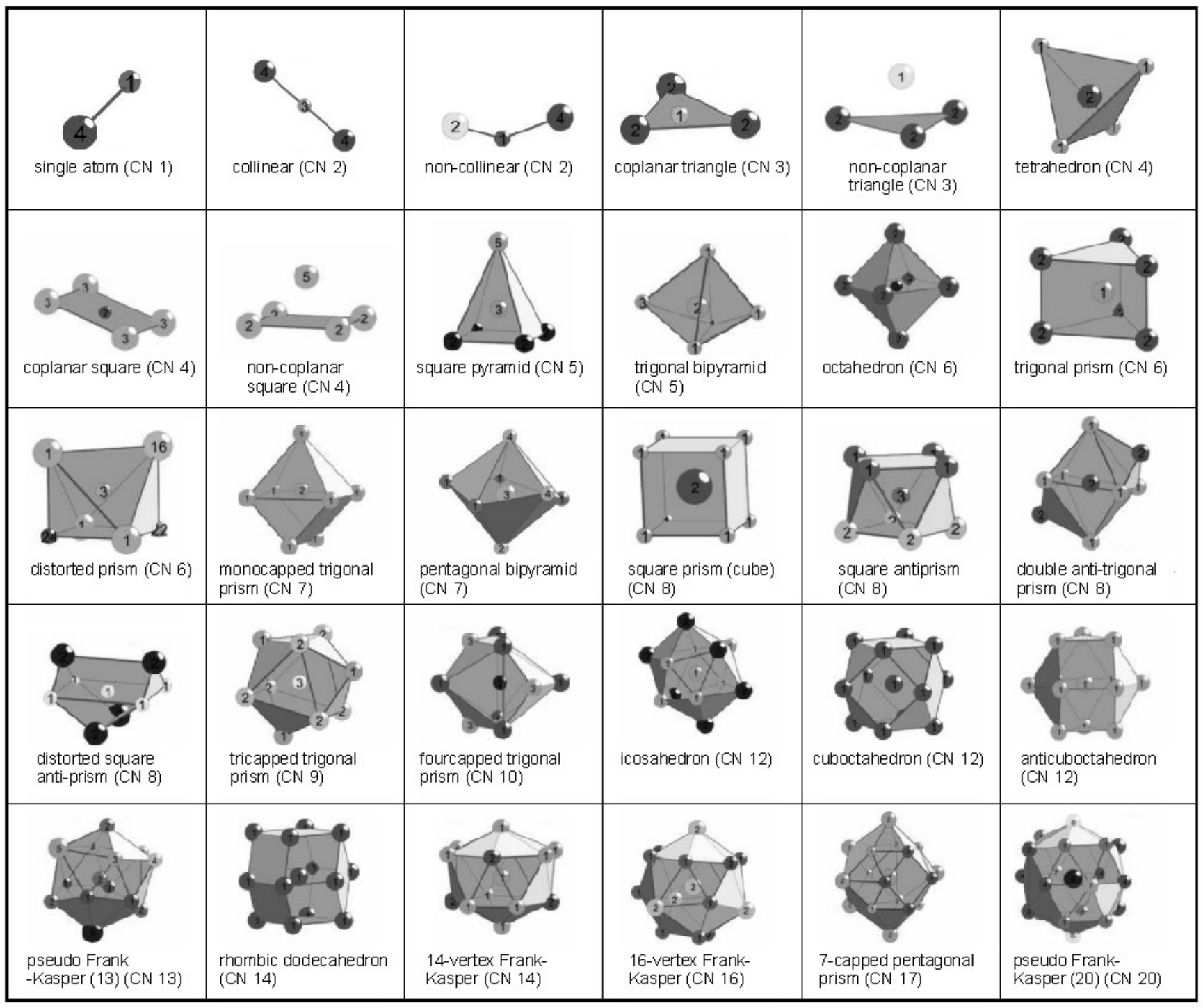

Fig. 24. The 30 most populous AET (see frequency in Table 8).

All the crystallographic structure entries with refined or fixed atom coordinates in the PAULING FILE are examined, with regard to the rules given above. 100 different AETs have been detected; the 30 most frequent of them are shown in Fig. 24 and given in Table 8. Coordination number (CN) and name of the coordination polyhedron define each AET; the number in the second column means the number of times this AET is present in Pearson's Crystal Data [34], release 2020/21.

| Frequency counts order | Counts | CN | *Name* |
|---|---|---|---|
| 1 | 361'674 | 1 | single atom |
| 8 | 32'013 | 2 | collinear |
| 2 | 291'022 | 2 | non-collinear |
| 10 | 24'868 | 3 | coplanar triangle |
| 5 | 135'350 | 3 | non-coplanar triangle |
| 4 | 210'239 | 4 | tetrahedron |
| 21 | 9'746 | 4 | coplanar square |
| 25 | 7'686 | 4 | non-coplanar square |
| 14 | 18'046 | 5 | square pyramid |
| 15 | 17'799 | 5 | trigonal bipyramid |
| 3 | 221'483 | 6 | octahedron |
| 19 | 10'615 | 6 | trigonal prism |
| 30 | 5'002 | 6 | distorted prism |
| 16 | 12'829 | 7 | monocapped trigonal prism |
| 22 | 9'520 | 7 | pentagonal bipyramid |
| 12 | 19'422 | 8 | square prism (cube) |
| 11 | 21'728 | 8 | square antiprism |
| 27 | 5'717 | 8 | double anti-trigonal prism |
| 28 | 5'260 | 8 | distorted square anti-prism |
| 7 | 34’764 | 9 | tricapped trigonal prism |
| 18 | 11’346 | 10 | fourcapped trigonal prism |
| 9 | 29’181 | 12 | icosahedron |
| 6 | 47’952 | 12 | cuboctahedron |
| 26 | 6’936 | 12 | anticuboctahedron |
| 24 | 8’338 | 13 | pseudo Frank-Kasper (13) |
| 13 | 19’355 | 14 | rhombic dodecahedron |
| 17 | 11’818 | 14 | 14-vertex Frank-Kasper |
| 20 | 9’879 | 16 | 16-vertex Frank-Kasper |
| 29 | 5’112 | 17 | 7-capped pentagonal prism |
| 23 | 8’697 | 20 | pseudo Frank-Kasper (20) |
| sum | 1’532’970 | | |

Table 8. The 30 most frequent occurring atomic environment types (AETs) with their counts (number of point-sets).

This merely geometrical approach elaborated for intermetallic substances does not separate the types of chemical bonding. AET is a purely topological instrument for grouping of crystallographic structures into geometrically similar ‘structure prototypes’ **[35]**, called ‘coordination types’ categorization **[27]**.

*4.3.8. Cell parameters from plots*

Values are drawn from plots of cell parameters (or functions of these) *vs.* temperature, pressure, or composition **[36]** and accumulated in the database. Three cases are possible: experimental points, fit to experimental points, linear dependence. The drawn values are converted to SI units and normally produce new figures illustrating phase transitions (Figs. 25a-b), thermal expansion (Fig. 25c) or compression (under pressure) (Figs. 25d).

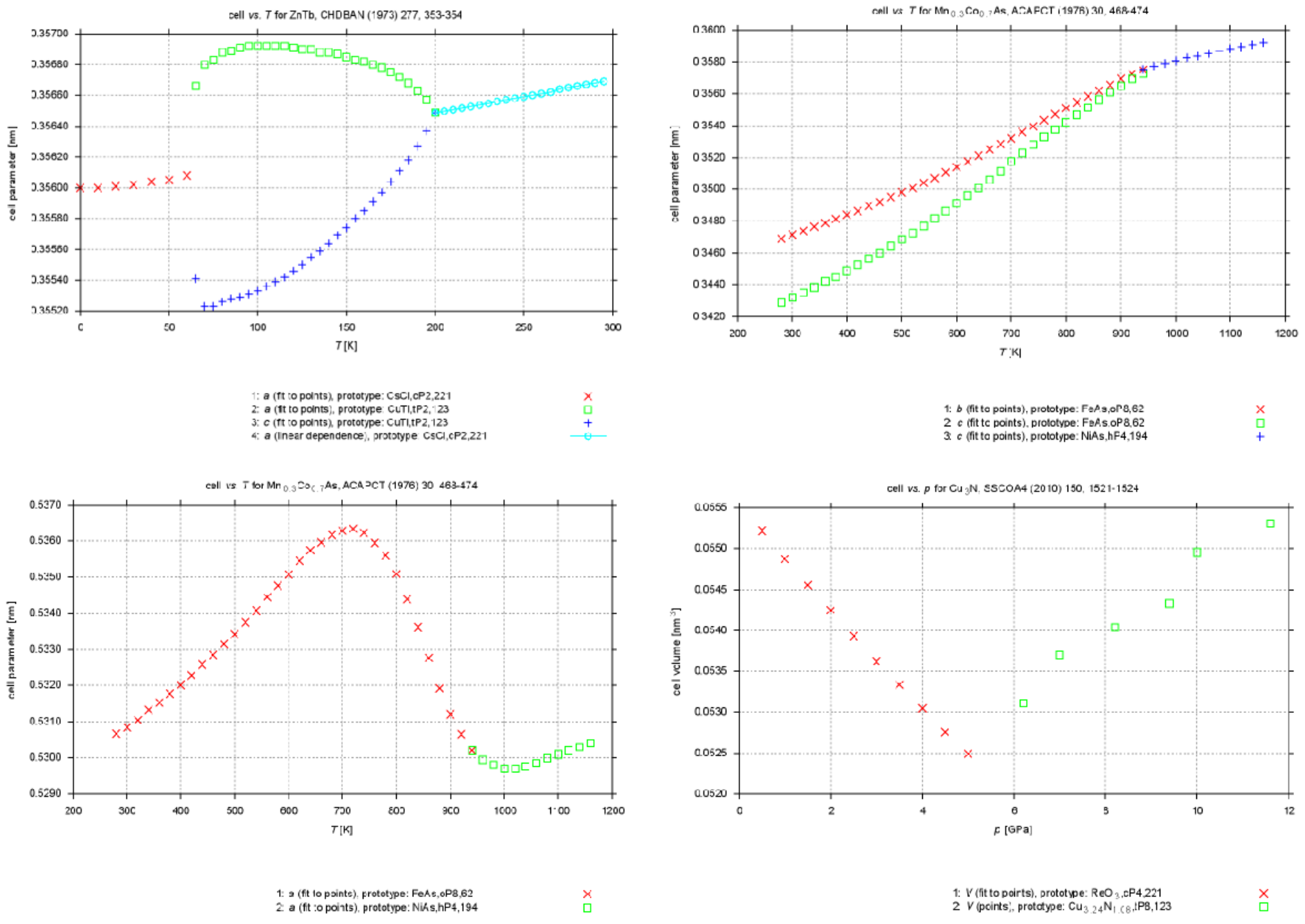

Fig. 25a-d. Cell parameters versus T for $ZnTb_2$ and $Mn_{0.3}Co_{0.7}As$; cell volume versus p for $Cu_3N$.

### 4.4. PAULING FILE – Phase Diagrams

The phase diagram part of the PAULING FILE comprises temperature-composition phase diagrams for binary systems, as well as horizontal and vertical sections and liquidus / solidus projections for ternary and multinary systems. Both kinds of experimentally determined and calculated diagrams are processed. Primary literature is considered in priority, but diagrams from several commonly known collections, such as the compendium of binary phase diagrams edited by Massalski *et al*. **[37]** and the series of books on ternary phase diagrams edited by Petzow and Effenberg **[38]**, have also been included. All the diagrams have been converted to at.% and °C and redrawn in a standardized version, so that different values for the same chemical system can be easily seen in comparison. Single-phase areas are colored in blue, two-phase areas in white, three-phase areas in yellow, four-phase areas in red, and five-phase in green. The *distinct phases* identified on the diagrams are named according to PAULING FILE conventions, but also the original names are posted in the database. Examples of phase diagrams redrawn for the PAULING FILE are shown in Figs. 26a-d: (a) phase diagram of a binary system (b) vertical section and (c) isothermal section of a ternary system, (d) liquidus projection of the phase diagram of a ternary system. Each phase diagram is linked to a database entry, which comprises the following database fields:

- *Classification*
- *Investigation*
- *Bibliographic data*
- *Original diagram*
- *Redrawn diagram*
- *List of distinct phases presented on the diagram*

For binary systems also the temperature and reaction type for the upper and (or) lower limit of existence of the *distinct phase* are stored.

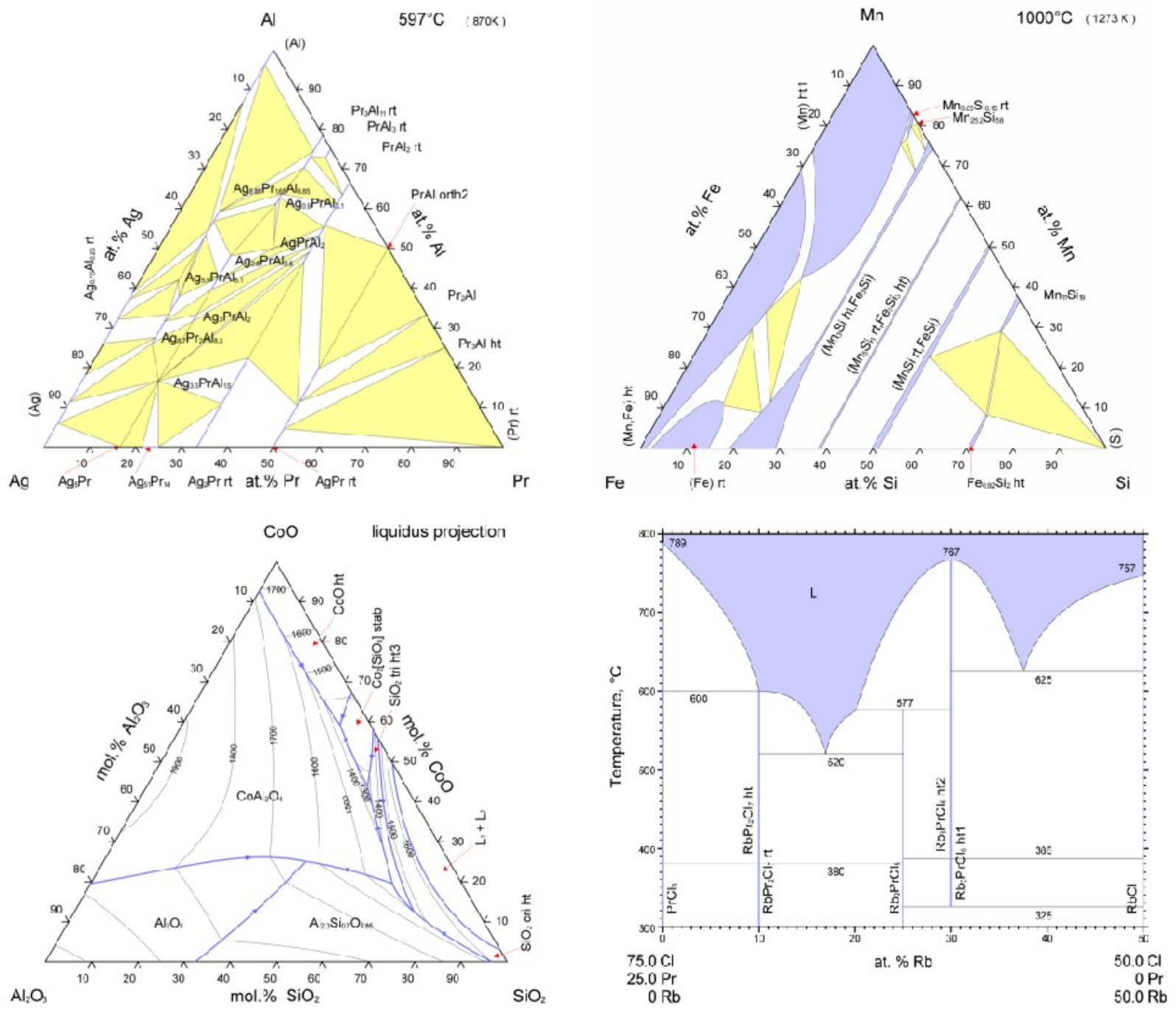

Fig. 26a-d. Examples of phase diagrams as redrawn for the PAULING FILE: (a,b) isothermal sections of two ternary systems, (c) liquidus projection of a multinary system, (d) vertical section of a ternary system.

### 4.5. PAULING FILE – Physical Properties

The physical properties part of the PAULING FILE stores experimental and (to a limited extent) simulated data for a broad range of physical properties of inorganic substances in the solid, crystalline state. Each database entry groups selected data extracted for a *distinct phase* in a publication. Focus is on the characterization of inorganic substances (single-phase samples). When published, the entries also contain information about synthesis and sample preparation, as well as information that help to establish the links to phase diagram and crystallographic structure entries, such as colloquial names, crystallographic data, and limits of stability of the phase with respect to temperature, pressure, or composition. The physical properties are stored in four different ways:

- Property class, such as superconductor, ferroelectric, *etc*.,
- Additional data, such as keywords indicating the existence of fragmentary data, *e.g.* spectra, *etc*.,
- Figure description (*y vs. x*),
- Numerical value.

The symbols for most common physical properties have been standardized, mainly based on the CRC Handbook of Chemistry and Physics **[39]**. Great flexibility is provided through the links to reference tables, thanks to them the physical properties may be selected, as well as their symbols, units, and ranges of magnitude can be controlled.

*4.5.1. Data selection*

Data are taken from primary literature. Each database entry corresponds to an inorganic substance, but can contain several numerical values, figure descriptions, and keywords. For an investigation of an inorganic substance through a temperature- or pressure-induced structural phase transition there will be two database entries, for instance one for the room-temperature modification and one for the low-temperature modification. Some simulated data from *ab initio* simulations with code identification are also included, in energy band structures, but focus is on experimentally measured data and values directly derived from measurements.

*4.5.2. Database fields*

In addition to the physical properties (in the form of property classes, additional data, figure descriptions, and numerical values), and compulsory items such as the chemical formula, large amounts of information concerning the sample preparation and experimental conditions are posted in the PAULING FILE. The following database areas may be present in the physical properties database entry:
- *Inorganic substance*
- *Bibliographic data*
- *Preparation*
- *Sample description*
- *Crystallographic data*
- *Numerical values*
- *Figures*
- *Keywords*
- *Property class*

The data extracted and stored for $TaS_2$ 1T ht (fictive example) are shown in Table 9 (external file).

*4.5.3. Physical properties considered in the PAULING FILE*

The physical properties considered in the PAULING FILE belong to one of the following domains: phase transitions, mechanical properties, thermal and thermodynamic properties, electronic and electrical properties, optical properties, magnetic properties, and superconductivity. Table 10 lists the main physical properties considered in the PAULING FILE for the domain 'thermal and thermodynamic properties'. Items in square brackets are keywords, for which numerical values are in principle not yet extracted.

Primary properties, to which a certain attention is paid for the extraction of numerical values, are emphasized with bold characters.

---

*thermal expansion*
**linear thermal expansion coefficient**
  pressure derivative of linear thermal expansion coefficient
  temperature derivative of linear thermal expansion coefficient
*for cell parameters vs. temperature see PAULING FILE – Crystal Structures*
magnetic contribution to linear thermal expansion coefficient
**volume thermal expansion coefficient**
  pressure derivative of volume thermal expansion coefficient
  temperature derivative of volume thermal expansion coefficient
*volume change at phase transition*
**relative volume change at melting**
relative volume change at structural phase transition
relative volume change at Curie point
relative volume change at Néel point
*enthalpy/energy*
**enthalpy of formation**
enthalpy of reaction
Gibbs energy of formation
Gibbs energy of reaction
Helmholtz energy of formation
Helmholtz energy of reaction
[ cohesive energy ]
[ total energy calculation data ]
*enthalpy/energy change at phase transition*
**enthalpy of fusion**
enthalpy change at eutectoid decomposition
enthalpy change at peritectoid decomposition
enthalpy of sublimation
enthalpy change at structural phase transition
*heat capacity (specific heat)*
**heat capacity at constant pressure**
**heat capacity at constant volume**
phonon heat capacity at constant pressure
phonon heat capacity at constant volume
electronic contribution to heat capacity
superconducting heat capacity
electronic contribution to superconducting heat capacity
**magnetic heat capacity**
Schottky heat capacity
hyperfine heat capacity
*heat capacity discontinuity*
heat capacity discontinuity at melting
heat capacity discontinuity at structural phase transition
heat capacity discontinuity at superconducting transition
*heat capacity coefficients*
**electronic heat capacity coefficient**
coefficient of Schottky term in heat capacity
coefficient of third-order term in heat capacity
coefficient of fifth-order term in heat capacity
coefficient of spin fluctuation term in heat capacity
*Debye/Einstein temperature*

**Debye temperature**
Einstein temperature
*entropy*
**entropy**
entropy of formation
magnetic entropy
*entropy change at phase transition*
entropy change at structural phase transition
**entropy change at melting**
sublimation entropy
*thermoelectric power*
**thermoelectric power (Seebeck coefficient)**
temperature derivative of thermoelectric power
**thermoelectric figure of merit**
*thermal conductivity*
thermal conductivity
  pressure derivative of thermal conductivity
  temperature derivative of thermal conductivity
electronic contribution to thermal conductivity
phonon contribution to thermal conductivity
Lorentz number
*diffusion*
[ hydrogen diffusion ]
[ diffusion ]
*etc.*

---

Table 10. Thermal and thermodynamic properties considered in the PAUING FILE.

Thanks to the flexible metadata construction of this relational database, new physical properties can easily be added. Fig. 27 shows the number of items in the physical properties part of the PAULING FILE according to the property domain and the data category; from bottom to top for each column: numerical value, figure description, keywords for additional data, and property class, distributed over the eight property categories considered in the PAULING FILE.

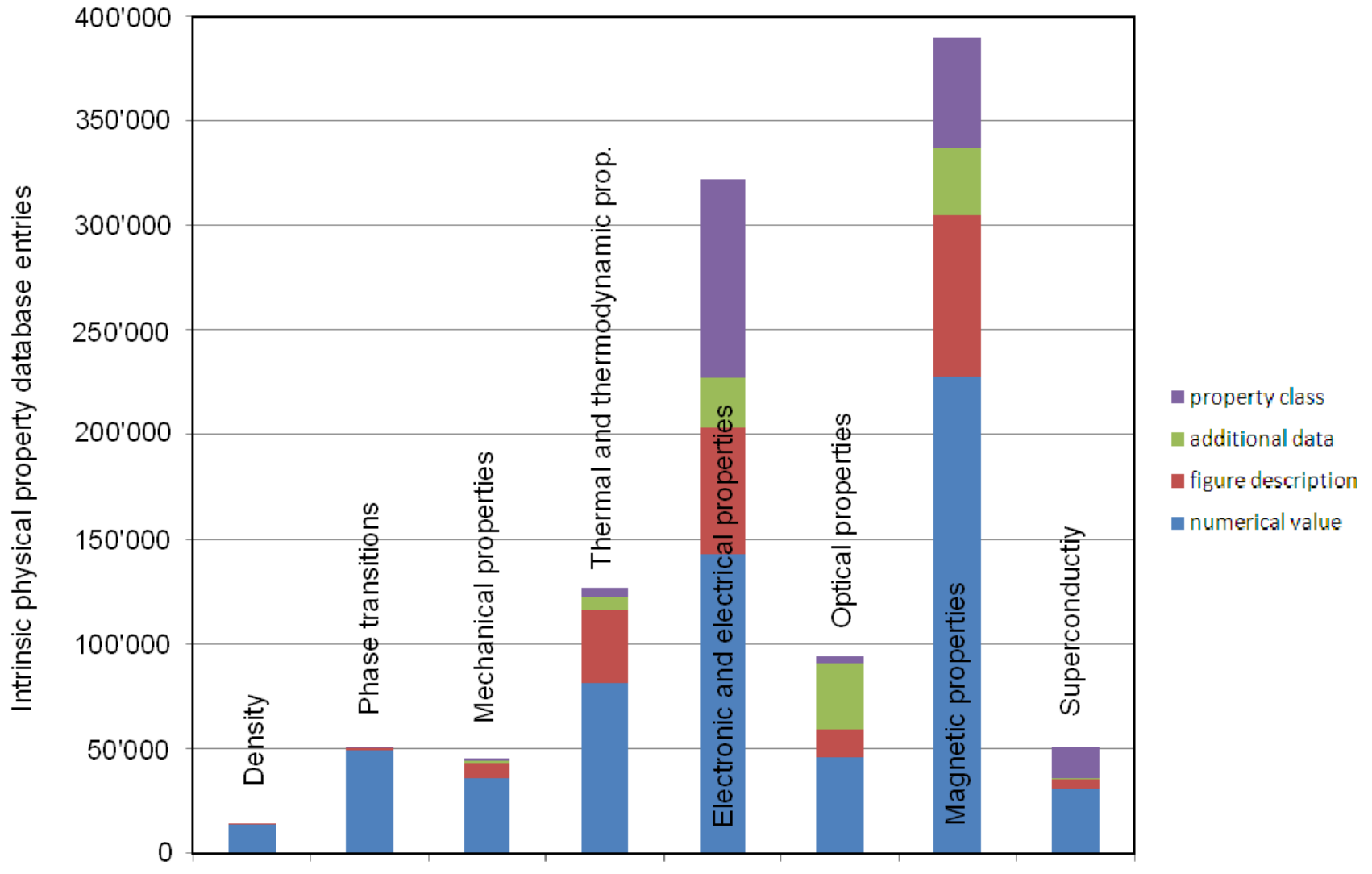

Fig. 27. Number of items in the physical properties part of the PAULING FILE according to the property category and the data category; from bottom to top for each column: numerical value, figure description, keyword for additional data, and property class.

### 4.6. Data quality

Only reliable data can be used for sensible data mining and great importance is given to the quality of the data in the PAULING FILE. The articles selected for processing are analyzed by scientists specialized in crystallography, phase diagrams, or solid-state physics, most of them with a doctor degree and own experience in solid-state chemistry or physics research.

#### *4.6.1. Computer-aided checking*

The PAULING FILE data are checked for consistency with the help of an original software package, ESDD (Evaluation, Standardization, Derived Data), containing more than 100 different modules [28]. The checking is carried out progressively, level by level.

*Checks on individual database areas are:*
- Formatting of numerical values,
- Units and symbols for physical properties,
- Hermann-Mauguin symbols, Pearson symbols,
- Consistency journal code–year–volume, first–last page for literature references,
- Formatting of chemical formulae,
- Usual order of magnitude,
- Spelling.

*Consistency within individual data sets: e.g.*
- Consistency atom coordinates-Wyckoff letters-site multiplicity

*Special checking of crystallographic data:e.g.*
- Comparison of interatomic distances with the sum of atomic radii

*Consistency within the database: e.g.*
- Comparison of densities

Wherever possible, misprints detected in the original paper are corrected, based on arguments explained in remarks. Since editing mistakes can never be completely avoided, all modifications of the originally published data and interpretations of ambiguous data are posted in remarks.

### 4.7. Distinct Phase Concept

The first part of the challenge is in building up a comprehensive database compiling the large amounts of data. However, to provide 'holistic views' of the database content and allow combined retrieval, it was also necessary to link the different database entries from the three parts of the PAULING FILE in a more efficient way than provided through the bibliographic information and the chemical system. The *Distinct Phases Concept* was introduced for this purpose. The linkage of the three different groups of data is achieved *via the Distinct Phases Table*, to which each individual crystallographic structure, phase diagram, and physical properties entry is linked with the *permanent distinct phase identifier.*
To make this table, each chemical system has been evaluated and the *distinct phases* identified using the information available in the PAULING FILE. As an example, the twelve phases reported in the Al–Hf system are listed in Table 11. A *distinct phase* is in the PAULING FILE defined by the chemical system, the structure prototype (when known), and (or) the domain of existence with respect to temperature, pressure, or composition. Each *distinct phase* has been given a unique name containing a representative chemical formula, when necessary followed by a specification such as "ht", "rt", "3R", "hex", *etc.* For not yet (fully) investigated crystallographic structures, partial structural information is given, if available, *e.g.* the complete Pearson symbol may be replaced by *t*** (tetragonal) or *cI** (cubic body centered). Information about colloquial names and stability with respect to temperature, pressure, or composition, collected in the three parts of the database, is used to assign a phase identifier to the physical properties and phase diagram entries with no crystallographic structure data.

| System | at.% Hf | **Distinct phase** | Structure prototype | Space group |
|---|---|---|---|---|
| Al-Hf | | (Al) | Cu,*cF4*,225 | *F*m-3m |
| Al-Hf | 25 | $HfAl_3$rt | $ZrAl_3$,*tI16*,139 | *I*4/mmm |
| Al-Hf | 25 | $HfAl_3$ ht | $TiAl_3$,*tI8*,139 | *I*4/mmm |
| Al-Hf | 33.32 | $HfAl_2$ | $MgZn_2$,*hP12*,194 | *P*$6_3$/mmc |
| Al-Hf | 40 | $Hf_2Al_3$ | $Zr_2Al_3$,*oF40*,43 | *F*dd2 |
| Al-Hf | 50 | HfAl | TlI,*oS8*,63 | *C*mcm |
| Al-Hf | 57.14 | $Hf_4Al_3$ rt | $Zr_4Al_3$,*hP7*,191 | *P*6/mmm |
| Al-Hf | 60 | $Hf_3Al_2$ rt | $Zr_3Al_2$,*tP20*,136 | *P*$4_2$/mnm |
| Al-Hf | 62.5 | $Hf_5Al_3$ stab | $Mn_5Si_3$,*hP16*,193 | *P*$6_3$/mcm |
| Al-Hf | 66.67 | $Hf_2Al$ rt | $Hf_2Al$,*tI12*,140 | *I*4/mcm |
| Al-Hf | 100 | (Hf) rt | Mg,*hP2*,194 | *P*$6_3$/mmc |
| Al-Hf | 100 | (Hf) ht | W,*cI2*,229 | *I*m-3m |

Table 11. The 12 distinct phases in the Al-Hf system.

There exist of course still parts of chemical systems that are little explored and reports in the literature are sometimes contradictory. The *distinct phase* assignment becomes here difficult, and the list of *distinct phases* may sometimes contain more *distinct phases* than there exist in reality. It turns out there is a certain amount of subjectivity when assigning a *distinct phase identifier*; however this approach represents a substantial advantage for the user, *and in addition it is a requirement to link the three major parts of the PAULUNG FILE*.

### 4.8. Chemical Formulae and Phase Names

The chemical formulae have been standardized so that the chemical elements are always written in the same order, an order that roughly corresponds to the order of the groups in the periodic system. Chemical units, such as water molecules or sulfate ions, are distinguished and written within square brackets. Deuterium and tritium are considered as separate chemical elements. In the crystallographic structure fragment of the PAULING FILE, whenever a structure prototype has been assigned to the published data, the chemical formula is written so that the number of formula units per cell is the same as for the type-defining inorganic substance. A *distinct phase* containing 50 at.% *A* and 50 at.% *B*, for example, will be called $A_{0.50}B_{0.50}$ if the structure prototype is Cu,*cF4*,225 (*Z*= 4), but *AB* if it is CuAu,*tP2*,123 (*Z*= 1) and $A_2B_2$ if it is $Cu_3Au$,*cP4*,221 (*Z*= 1). A two-phase sample of the same composition would be written $A_{50}B_{50}$. Such conventions imply a certain hypothesis on the atom distribution in the case of off-stoichiometric formulae. Each *distinct phase* is assigned with the name, which in general case is a representative chemical formula, written as described above. If several *distinct phases* are known for the same chemical composition, a short code specifying the modification is added. Preference is given to terms such as 'rt' (room-temperature), 'ht' (high-temperature), 'lt' (low-temperature), or 'hp' (high-pressure), possibly followed by a digit when a series of temperature- or pressure-induced phase transitions are known. If only one modification, stable at room temperature, is known, the field modification is left blank. Ramsdell notations are used for polytypic inorganic substances such as $CdI_2$. Mineral names can also be used as specifications, and are then abbreviated to the first three letters. Special notations are used in the phase diagram fragment, where a chemical element in parentheses indicates a terminal solid solution based on this element. For complete solid solutions two or more chemical elements, or chemical formulae (if relevant, with specifications) are written within parentheses, separated by commas, *e.g.* (LiBr,AgBr) or ($Ag_2La$,$Ag_2Ce$ rt).

### 4.9. Phase Classifications

A certain number of characteristics, attributed to the *distinct phases*, are posted in the *Distinct Phases Table*.

• *Inorganic substance classes*
• *Crystallographic structure classes*: e.g. The nomenclature of zeolites, using 3-letter codes to characterize different frameworks, is taken from the Database of Zeolite Structures **[40]**.
• *Physical property classes: e.g.* Such as antiferromagnet, ferroelectric, semiconductor, metal, ionic conductor, superconductor, etc. are distinguished based on data available in the physical properties part of the PAULING FILE.
• *Mineral names:* The names reported in the original publications have been checked by consulting Strunz Mineralogical Tables **[41]** and the list of minerals approved by the International Mineralogical Association **[42]**, and updated consequently.
• *Color:* Color has tentatively been assigned also at the phase level **[43]**, but in some cases it is strongly composition-dependent, or due to small amounts of impurities.

### 4.10. Towards a super-database

After about 30 years of existence, the PAULING FILE has reached a respectable size in the areas of crystallographic structures and phase diagrams of inorganic substances (systems). Focus is here on the yearly update, and old, not yet processed publications represent a few percent. On the contrary, in spite of the relatively high number of database entries, the coverage of physical properties is still at a lower level, considering the huge amount of data published in this field.

At the present stage of development, the PAULING FILE comprises about 350,000 crystallographic structure entries (including atom coordinates and displacement parameters, when relevant) for some 170,000 *distinct phases*, more than 50,000 phase diagrams (with updated *distinct phase* assignment) for more than 10,000 chemical systems, and about a million of physical property entries (500,000 numerical values and 200,000 figure descriptions) for about 150,000 *distinct phases*. To reach this result, about 310,000 scientific publications have been processed, from more than 1,500 different journals. Some 250 scientific journals are browsed from cover to cover for the yearly updates.

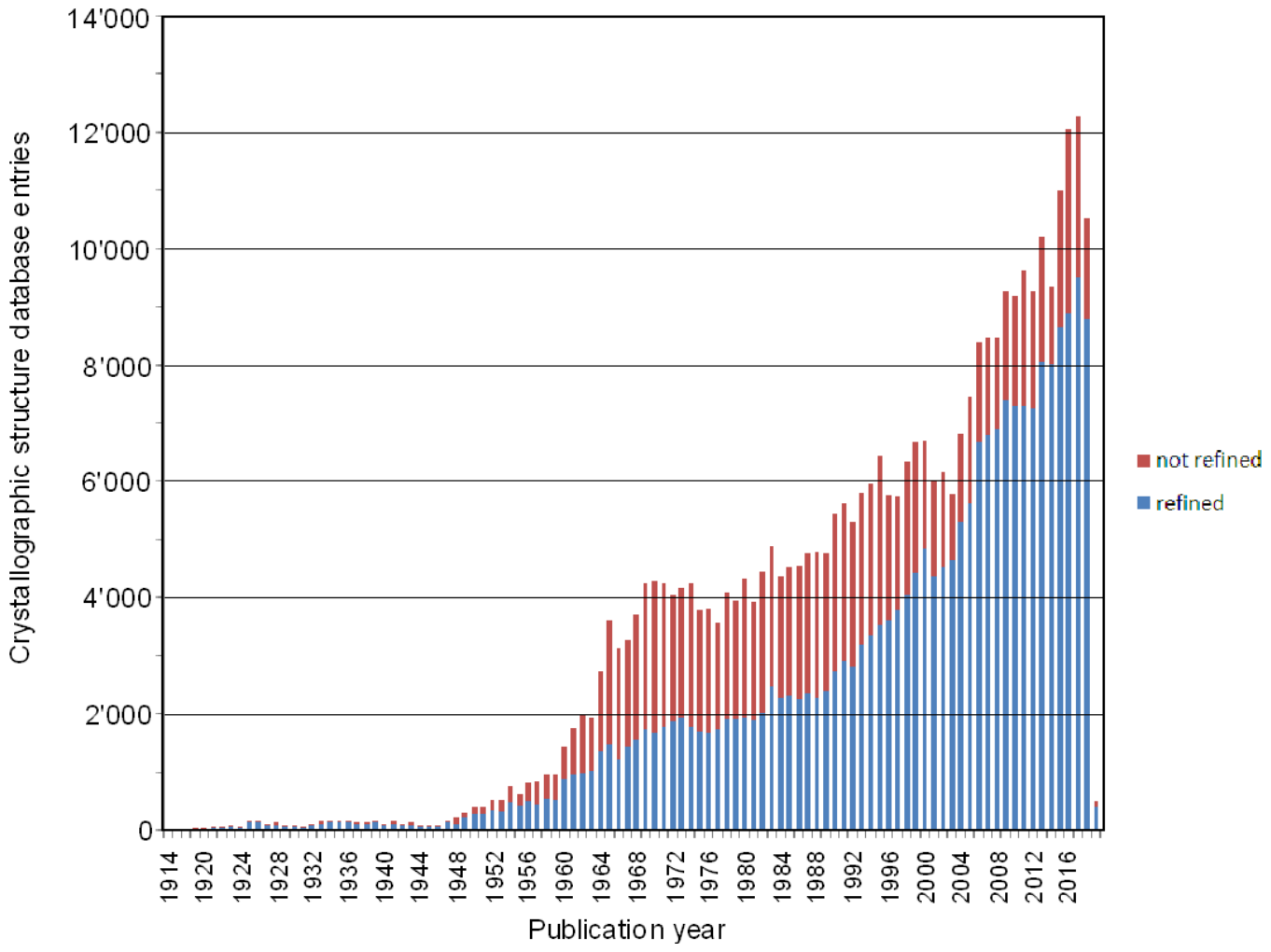

Fig. 28. Distribution of the database entries in the PAULING FILE according to the publication year for crystal structure. The blue part of the bars represent the refined structures, the red part the not refined structures.

Crystallographic structure database entries in Fig. 28 show the distribution of the database entries per publication year. The regular shape of the diagrams for crystallographic structure and phase diagram entries confirms the good coverage of the world literature for these two sections of the PAULING FILE. Fig. 29 confirms that the number of experimental investigations of phase diagrams per year is decreasing, whereas the number of thermodynamic assessments is increasing. Fig. 30 shows the number of physical property entries as a function of publication year.

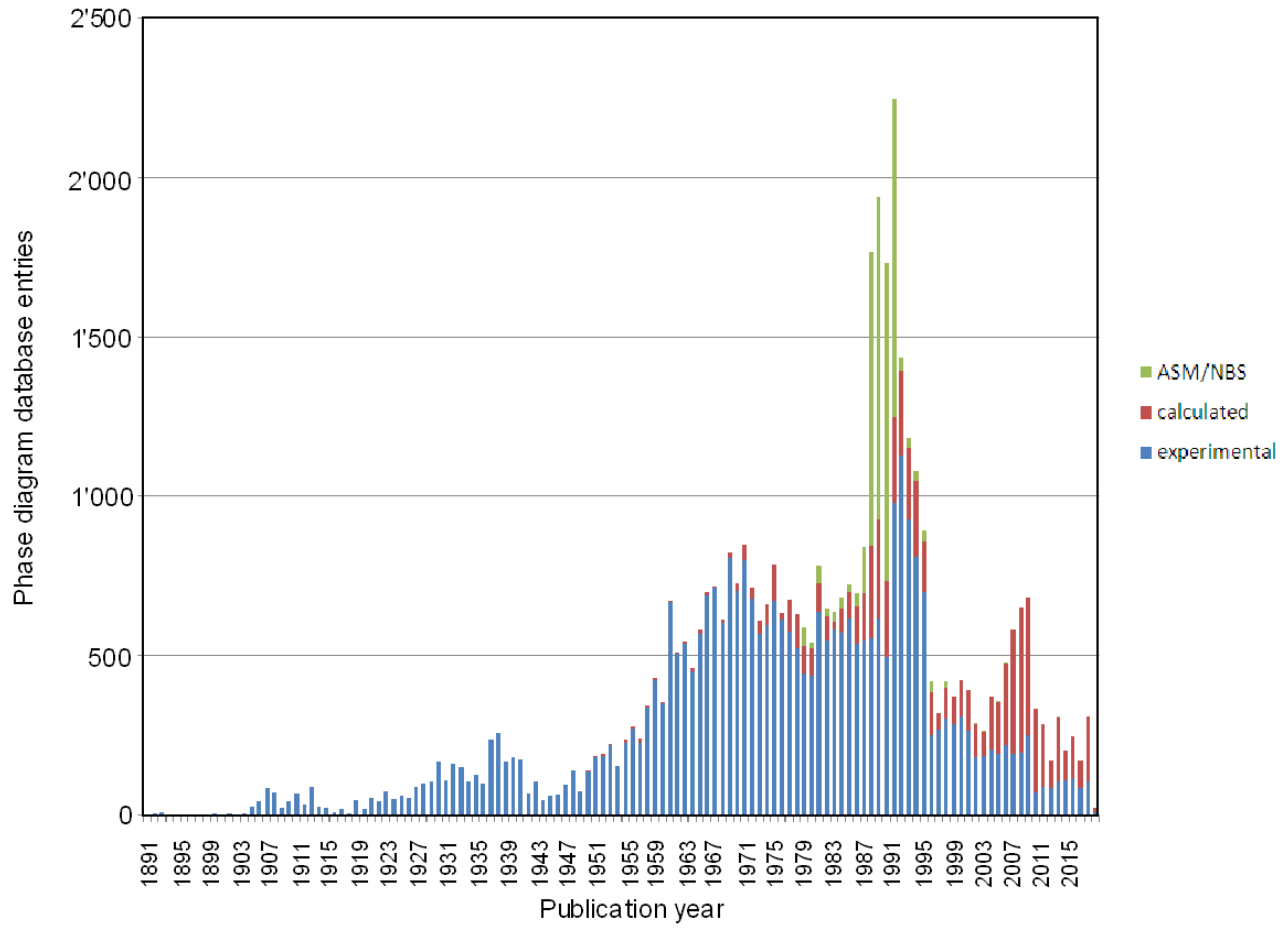

Fig. 29. Distribution of the database entries in the PAULING FILE according to the publication year for phase diagrams. Here the differential is made between evaluated (ASM/NBS), experimental and calculated.

The second overview, shown in Fig. 31, proves that, contradicting common ideas based on earlier works by one of the author (P.V., e.g. **[59]**), the PAULING FILE is not limited to intermetallics. On the contrary, except for the phase diagram fragment, oxides dominate. Tables 12a+b give some numbers from the main product for the crystallographic structure data, Pearson's Crystal Data **[60]**, release 2020/21.The first table shows the distribution according to the number of chemical elements, and the second one the distribution according to the level of structural investigation. These numbers are in Table 13 comparing the PCD (included in the MPDS platform) and the ICSD [61]. One can see:

1) PCD-2020/21 [8] comprises 60% more entries than the ICSD-2020 [61], and therefore covers the world literature more comprehensively. In addition each entry comprises more than twice database areas within one entry.
2) PCD-2020/21 [62] comprises 39,990 structure prototypes, four times more than ICSD-2020 [61] with about 10,000 structure prototypes. In addition ICSD uses a mixture in the structure prototype assignment between the 'old' definition from my (P.V.) first handbook [68] and the 'new' very rigid definition used in the PAULING FILE.

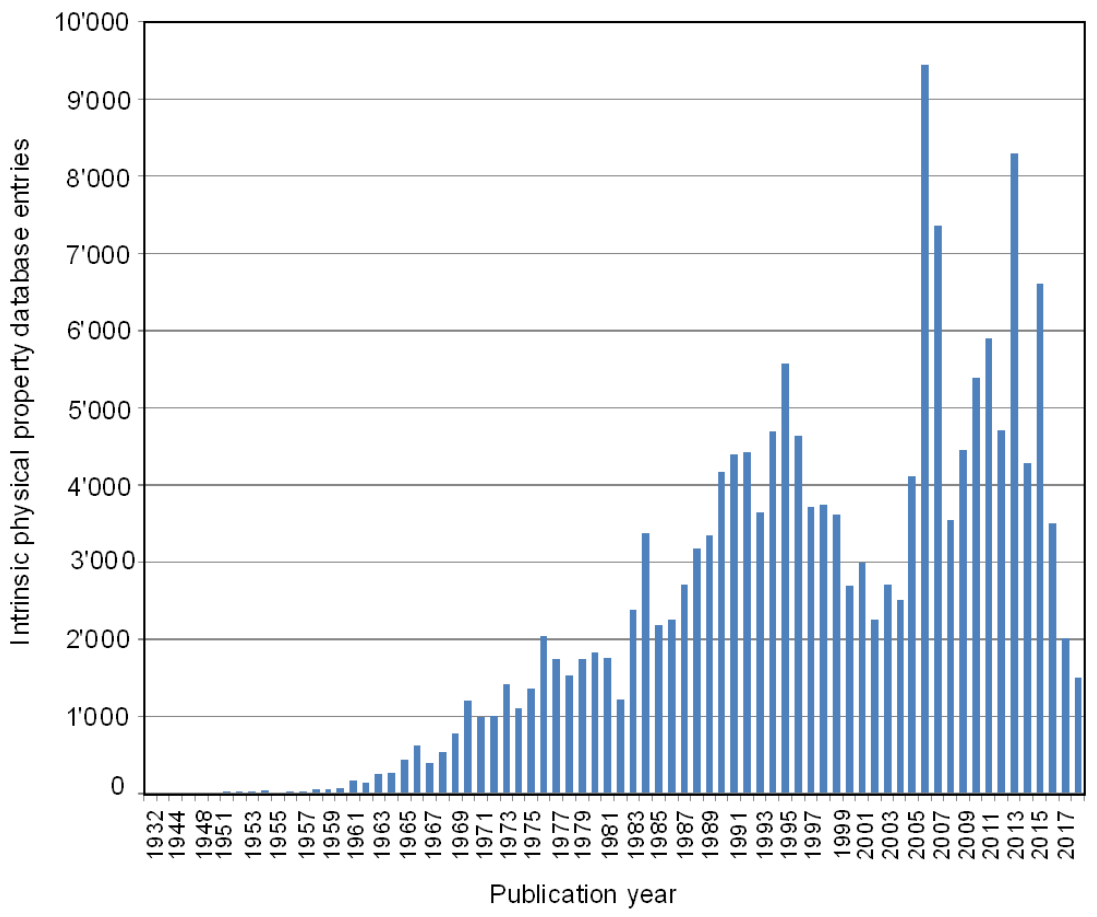

Fig. 30. Distribution of the database entries in the PAULING FILE according to the publication year for physical properties.

| Feature | MPDS | ICSD Web |
|---|---|---|
| Current release | 2021 | 2020.2 |
| Entries | 349,046 | 232,012 |
| Theoretical structures | — | 18,057 |
| Structure types | 39,990 | 9,490 |
| Assigned to structure type | 331,594 | 170,567 |
| Unaries | 3,321 | 2,995 |
| Binaries | 59,356 | 41,832 |
| Ternaries | 132,714 | 78,064 |
| Quaternaries | 93,489 | 53,231 |
| Quinternaries | 41,350 | 30,684 |
| Articles | 309,460 | 87,676 |
| Journals and books | 2,599 | 1,744 |
| Available since | 2017 | 2009 |
| Distinct phase concept (data linked for the same compound) | yes | no |
| Phase diagrams & physical properties linked to structures | yes | no |
| Atomic environment & polyhedra searches | yes | no |
| Data from the old Soviet & Japanese journals (never online) | yes | no |
| Programmatic broadband access (API) for data-mining | yes | no |
| Data export in CIF, PDF, etc. | yes | yes |
| Data export for ab initio simulation inputs | yes | no |

Table 13. Comparison of the MPDS and ICSD Web.

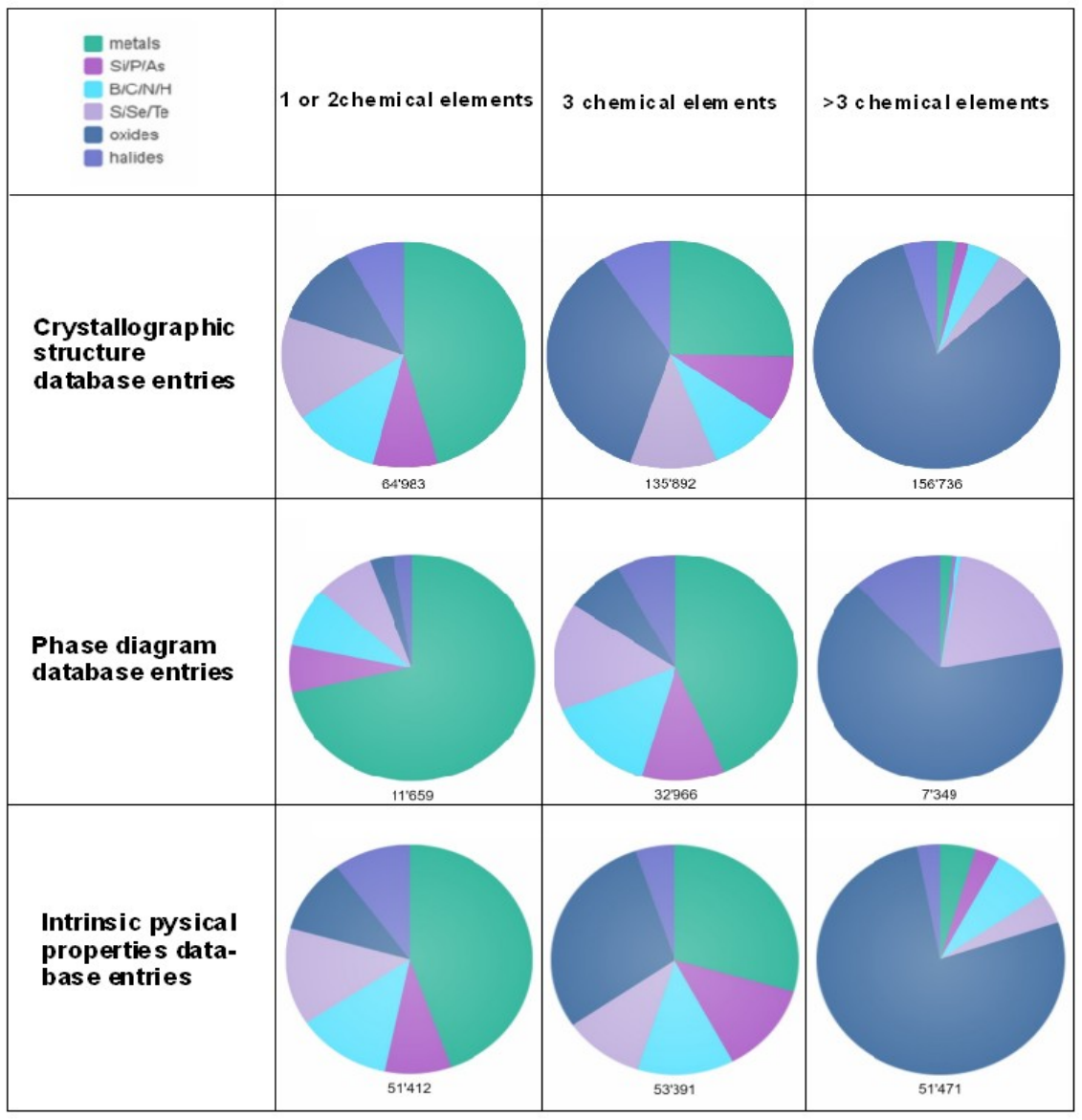

Fig. 31. Distribution of database entries in the PAULING FILE according to the chemical class. The order in the legend corresponds to clockwise order, starting from the top, on the diagrams.

| Number of chemical elements | Distinct chemical systems | Distinct chemical formula | Number of entries | Average entry per chemical entry |
|---|---|---|---|---|
| Unaries | 97 | 441 | 3,323 | 7.5 |
| Binaries | 2,591 | 19,392 | 59,393 | 3.1 |
| Ternaries | 18,982 | 68,544 | 132,800 | 1.9 |
| Quaternaries | 23,970 | 58,904 | 93,561 | 1.6 |
| Quinternaries and higher | 24,137 | 47,899 | 60,404 | 1.3 |
| **Total** | **69,777** | **195,180** | **349,481** | **1.79** |

Table 12a. Numbers of distinct chemical systems, chemical formula, and entries in PCD-2020/21, subdivided into 1, 2, 3 and more than 3 chemical elements, and the total numbers.

| | Number of entries |
|---|---|
| atom coordinates, data set defining structure prototype | 39,990 |
| atom coordinates, structure prototype assigned | 183,735 |
| part of atom coordinates (protons ignored) | 873 |
| atom coordinates and structure prototype assigned (by editors), no atom coordinates published by authors | 102,675 |
| no atom coordinates, parent type assigned for filled-up derivative | 1,508 |
| no atom coordinates, no structure prototype assigned | 20,700 |
| **Total** | **349,481** |

Table 12b. Numbers of entries in PCD-2020/21 according to the level of structural investigation.

## 4.11. Products based on the PAULING FILE

The hundreds of related database fields can be used as LEGO bricks. To create different product requires strictly standardized digital system with the interoperability and modularity. The PAULING FILE data are included as a core component in several on-line, off-line, and printed products: some of them are listed below.

• *ASM Phase Diagram Database (APD), ASM International (on-line)* **[63]**
The Phase Diagram Database offers easy viewing of phase diagram details, crystallographic and reaction data. The content was updated on an annual basis until 2018/19 and brings the database to more than 50,000 on-line phase diagrams for binary and ternary systems.

• *Inorganic Material Database (AtomWork), NIMS (on-line)* **[18]**
The data part of AtomWork is the result of collaboration between Japanese organizations (JST, NIMS) and MPDS. The Inorganic Material Database aims to cover all crystallographic structure, X-ray diffraction, physical properties and phase diagram data of inorganic substances from main literature sources. AtomWork covers mainly binary systems and contains 82,000 crystallographic structures, 55,000 physical properties and 15,000 phase diagrams.

• *Inorganic Material Database (AtomWork-advanced), NIMS (on-line)* **[20]**
AtomWork-adv. was released the first time in 2019 and covers binary, ternary and multinary systems.

• *PAULING FILE – Binaries Edition, ASM International (off-line)* **[8]**
The Binaries Edition of the PAULING FILE, which is limited to binary inorganic substances, was published in 2002. It contains 8,000 phase diagrams covering 2,300 binary systems, 28,300 crystallographic structure data sets for more than 10,000 different *distinct phases*, and around 17,300 physical property entries (with about 43,100 numerical values and 10,000 figure descriptions) for some 5,000 *distinct phases*. To reach this result, 21,000 original publications had been processed. The user-friendly retrieval program offers numerous possibilities, including some data-mining options.

• *Pearson's Crystal Data (PCD), ASM International (off-line)* **[62]**
The 14$^{th}$ release of Pearson's Crystal Data (2020/21) contains 349,481 structural data sets for 143,000 different *distinct phases*. Differently from the similarly named Pearson's Handbook, the electronic product contains data for all classes of inorganic substances (approx. 50% oxides). All data sets with published coordinates, and 80% of the data sets where only cell parameters were published, have been assigned a structure prototype: 224,598 data sets with published atom coordinates, 102,675 data sets with assigned atom coordinates, 22,208 data set with only cell parameters. Atomic Environment Types (AETs) have been defined for the first category. The software offers numerous possibilities to retrieve and process the data. In its turn, it is also integrated in the other popular software, such as MedeA from the Materials Design, Inc. **[76]**.

• *Powder Diffraction File PDF-4$^+$, ICDD (off-line)* **[64]**
Since 1940 ICDD provides tools in the form of experimental and calculated powder patterns for phase analysis based on diffraction methods. PDF-4$^+$ (inorganic substances) and PDF-4 Minerals also include atom coordinates, which can be used to perform Rietveld refinements. Over 80% of the inorganic substances in the current edition of PDF-4$^+$ originate from the PAULING FILE.

• *SpringerMaterials, Springer (on-line)* **[19]**
Based on the well-known series of Landolt-Börnstein Handbooks, SpringerMaterials allows materials scientists to identify materials and their properties by offering access to physical and chemical data in materials science on an on-line platform. The PAULING FILE provides crystallographic structure, phase diagram, and physical properties entries on a yearly basis to the section *Inorganic Solid Phases*.

• *Materials Platform for Data Science MPDS (on-line)* **[17a]** *and ASM International-branded Materials Platform for Data Science MPDS (on-line).* **[17b]**
MPDS is a web platform, presenting on-line all the three parts of the PAULING FILE data. The 2020/21 release contains 73,365 phase diagrams, 489,365 crystallographic structures, and 1,001,516 physical properties entries. About 80% of the data can be requested programmatically in a developer-friendly format, ready for any external data-mining applications. The remaining 20% can be obtained as references to the original publications. Altogether 309,460 scientific publications in materials science, chemistry, physics, *etc.* serve as starting points for the platform data, and this number increases. In addition, 1,102,333 values for the 10 most commonly reported physical properties were derived by machine-learning (random forest regression model) using the PAULING FILE data as an input. And, finally, 82,874 physical properties have been simulated via the *ab initio* Gaussian electron basis sets and hybrid Hartree-Fock DFT approximation, again using the peer-reviewed PAULING FILE data as the starting point. A web-browser (without any plugins) and internet connection allows comfortable work with the scientific data, be it for a literature overview, evaluation of hypotheses, or design of new materials. For the further details see Chapter 10.

• *Materials Landolt-Börnstein handbook series Inorganic Crystal Structures* **[65]** and the *Handbook of Inorganic Substances* **[43]**, also contain PAULING FILE crystallographic structure data. The former describes structure prototypes in space groups 123-230, whereas the most recent edition of the latter lists crystallographic data for 157,000 inorganic substances (*distinct phases*). The electronic book *Inorganic Substances Bibliography* **[66]**, lists publications selected for processing in the PAULING FILE, ordered according to the chemical systems considered in the papers.

## 5. Holistic views: Crystallographic systematization of inorganic substances for 'inorganic substance former systems'

*Holistic view stands for 'The whole is greater than the sum of its parts'.*

A remarkable example is given in Fig. 32, which shows a holistic view of all Fe-containing binary systems in Meyer periodic system representation. One can e.g. easily discover that the Fe-d$^{7-9}$ element systems behave completely different from the Fe-RE systems in respect of inorganic substance formation (formers vs. non-formers).

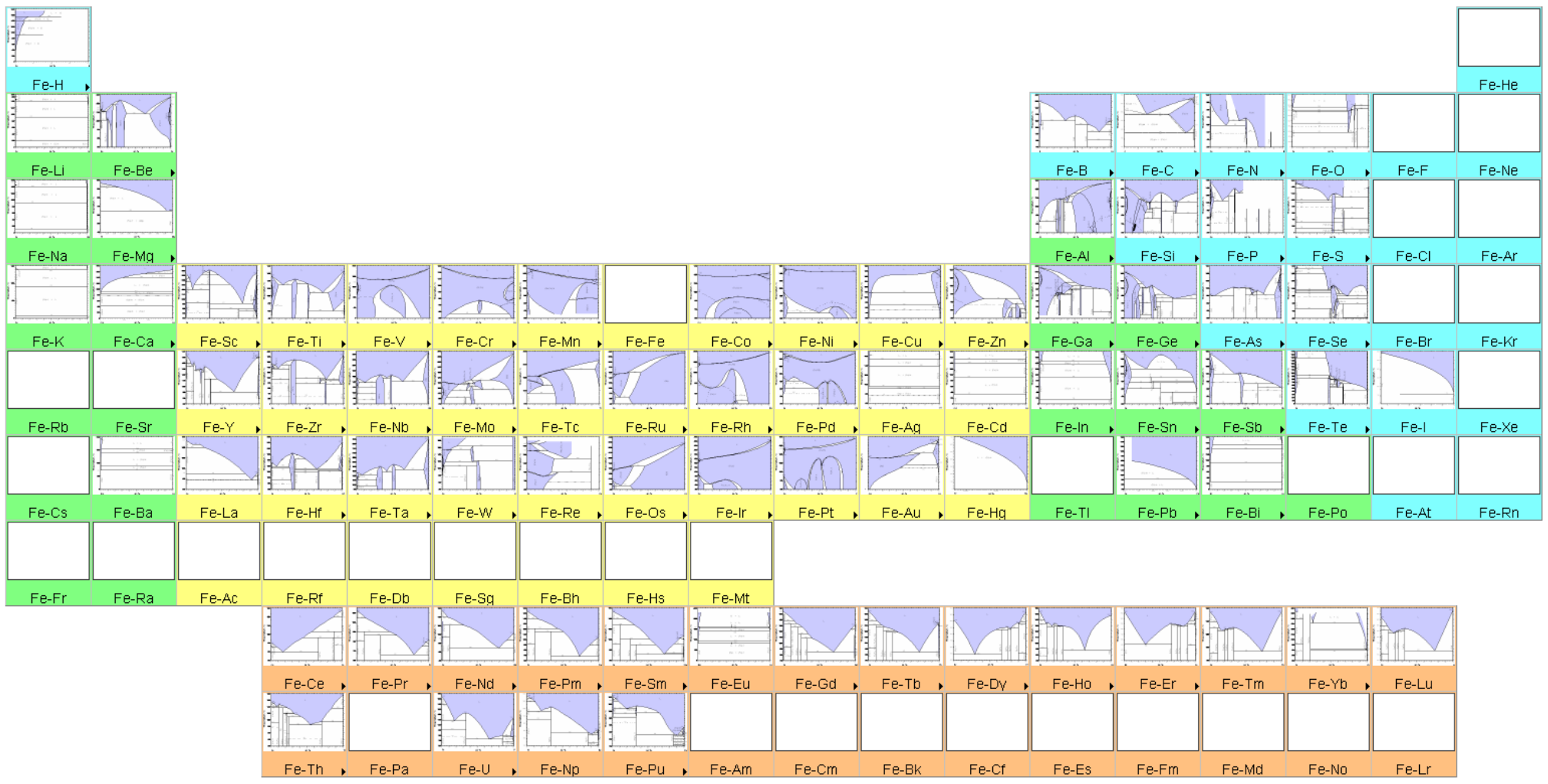

Fig. 32. Example of the constitution browser in the PAULING FILE – Binaries Edition **[8]**, **s**howing phase diagrams of binary systems containing Fe, giving a simple 'holistic view'.

There exist three crystallographic systematization schemes based on the structure stability maps:

1. Structure stability maps using the structure prototype classification
2. Structure stability maps using the coordination type classification
3. Generalized 'coordination number ranges' respectively 'AET' stability maps using the AET classification (which is independent of its stoichiometry and its number of involved chemical elements).

### 5.1. Structure stability maps using the structure prototype classification

The major aim of structure stability maps using the structure prototype classification is to correlate them with APP or APPE of its constituent chemical elements. If groupings can be achieved within such structure stability maps, its crystal chemistry can be better understood, and in the best case predictions can be made for chemical systems, where no or limited information is experimentally known.

Table 14 (external file) lists in summary for the binary AB inorganic substances the published structure stability maps together with some significant numbers: class of inorganic substances, map coordinates used, structure prototypes considered (number of inorganic substances included), total number of inorganic substances included, and number of violations. Most of the above-mentioned structure stability maps were also done for $AB_2$, $AB_3$ $A_2B_3$, and $A_3B_5$ inorganic substances; here are the most outstanding references **[44-58]**. Analyzing APP of all coordinates listed in Table 14 that are successful in separating experimentally determined structure prototypes of binary AB inorganic substances into separate domains (including also the coordinates used for ternary substances) shows that APP of four atomic property parameter pattern groups are involved:

- Atomic number pattern group (main quantum number QN),
- Periodic number pattern group (number of electron vacancies per atom, number of valence s+d electrons, and Pettifor's Mendeleyev number),
- Atomic size pattern group (covalent radii $R_c$, ionic radii $R_i$, metallic radii $R_m$, and pseudo-potential radii $R_{s+p}$),
- Atomic reactivity pattern group (electronegativities EN, Herman-Skillman term values, and s-p parameters).

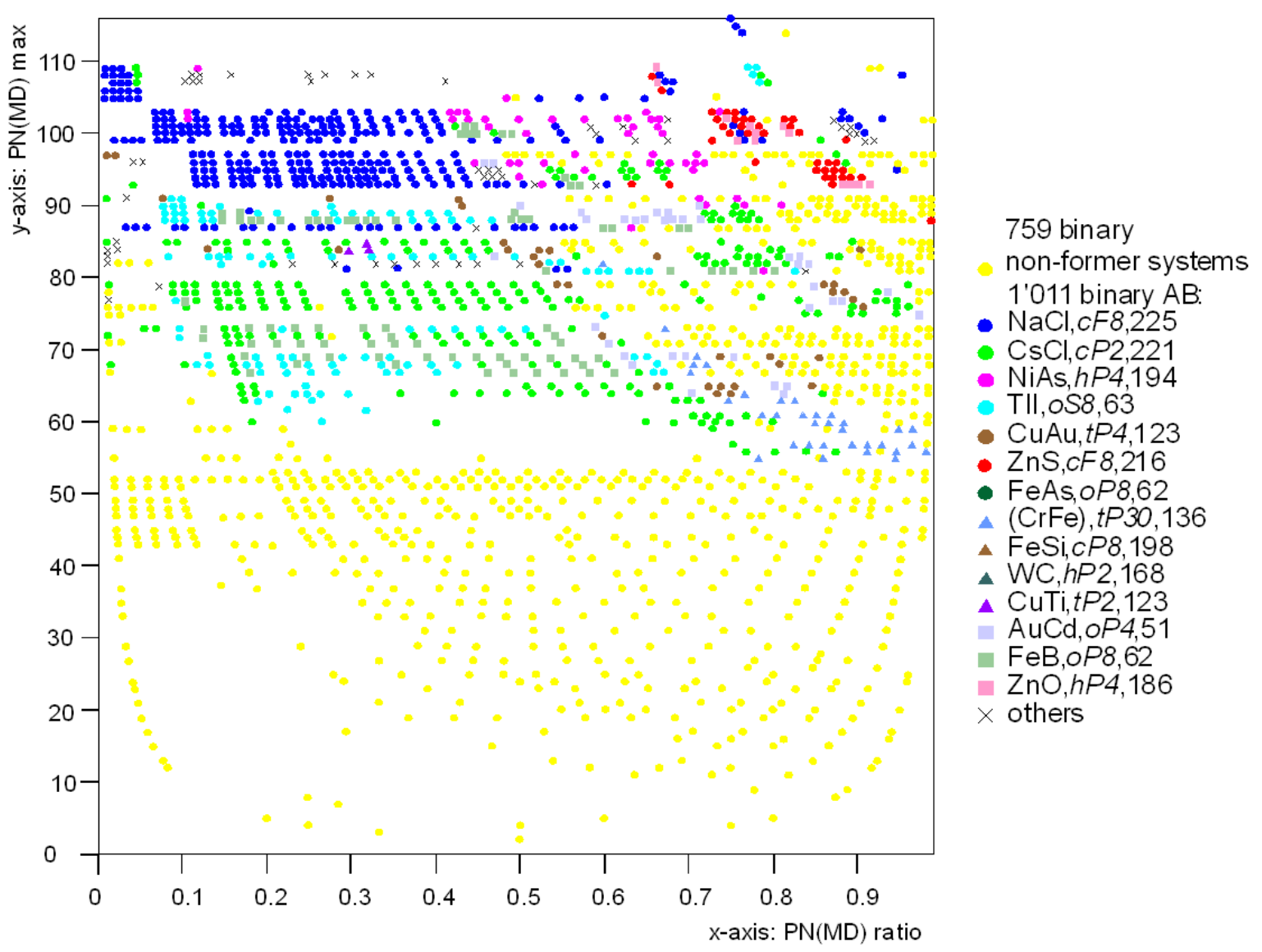

Fig. 33. Structure prototype map showing $[(PN_{MD})_A,(PN_{MD})_B]_{max}$ (y-axis) versus $(PN_{MD})_A/(PN_{MD})_B$ (x-axis) for AB inorganic substances focusing on the 24 most populous 1:1 structure prototypes.

Using the MPDS, release 2020/21 **[17a]**, we created two 2 binary structure stability maps focusing on the 15(24) most populous structure prototypes for AB inorganic substances. Fig. 33 using as APPE $[(PN_{MD})_A, (PN_{MD})_B]_{max}$ (y-axis) vs. $(PN_{MD})_A/(PN_{MD})_B$ (x-axis). The separation between the different structure prototype domains are even simpler in Fig. 34 using as axis only APP: $(PN_{MD})_A$ (y-axis) vs. $(PN_{MD})_B$ (x-axis). Fig.35 shows its analogous to Fig. 33 for the 16(25) most populous $AB_2$ structure prototypes using as APPE *concentration-weighted* $[(PN_{MD})_A,(PN_{MD})_B]_{max}$ (y-axis) versus $(PN_{MD})_A/(PN_{MD})_B$ (x-axis). The concentration-weighting has the advantage that one can include $AB_2$ and $A_2B$ in the same structure map (analog to its AB structure map). Fig. 36 shows its analogous to Fig. 34 for $AB_2$ substances, in this case also using the simpler APP: $(PN_{MD})_A$ vs. $(PN_{MD})_B$. Even the overall separation is complex it is far from a statistical distribution, and clear structure prototype stability domains can be seen. As a consequence of this one can predict the potential structure prototype (or a limited number of competing structure prototypes) for yet not investigated systems. Analog behavior is also seen for all other binary stoichiometric ratios. An important fact is that by using $PN_{MD}$ or $PN_{ME}$ as APP or APPE within the same maps the inorganic substance non-formers area is clearly separated from the inorganic substance former area, and within the inorganic substance former area the different structure prototype domains are nicely grouped. This is unique and was never considered in published structure stability maps, representing a big improvement, as it highlights the 'parameterization power' of the APP: $PN_{MD}$ *or* $PN_{ME}$ touching upon inverse problem solving on models for extrapolation rather than simple fittings of parameters for interpolation.

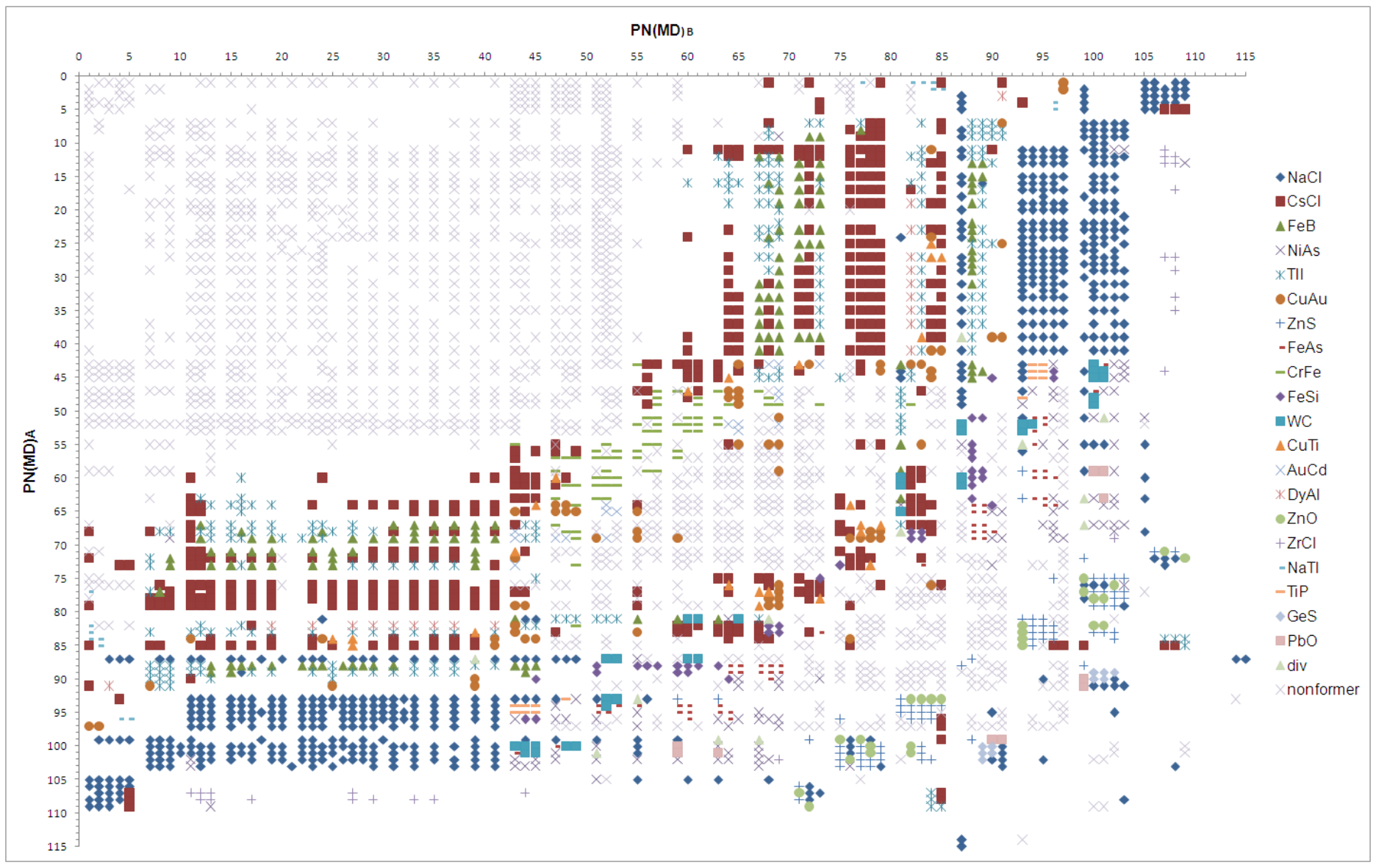

Fig. 34. Structure prototype map showing $(PN_{MD})_A$ (y-axis) versus $(PN_{MD})_B$ (x-axis) for AB inorganic substances focusing on the 24 most populous 1:1 structure prototypes.

The extension from binary to ternary (multinary) systems leads to two serious additional issues. First, while 13,750 distinct different inorganic substances (about 64% of those possible on the chemical system level) are known for the binary case, we are confronted with 53,619 distinct different inorganic substances for ternaries. Although the number 53,619 inorganic substances looks overwhelming, in fact it is only a small fraction of the still not yet investigated inorganic substances. If we consider 100 chemical elements we can define 161,700 ternary chemical systems, and assuming the existence of 4 inorganic substances per one ternary chemical system (see Fig. 47), we end up with about 646,800 ternary inorganic substances (about 8 % on the inorganic substances level).

Second, in the binary case we could restrict ourselves to fewer than 100 structure prototypes and still include 70% of all binary inorganic substances, but inclusion of ternary inorganic substances requires that about 1,000 of the most populous structure prototypes to be included to treat 70% of all ternary inorganic substances. The past has shown us that it is impossible to reduce this number significantly by taking only ternary basic structure prototypes, as illustrated by the following example. In the 'old' Pearson's Handbooks **[67]** we find the tetragonal $BaAl_4$,*tI10*,139 structure prototype with 14 exclusively binary representatives. Its successor **[68]**, 20 years later, lists 17 binary $BaAl_4$,*tI10*,139 and 450 ternary $CeAl_2Ga_2$,*tI10*,139, mainly 1:2:2 representatives, e.g. $BaMn_2Sb_2$, $NdRh_2Si_2$, and $LaNi_2Sn_2$. Assuming one would need five different structure stability maps (e.g. each treating one class of iso-stoichiometric 1:1:1, 1:1:2, 1:1:3, 1:2:2, and 1:2:3 inorganic substances), then each structure stability map would have to be able to separate 200 different structure prototype domains within the limited two-dimensional (three-dimensional) APP or APPE area (respectively space) available. It is very unlikely that it would be possible to separate correctly those inorganic substances, based on so few experimentally determined data, into 200 separately different structure prototype domains bounded by demarcation lines (surfaces). According to the experience of one author (P.V.) with binaries, at least 50% of all ternary information should be experimentally known to be able to draw such 'finely tuned' complex demarcation lines (surfaces) that the predictions would be trustworthy.

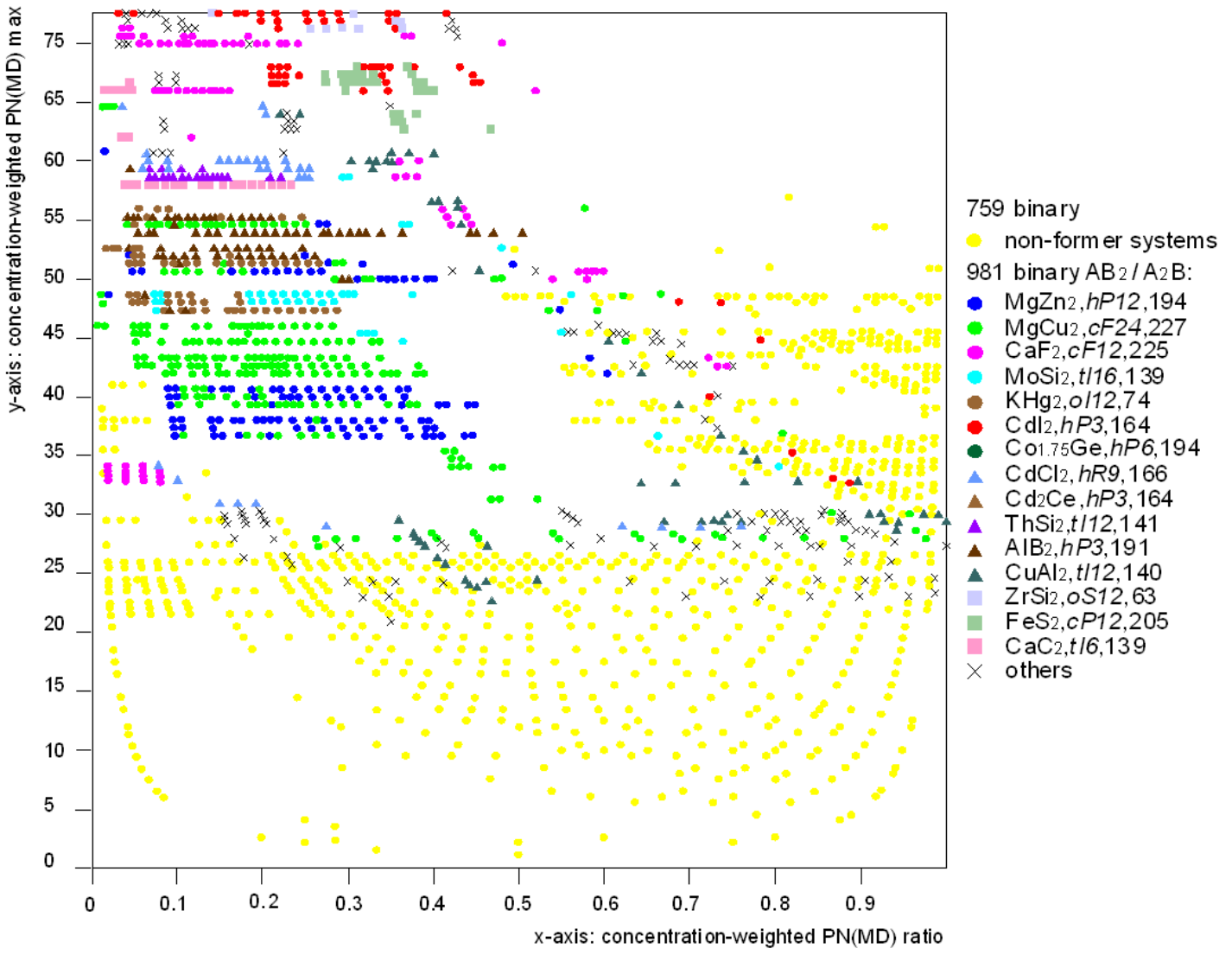

Fig. 35. Structure prototype map showing $[(PN_{MD})_A,(PN_{MD})_B]_{max}$ (y-axis) versus $(PN_{MD})_A/(PN_{MD})_B$ (x-axis) for $AB_2$ inorganic substances focusing on the 16 most populous 1:2 structure prototypes.

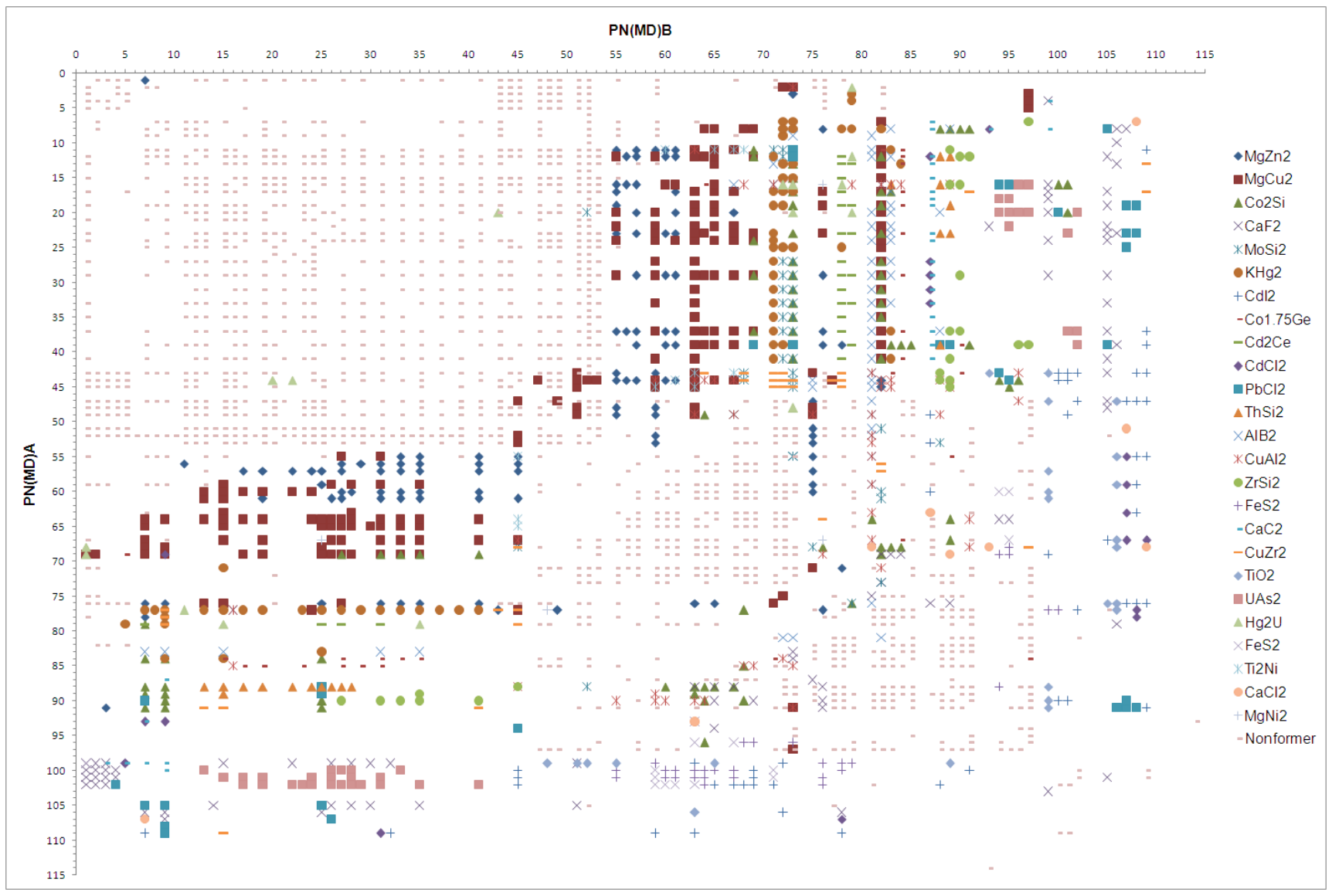

Fig. 36. Structure prototype map showing $(PN_{MD})_A$ (y-axis) versus $(PN_{MD})_B$(x-axis) for $AB_2$ inorganic substances focusing on the 25 most populous 1:2 structure prototypes.

Even with the above mentioned two principal restraints, which cannot be surrounded without having substantial more data (multinary system), one can get surprisingly nice separations between iso-stoichiometric ternary or quaternary inorganic substances and the concept of using 2D- and 3D- APP and respectively APPE features sets (of course always focusing on the most populous iso-stoichiometric ratios as well as its most populous structure prototype). See further details and examples under 7.3.4.2.

### 5.2. Structure stability maps using the coordination type classification

The AET is not a simple inorganic substance property in the sense that different AETs are generally observed in the same inorganic substance (details see under 4.3.). Within the PAULING FILE data **[17a,b]** the number of different AETs realized within a specific inorganic substance is as follows:

| | |
|---|---|
| *Single-environment type* | 6 % |
| *Two-environments type* | 29 % |
| *Three-environments type* | 21 % |
| *Four-environments type* | 11% |
| *Multi-environments type* | 33% |

In a *single-environment type* inorganic substance all atoms present in that specific structure prototype have the same AET, in a *two-environments type* inorganic substance the atoms adopt two different AETs, and so on. It is also worth mentioning that structure prototypes belonging to *single-environment type* are not limited to the simple chemical elements. They are also found in binary, ternary, and multinary inorganic substances. On the other hand, *two-environments type* inorganic substances occur in unaries (chemical element structure prototypes), as well as in binaries, ternaries, and multinaries. The same is valid for the *three-* and *higher-environments* types. In other words this means that in most inorganic substances the same chemical element realizes one or two different AETs. Experience shows, however, that the number of different AETs realized by one chemical element within the same inorganic substance with a specific structure prototype is generally low. The AET, or coordination polyhedra, of the atoms, i.e. the number of nearest neighbors and the geometrical arrangement formed by these, is an important characteristic of each structure prototype. A classification scheme of crystallographic structures based on the coordination of the smallest atoms in the structure, where the CN range from 2-12, was proposed by Kripyakevich **[35]**.

#### *5.2.1. Observed AETs within the coordination type classification*

The AET of each crystallographic point-set within all binary AB inorganic substances were examined. In total the AETs of the two elements in 1,126 binary inorganic substances AB for dataset 1 (using basic definition of non-formers) and 1,384 for dataset 2 (including simple chemical elements and its solid solutions) were investigated. The results show that 31 different AETs are realized in AB inorganic substances, but coordination polyhedra such as tetrahedron, trigonal prism, octahedron, cube, tricapped trigonal prism, cuboctahedron, rhombic dodecahedron, heptacapped pentagonal prism, largely predominate. The 18 most populous AETs are shown in Fig. 37 and their occurrences are given in Table 15b. Usage of the MPDS platform shows, that 20 years later the numbers in Tables 15a+b did almost not change, therefore we did not repeat the work with the 2020/2021data-sets. This is completely consistent with the general research tendencies; binaries were mainly investigated up to 1980, from then on scientists focused mainly on ternaries, and from about 2000 moving more and more towards multinaries.

| Chemical systems | Dataset 1 | Dataset 2 |
|---|---|---|
| not yet investigated | 1,480 (34.5 %) | 1,481 (33.9 %) |
| no compounds formed (non-former) | 760 (17.8 %) | 540 (12.3 %) |
| no AB compounds formed (non-AB former) | 839 (19.6 %) | 794 (18.2 %) |
| AB compound reported but no crystallographic structure data | 73 ( 1.8 %) | 79 ( 1.8 %) |
| AB compound with structure prototype assignment (AB former) | 1,126 (26.3 %) | 1,384 (31.7 %) |
| chemical elements | - | 93 (2.1 %) |
| total (considering 93 chemical elements) | 4,278 (100 %) | 4,371 (100 %) |

Table 15a. AB former and non-former statistic for dataset 1 (basic definition of non-formers) and dataset 2 (including simple chemical elements and solid solutions).

| AET | CN | Dataset 1 | Dataset 2 |
|---|---|---|---|
| single atom | 1 | 27 | 36 |
| collinear | 2 | 9 | 11 |
| non-collinear | 3 | 20 | 25 |
| (non) coplanar triangle | 3 | 61 | 70 |
| tetrahedron | 4 | 103 | 108 |
| (non coplanar square | 4 | 12 | 12 |
| square pyramid | 5 | 8 | 10 |
| coctahedron | 6 | 628 | 637 |
| trigonal prism | 6 | 120 | 121 |
| square prism (cube) | 8 | 226 | 227 |
| tricapped trigonal prism | 9 | 194 | 194 |
| icosahedron | 12 | 16 | 18 |
| (anti)cuboctahedron | 12 | 104 | 525 |
| rhombic dodecahedron | 14 | 326 | 445 |
| heptacapped pentagonal prism | 17 | 168 | 169 |
| Others | | 268 | 225 |

Table 15b. Particular AETs observed in AB compounds, their CNs and occurrences for dataset 1 (basic definition of non-formers) and dataset 2 (including simple chemical elements and solid solutions).

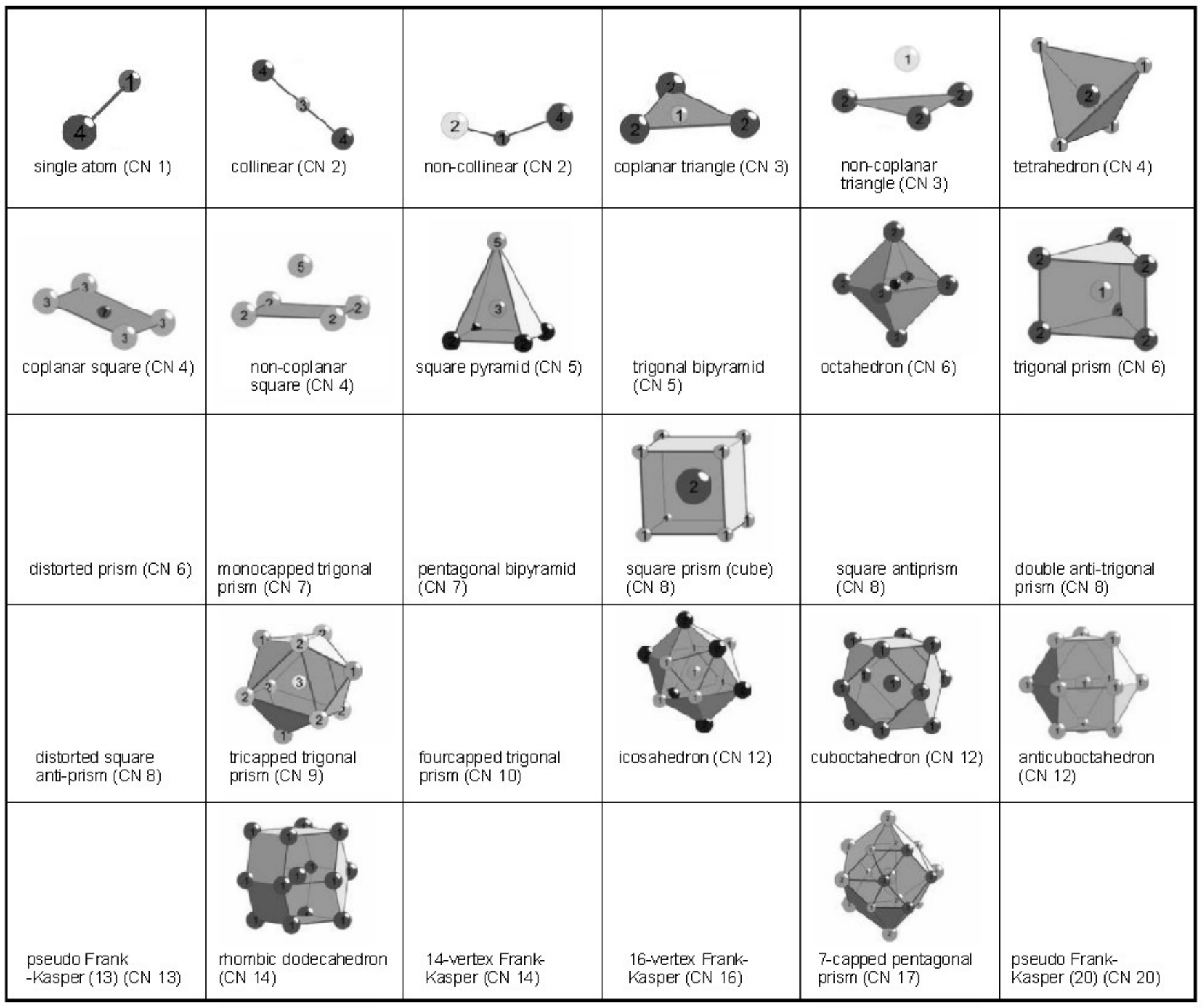

Fig. 37. The 18 most populous AET's for AB inorganic substances.

### *5.2.2. AET - $(PN_{MD})_{A(central\ atom)}$ versus $(PN_{MD})_{B(coordinating\ atoms)}$ stability maps (matrices)*

Fig. 38 a shows the distribution of the AETs observed for binary AB inorganic substances in dataset 1, including information about non-AB (1:1) former (comprises binary inorganic substances, but not at the composition AB) and non-former systems. The chemical element occupying the center of the polyhedron, hereafter referred to as A, is given on the *y*-axis and the other element in the binary inorganic substance, B, on the *x*-axis. It should, however, be noted that both A and B atoms can be fragment of the AET. On both axes the chemical elements are ordered according to the periodic number $PN_{MD}$ (using Mendeleyev periodic system). Each chemical system is thus present twice in the map, the AETs of the chemical element with the highest periodic number $PN_{MD}$ being indicated in the areas situated below the diagonal corresponding to the simple chemical elements. The diagonal is left empty for dataset 1. It should be noted that an AET map is different from the better-known structure prototype maps, where each AB inorganic substance is characterized by its structure prototype. The number of homonuclear bonds per central atom is indicated by a point pattern in Fig. 38a. The density of the pattern increases when going from 1 (e.g. KO) to three homonuclear bonds (e.g. NaSi) per atom. Some of the information included in the AET map $(PN_{MD})_{A(central\ atom)}$ versus $(PN_{MD})_{B(coordinating\ atoms)}$ in Fig. 38a is by definition symmetric with respect to the diagonal. This is the case for the codes for non-formers, non-AB formers and AB formers with no structure data. The AET of the chemical elements A and B do not have to be identical, however, the overall picture presents a strong symmetry with respect to the diagonal, due to the high occurrence of simple structure prototypes like CuAu,*tP2*,123, AuCd,*oP4*,51, NaCl,*cF8*,225, ZnS,*cF8*,216 sphalerite, CsCl,*cP2*,221 (see restrictions above) where the two atom sites are interchangeable. This situation is common for the equi-atomic composition, where, among the about 50 known AB structure prototypes, most are single- environment types, meaning that the AETs of both elements are the same. A relatively good separation of the different chemical systems into separate AET and non-former domains is achieved using this simple classification. No attempt was made to optimize the periodic number $PN_{MD}$ (using the Mendeleyev periodic system) by modifying the order of the rows and columns in the table. It is, however, worth noting that when the periodic number using the periodic system from Mendeleyev ($PN_{MD}$) is replaced by an APP from one of the other four APP pattern groups (e.g. atomic number, atomic size, atomic reactivity, and atomic affinity) the resulting AET maps show a lower degree of local ordering. Well-defined regions of identical colors are observed also for dataset 2 where solid solutions and the chemical elements are included (see Fig. 38b). In particular, the categories of (anti-)cuboctahedra (close packed structures) and rhombic dodecahedra (W,*cI2*,229 / CsCl,*cP2*,221) are substantially increased. Fig. 38c has been generated using the MPDS.

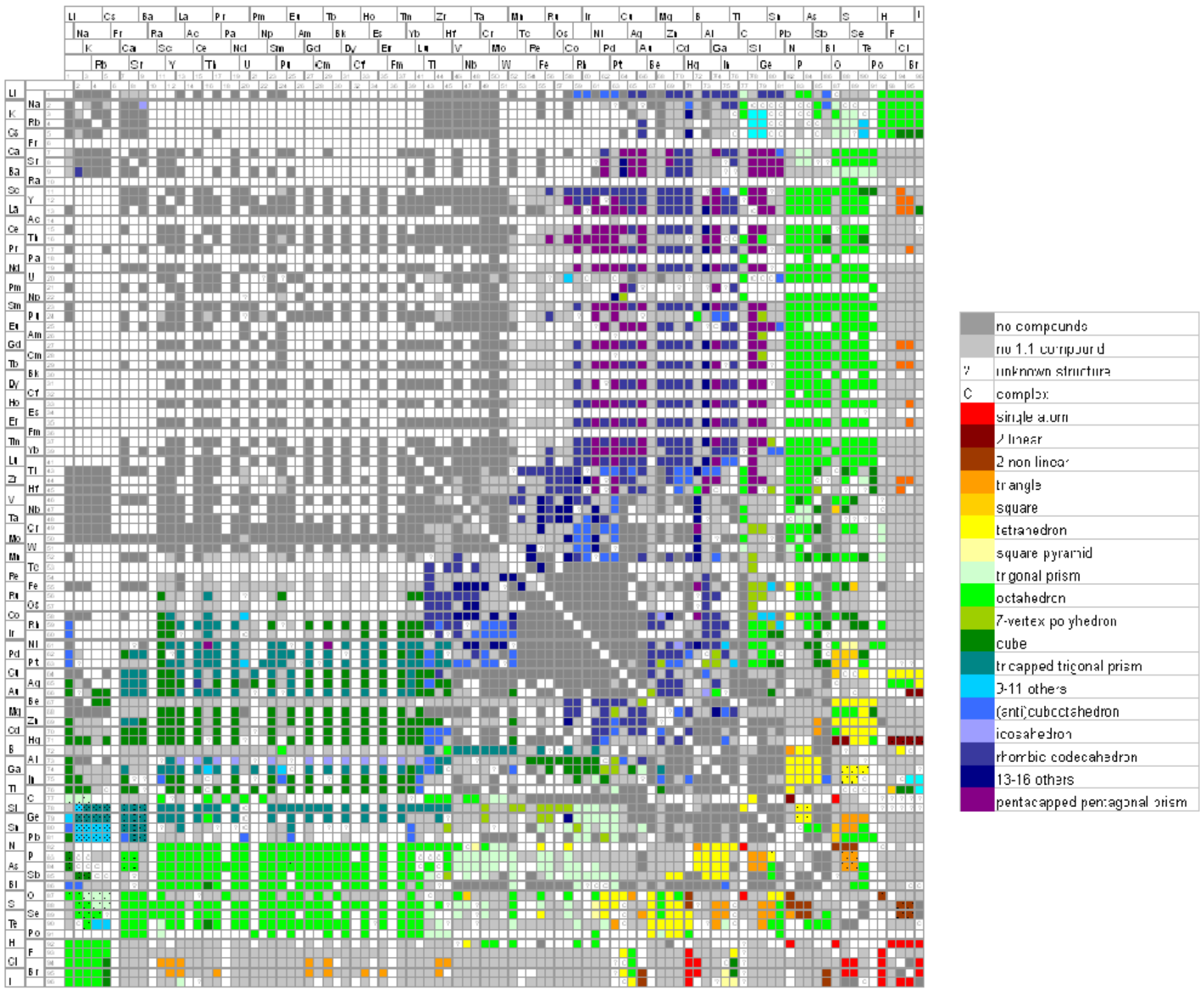

Fig. 38a. The atomic environment type (AET) map showing $PN_{MD(central\ atom)}$ (y-axis) versus $PN_{MD(coordinating\ atoms)}$ (x-axis) for AB inorganic substances dataset 1 (AET's being limited to thermodynamically defined AB former systems). Element A is defined as the central atomic environment type, so that AET's centered by the same element in different inorganic substances are found along the same row. Dotted patterns indicate the existence of covalent homonuclear bonds.

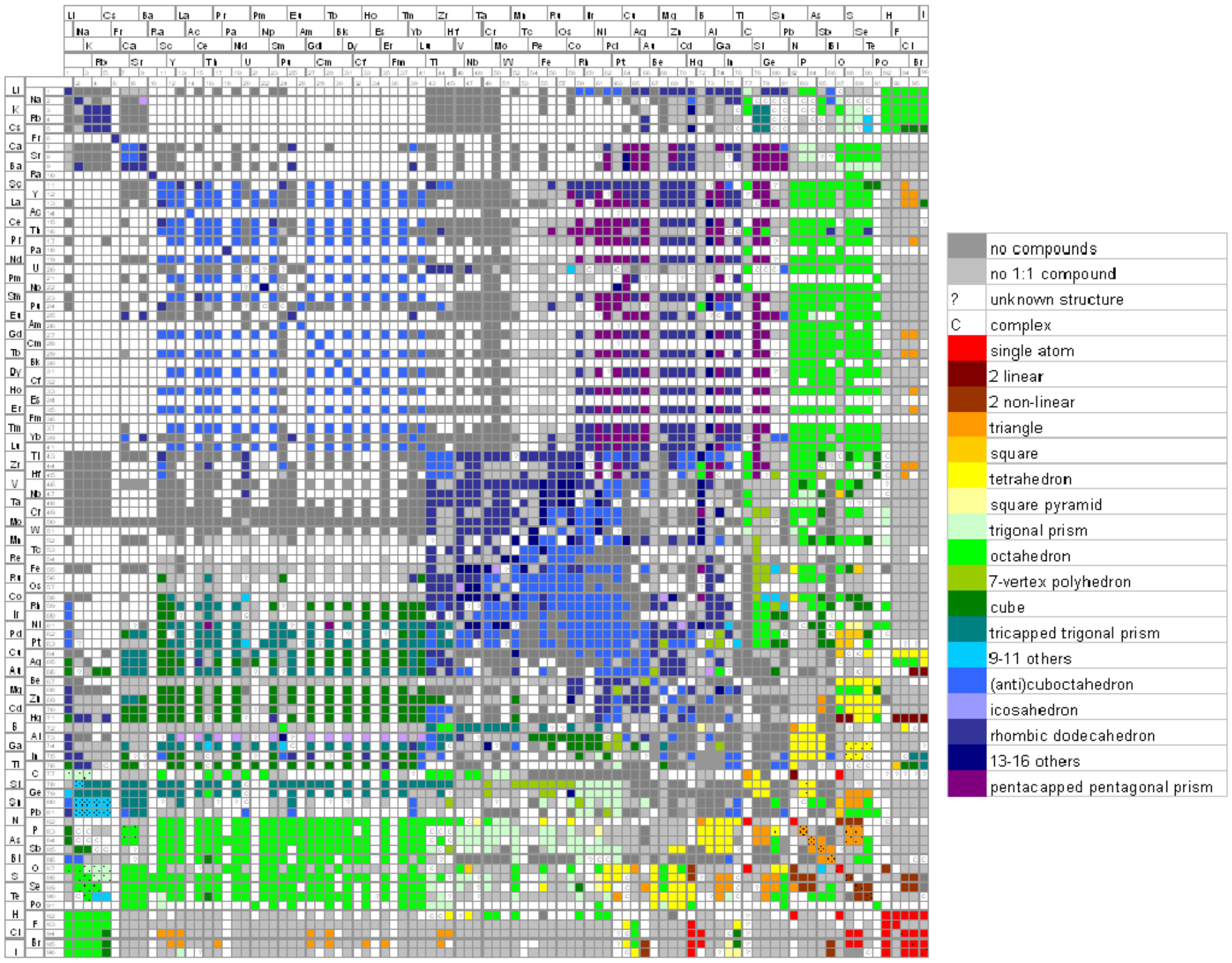

Fig. 38b. Atomic environment type (AET) map: $PN_{MD(coordinating\ atoms)}$ ($x$) vs. $PN_{MD(central\ atom)}$ ($y$) or AB inorganic substances dataset 2 (AET's with chemical elements and extended terminal solid solutions). Element A is defined as the central atom, AET's centered by the same element in different inorganic substances are found along the same row. Dotted patterns show the existence of the covalent homonuclear bonds.

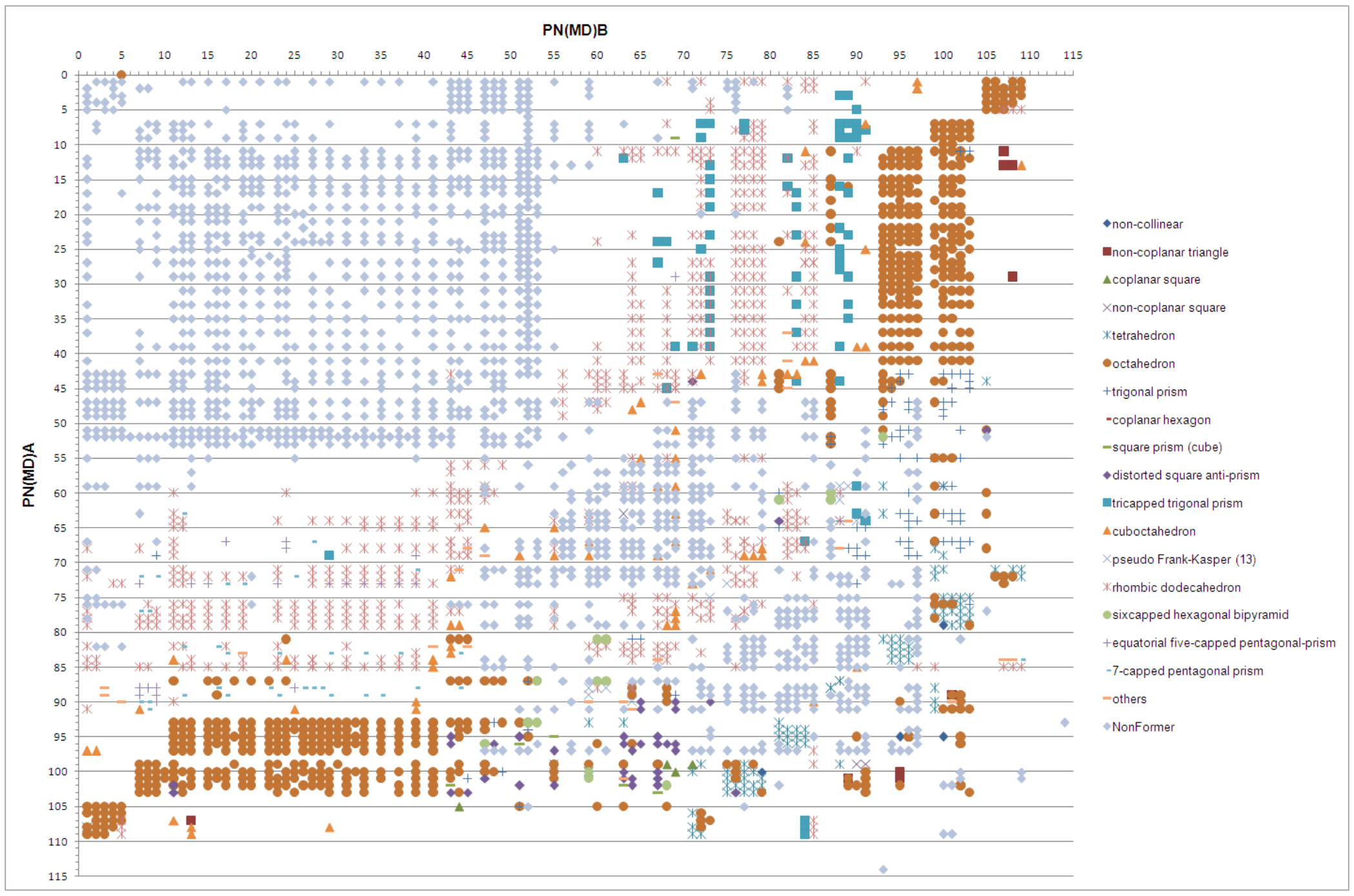

Fig. 38c. Using the MPDS, an analogue of Figure 38a has been generated automatically.

### *5.2.3. AET - $[(PN_{MD})_A,(PN_{MD})_B]_{max}$ versus $(PN_{MD})_{max}/(PN_{MD})_{min}$ stability maps*

The pattern observed in the AET - $(PN_{MD})_{A(central\ atom)}$ versus $(PN_{MD})_{B(coordinating\ atom)}$ stability map encouraged us to search for more strongly defined correlation by combining APP. Our aim was to try to define stability conditions for each AET, e.g. identify the region of binary inorganic substance where the chemical element Zr forms equi-atomic inorganic substances where each Zr atom is surrounded by six atoms forming an octahedron. If this is achieved, it is possible to quantify the conditions for the formation of an AET. As the next step it is easy to calculate the appropriate parameters for a non-investigated system and see if the conditions for the formation of an AET are fulfilled. In order to find such correlation we combined APPs of its constituent elements ($APP_A$ and $APP_B$) by applying simple mathematical operators (+, -, *, /, min, max). The following APPEs were used, where the APPs can be the atomic number, group number, size factors etc.:

Sum = $APP_A+APP_B$ (7)
Difference = $|APP_A- APP_B|$ (8)
Product = $APP_A*APP_B$ (9)
Ratio = $APP_A/APP_B$, with $APP_A<APP_B$ (10)
Maximum = $[APP_A,APP_B]_{max}$ (11)
Minimum = $[APP_A,APP_B]_{min}$ (12)

As noted above, AET represents the special case since for each inorganic substance we have defined not one but two inorganic substance properties (the AET of each of the chemical elements A, respectively B). For that reason each dataset was thus divided into two sub-sets:

i) central atom of the AET being the chemical element with highest periodic number $PN_{MD}$ in the inorganic substance, the $AET(PN_{A(central\ atom)\ MD})_{max}$,
ii) central atom of the AET being of the chemical element with lowest periodic number $PN_{MD}$, $AET(PN_{A(central\ atoms)\ MD})_{min}$.

All possible 2D- and 3D- feature sets, e.g. sum = $APP_A + APP_B$ versus product = $APP_A - APP_B$, were investigated for all APPs and the correlation evaluated by the closest neighbor (domain) method. The best separation into domains was achieved for the AET stability map $(PN_{MD})_{max}$ versus $(PN_{MD})_{min}/(PN_{MD})_{max}$. The two sub-sets (i) and ii)) of dataset 1 are presented together in Fig.39a with a pseudo-mirror plane at $(PN_{MD})_{min} = (PN_{MD})_{max}$, i.e. the vertical line corresponding to the chemical elements. On the left side $(PN_{MD})_{max}$ versus $(PN_{MD})_{min}/(PN_{MD})_{max}$ where the central atom is the chemical element with $(PN_{MD})_{max}$, and on the right side $(PN_{MD})_{max}$ versus $(PN_{MD})_{min}/(PN_{MD})_{max}$ where the central atom is the chemical element with $(PN_{MD})_{min}$.

The separation of the different categories of chemical systems into separate stability domains is very good and several observations can be made. First of all, the non-former systems are clearly separated from the AB former systems. With increasing values of the periodic number $PN_{MD}$ one goes from non-formers to AB formers, and within the AB formers the coordination numbers decrease from 15 to 1. The ratio $(PN_{MD})_{min}/(PN_{MD})_{max}$ ($x$-axis) is, however, necessary to achieve a good separation. Inorganic substances formed by elements with similar periodic numbers $PN_{MD}$ are situated close to the vertical line at $x = 1$.

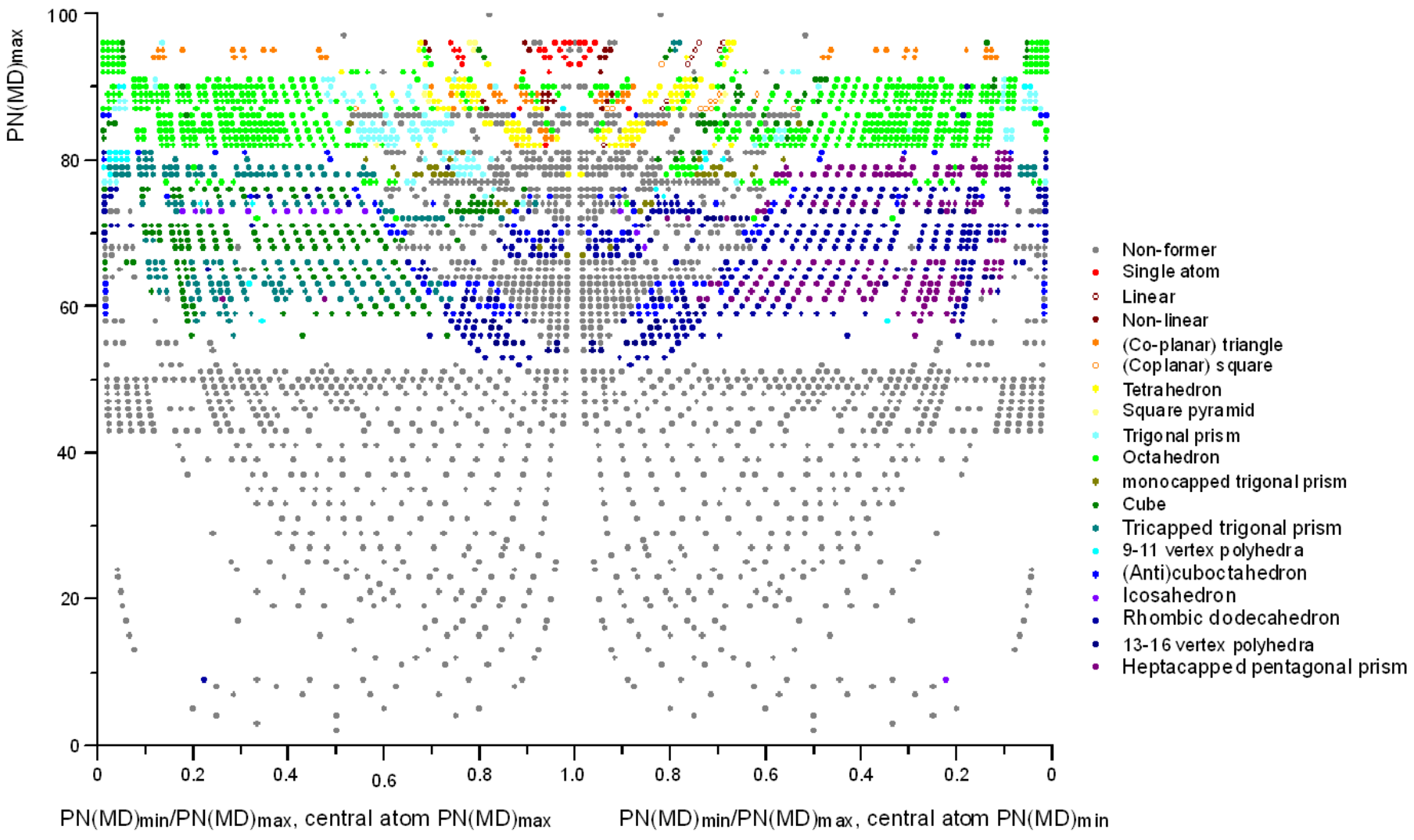

Fig. 39a. Atomic environment type (AET) stability maps showing periodic number $(PN_{MD})_{max}$ vs. $(PN_{MD})_{min}/(PN_{MD})_{max}$ for AB inorganic substances for the dataset (AETs) being limited to thermodynamically defined AB former systems (dataset 1). AETs of the elements with the highest Mendeleyev number in the AB inorganic substances are given on the left-hand side of $(PN_{MD})_{min}/(PN_{MD})_{max} = 1$, AETs of the elements with the lowest Mendeleyev number on the right hand-side.

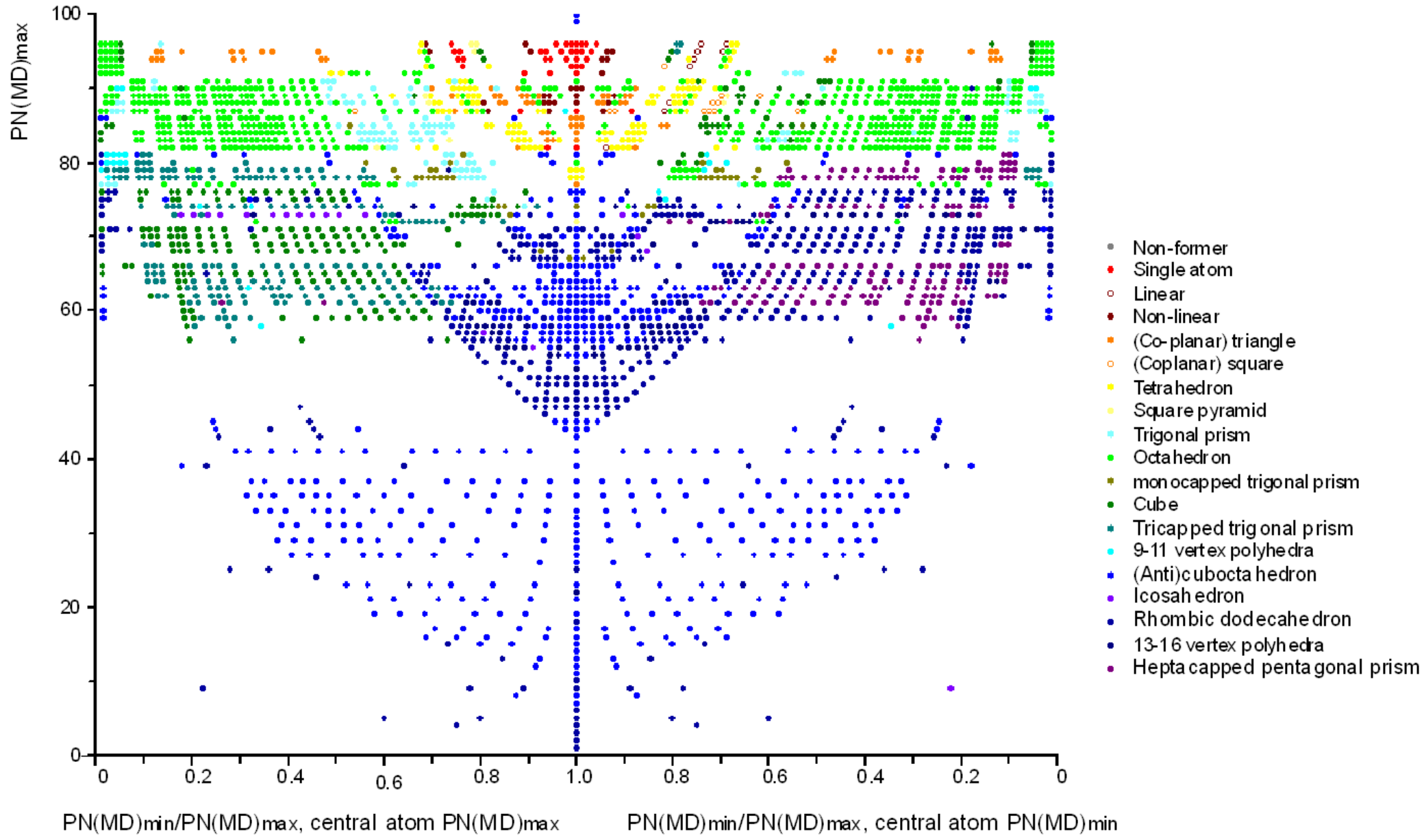

Fig. 39b. Atomic environment type (AET) stability maps showing periodic number $(PN_{MD})_{max}$ vs. $(PN_{MD})_{min}/(PN_{MD})_{max}$ for AB inorganic substances for the dataset 2 (AETs including simple chemical elements and extended terminal solid solutions).

In the corresponding graph made for dataset 2 (Fig. 39b), about one third of the non-formers have been replaced by AET codes, mainly located next to the pseudo-mirror plane at $(PN_{MD})_{min}/(PN_{MD})_{max}= 1$. It is satisfactory to see that the additional AETs, including those of the simple chemical elements, fit well into the general pattern. The chemical systems where substances form, but not at the equi-atomic compositions (non-AB formers), are not shown in Fig. 39a+b. They are not grouped together in clearly defined domains, nevertheless some tendencies are observed: very few non-AB formers are located in the non-former domains. Very few non-AB formers are also located in the domains where $(PN_{MD})_{min}/(PN_{MD})_{max}<0.5$ and octahedra are observed. For certain $(PN_{MD})_{max}$ values (e.g. 56, 58, 67, 72 and 93, corresponding to Ru, Co, Be, B and F, respectively) a high density of non-AB formers is noted. Non-AB formers are also common at the boundary between the non-formers and AB formers.

Chapters 5.1 and 5.2. mainly focusing on the binary inorganic substances reveal the following facts:

• The chemical elements prefer a limited number of AETs (tetrahedron, trigonal prism, octahedron, cube, tricapped trigonal prism, cuboctahedron, rhombic dodecahedron, and heptacapped pentagonal prism).
• The chemical elements prefer to realize one single kind of AET, even in structure prototype having several different crystallographic sites.
• The distribution of A and B atoms within a 'mixed' coordination polyhedron is in general limited to the highest symmetrical distribution, although there exist in principle many ways to distribute them, e.g. eight B and six A atoms at the vertices of a rhombic dodecahedron.
• The maximum number of coordinating atoms B observed in the AET of an atom A is nine, although the coordination numbers (CNs) vary from 1 to 18.
• For the lower coordination numbers (CN$\leq$ 6), all coordinating atoms are in general of the other kind. For graphical presentations, the less frequently observed AETs were sub-divided into four categories: 7-vertex polyhedra, 9–11-vertex polyhedra, 13–16-vertex polyhedra and complex.
Stability domains are clearly distinguished in AET stability maps based on a combination of the periodic number $PN_{MD}$ of the constituent elements AET- $(PN_{MD})_{max}$ versus $(PN_{MD})_{min}/(PN_{MD})_{max}$, as well AET-$(PN_{MD})_A$ versus $(PN_{MD})_B$ stability maps. It is now possible to predict the existence of equi-atomic inorganic substances in binary systems that have not yet been studied and to assign a probable AET, or to search for equi-atomic inorganic substances with an AET.

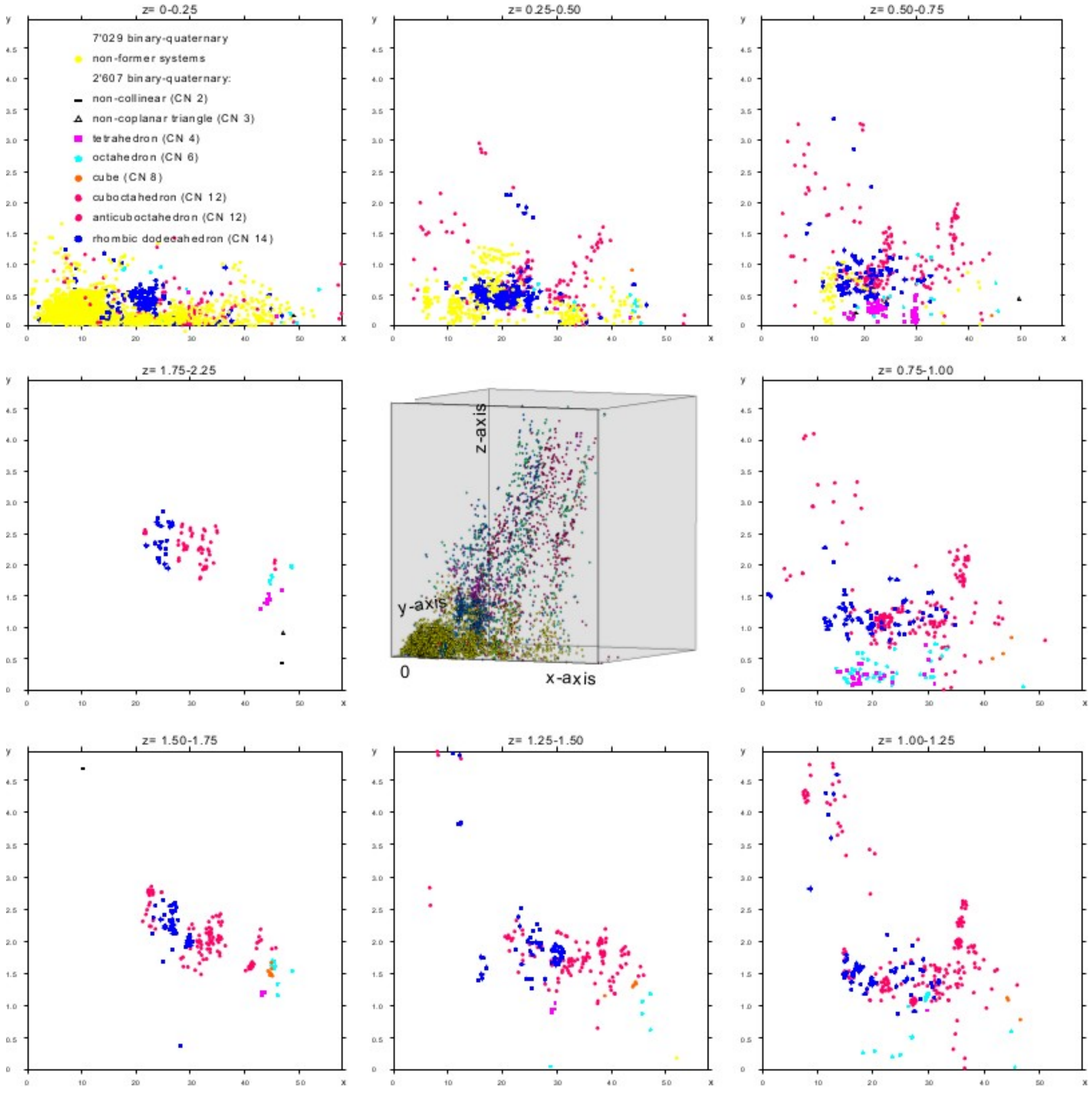

Fig. 40. The 3D-AET classification maps for single environment type including binary to multinary inorganic substances. The following feature sets are used: Mendeleyev number $PN_{MD}$: sum (x-axis), pseudo-potential radii (Zunger) R(Z): difference (y-axis) and electronegativity (Martynov + Batsanov) EN(MB): difference (z-axis). To demonstrate its separation ability we have in addition to the 3D-graph given eight 2D-graphs for defined z ranges: z=0-0.25, z=0.25-0.50, etc.

The results prove that the analysis of critically evaluated datasets can lead to the discovery of so far 'hidden' patterns, which can then be used to develop rules for semi-empirical inorganic substances design. The effect of the concentration on the AET stability criteria can be investigated by investigating $AB_2$, $AB_3$, $A_2B_3$, $A_3B_5$, and etc. inorganic substances.

In addition this approach can be easily extended to ternaries and quaternaries. Fig. 40 shows such a three-dimensional structure stability map using the 'coordination type' classification focusing on single-environment binary, ternary, and quaternary inorganic substances (single-environment type inorganic substances show within each inorganic substance just 1 kind of AET, e.g. all point-sets have as AET the tetrahedron). The following feature sets where used: Mendeleyev number $PN_{MD}$: sum (x-axis), pseudo-potential radii (Zunger) $R_Z$: difference (y-axis) and electronegativity (Martynov + Batsanov) $EN_{MB}$: difference (z-axis). To demonstrate its separation ability we have in addition to the 3D-graph given eight 2D-graphs for defined z ranges: z=0-0.25, z-0.25-0.50, etc. As the separation of the AET domains require three different APPE the domains have to be considered as volumes, some show clear dominance of one specific AET, other region there exists an overlapping between 2 different AETs. The situation gets of course more complex focusing on two-, multi-environments type inorganic substances.

## 5.3. Structure prototype classification versus coordination type classification

The two different general applicable crystallographic classifications are pictured in Fig. 41, as well as their complementary effect. By comparing the advantages respectively disadvantage of each classification in context with structure stability maps for binary, as well as multinary, the following facts are seen:

| Structure prototype classification *(classical view)* | Coordination type classification *(non-conventional view)* |
|---|---|
| 39,990 different structure prototypes | 100 distinct AETs |
| 1,000 most populous structure prototypes (70%) | 30 most populous AETs (70%) |
| 170,000 distinct different inorganic substances | 1,600,000 fully occupied point-sets |
| About 500 different stoichiometric ratios are realized by daltonide inorganic substances belonging to basic structure prototypes | *Single-environment type* (6 %)<br>*Two-environments type* (29 %)<br>*Three-environments type* (21 %)<br>*Four-environments type* (11 %)<br>*Multi-environments type* (33 %) |
| In average: about 4 inorganic substances per structure prototype | In average: about 53,300 point-sets per AET |

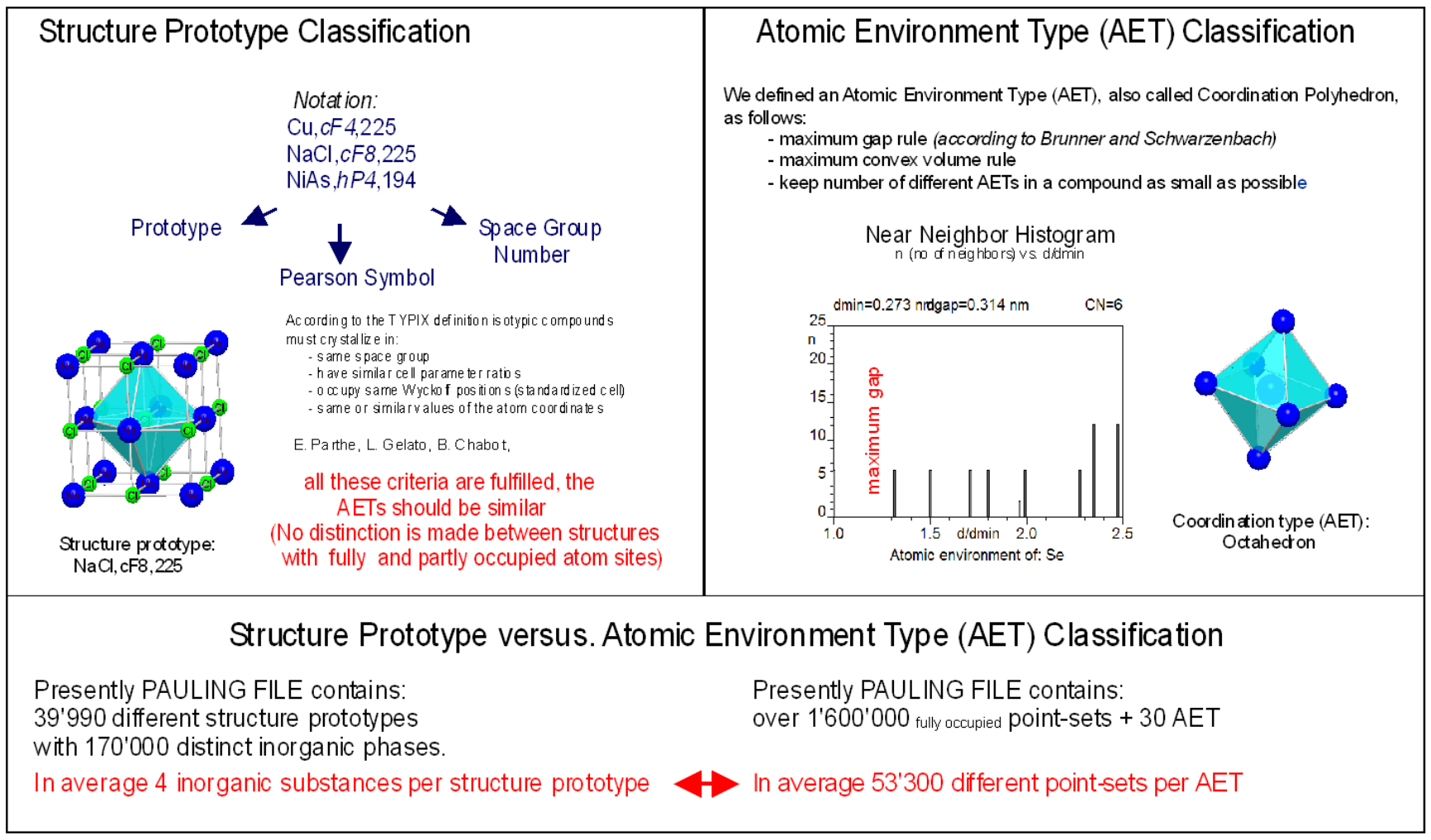

Fig. 41. Structure prototype versus atomic environment type classifications (advantages / disadvantages).

The conclusion is best to use both classifications, and judge case by case which leads to a simpler solution focusing on a specific problem. Another important help is to take advantage of the many structural relationships. It would exceed the aim of this review to go here into details, but one can easily derive thousands of such structural relations.

## 5.4. Generalized coordination number ranges and AETs stability maps using the AET classification

### *5.4.1. Observed AETs within the AET classification*

On the MPDS platform **[17a,b]** the following information was taken into consideration: AET, the nature of the central atom, hereafter referred to as $A$, and the nature of the coordinating atoms ($B_xC_yD_z$), hereafter referred to as $B$, e.g. octahedron, Na,$F_2O_4$; tetrahedron, Na,$As_4$.
100 different AETs are realized in 223,635 refined inorganic substance (representing about 143,000 *distinct* phases), but AETs such as non-collinear, non-coplanar triangle, tetrahedron, octahedron, pentagonal pyramid, icosahedron, cuboctahedron, and rhombic dodecahedron largely predominate. The 16 most populous AETs, representing 1,136,625 sites are shown in Fig. 42 and their counts are given in Table 16.

Table 17 shows the ‘Zintl line’ considered, and gives the counts per chemical element focusing on all fully occupied sites (in total 1’532’970 sites). The following observations within the AETs are made:

• By far the highest occurrence counts of *[central atom–coordinating atoms] combinations* are realized when chemical elements located to the right of the ‘Zintl line’ act as central atom. The top-10 with decreasing counts are O, H, S, F, C, Si, P, N, and Cl, as well as the $s^1$-element Na (see blue areas in Table 17). Oxygen is by far the most populous chemical element, 8 times more the second top chemical element hydrogen.
• The 16 most populous AETs are highly symmetrical, and consequently realize a ‘spherical’ distribution of the coordinating atoms within the AET (see Fig. 42).

| CN | Occurrence | Name |
|---|---|---|
| 2 | 32’013 | collinear |
| | 291’022 | non-collinear |
| 3 | 24’868 | coplanar triangle |
| | 135’350 | non-coplanar triangle |
| 4 | 210’239 | tetrahedron |
| 6 | 221’483 | octahedron |
| | 10’615 | trigonal prism |
| 8 | 19’422 | square prism (cube) |
| | 21’728 | square antiprism |
| 9 | 34’764 | tricapped trigonal prism |
| 12 | 29’181 | icosahedron |
| | 47’952 | cuboctahedron |
| | 6’936 | anticuboctahedron |
| 14 | 19’355 | rhombic dodecahedron |
| | 11’818 | 14-vertex Frank-Kasper |
| 16 | 9’879 | 16-vertex Frank-Kasper |
| total | 1’136’625 | represents about 70% all considered sites |

Table 16. The 16 most frequent occurring atomic environment types (using refined structures).

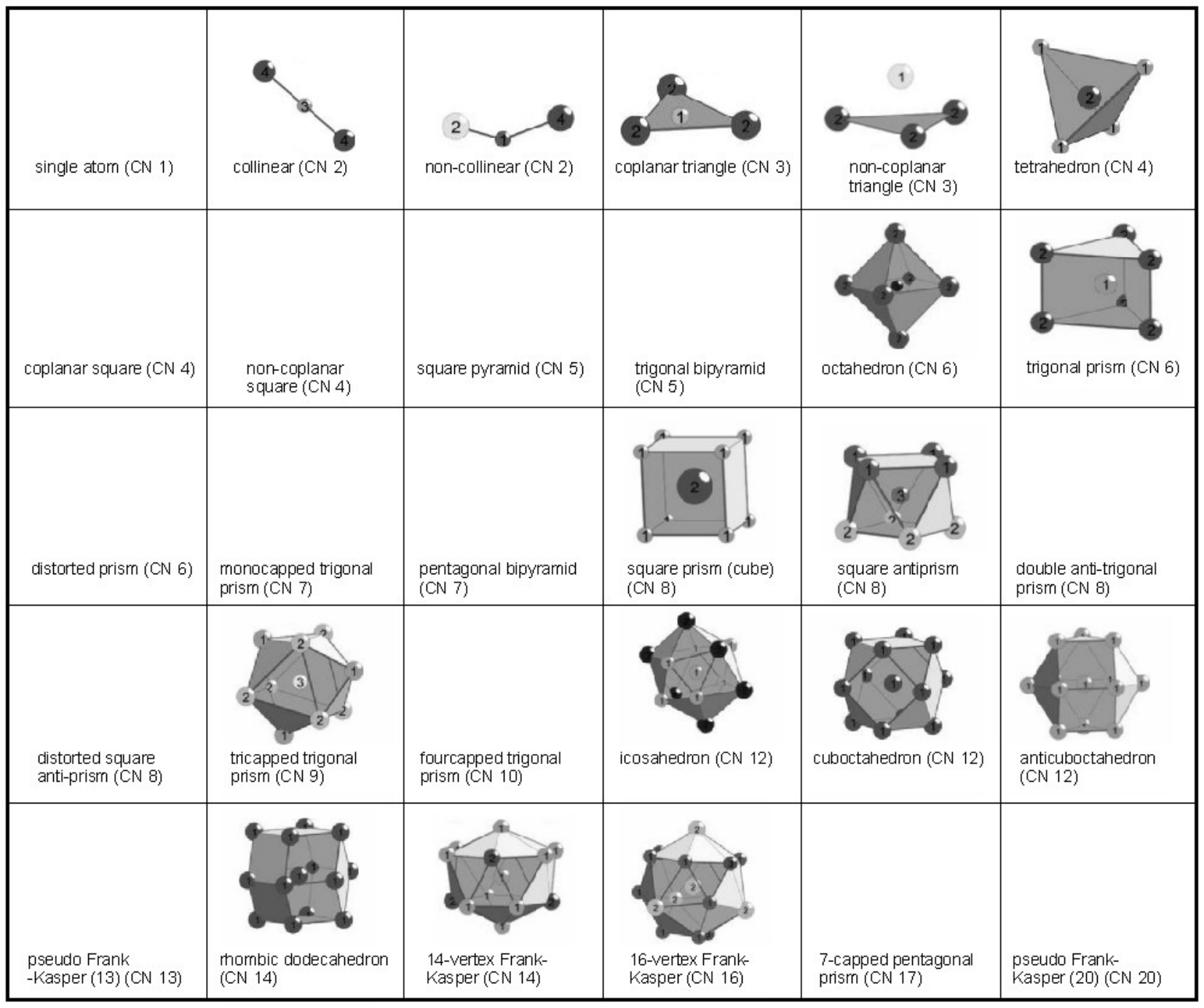

Fig. 42. The 16 most populous AETs of about 223,000 refined inorganic substances published.

AN PN$_{MD}$ — number of counts for central atom positions

total number of point-sets occupied by a single chemical element with occupancy= 1: 1'532'970

'Zintl line'

| | IA | IIA | IIIB | IVB | VB | VIB | VIIB | VIIIB | | | IB | IIB | IIIA | IVA | VA | VIA | VIIA | VIIIA |
|---|---|---|---|---|---|---|---|---|---|---|---|---|---|---|---|---|---|---|
| GN | 1 | 2 | 3 | 4 | 5 | 6 | 7 | 8 | 9 | 10 | 11 | 12 | 13 | 14 | 15 | 16 | 17 | 18 |
| 1 | | | | | | | | | | | | | | | | | 1 H 105<br>79'346 | 2 He 112<br>44 |
| 2 | 3 Li 1<br>11'348 | | | | | | | | | | | 4 Be 75<br>1'518 | 5 B 81<br>22'755 | 6 C 87<br>57'801 | 7 N 93<br>40'899 | 8 O 99<br>586'661 | 9 F 106<br>39'393 | 10 Ne 113<br>16 |
| 3 | 11 Na 2<br>19'568 | | | | | | | | | | | 12 Mg 76<br>9'153 | 13 Al 82<br>20'736 | 14 Si 88<br>33'540 | 15 P 94<br>27'791 | 16 S 100<br>46'893 | 17 Cl 107<br>22'218 | 18 Ar 114<br>14 |
| 4 | 19 K 3<br>18'599 | 20 Ca 7<br>14'258 | 21 Sc 11<br>2'760 | 22 Ti 43<br>11'141 | 23 V 47<br>9'794 | 24 Cr 51<br>5'678 | 25 Mn 55<br>16'394 | 26 Fe 59<br>23'421 | 27 Co 63<br>13'321 | 28 Ni 67<br>13'220 | 29 Cu 71<br>22'031 | 30 Zn 77<br>12'869 | 31 Ga 83<br>7'684 | 32 Ge 89<br>12'812 | 33 As 95<br>10'362 | 34 Se 101<br>20'174 | 35 Br 108<br>7'555 | 36 Kr 115<br>43 |
| 5 | 37 Rb 4<br>6'174 | 38 Sr 8<br>11'004 | 39 Y 12<br>6'711 | 40 Zr 44<br>5'267 | 41 Nb 48<br>7'652 | 42 Mo 52<br>13'042 | 43 Tc 56<br>266 | 44 Ru 60<br>3'490 | 45 Rh 64<br>3'067 | 46 Pd 68<br>5'277 | 47 Ag 72<br>5'216 | 48 Cd 78<br>6'720 | 49 In 84<br>7'232 | 50 Sn 90<br>9'773 | 51 Sb 96<br>11'566 | 52 Te 102<br>12'092 | 53 I 109<br>12'531 | 54 Xe 116<br>336 |
| 6 | 55 Cs 5<br>9'057 | 56 Ba 9<br>19'002 | 6' | 72 Hf 45<br>1'702 | 73 Ta 49<br>4'008 | 74 W 53<br>24'593 | 75 Re 57<br>2'288 | 76 Os 61<br>1'302 | 77 Ir 65<br>2'342 | 78 Pt 69<br>3'954 | 79 Au 73<br>3'125 | 80 Hg 79<br>2'800 | 81 Tl 85<br>2'817 | 82 Pb 91<br>8'151 | 83 Bi 97<br>9'496 | 84 Po 103<br>41 | 85 At 110 | 86 Rn 117 |
| 7 | 87 Fr 6 | 88 Ra 10<br>5 | 7' | 104 Rf 46 | 105 Db 50 | 106 Sg 54 | 107 Bh 58 | 108 Hs 62 | 109 Mt 66 | 110 Ds 70 | 111 Rg 74 | 112 Uub 80 | 113 Uut 86 | 114 Uuq 92 | 115 Uup 98 | 116 Uuh 104 | 117 Uus 111 | 118 Uuo 118 |

| QN | | | | | | | | | | | | | | | |
|---|---|---|---|---|---|---|---|---|---|---|---|---|---|---|---|
| 6' | 57 La 13<br>10'169 | 58 Ce 15<br>6'644 | 59 Pr 17<br>4'417 | 60 Nd 19<br>5'636 | 61 Pm 21<br>10 | 62 Sm 23<br>3'380 | 63 Eu 25<br>3'282 | 64 Gd 27<br>4'963 | 65 Tb 29<br>3'755 | 66 Dy 31<br>3'134 | 67 Ho 33<br>3'290 | 68 Er 35<br>3'595 | 69 Tm 37<br>1'563 | 70 Yb 39<br>3'290 | 71 Lu 41<br>1'767 |
| 7' | 89 Ac 14<br>5 | 90 Th 16<br>1'403 | 91 Pa 18<br>44 | 92 U 20<br>7'202 | 93 Np 22<br>629 | 94 Pu 24<br>554 | 95 Am 26<br>117 | 96 Cm 28<br>133 | 97 Bk 30<br>39 | 98 Cf 32<br>39 | 99 Es 34<br>1 | 100 Fm 36 | 101 Md 38 | 102 No 40 | 103 Lr 42 |

Table 17

Table 17. Mendeleyev type periodic system including the atomic numbers AN and periodic numbers PN$_{MD}$, as well as the 'Zintl line' considered. The occurrence count (number of times the chemical element occurs as a central atom) for each element. The blue color highlighted elements have an outstanding high occurrence frequency as the central atom.

*5.4.2. Generalized 'Coordination Number Ranges' – $(PN_{MD})_A$ versus $(PN_{MD})_B$ Stability Matrix*

We made the restraints that the considered AET have an occurrence count larger than 10 per [central atom–coordinating atoms] combination. This additional criterion was necessary to exclude special, as well as exotic cases. Sites with mixed occupation, disorder such as site spitting or partial vacancies were not taken into consideration. Incorrect or incomplete structure determinations, which often generate 'strange asymmetrical' AETs, are excluded from our considerations. In addition we focused on atom sites which belong to single- or two-atomic environments type's inorganic substances. Furthermore we included information about binary non-former systems [1]. This leads to the idealized square- or rectangular-shaped *non-former* domains, the following 30 binary systems represent violations having stable binary inorganic substances: Ba-Na, Bi-In, Bi-Mn, Bi-Pt, Bi-Rh, Br-H, Br-I, C-Si, Cd-Hg, Cd-Mg, Cl-H, Cl-I, F-H, Fe-Tc, Ga-Mg, H-I, Hg-In, Hg-Mg, Hg-Mn, In-Mg, In-Ni, In- Pd, In-Rh, Ir-Pb, Mg-Tl, Mn-Zn, Pb-Pd, Pb-Pt, Pb-Rh, and Pt-Tl.

The chemical element A occupying the center of the AET is given on the *y*-axis and the chemical elements B acting as coordinating atoms on the *x*-axis. On both axes the chemical elements are ordered according to the periodic number $PN_{MD}$. Each chemical element combination is present twice in the stability matrix, once the considered element is acting as central atom A(y-axis), and once it is acting as coordinating atom B(x-axis). The [central atom– coordinating atoms] combinations with $(PN_{MD})_{A(central\ atom)}$ larger than $(PN_{MD})_{B(coordinating\ atoms)}$ are situated to the left of the diagonal 1/1-118/118, whereas the [central atom–coordinating atoms] combinations with $(PN_{MD})_{A(central\ atom)}$ smaller than $(PN_{MD})_{B(coordinating\ atoms)}$ are situated to the right of the stability matrix.

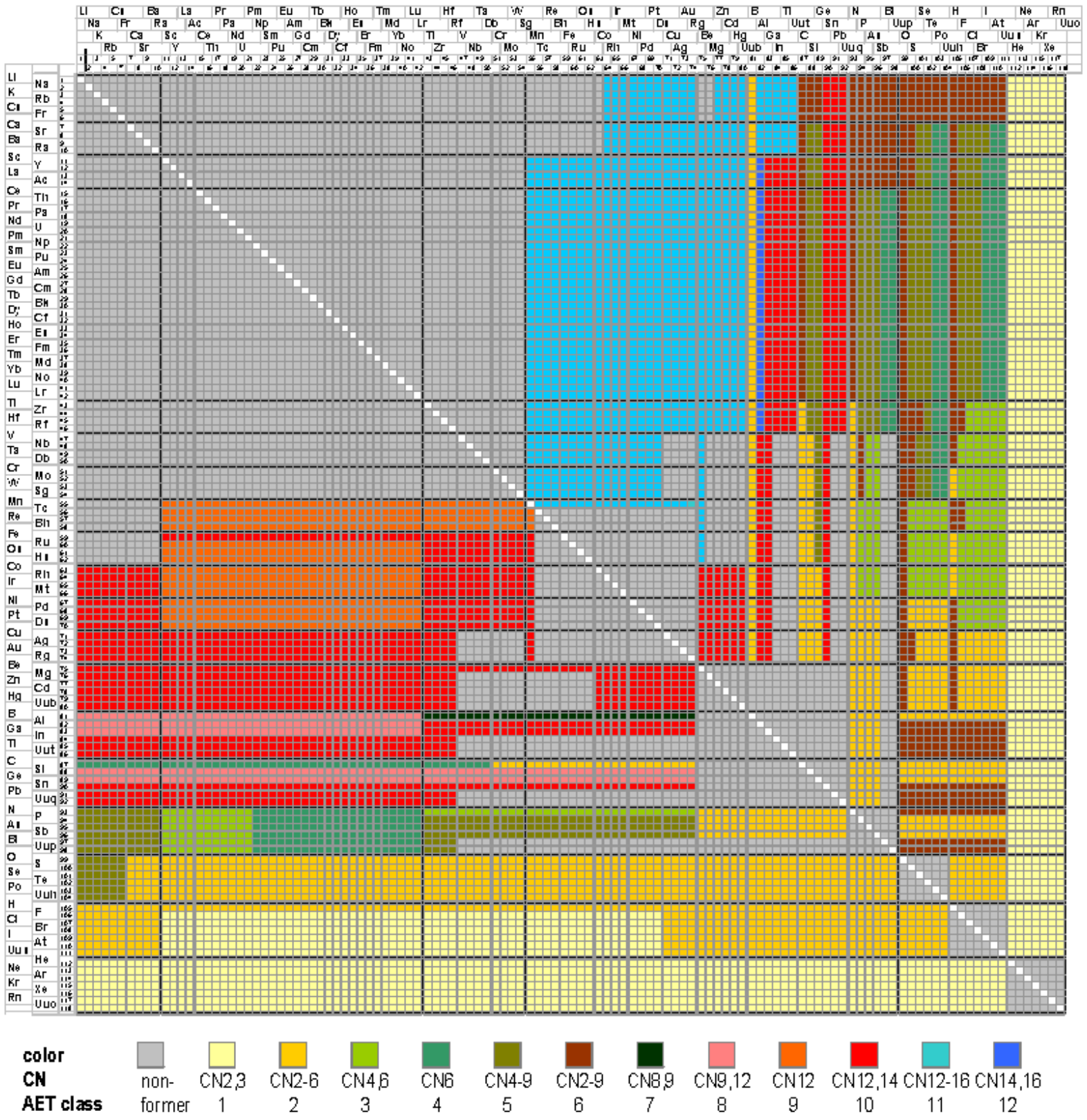

Fig. 43. Generalized 'Coordination number range' - $(PN_{MD})_A$ versus $(PN_{MD})_B$ stability matrix, which is independent of the stoichiometry and the number of chemical elements in the inorganic substance, prediction-completed for all possible [central atom-coordinating atoms] combination. The element A occupying the center of the AET is given on the y-axis and the coordinating elements B on the x-axis. For each element B as coordinating atom the total CN is considered.

We were able to generate a 'prediction-completed' generalized stability matrix valid for all potential [central atom - coordinating atoms] combination possibilities, shown in Fig. 43. Several well-defined 'stability domains', which can be simply formulated as '$(PN_{MD})_{A(\text{central atom})}$ range - $(PN_{MD})_{B(\text{coordinating atoms})}$ range' rectangular stability domains, are observed within the stability matrix. In order to obtain this simple separation into separate stability domains (see Figs. 43-44) we had to release the separate most populous 16 AET classes (see Fig. 42) into 12 'coordination number ranges' classes (see Table 18). The necessity to introduce these 'coordination number ranges' classes is reinforced by the fact that only about 1/3 of all atom sites considered here belong to single- or two-atomic environments types, whereas the remaining 2/3 belong to three- or higher-atomic environments types. This is related to the fact that we had to reduce the problem to a simplified situation where the coordinating atoms, which can be at most three different chemical elements, are plotted along one axis, $(PN_{MD})_{B(\text{coordinating atoms})}$. In the extreme case the same 'coordination number ranges' has been indicated for three different element combinations, A-B, A-C and A-D. When considerably more data will be available it might be possible to 'separate' specific AET into separate AET domains by taking in a different way the additional chemical elements C and D into account.

| AET class | Coordination Number CNs | Atomic Environment Types (AETs) |
|---|---|---|
| 1 | 2,3 | (non-)collinear, (non-)coplanar triangle |
| 2 | 2-6 | (non-)collinear, (non-)coplanar triangle, tetrahedron, octahedron, distorted prism |
| 3 | 4,6 | tetrahedron, octahedron, distorted prism |
| 4 | 6 | octahedron, distorted prism |
| 5 | 4-9 | tetrahedron, octahedron, distorted prism, square prism, square antiprism, tricapped trigonal prism |
| 6 | 2-9 | (non-)collinear, (non-)coplanar triangle, tetrahedron, octahedron, distorted prism, square prism, square antiprism, tricapped trigonal prism |
| 7 | 8,9 | square prism, square antiprism, tricapped trigonal prism |
| 8 | 9,12 | tricapped trigonal prism, icosahedron, cuboctahedron, anticuboctahedron |
| 9 | 12 | icosahedron, cuboctahedron, anticuboctahedron |
| 10 | 12,14 | icosahedron, cuboctahedron, anticuboctahedron, 14 Frank-Kasper, rhombic dodecahedron |
| 11 | 12-16 | icosahedron, cuboctahedron, anticuboctahedron, rhombic dodecahedron, 16 Frank-Kasper |
| 12 | 14,16 | 14 Frank-Kasper, rhombic dodecahedron, 16 Frank-Kasper |

Table 18. Coordination number ranges (and its atomic environment type (AET) classes).

The [central atom–coordinating atoms] combinations that do not occur are located symmetrically with respect to the diagonal 1/1- 118/118. The overall picture reveals a strong pseudo symmetry with respect to the diagonal, due to the dominating influence of the valence electron configuration of the central atom A on the choice of 'coordination number ranges' class when $(PN_{MD})_{A(\text{central atom})} > (PN_{MD})_{B(\text{coordinating atom})}$. On the contrary, for combinations with $(PN_{MD})_{A(\text{central atom})} < (PN_{MD})_{B(\text{coordinating atom})}$, the 'coordination number ranges' class is mainly determined by the valence electron configuration of the chemical elements acting as coordinating atoms. Chemical elements with $PN_{MD} > 54$ determine the AET, regardless of whether they act as central or as coordinating atoms.
The fact that the 'non-existence' of [central atom–coordinating atoms] combinations is consistent with the binary non-former systems demonstrates that if the 'reactivity difference' is too small in the binary case, it is also too small for the realization of [central atom-coordinating atoms] combinations within any AET, even including other chemical elements. The domains of not observed [central atom–coordinating atoms] combinations are located along the diagonal in the generalized 'coordination number ranges' stability matrix, and their size decreases from $(PN_{MD})_{A(\text{central atom})} = (PN_{MD})_{B(\text{coordinating atoms})} = 1$ to $(PN_{MD})_{A(\text{central atom})} >$ $(PN_{MD})_{B(coordinating\ atoms)}$= 118. This is consistent with the behavior of the atomic reactivity $RE_a$ as a function of $PN_{MD}$.

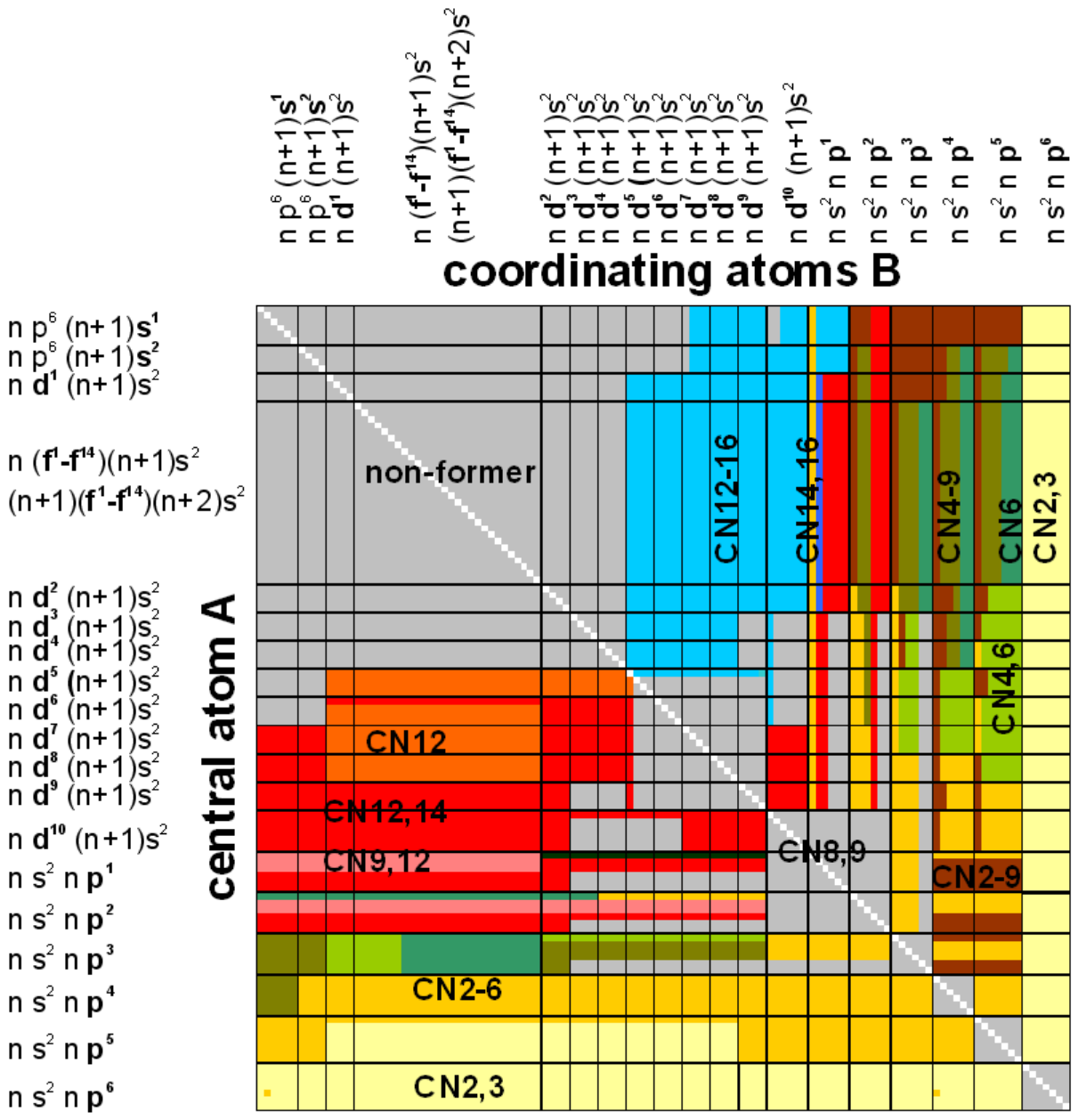

Fig. 44. Schematic generalized 'Coordination number range' - $(PN_{MD})_A$ versus $(PN_{MD})_B$ stability matrix highlighting the correlation to the valence electron configurations of the constituent chemical elements. The element A occupying the center of the AET is given on the y-axis and the coordinating element B on the x-axis, For each element B acting as coordinating atom the total CN is given.

A relatively good separation of all potential [central atom–coordinating atoms] combinations – 'coordination number ranges', including domains where no such combinations are observed, is achieved by this simple generalized 'coordination number ranges' – $(PN_{MD})_{A(central\ atom)}$ versus $(PN_{MD})_{B(coordinating\ atoms)}$ stability matrix. Also here no attempt was made to optimize the periodic number $PN_{MD}$ by modifying the order of the rows and columns in the Mendeleyev periodic system. It is, however, worth noting that when the periodic number $PN_{MD}$ is replaced by any of the other 40 selected APP listed in Table 1 the resulting generalized 'coordination number ranges' $(APP)_{A(central\ atom)}$ versus $(APP)_{B(coordinating\ atoms)}$ stability matrixes show a lower degree of local 'coordination number ranges' class ordering. Focusing on the electron configurations of the chemical elements shown in a schematic representation of 'coordination number ranges' classes – $(PN_{MD})_{A(central\ atom)}$ versus $(PN_{MD})_{B(coordinating\ atoms)}$ stability matrix (see Fig. 44), it can be concluded that the 'coordination number ranges' classes are in first priority controlled by the valence electron configurations of the constituent elements. The sizes of the atoms, often considered as being important in this context, appear to be almost irrelevant. The relative size of the atoms is, however, responsible for the frequent distortions observed within the AETs.

### *5.4.3. Generalized 'Periodic System – Coordination number ranges' graphs for the 18 (21) different GNs*

The second way to present exactly the same results is given by the fact that chemical elements with the same group number GN are behaving in a very similar way (except the $p^1$- to $p^3$- elements), and cannot be distinguished from each other. From Table 19 it can be seen that the dominant factor is the group number GN (valence electron configuration), which is contained in the periodic number $PN_{MD}$ ($PN_{ME}$), but not in the atomic number AN. The periodic system – 'coordination number ranges' graphs shown for e.g. $d^9$-elements respectively $s^1$-elements in Table 20a+b have the advantage that they are very easy to read. The reference chemical elements (blue shaded areas) represent the central atom A, the remaining chemical elements the coordinating atoms B. The 118 chemical elements can be separated into 18(21) major groups, which behave differently in context with the possible ‘coordination number ranges’ classes, respectively the ‘non-existence’ of [central atom–coordinating atoms] combinations (Table 18).

| Group Number GN | Valence Electrons VE | Chemical elements | Special cases |
|---|---|---|---|
| 1 (IA) | n $s^1$ | Li, Na, K, Rb, Cs, Fr | |
| 2 (IIA) | n $s^2$ | Ca, Sr, Ba, Ra | |
| 3 (IIIB) | n $d^1$ | Sc, Y, La, Ac | |
| 3 (IIIB) | n $f^{1-14}$ | Ce, Th, Pr, Pa,…Yb, No, Lu , Lr | |
| 4 (IVB) | n $d^2$ | Ti, Zr, Hf, Rf | |
| 5, 6 (VB, VIB) | n $d^3$,n $d^4$ | V, Nb, Ta, Db; Cr, Mo, W, Sg | |
| 7 (VIIB) | n $d^5$ | Mn, Tc, Re, Bh | Mn 12-16 for GN = 8-11 |
| 8 (VIIIB) | n $d^6$ | Fe, Ru, Os, Hs | |
| 9 (VIIIB) | n $d^7$ | Co, Rh, Ir, Mt | |
| 10 (VIIIB) | n $d^8$ | Ni, Pd, Pt, Ds | |
| 11 (IB) | n $d^9$ | Cu, Ag, Au, Rg | |
| 12 (IIB) | n $d^{10}$ | Be, Mg, Zn, Cd, Hg, Cn | Be 12,14 for GN = 5-8 |
| 13 (IIIA) | n $p^1$, n = 2-4 | B, Al, Ga | B 8,9 for GN = 4-11; B 2-6 for GN = 16-17 |
| 13 (IIIA) | n $p^1$, n = 5-7 | In, Tl, Nh | |
| 14 (IVA) | n $p^2$, n = 2-4 | C, Si, Ge | C 6 for GN = 1-5; C 2-6 for GN = 6-9 |
| 14 (IVA) | n $p^2$, n = 5-7 | Sn, Pb, Fl | Sn 12,14 for GN = 5-11 |
| 15 (VA) | n $p^3$, n = 2-5 | N, P, As, Sb | N 4,6 for GN = 4-11; N 2-9 for GN = 16-17 |
| 15 (VA) | n $p^3$, n = 6, 7 | Bi, Mc | |
| 16 (VIA) | n $p^4$ | O, S, Se, Te, Po, Lv | |
| 17 (VIIA) | n $p^5$ | H, F, Cl, Br, I, At, Ts | H 2-6 for GN = 3-10 |
| 18 (VIIIA) | n $p^6$ | He, Ne, Ar, Kr, Xe, Rn, Og | |

Table 19. Some details for the 21 different generalized periodic system – coordination number ranges respectively AET groups.

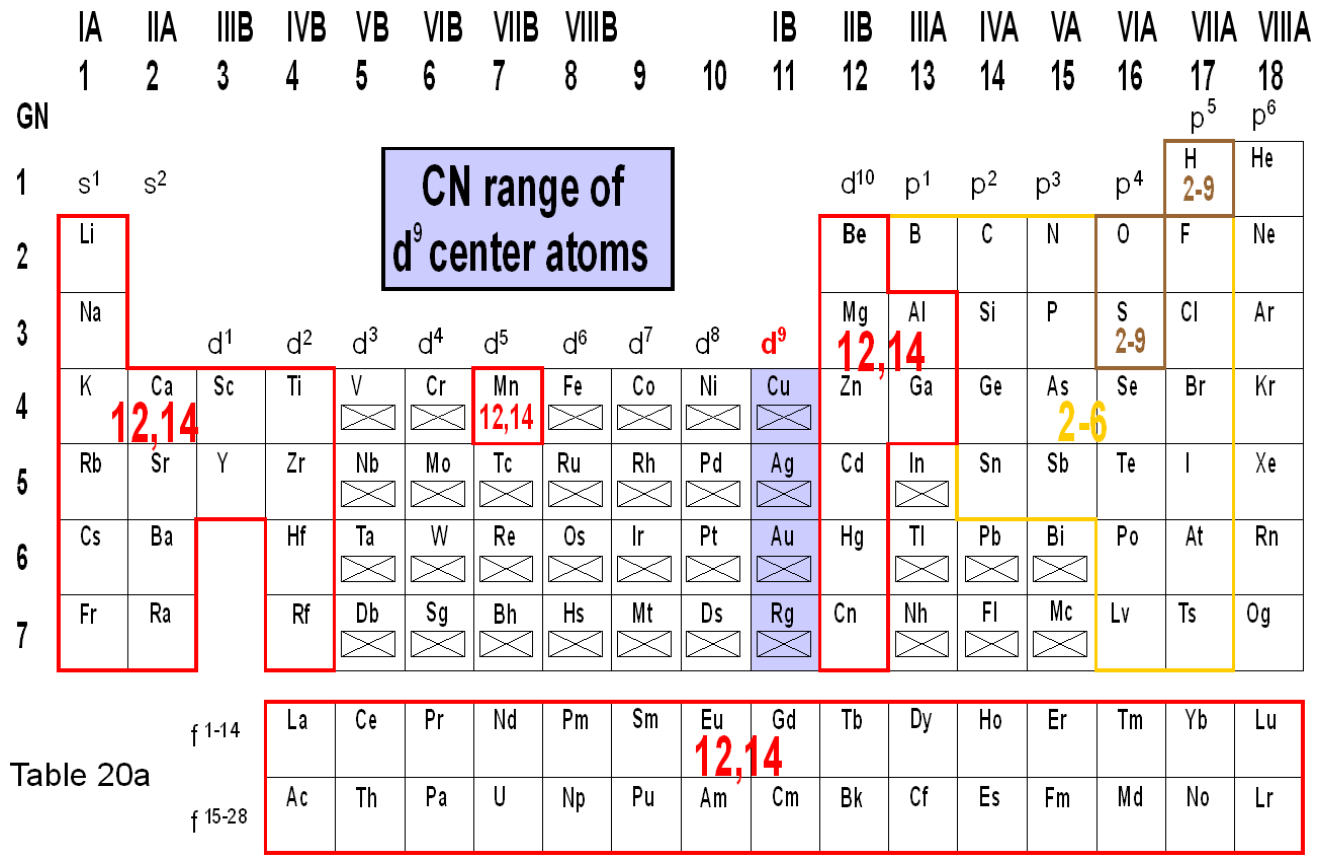

Table 20a. Generalized periodic system – coordination numbers ranges graph for $d^9$-elements as central atoms [central atom – coordinating atoms] combinations. Coordination numbers ranges are defined for the substance formers, the non-formers are marked by a ‘cross in a box’. The coordination numbers and their ranges are consistent within the AET classes defined in Table 18.

The ‘coordination number ranges’ classes for the elements acting as coordinating atoms B located to the left of the column of the reference central atoms A (e.g. $d^9$- elements Cu, Ag, Au, Rg) in the periodic system – ‘coordination number ranges’ graphs are valence electron governed by the central atom A. This is a consequence of the fact that the periodic number $PN_{MD}$ of these coordinating atoms B is lower than the periodic number $PN_{MD}$ of the central atom A. In contrast, the ‘coordination number ranges’ classes formed by coordinating atoms B located to the right of the column of the reference central atoms A are valence electron-governed by the coordinating atoms B, because here $(PN_{MD})_{B(\text{coordinating atoms})} > (PN_{MD})_{A(\text{central atom})}$.

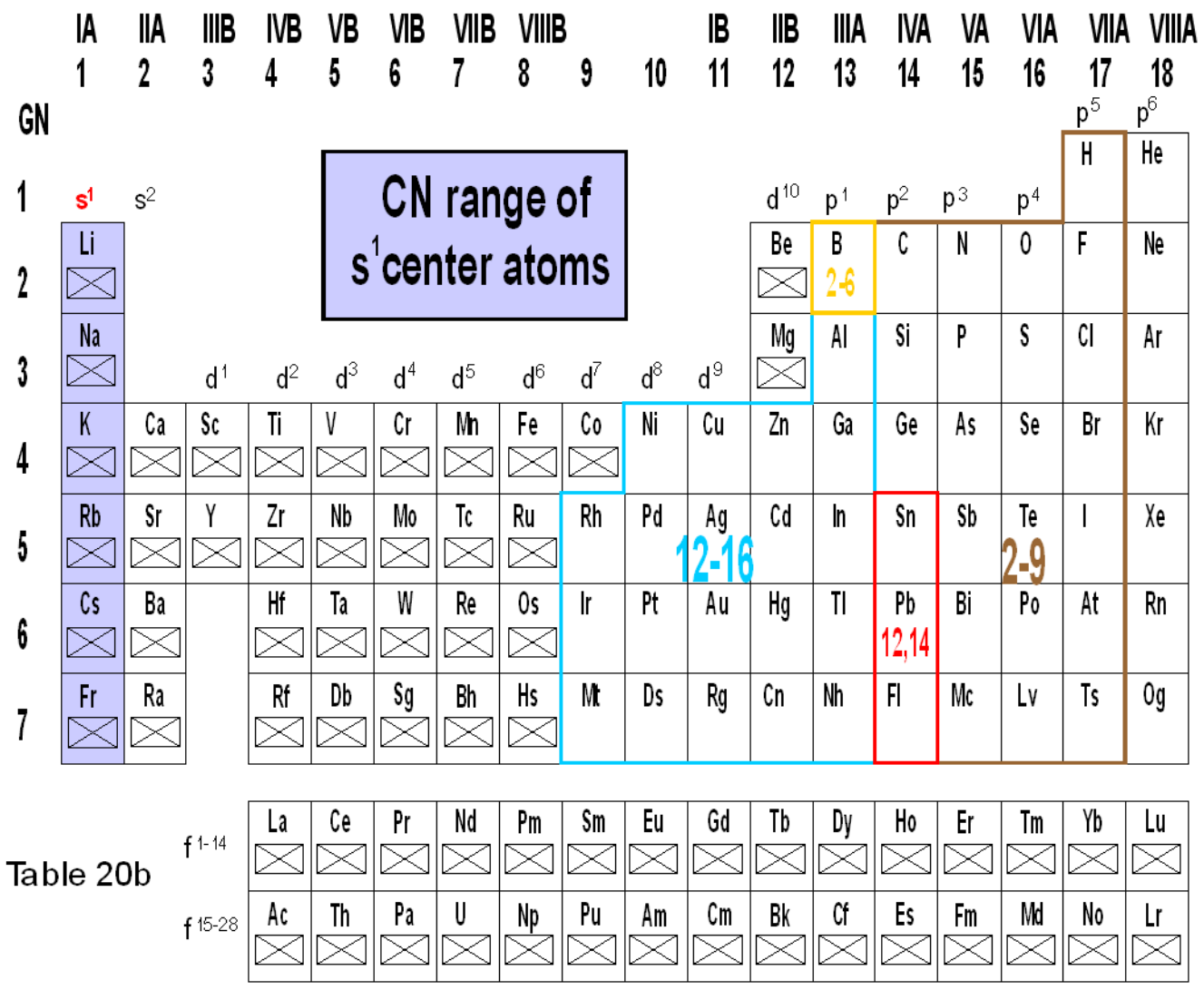

Table 20b. Generalized periodic system – coordination numbers ranges graph for $s^1$-elements as central atoms [central atom – coordinating atoms] combinations. Coordination numbers ranges are defined for the substance formers, the non-formers are marked by a ‘cross in a box’. The coordination numbers and their ranges are consistent within the AET classes defined in Table 18.

*5.4.4. Generalized ‘Periodic System – AETs’ graphs for the 18(21) different GNs (work in progress)*

As the generalized periodic system - AET graphs are much easier to read than the generalized AET - $(PN_{MD})_{A(\text{central atom})}$ versus $(PN_{MD})_{B(\text{coordinating atoms})}$ stability matrix we focus in this review on the ‘periodic system representation’, stressing that they give the identical information. In a most recent and ongoing research activity we found that the big disadvantage of having to group ‘AETs’ to ‘coordination number ranges’, and therefore having to a certain degree limited systematization of the APP or APPE - AET. Analog to Tables 20a+b we created periodic system – AET graphs (Tables 21a+b) by introducing a third dimension to the periodic system of Mendeleyev, this by giving to each of the 18 (21) group numbers (GN= 1-18) a different color, which correlates the 18 lines of each group number GN, sorted according with increasing periodic number $PN_{MD}$. As this is an ongoing research activity and for simplicity we intentionally differentiate here between the 4 most populous AET within each CN: tetrahedron (CN 4), octahedron (CN 6), cuboctahedron (CN 12) and rhombic dodecahedron (CN 14).

The ‘crossed’ chemical elements within the periodic system mean that those chemical elements do not contribute as coordination atoms to e.g. the considered $d^9$-central elements A (Cu, Ag, Au, Rg), e.g. Ag is never surrounded by either V and (or) Ru. The ‘uncrossed’ chemical elements B being first kind of coordination atoms B, e.g. $s^1$- elements Li, Na, K, Rb, Cs, Fr can have in addition to themselves a second (third) kind of coordinating atoms C, D (e.g. Cu ($PN_{MD}$= 71) to Hp ($PN_{MD}$ =92). All the remaining crossed chemical elements within the brown row cannot act as the second (third) kind of coordinating atoms. E.g. Cu with coordinating atom Zr cannot have as additional second coordination atom e.g. K, Dy, or Te. And finally the uncrossed chemical elements can contribute as the second (third) kind of coordinating atom.

With 21 such periodic system – AET graph we can systematize any chemical element through its GN to the relevant periodic system – AET graph. Therefore for any considered chemical element acting as the central atom we can verify or predict which chemical elements can act as coordinating atoms, and which AET they will form. This all is independent of its stoichiometry and its number of involved chemical elements. Any through simulation created hypothetical crystal structure (e.g. *ab initio*) can be checked if it obeys the required AET stability condition. This means the kind of the AETs, as well as its chemical element combinations between the central atom and its coordinating atoms forming the required AET.

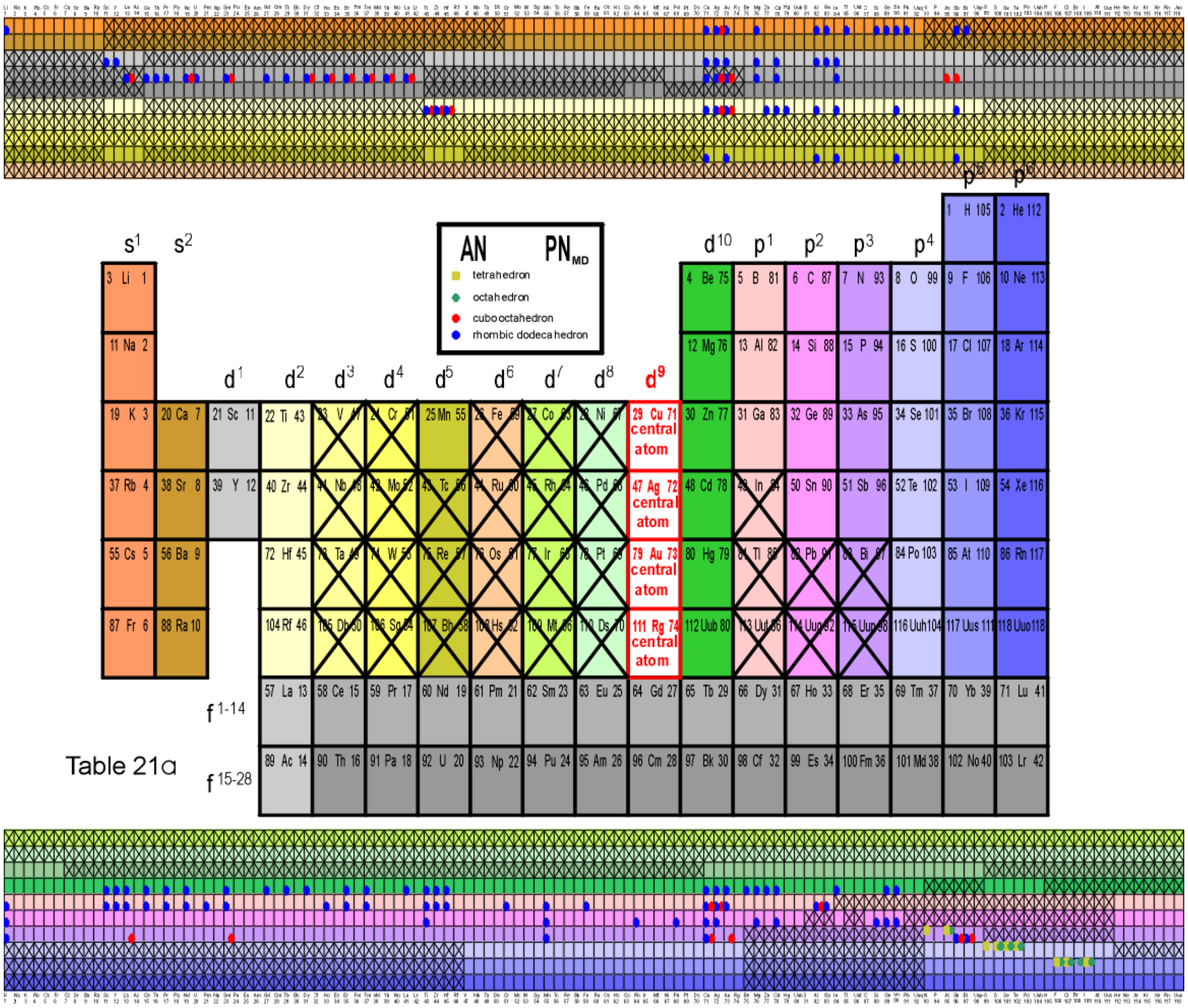

Table 21a. Generalized periodic system – AET graph for $d^9$-elements as central atoms [central atom – coordinating atoms] combinations. AETs are defined for the substance former, the substance non-formers are marked by a ‘cross in a box’.

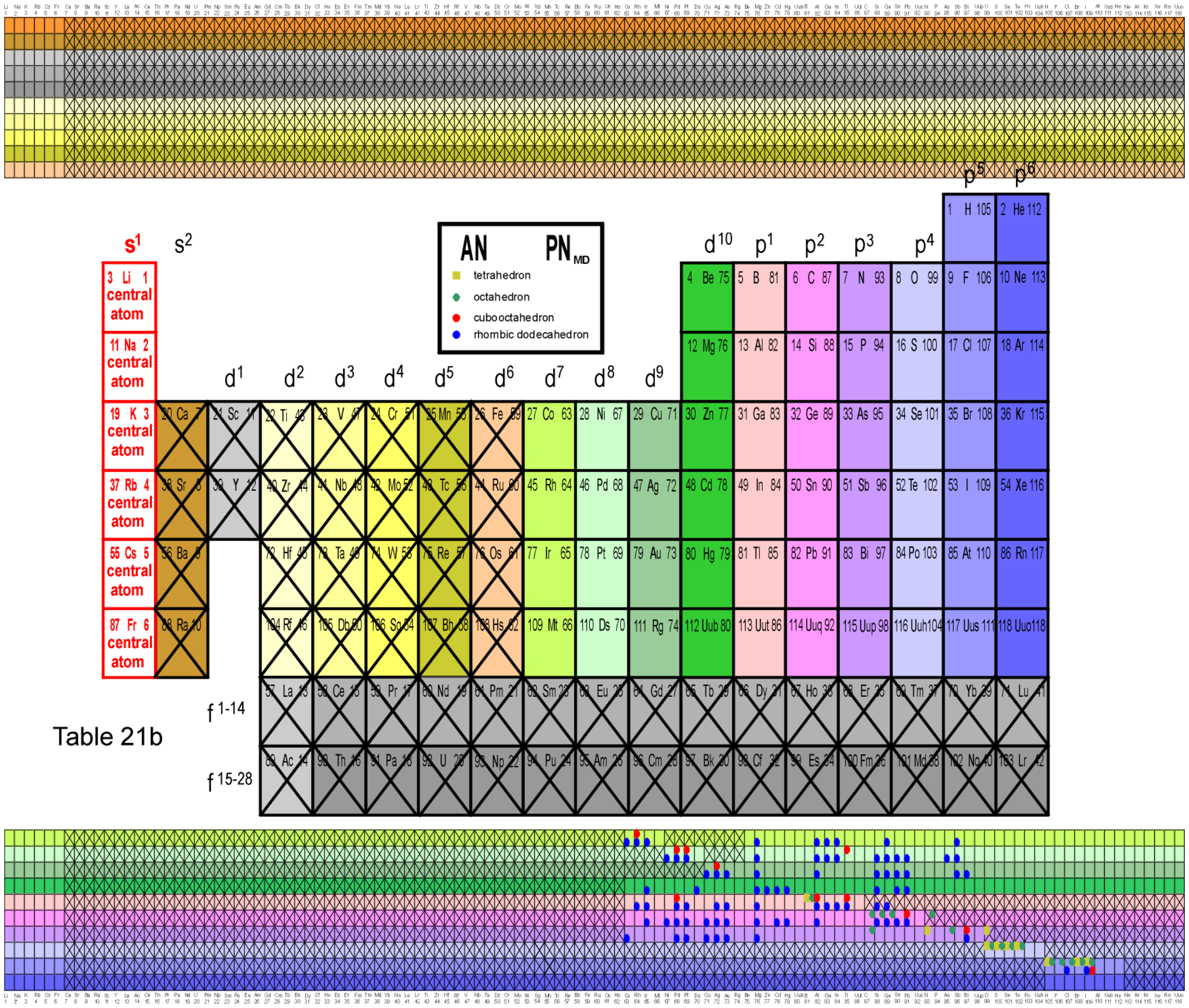

Table 21b. Generalized periodic system – AET graph for s[1]-elements as central atoms [central atom – coordinating atoms] combinations. AETs are defined for the substance former, the substance non-formers are marked by a 'cross in a box'.

Focusing on the binary, ternary, and multinary inorganic substances, chapter 5.4 reveals the following facts:

• The chemical elements prefer a limited number of AETs such as (non)-collinear, (non)-coplanar triangle, tetrahedron, octahedron, pentagonal pyramid, icosahedron, cuboctahedron, and rhombic dodecahedron.

• There exist (18)21 different groups of chemical elements acting as central atom A, which prefer AET's for different coordinating elements B, or do not form the [central atom–coordinating atoms] combinations with B.

• Chemical elements with $PN_{MD}$> 54 define the AETs they are part of, regardless of whether they act as central or as coordinating atoms.

• The observation that the non-existence of the [central atom–coordinating atoms] combinations is consistent with the binary non-former systems formed by the same elements demonstrates that if the 'reactivity difference' is too small in binary case, it is also too small for the [central atom-coordinating atoms] combinations within any AET, even when additional coordinating atoms would be included.

• 'Coordination number ranges' class domains are distinguished in the generalized 'coordination number ranges' - $(PN_{MD})_{A(central\ atom)}$ vs. $(PN_{MD})_{B(coordinating\ atoms)}$ stability matrix. It is now possible to predict the existence or 'non-existence' of the [central atom– coordinating atoms] combinations in not yet studied systems and to assign probable 'coordination number ranges', or to search for inorganic substances with a particular 'coordination number ranges'. The identical results can be achieved using its periodic system representation (which is easier to read).

• In our most recent and ongoing research activity we found that the disadvantage of having to group different AETs to 'coordination number ranges' could be resolved by introducing a third dimension, and therefore being able to create 18(21) different (for each GN) generalized periodic system – AET graphs.

• The following simplifying three observations are made:

1) The AETs are independent of the stoichiometry,
2) The AETs are independent of the number of chemical elements,
3) The AETs are independent on the number of different AET(s) (this means also independent of belonging to 1-, 2-, 3-, 4- or multi-environment types), within an inorganic substance.

This opens a new AET classification (in contrast to the coordination type classification), which has extreme simplifying consequences in context of systematization of crystallographic structures in respect to formulate AET stability criteria. *In essence they depend only on the [central atom A – coordinating atoms B,C] combinations.*

On the example of structure stability maps, chapter 5 demonstrates the discovery of several so far 'hidden' patterns, which can be used to develop rules for semi-empirical inorganic substances design. To our own surprise it is possible to develop *a concentration-independent and number of chemical-element-independent generalized AET stability matrix.* The identical results are even simpler shown in its 18(21) generalized periodic system AET graphs, that can be used for the verification of the most probable AET(s) within any hypothetical simulated inorganic substance, or to verify its conformity of an experimentally discovered novel inorganic substance. Inorganic substances following such generalized periodic system – AET graphs are behaving like the majority of experimentally known inorganic substances. In other words we claim that predicted hypothetical inorganic substances can only be synthesized if for each atom (chemical element) within the considered inorganic substance the "AET conditions" are fulfilled. Otherwise the hypothetical inorganic substance is not stable or very 'exotic'.

## 6. Holistic views: Physical property systematization for inorganic substance former systems

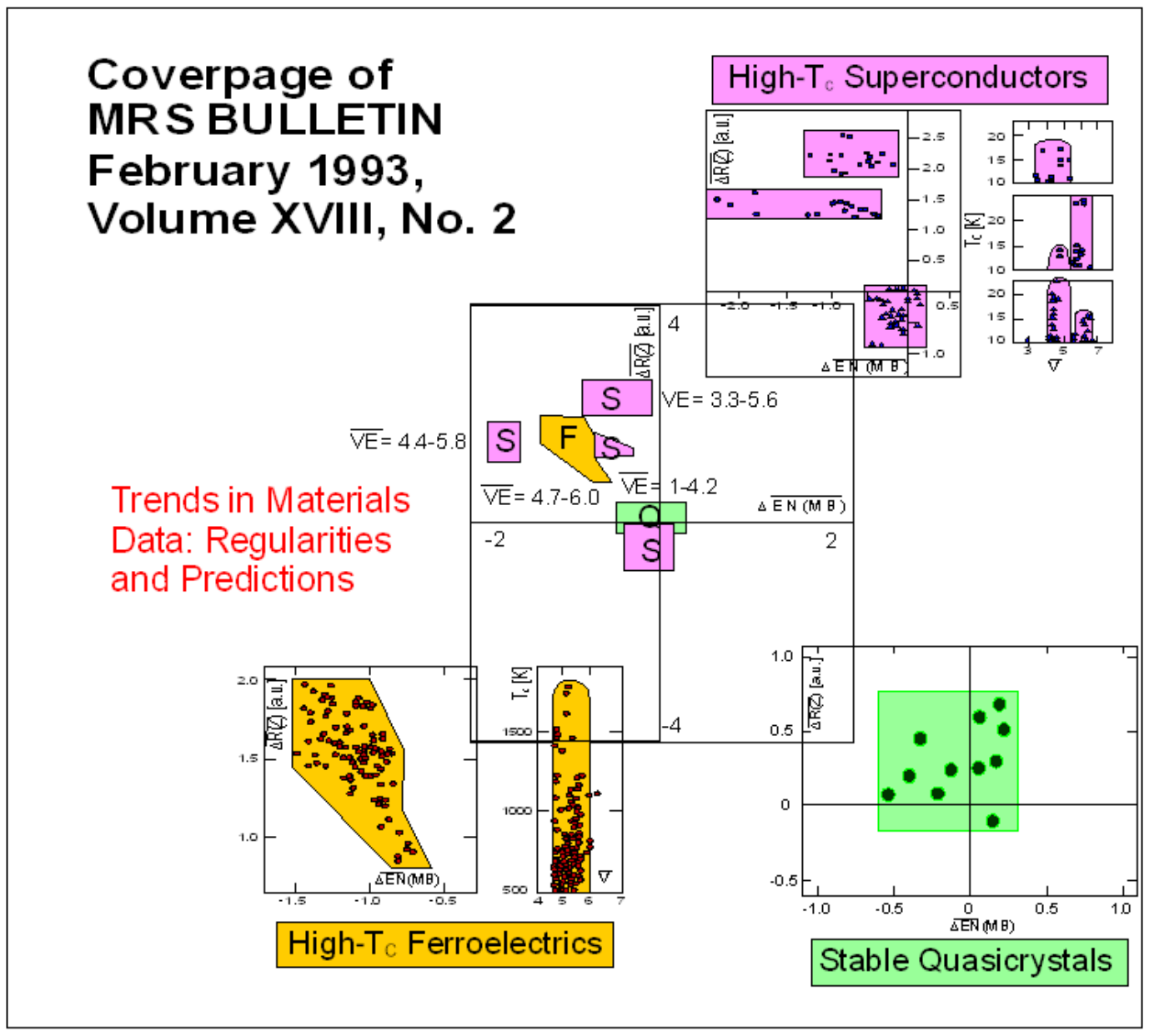

Fig. 45a. Quantum structure diagrams (QSD) focusing on High-T$_c$ superconductors, ferroelectrics, and stable quasicrystals showing trends in materials data: regularities and prediction [69].

In 1993 Rabe [69] reviewed the quantum diagram technique (it includes quantum structural diagrams, as well as quantum stability diagrams) as follows: The complexity of the systems under considerations dictates the use of the by Villars proposed quantum diagram technique, which is a procedure for displaying data graphically and identifying trends in the full database of 22,000 intermetallic substances. This global organization of structure prototypes and stability data involves 3D-diagrams in which each system is represented by a point whose coordinates are determined by its composition and its APP or APPE. Closely related systems have similar values of diagrammatic coordinates; in general, simple surfaces can be drawn to separate different classes of system.
The goal of the 'quantum structural diagrams' (QSD) is to systematize the relationship between composition and its APPE and structure prototype. It has been shown that binary inorganic substances can be effectively classified according to their structure prototype with the diagrammatic coordinate definitions:

$$\text{'mean } \Delta EN_{MB}\text{'} = |(EN_{MB})_A - (EN_{MB})_A|, \quad (13)$$
$$\text{'mean } \Delta R_Z\text{'} = |(R_Z)_A - (R_Z)_B|, \quad (14)$$
$$\text{'mean } \sum VE\text{'} = (VE)_A + (VE)_B, \quad (15)$$

where the APP: $EN_{MB}$ is Martynov-Batsanov electronegativity, $R_Z$ is Zunger pseudo-potential core radius, and VE is the valence electron number.

These diagrammatic coordinates (in the literature also called 'golden coordinates') are based on a relatively large set a crystallographic data (even 30 years ago). Therefore it was at that time worth trying to apply the same diagrammatic coordinates to inorganic substances with specific physical properties. The bottleneck is for all physical properties the same: there exists a very limited number of experimental data.

From the previous chapters we learned: 'Structure prototype-sensitive physical properties of inorganic substances are quantitatively described by its APPs of its constituent chemical elements.' As shown in Chapter 2, we obtained a direct link between the fundamental APP AN and PN and its derived APPs $SZ_a$, $RE_a$, and $AF_a$. Therefore we can generalize our postulate to the following statement: 'Structure prototype-sensitive physical properties are quantitatively determined by the APPs AN and PN (or simple mathematical functions of them) of its constituent chemical elements.' This generalization is an important link to strategically explore structure prototype-sensitive physical properties of inorganic substances. *Nevertheless those conditions are necessary but not sufficient.*

Already in 1993 it was possible with the help of such QSD to separate the following three, below listed, different classes of inorganic substances into three different domains (see Fig. 45a cover page of MRS Bulletin, 1993, 18, no.2).

67 high-$T_c$ (> 10 K) superconductors,
153 high-$T_C$ (> 500 K) ferroelectrics and anti-ferrolelectrics,
13 stable quasi-crystals

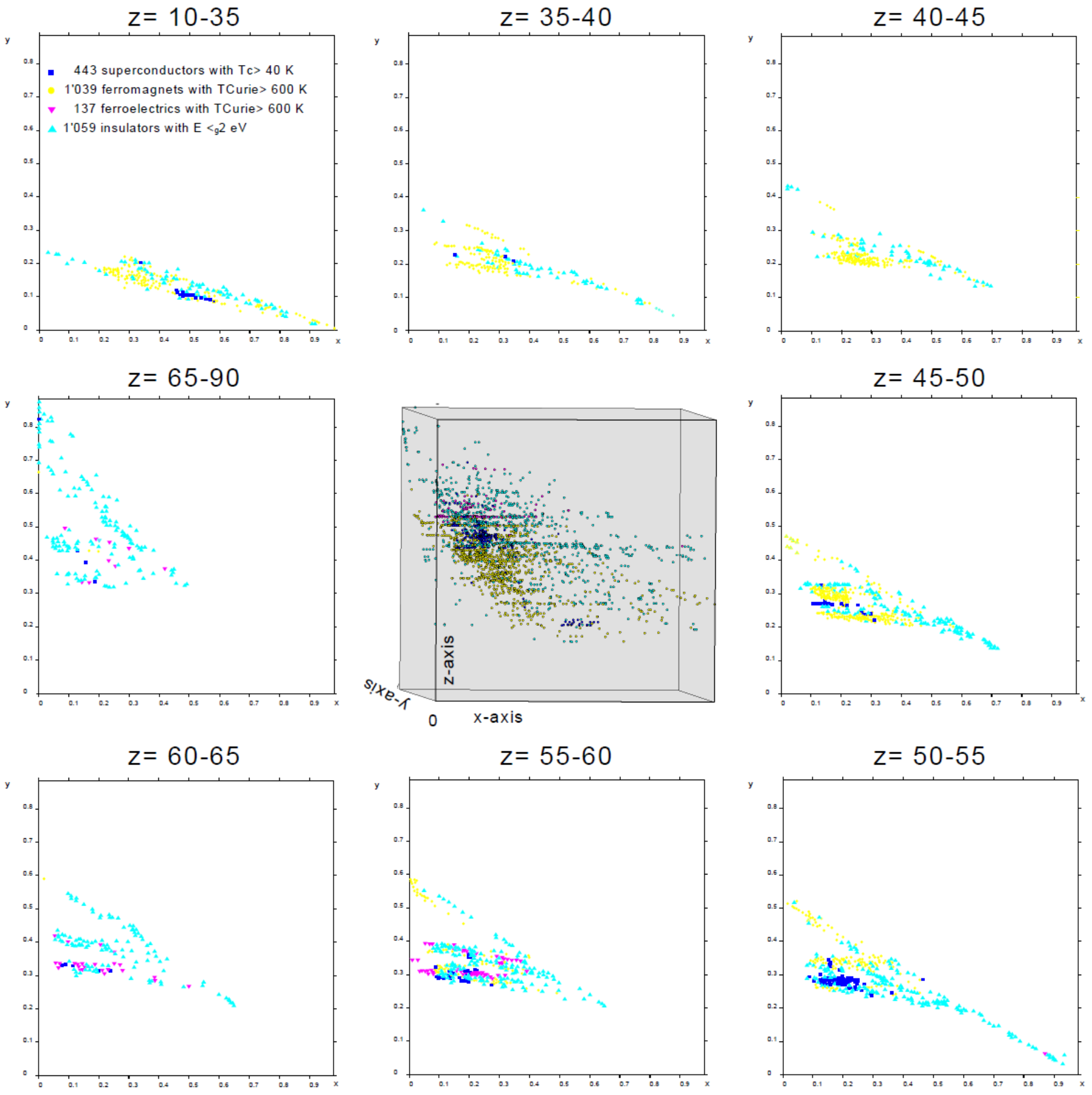

Fig. 45b. The 3D-APPE space using as x-axis: concentration-weighted $PN_{MD}$ ratio, y-axis: concentration-weighted $PN_{MD}$ difference, and z-axis: concentration-weighted $(PN_{MD})_{max}$ used to separate inorganic substances with different physical properties: high-$T_c$ superconductors, ferromagnets, ferroelectrics, and insulators (which are structure prototype-sensitive). To demonstrate its separation ability we have in addition to the 3D-graph given eight 2D-graphs for defined z ranges: z=10-35, z=35-40, z=40-45, etc.

About 30 years later with 2,678 different inorganic substances (summary listed below) we reach an analogous high quality of separation (see Fig. 45b) using as x-axis: concentration-weighted $PN_{MD}$ ratio, y-axis: concentration-weighted $PN_{MD}$ difference, and z-axis: concentration-weighted $PN_{MD}$ max. Notably we use only one APP, the periodic number (i.e. just integers from the Mendeleyev periodic system). Focusing on the structure prototypes and specific physical property (e.g. ferromagnets with paramagnetic $T_{Curie}$ > 600 K) we see it is realized in 102 different structure prototypes. This means the structural situation is very complex, nevertheless we are convinced that the materials design can be successful only through that "door". Here again the knowledge of structural relationships is playing a key position.

| Physical properties | Number of distinct inorganic substances | Number of different structure prototypes |
|---|---|---|
| superconductors with $T_c$> 40 K | 443 | 66 |
| ferromagnets with paramagnetic $T_{Curie}$> 600 K | 1,039 | 102 |
| ferroelectrics with $T_{Curie}$ > 600 K | 137 | 65 |
| semiconductors with $E_g$< 2 eV | 1,059 | 448 |
| Total number of considered inorganic substances | **2,678** | |

By adding two additional properties (ferromagnets and insulators) we have 12 times more data point in the 'QSD' (now called 'physical property' – 2(3)D-APP or APPE graphs).

Chapter 6 reveals the following fact. The quantum diagram technique (including quantum structural diagrams, as well as quantum stability diagrams), nowadays called (e.g. Structure stability maps using the structure prototype classification; Structure stability maps using the coordination type classification; Generalized 'coordination number ranges' respectively AET stability maps using the AET classification independent of its stoichiometry and its number of involved chemical elements) are very powerful to link 'structure prototype-sensitive physical properties of inorganic substances' with its crystallographic structure (either using the structure prototype-, or coordination type-, or AET-classifications), as well as to its APPs or APPEs.

## 7. Holistic views: PAULING FILE suggests 12 principles in materials science — Four Cornerstones given by Nature

### 7.1. Introduction

Confronted with the explosion of computing power, as well as of materials data, J. Gray proposed in 2009 the *Fourth Paradigm of Science: Data-Intensive Discovery through Data Exploration* (eScience) **[70]**, which means electronically unify experiment, theory and computation by data. The same agendas have been declared many times since 1980s, also even by the advocate Kenneth Wilson chairman of the third paradigm computational science, as a keynote of 11$^{th}$ CODATA **[118]**. After decades of incubation period for establishing a knowledge infrastructure of ICT, computational models and data, a kind of 'Cambrian Explosion' towards materials the *digital transformation* is being started in this decade.

The Executive Office of the President of the United States, National Science and Technology Council, launched mid-2011 the white paper *Materials Genome Initiative for Global Competitiveness* **[71]**, having as major aim to shorten the time between discovery of advanced materials and their industrial application by at least a factor two.

Reflecting these trends, many ideas have been proposed in the world to explore new dimensions, trying to derive knowledge from a collection of big data in full swing regardless of quality. To show a clear direction for such trends, we draw here another roadmap by taking advantage of peer-reviewed scientific data for inorganic substances (materials). Three key developments during the past decades have opened up unprecedented opportunities, namely:

1) The power of high-speed computers has reached incredible levels, and their price has decreased. High throughput computing clusters are forming a new virtual environment in the world for materials science and engineering, waiting for useful computational tools of higher accuracy and efficiency and inspiring data for materials development.

2) Computational materials science, algorithmic developments, and sophisticated software systems for simulation are advancing at increasing rates. In particular, the *ab initio* methods are now established that can predict the structure prototype and a big range of physical properties of inorganic substances by the use of the fundamental principles only. The computed results are often comparable with the experimental data and can sometimes compensate the experiments with higher resolutions. In a similar way, the CALPHAD approach has reached maturity, and is able to calculate phase diagrams that can be compared with experimentally determined phase diagrams, compensating difficult experiments at extreme temperature, pressure and time-equilibrium and transient conditions.

3) High bandwidth communication at extremely low cost has revolutionized global collaboration and knowledge exchange, which enable a creative ecology for the collective knowledge. Reusable components such as algorithms, datasets, models, frameworks, scripts, experimental results, papers, standards, etc. can be easily taken and used for innovation. Extensible metadata and knowledge descriptions based on FAIR principles (findability, accessibility, interoperability and reusability) have been under development globally for international collaborations and locally for large-scale businesses **[119]**. The final goal is to facilitate the data and knowledge acquisition for researchers, practitioners and other users in production of values through R&D practices.

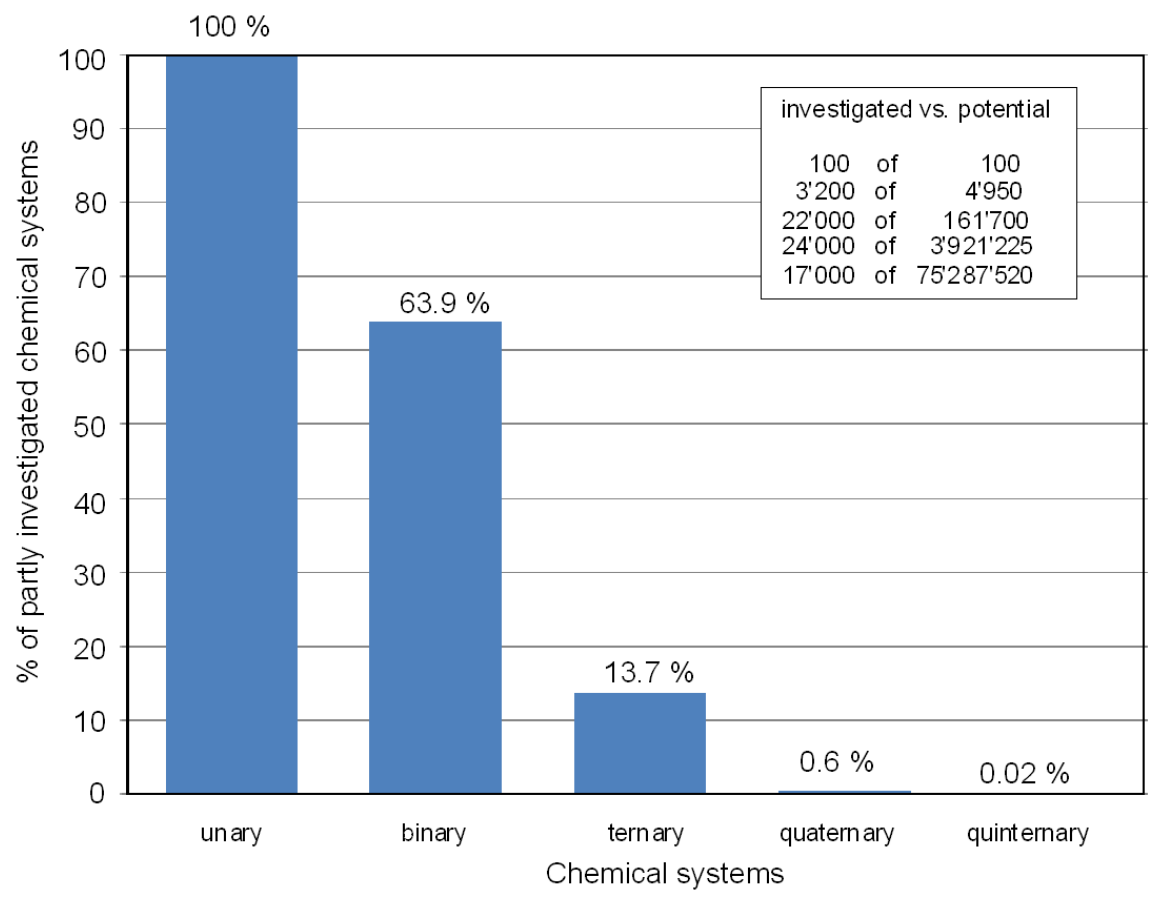

Fig. 46. Percentage of partly or fully investigated chemical systems grouped into unary, binary, ternary, quaternary and quinternary systems. The inset shows the actual numbers of investigated versus potential systems.

However, despite the enormous importance of inorganic substances for our industrialized society in areas such as housing, energy, transportation, civil engineering, communication, and health, mankind's knowledge of inorganic substances is astoundingly sparse. As seen from Fig. 46 for example, only about 14 % of all possible ternary chemical systems have been at most partly characterized. In the case of inorganic substances containing four or more chemical elements, this fraction drops to 0.6% or less **[9]**. In fact, the remarkable accomplishments in the development of the advanced technologies (such as aircraft engines, computer processors, magnetic recording devices, battery materials, or chemical catalysts) rely on the optimization of physical properties of inorganic substances. In this review we focus on inorganic substances. On the one hand it is worth noticing that the number of experimentally investigated inorganic substances is very low. But on the other hand the ability to simulate crystallographic structures and its physical properties for inorganic substances from first-principles methods, and phase diagrams with CALPHAD method is rapidly growing.

These two facts motivated us to build up the Materials Platform for Data Science (MPDS) [17a,b], an online database system consisting of two major inter-linked data parts: the experimentally determined data part (peer-reviewed), to be used as reference (from the world literature, see Chapter 4), and the calculated data (using first-principle simulations or machine learning high-throughput calculations, see Chapters 8 and 9) data part.

The engineering materials are typically the multi-phase materials, which are further affected by defects, interfaces, and the other microstructure features. Nevertheless the fundamental basis of all these engineering materials is made from the individual inorganic substances, which can explore a new dimension of self-assembling of structural primitives prepared by bottom-up and top-down approaches powered by ICT. Verifications of equivalence between representations of experimental data and calculated data and bridging gaps for identified differences are to be organized systematically and practically to satisfy each requirement of materials utilization. There are many unsolved theoretical issues on the equivalence and bridging gaps, namely, analogue and digital, continuous mathematics and discrete mathematics, data and models, parts and whole, fact and logic, intention and extension, element and set. One of trials in this context is to bridge a dataset and structural characterizations by calculated parameters of *ab initio* calculations, verifying the rewriting of the data set by words and logics as trustworthy [120]. Another challenge concerns are developing transfer models between materials complexities by iterations of the above direct (assembling) and inverse (disassembling) methods to converge into a solution [121], which is becoming feasible by the high quality database as explained in the following chapters.

Neither the Fourth Paradigm of Science [70] nor the Materials Genome Initiative [71] can be realized without the integration of restraints obtained by *'inorganic substances data exploration searching for principles (governing factors) with the aim to formulate restraints'*, as will be shown in this chapter. The infinite number of potential chemical element combinations forces us to develop approaches that are able to reduce this infinite number to a practicable number of the most probable potential inorganic substances (chemical systems), to be theoretically and experimentally investigated. In addition, it is essential to note that the realization of the Materials Platform for Data Science (MPDS) requires two preconditions:

*i) The first requirement is the introduction of the Distinct Phases Concept to link different kinds of inorganic substances data.*
This concept was introduced for the PAULING FILE material database system [8,9,18], and is implemented in its derived products [19,62,64]. A *distinct phase* is defined by the chemical system and the structural prototype, and has been given a unique name by a representative chemical formula and, when relevant, a modification specification. As the linkage of different groups of data was considered as the most important requirement, the PAULING FILE was designed as *a distinct phase-oriented* database, using as a key of a relational database system. This was achieved by the creation of a *Distinct Phases Table*, as well as the required internal links. In practice this means that each chemical system has been evaluated and the *distinct phases* identified based on all available information. Finally every database entry has been linked to such a *distinct phase*. This is also a requirement for linking materials databases created by different teams.

*ii) The second requirement is the existence of a comprehensive, critically evaluated inorganic substances database system of experimentally determined single-phase inorganic substances data from the world literature, to be used as reference.*
The PAULING FILE was launched about 30 years ago and represents meanwhile the sole and therefore world-largest data collection of its kind with 1,563,946 entries (subdivided into crystallographic structure, diffraction pattern, constitution, and physical properties) extracted from 309,460 scientific publications, covering all the inorganic substances (i.e. without C-H bonds). It is now becoming feasible to use it as a starting reference. The PAULING FILE data have been carefully checked, and fully standardized in a self-explanatory manner [23,28,31].

### 7.2. Exploration of materials data searching for principles (governing factors) to formulate restraints

One of the most challenging tasks in materials science is the design of novel inorganic substances with beforehand-defined physical and (or) chemical properties. In order to reach this objective, as already many times mentioned, in general, two different approaches are explored, the bottom-up and top-down [8]. For both the approaches, to make predictions it is necessary to start from experimental variables such as the selection of the chemical elements to be combined, their concentrations, temperature, and pressure. For the top-down approach we use, in addition, the laws of quantum mechanics, including the APP: atomic number (AN) (as well as some adjustable computation parameters), whereas for the bottom-up approach we use tabulated APPs (or APPEs), such as the atomic number (AN), periodic number (PN), atomic size $SZ_a$(AN,PN), atomic reactivity $RE_a$(AN,PN), and atomic affinity $AF_a$(AN,PN) [8].

For both approaches the key-position is the structure prototype (and (or) coordination type, AET) classification of the inorganic substances, which represents our 'window' to view the electronic interactions of the atoms within a specific inorganic substance. To simplify our discussions we use here, in first priority the structure prototype classification, which is the classical view. Presently 39,990 different structure prototypes have been experimentally established for inorganic substances. In other words, nature realizes from the symmetry point of view about 40,000 different geometrical arrangements of atoms, for inorganic substances without C-H bonds.

During the past 30 years several strong patterns in inorganic substances overviews – APP or APPE maps [1,3,8], considering thousands of data sets of different chemical systems / inorganic substances (summarizing the content of thousands of scientific publications), have been published. This proves that the underlying quantum mechanical laws can be parameterized by the use APP or APPE of its constituent chemical elements. The search for optimal APP or APPE led to relatively simple maps with well-defined stability domains, giving excellent 'holistic views' of experimental data for known inorganic substances. As a direct consequence the maps give some prediction ability.

Before initiating the PAULING FILE project, one of us (P.V.) reviewed the world literature, focusing on intermetallics and alloys, in context with the topic 'Factors Governing Crystallographic Structures' [2], and came up with nine principles. The validity of these principles was tested on a fair number of inorganic substances in well-defined groups of experimentally determined data sets and showed accuracy in the range of 90-100%. It was concluded, we can rely on the predictions based on those nine principles with considerable confidence. Applying the nine principles will both reduce the number of systems (or samples) to be investigated and remarkably increase the success rate of finding new inorganic substances. Now, about 30 years later, having access to five times more experimental data with coverage extended from intermetallics and alloys to include all other inorganic substances such as *e.g.* halides and ceramics, the situation has significantly improved. The PAULING FILE contains structural information for over 170,000 *distinct phases* (of which about 50% contain oxygen), compared to Pearson's Handbook of Crystallographic Data for Intermetallic Phases [59], which covered about 28,000 intermetallics and alloys (including sulfides and selenides).

In this chapter we focus on the 1,000 most populous structure prototypes and their representatives. This covers 75% of the entries in the MPDS release 2020/21 [17a,b]. Tables 12a + b give some details about the content of the crystallographic structure part of the MPDS release 2020/21, also called PCD [62].

The main aim of this chapter is on the one hand to test the nine principles proposed in [2] (based on about 17% of the by now available facts) and, if necessary, to modify them, and on the other hand to discover additional principles. Finally we ended up with in total twelve principles, derived either from statistical plots or from inorganic substances overview APPs or APPEs maps. Each of them reveals governing factors, which could be used to formulate restraints.

### 7.3. Nature defines cornerstones providing a marvelously rich, but still very rigid systematic restraint framework

Below we outline four fundamental cornerstones defined by nature. The first cornerstone is responsible for the fact that we are confronted with infinitely many chemical element combinations. The second to fourth cornerstones provide a very rigid systematic framework of restraints. Each of the latter is supported by four principles, *i.e.* in total twelve principles. Since the cornerstones reflect underlying natural laws, they have general validity. For comparison purposes, we will briefly outline the nine principles proposed earlier [2] and, when relevant, their extensions, and describe the newly discovered three principles.

### 7.3.1. First cornerstone: Infinitely many chemical element combinations

The experimental data in the PAULING FILE shows that the materials knowledge rate, by comparing the number of potential chemical element combinations with the number of chemical systems where we have at least partial information, is low. This is shown in Fig. 46 for unary to quinternary systems, and it is clearly seen that we have robust knowledge only for unary and binary systems. For ternary systems we are still at the very beginning and for higher systems we have close to no knowledge. This is surprising as, since 1980, materials scientists are mainly working on ternaries and higher systems. Going to higher-order systems leads to an astronomically large number of potential chemical systems to be considered. Fig. 47 shows that the average number of inorganic substances per binary chemical system is 5.3, for ternaries 3.9 and for quinternaries 2.3 and higher systems 1.2 (over 200 common stoichiometric ratios occur). In the following we will use the expression *basic structure prototype* for a structure prototype that has no mixed occupancy. Basic structure prototypes and their representatives are here referred also as daltonide inorganic substances (no significant homogeneity range), which may not always be true. Experimentally investigated systems reveal up to maximum 30 daltonide inorganic substances per ternary chemical system. Fig. 48 shows for all 36,745 basic (no mixed site occupancy) ternary inorganic substances, which belong to one of 490 distinct different stoichiometric ratios represented in a ternary concentration triangle.

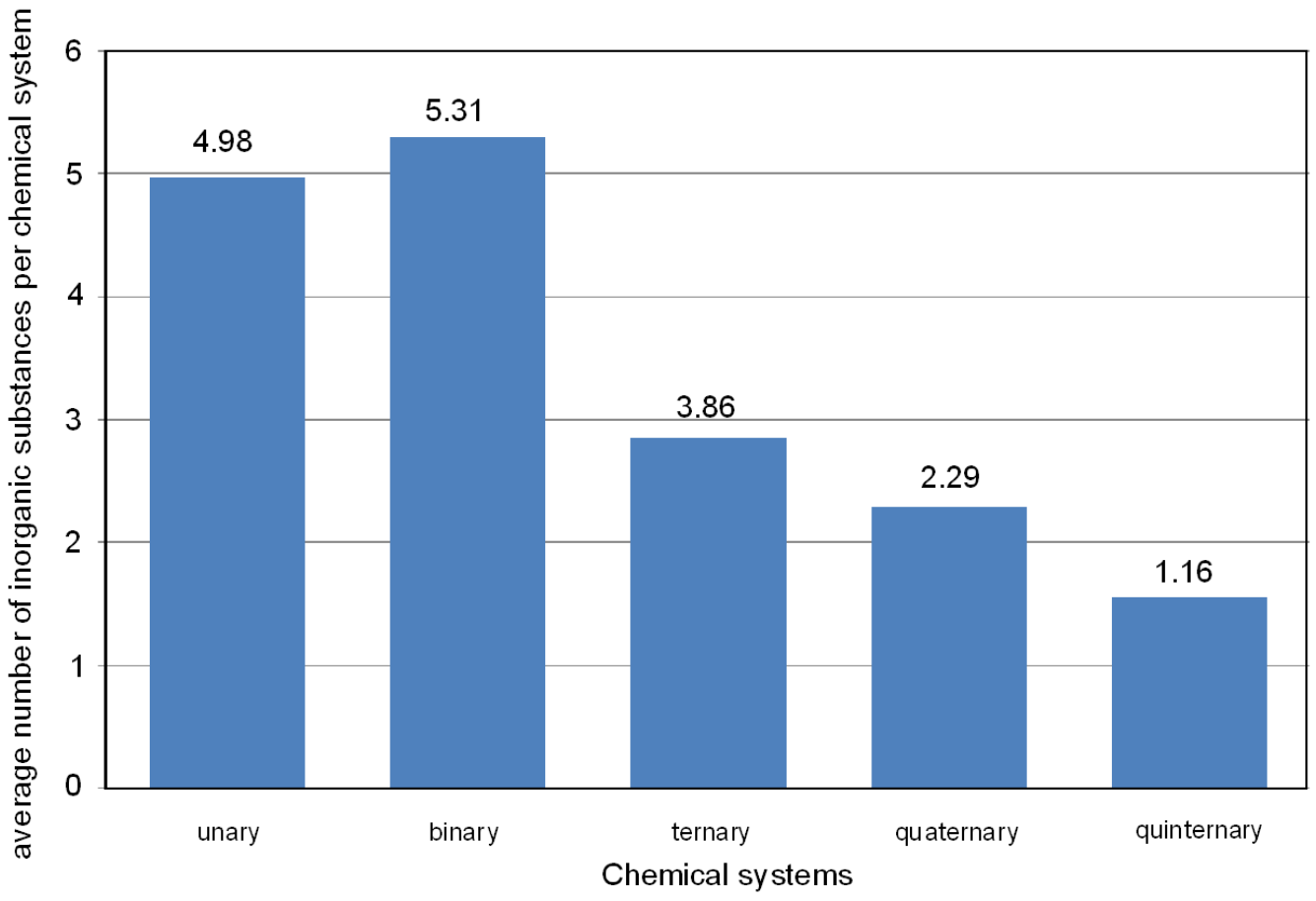

Fig. 47. Average number of inorganic substances per chemical system grouped into unary, binary, ternary, quaternary and quinternary systems.

Focusing on potential inorganic substances, going to higher-order systems leads in practice, with an average of about 2 inorganic substances per system, to an infinitely large variety of potential inorganic substances to be considered. Fig. 49 shows that the number of structure prototypes per literature year has increased approximately linearly from 100 structure prototypes per year in 1960 to 1,100 structure prototypes per year in 2010. Nature has so far realized about 40,000 structure prototypes for inorganic substances, and continued increase will lead to a huge variety of 3D-ways to arrange the atoms within a specific inorganic substance. Figs. 46-49 support, based on the data of [19], the overall conclusion of the first cornerstone: '*Nature provides us with 100 chemical elements as well as their combinations. A direct consequence of this fact is that there exist an infinite number of chemical element combinations. Furthermore, nature has worked out a huge number of three-dimensional ways of ordering the chemical elements (atoms) within inorganic substances. It may be added that the magnetic moments of the chemical elements can be ordered in an even higher number of four-dimensional ways.*' This leads, in practice, to a hopeless situation when trying to develop overall-valid efficient experimentation and calculation exploration strategies.

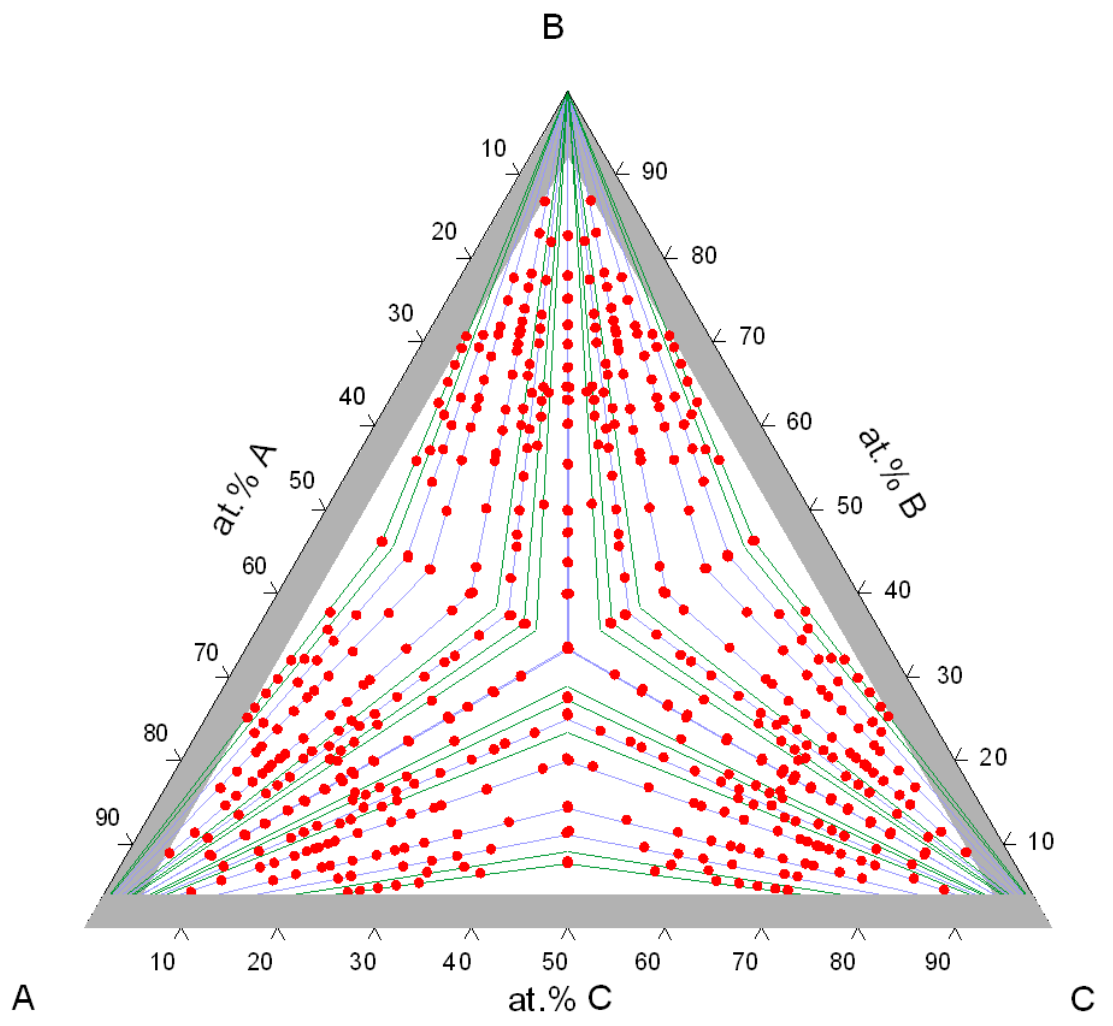

Fig. 48. 490 different stoichiometric ratios of the 36,745 daltonide ternary inorganic substances shown in a ternary concentration triangle.

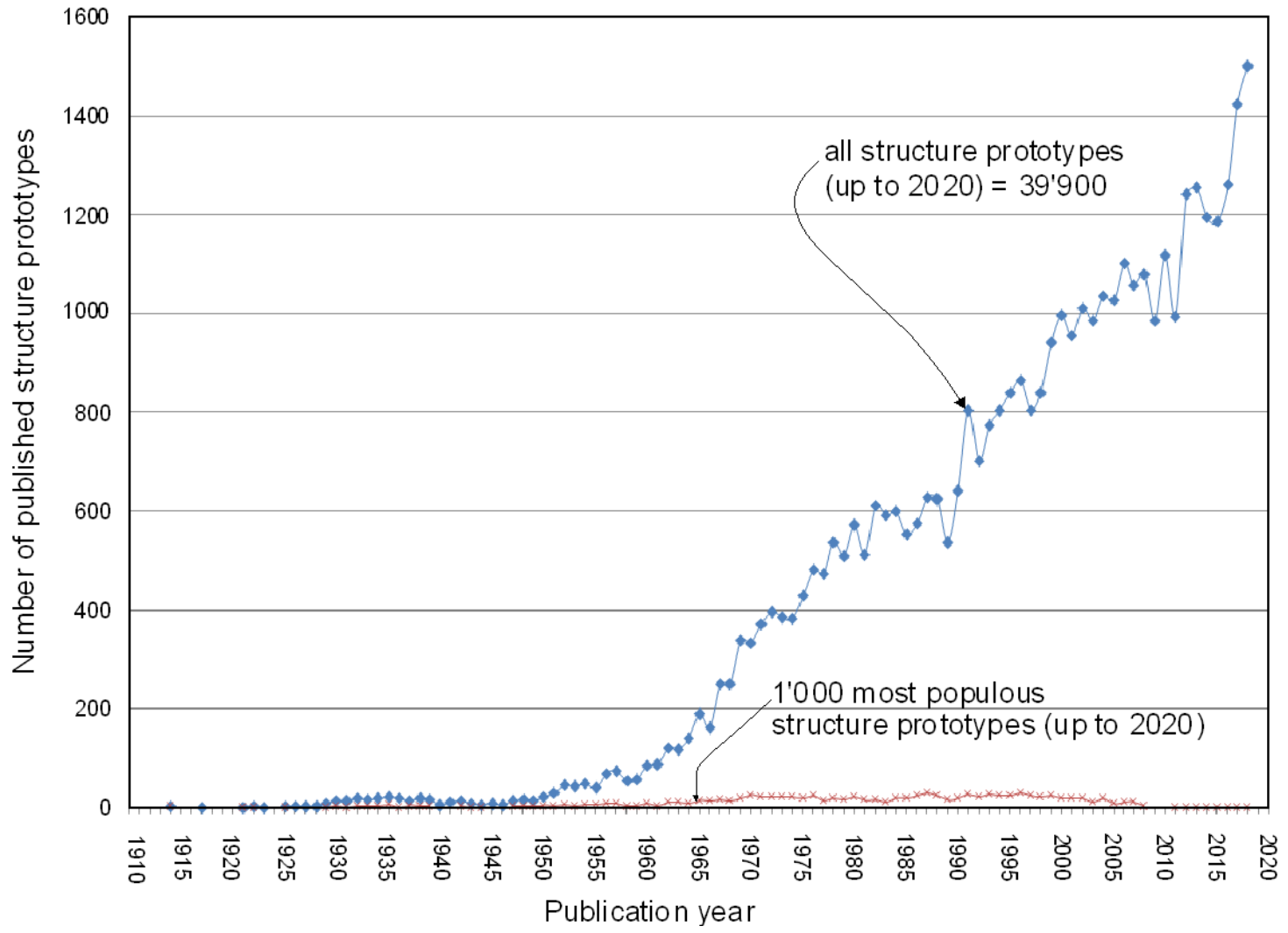

Fig. 49. Number of structure prototypes as a function of the publication year. The blue curve considers all prototypes, the brown curve consider the 1,000 most populous prototypes in PCD-2020/21 [62].

### 7.3.2. Second cornerstone: Laws that define inorganic substance formation

The second cornerstone of nature sets strict restrictions for the formation of inorganic substances. The enthalpy of formation has to be negative; otherwise the chemical element combination will not lead to inorganic substance. In addition, at constant pressure, Gibbs' phase rule P = C-F+1 defines the relation between the number of phases (P, here potential inorganic substances), the number of components (C, here chemical elements) and the degree of freedom (F) of intensive properties such as temperature and composition. It is indirectly possible, with the help of inorganic substances 'holistic views' – APP or APPE maps and statistical plots to formulate four principles. Thanks to these four principles it is possible to derive restraints, which can exclude non-former systems (chemical systems where no inorganic substance form), and some pre-conditions to be fulfilled for inorganic substance formation.

*7.3.2.1. Inorganic substance-formation map principle (first principle)*

Our work from 2008 [8] showed that the periodic number ($PN_{ME}$, $PN_{MD}$) is by far the most efficient APP in separating formers (chemical systems where at least one distinct inorganic substance exists) from non-formers. PN represents a different enumeration of the chemical elements within the periodic system (see Tables 2 and 3). In contrast to the atomic number (AN), PN emphasizes the number of valence electrons, *i.e.* the periodicity. Fig. 50 shows an idealized two-dimensional inorganic substance former/non-former map for binary A–B systems, using as axes the periodic number of the chemical element A($PN_A$) *versus* the periodic number of the chemical element B($PN_B$). PN as APP works well for binary, ternary, and quaternary systems. However, it was shown in the same publication [8] that the atomic size $SZ_a$, the atomic reactivity $RE_a$, as well as the atomic affinity $AF_A$ (in this work), are a direct function of PN and AN and strongly reflects the periodicity (valence electrons) of the periodic system.

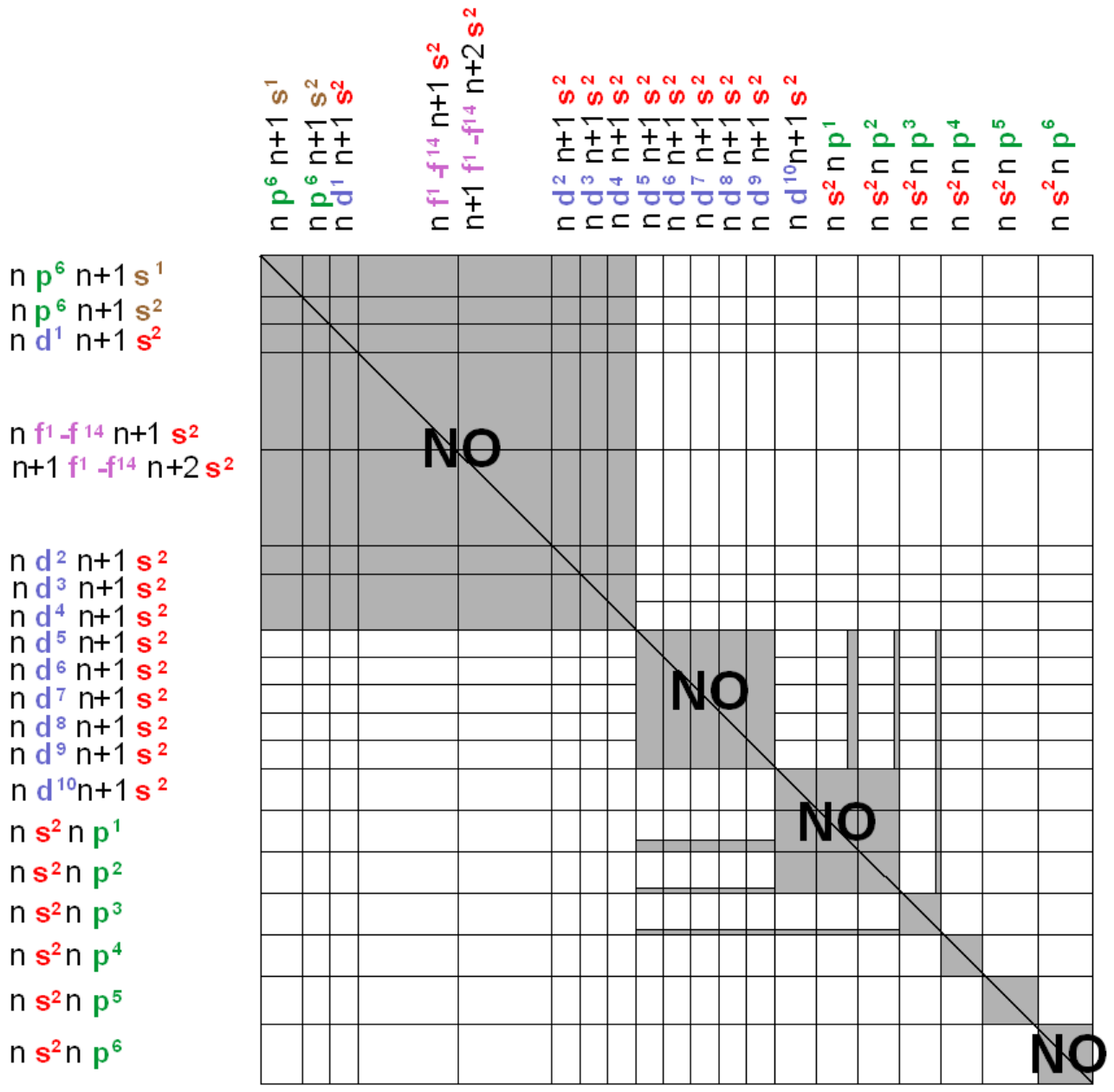

Fig. 50. Schematic two-dimensional inorganic substance-formation map for binary A-B systems using as axis: $PN_A$ versus $PN_B$.

Therefore, $SZ_a$, $RE_a$, and $AF_a$ can also be used as APPs. Compared with the earlier work of Miedema for binaries [72], and the work of Villars for ternaries [73], the maps using PN separate the domains more clearly, with an overall accuracy of more than 98%. The advantage of the APP PN with respect to parameters such as radius or reactivity is that it is an integer number.

The inorganic substance former/non-former map principle was formulated in 1994 [2] as follows: '*The valence electron, size, electrochemical factors are the factors governing inorganic substance formation*.' The inorganic substance former/non-former maps predict inorganic substance formation for binary [72], and for ternary [73] systems. This principle has been fully verified, and in addition, during the last 30 years, has been significantly improved in its simplicity and accuracy in separating former from non-former systems. Overall one can predict that about 30% of all chemical element combinations will form no inorganic substances (non-formers), this being true for binary, ternary, and quaternary systems.

*7.3.2.2. Number of chemical elements – AET correlation principle (second principle)*

Fig. 51, a 3D-plot, shows the distribution of point-sets in % (within unary, binary, ternary, quaternary, quinternary, sixternary) focusing on the 27 most populous AETs (sorted by increasing coordination number (CN)). In Fig. 51 (in blue color) we see that overall the quaternary, quinternary, as well sixternary inorganic substances behave very similar, this means in other words going higher than quaternaries will not lead to something structurally fundamental new. Focusing on quaternary to sixternary inorganic compounds Fig. 52 shows the distribution of point-sets in % *versus* the 12 most populous atomic environment types (AETs) with CN< 7 (sorted by increasing coordination numbers (CN)) and *versus* the valence electron groups: $p^1$, $p^2$, $p^3$, $p^4$, $p^5$, $p^6$, $s^1$, $s^2$, $d^{1\text{-}10}$, $f^{1\text{-}14}$- elements. In quaternary and higher-order inorganic substances p-elements (especially $p^3$- $p^5$ elements) occupy the majority of the point-sets. $s^{1+2}$, $d^{1\text{-}10}$, as well as $f^{1\text{-}14}$ elements show an overall low frequency, with the exception of point-sets having as AET the tetrahedron or the octahedron for $d^{1\text{-}10}$ elements as central atoms.

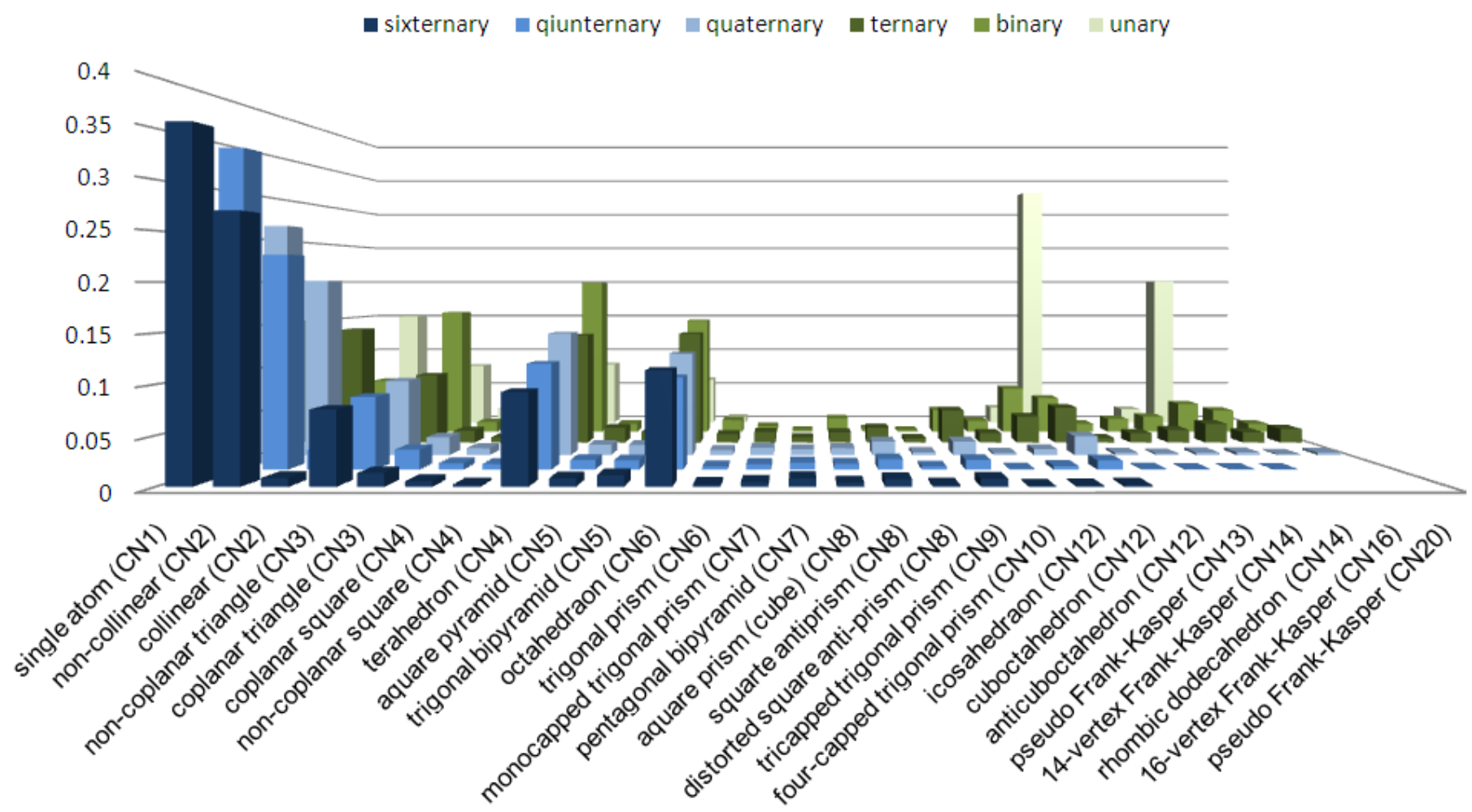

Fig. 51. Frequency of the number of point-sets in percentage (grouped in unary, binary, ternary, quaternary, quinternary, and sixnary phases) versus the 27 most populous atomic environment types (AETs), sorted by increasing coordination number (CN).

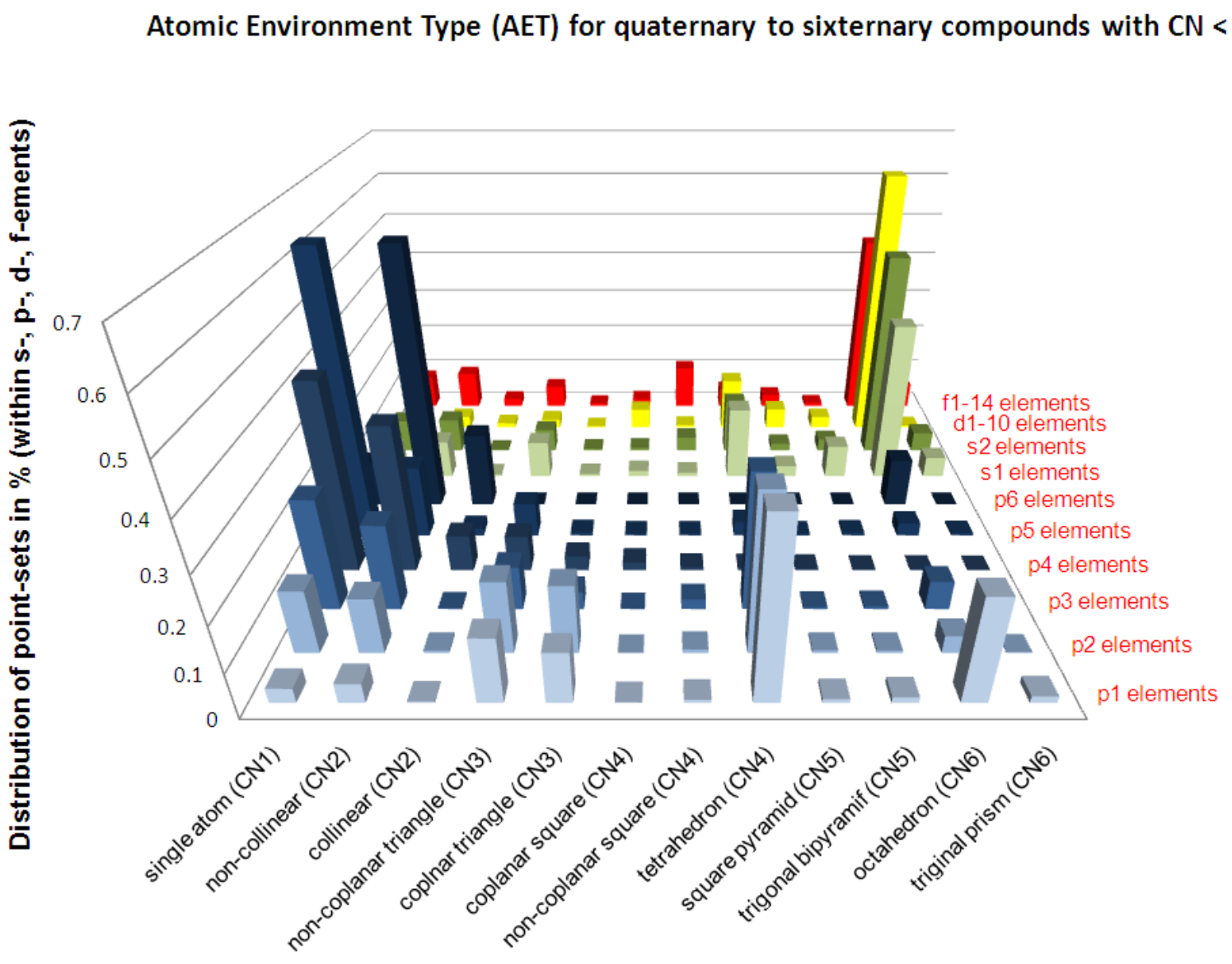

Fig. 52. Number of point-sets (grouped in quaternary, quinternary, sixnary phases) versus the 9 most populous atomic environment types (AETs), sorted by increasing coordination numbers (CN) versus element group: $p^1$, $p^2$, $p^3$, $p^4$, $p^5$, $s^1$, $s^2$, $d^{1-10}$, $f^{1-14}$-elements.

Thus, the summary of this newly discovered principle is as follows. The maximal AET diversity is reached within binary and ternary inorganic substances. Unaries (chemical elements) prefer AETs with high coordination number (CN= 12 or higher). Quaternary and higher-order inorganic substances strongly prefer AETs with low coordination number (CN= 6 or lower) with the central atoms being mainly $p^1$- to $p^6$-elements preferring single atom (CN= 1) and non-collinear (CN= 2), whereas $s^1$-, $s^2$- ,$d^{1-10}$ and $f^{1-14}$ elements predominately have as AET tetrahedron (CN= 4) or/and octahedron (CN= 6).

In other words geometrically different structure prototypes can be achieved with two, three or four different chemical elements within inorganic substances. Surprisingly this diversity going from four to six does no increase the diversity.

*7.3.2.3. Active concentration range principle (third principle)*

This principle was formulated in 1994 [2] as follows: '*Active composition range means that at least 5 at.% of minority elements is needed to form an inorganic substance.'* This is valid for binary, ternary and quaternary inorganic systems. As a consequence, in binary systems 10%, in ternary systems 15% and in quaternary systems 38.6% of the available concentration range (area, volume) is 'inactive' for the formation of daltonide substances (see gray area in Figs. 53 and 54). This principle has been fully confirmed.

*7.3.2.4. Stoichiometric ratio condition principle (fourth principle)*

By now 64% of all binary systems have been investigated and we can state that 95% of all binary daltonide phases crystallize in one of the following 10 stoichiometric ratios: $AB_x$ where $x$ = 6, 5, 4, 3, 2, 1.67, 1.5, 1.33, 1.25, and 1 (see Fig. 53). For ternaries, 30 years ago it was only possible to list 7 most often occurring stoichiometry ratios. Meanwhile 99% of the 36,745 daltonide ternary phases respect the following ternary stoichiometric ratio condition (see Fig. 54): $AB_xC_y$, where $x$ takes one of the 10 above listed values for the most frequent binary stoichiometric ratios and $y$ is equal to $x$ or an integer divided by 1, 2, 3, or 4, and larger than $x$. In combination with the active composition range principle, this leads to 1 stoichiometric ratio for $AB_xC_y$ ($x$=$y$=1) inorganic substance, 9 possible stoichiometric ratios for $AB_xC_y$ ($x$=$y$) and 819 possible stoichiometric ratios for $AB_xC_y$ ($x$<$y$). For the stoichiometry 1:1:1 there is only one possibility within a given ternary system *A–B–C*, but for 1:*x*:*y*, *x*= *y* there are three possible substances: $AB_xC_x$, $A_xBC_x$, and $A_xB_xC$, and for 1:*x*:*y*, *x*≠*y* six possibilities. This leads in total to 4,636 stoichiometric ratios assuming *A*≠*B*≠*C*, of which 490 have so far been found experimentally. It is worth mentioning that stoichiometric ratios following certain additional conditions are highly preferred: $AB_xC_y$, *x* equal one of the 10 above listed most frequent values for binary stoichiometric ratios, *y* equal *x* or an integer larger than *x*. With these conditions we end up with 769, compared with the 490 (64%) so far experimentally found stoichiometric ratios. The stoichiometric ratio condition principle was formulated in 1994 **[2]** as follows: '*The vast majority of daltonide substances have each point-set occupied by just one kind of chemical element. This leads for binaries to the following highly preferred stoichiometric ratios: 1:1, 4:5, 3:4, 2:3, 3:5, 1:2, 1:3, 1:4, 1:5, and 1:6.'* This has been fully verified in this work. For ternary intermetallic substances, the stoichiometric ratios 1:1:1, 1:1:2, 1:1:3, 1:1:4, 1:2:2, 1:2:3, and 1:2:4 predominate. In this work the above given ternary stoichiometric ratio condition was discovered.

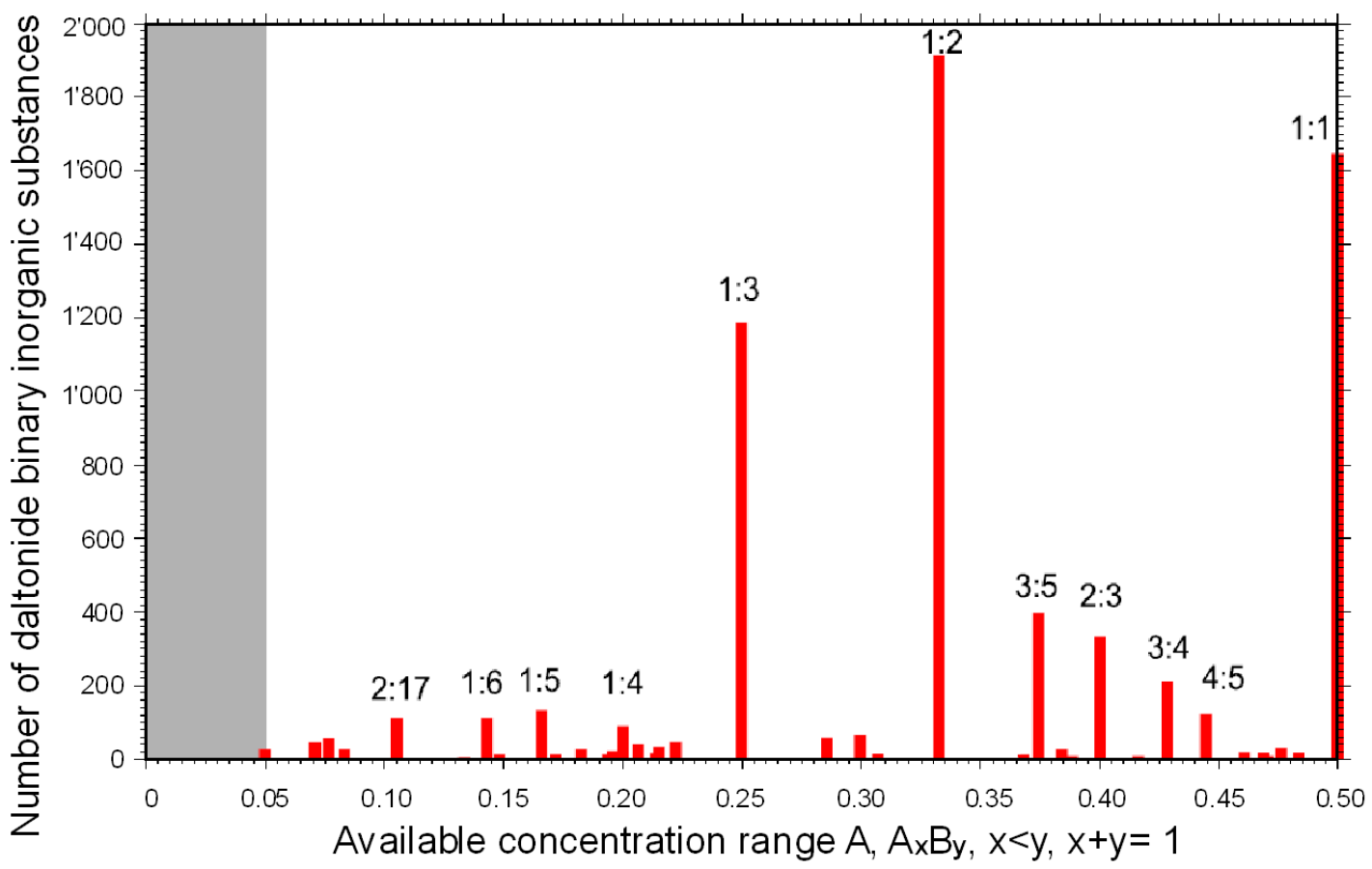

Fig. 53. Number of binary daltonide inorganic substances $A_xB_y$ (x< y, x+y= 1) versus x- concentration indicate frequent m:n ratios ($A_mB_n$ where m and n are integers); the gray color shows the composition range where no binary phases occur.

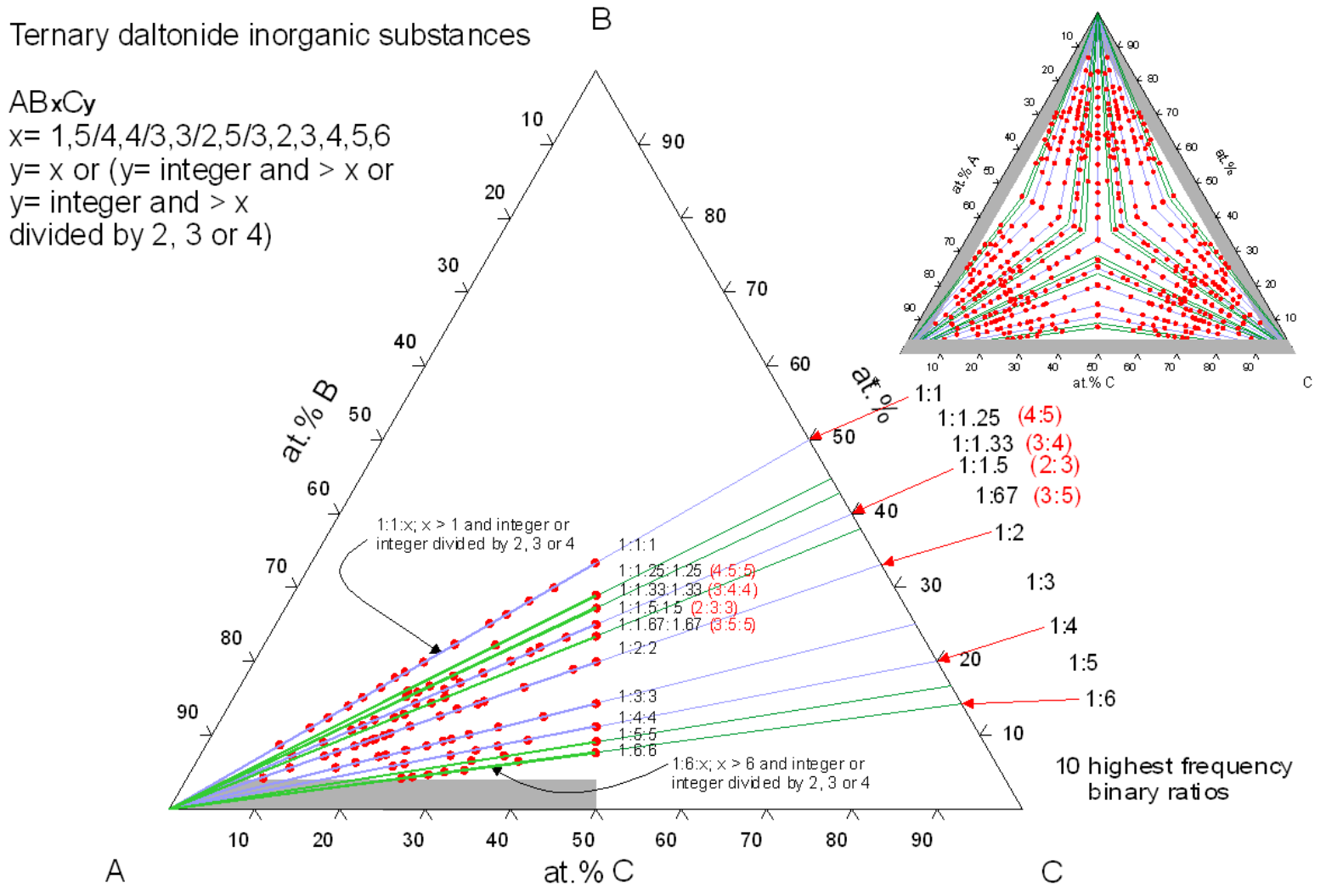

Fig. 54. Occurrence of daltonide non-organic phases in the available concentration range for ternary systems ($A_xB_yC_z$, x<y< z, x+y + z= 1): the gray area shows where no ternary phases occur.

The four principles demonstrate with the help of Figs. 50-54, that the overall conclusion of the second cornerstone is correct: '*nature provides us with a very rigid systematic framework of restraints, such as laws that define inorganic substance formation.'* Most efficient in reducing the number of potential chemical element combination is the fact that we can focus on unary to quaternary systems, since with them the full diversity of AETs (as well as structure prototypes) is achieved. *Quinternary and higher-order inorganic systems will not lead to inorganic substances with greater diversity, and therefore not to the discovery of additional principally different geometrical atom arrangements. In other words, all basic structure prototypes are realized within inorganic substances having four or less chemical elements.*

The number of potential chemical element combinations (chemical systems) for unaries to quaternaries is 4,087,975. About 30% of these systems are non-formers, which leaves about 2,900,000 combinations. Taking the average number of inorganic phases for unaries as 5, for binaries as 6, for ternaries as 4 and quaternaries as 2, leads to less than 6,000,000 inorganic substances, preferring one of 10 stoichiometric ratios for binaries, or obeying the stoichiometric ratio condition for ternaries. Taking into consideration the fact that there exist from the electronic point of view only four major groups of chemical elements (*s*-, *p*-, *d*- and *f*-elements) supports the above said. From the chemical point of view the binaries to quaternaries should represent the maximum of potential diversity that can be achieved. The basic structure prototypes and their basic representatives among quinary and higher-order inorganic substances cannot be built up with chemical elements from five (or more) distinct different chemical element groups, since there exist only four groups, therefore one or more chemical elements will have to belong to the same group of elements.

### 7.3.3. Third cornerstone: Laws that define ordering within inorganic substances

The third cornerstone of nature sets strict restrictions on the ordering of the chemical elements within a particular inorganic substance. When chemical elements combine to form solids, their crystallographic structures are beautifully rich, yet systematic patterns underlie this process. The most striking manifestation of this fact is the existence of so-called structure prototypes of inorganic substances, which can be understood as geometrical templates for large groups of inorganic substances, e.g. the NaCl,*cF8*,225 structure prototype has presently 1,550 different representatives. In other words, different inorganic substances belonging to the same structure prototype are geometrically very similar to each other. As discussed in the previous chapter 5. there exist, in principle, two main approaches to the classification of crystallographic structures, one considers as first criterion the overall symmetry (*e.g.* the Wyckoff sequence) **[31]**, the other focuses on each site (leads to the AET- or 'coordination type' classification) **[27]**. The first classification requires that the published crystallographic data are fully standardized **[23]**, which is consequently done in the PAULING FILE **[8,9,18]** and the products derived from it, *e.g.* Pearson's Crystal Data **[62]**. The following four principles demonstrate, with the help of four statistical plots and one fundamental crystallographic consideration, the correctness of the third cornerstone.

#### *7.3.3.1. Simplicity principle (fifth principle)*

Fig. 55 shows that the large majority of all structure prototypes of inorganic substances have less than 50 atoms per unit cell, with a maximum of types having around 20 atoms per cell. The number of point-sets per structure prototype is for the majority lower than 6, with a maximum around 2-3 point-sets per structure prototype (see Fig. 56). This principle was formulated in 1994 **[2]** as follows: '*The vast majority of the intermetallic substances have less than 24 atoms per unit cell.'* By the inclusion of other inorganic substances 24 has become 50. '*In addition the majority of all crystal structures have three or fewer AETs within the crystallographic structures (single-, two-, and three-environments types).'* This is still supported. An analog observation can also be made focusing on the number of point-sets instead of the number of different AETs.

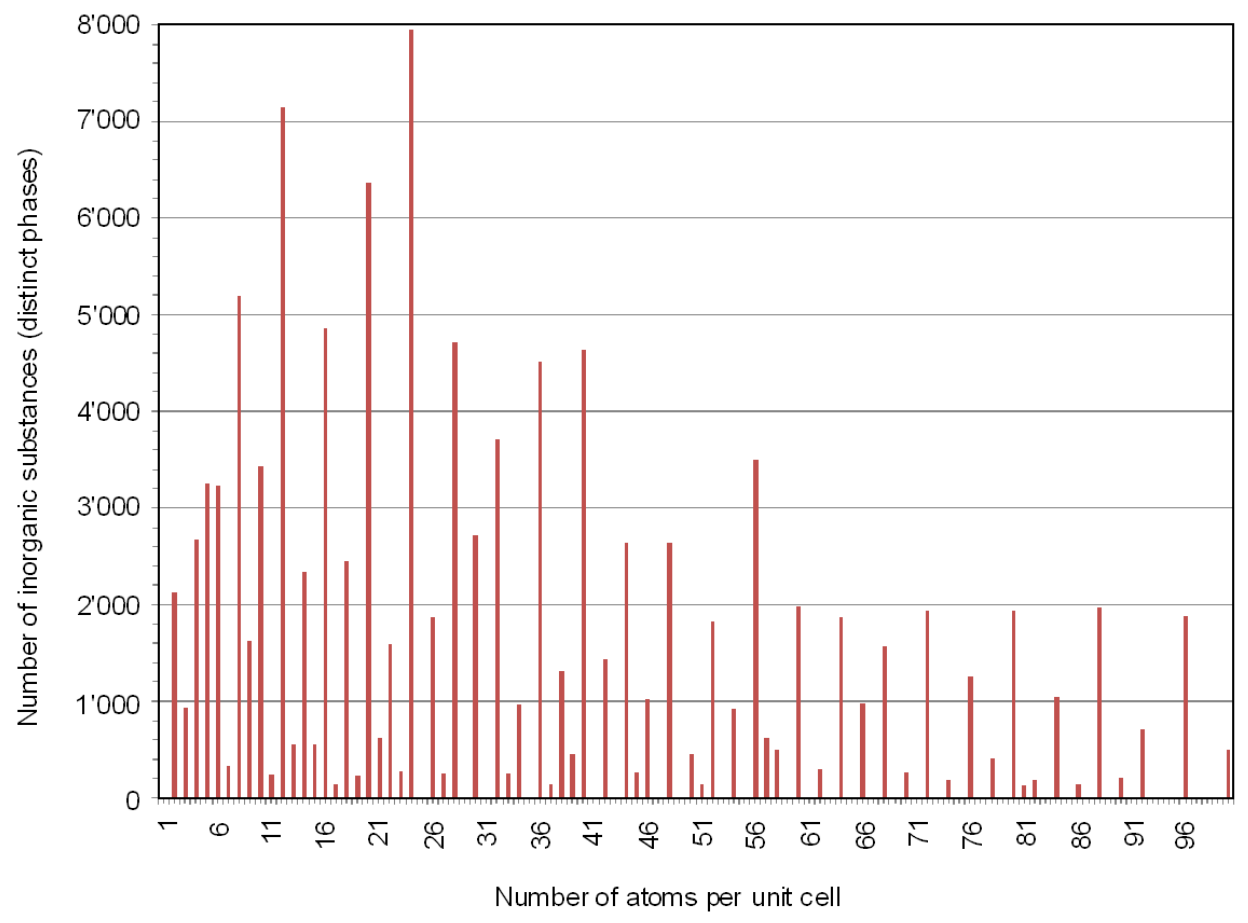

Fig. 55. Number of inorganic substances (distinct phases) versus number of atoms per unit cell considering the 1,000 most populous prototypes and their representatives.

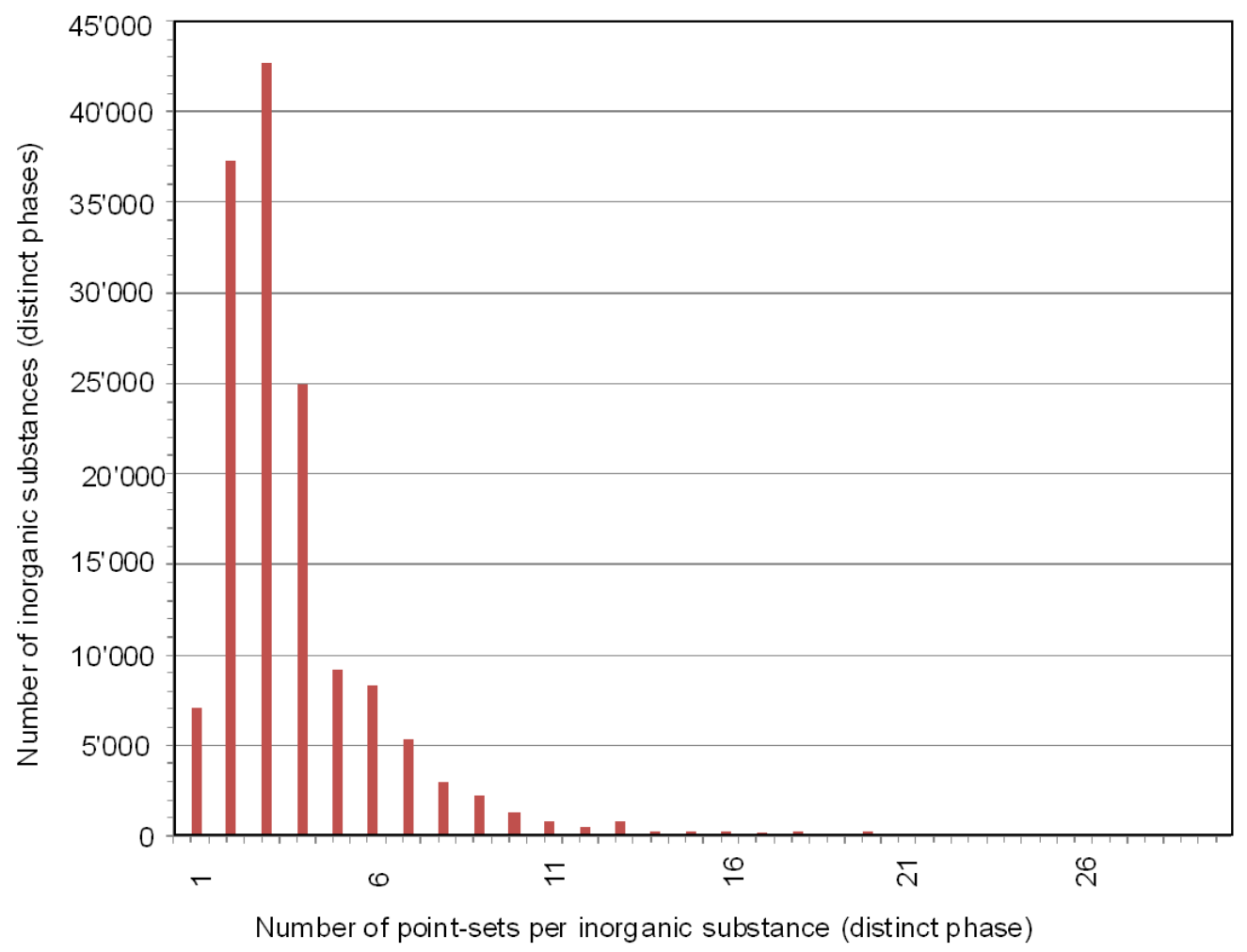

Fig. 56. Number of inorganic substances (distinct phases) versus number of point-sets in the structure, considering the 1,000 most populous prototypes and their representatives.

*7.3.3.2. Symmetry principle (sixth principle)*

The symmetry principle, which was formulated in 1994 [2] as follows: '*The vast majority of all intermetallic substances and alloys crystallize in one of the following 11 space groups: 12, 62, 63, 139, 166, 191, 194, 216, 221, 225, and 227.'* By extending to all other inorganic substances the following space group numbers have to be added: 2, 14, 15, 123, 129, 136, 140, 148, 164, 167, 176, and 229. Within each crystal system a few space groups (the ones with higher symmetry) are preferred. 10% of the space groups cover 67% of the inorganic substances (235,000 of 350,000 inorganic substances). Fig. 57 gives the number of inorganic substances distributed by space group number.

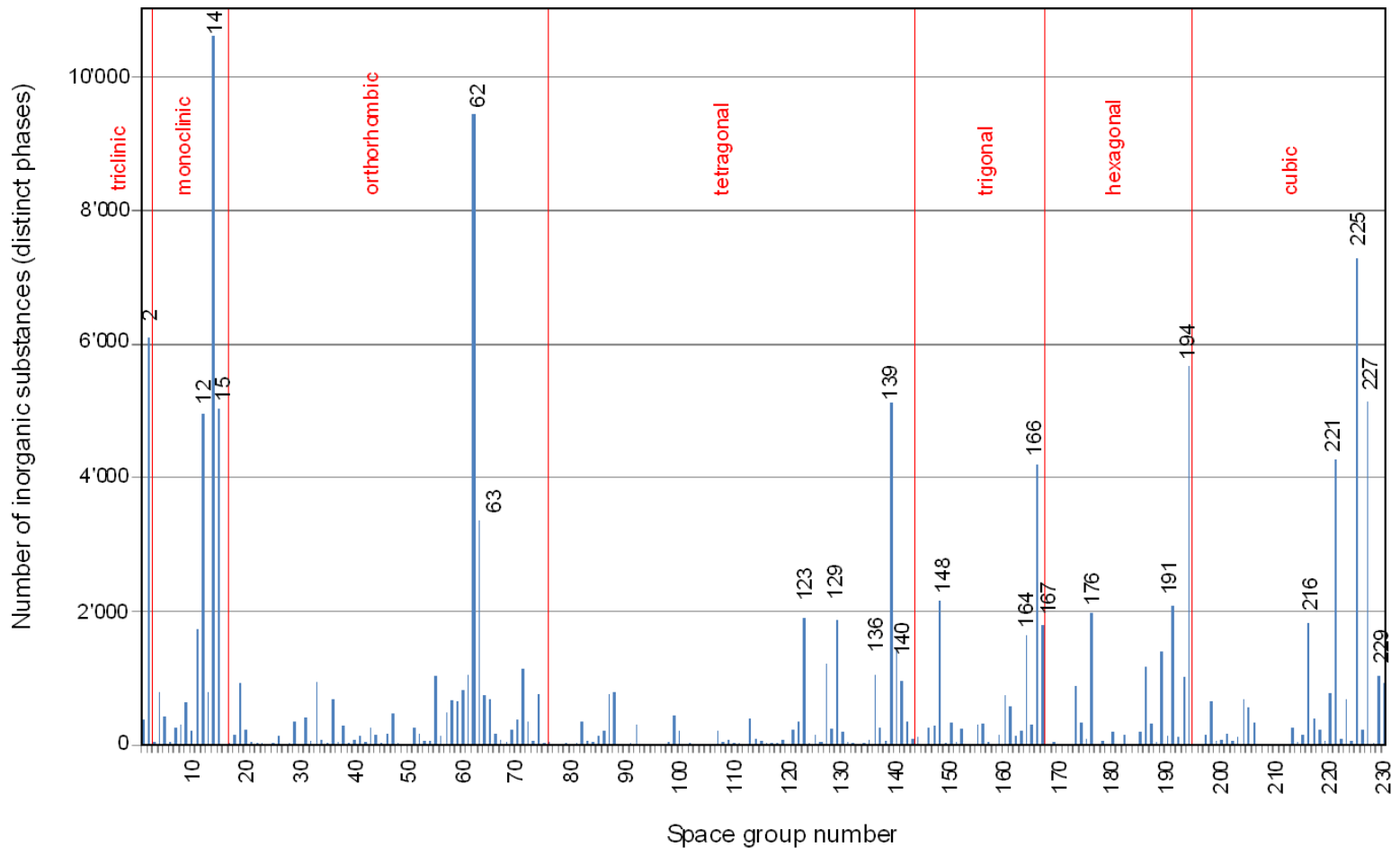

Fig. 57. Number of inorganic substances (distinct phases) distributed by space group numbers considering all entries in PCD-2020/21 **[62]**.

*7.3.3.3. AET principle (seventh principle)*

Fig. 58 shows a frequency plot for the 24 most populous AETs considering the 1,000 most populous prototypes and their representatives. Table 8 lists the 30 most populous AETs among all the refined inorganic substances.

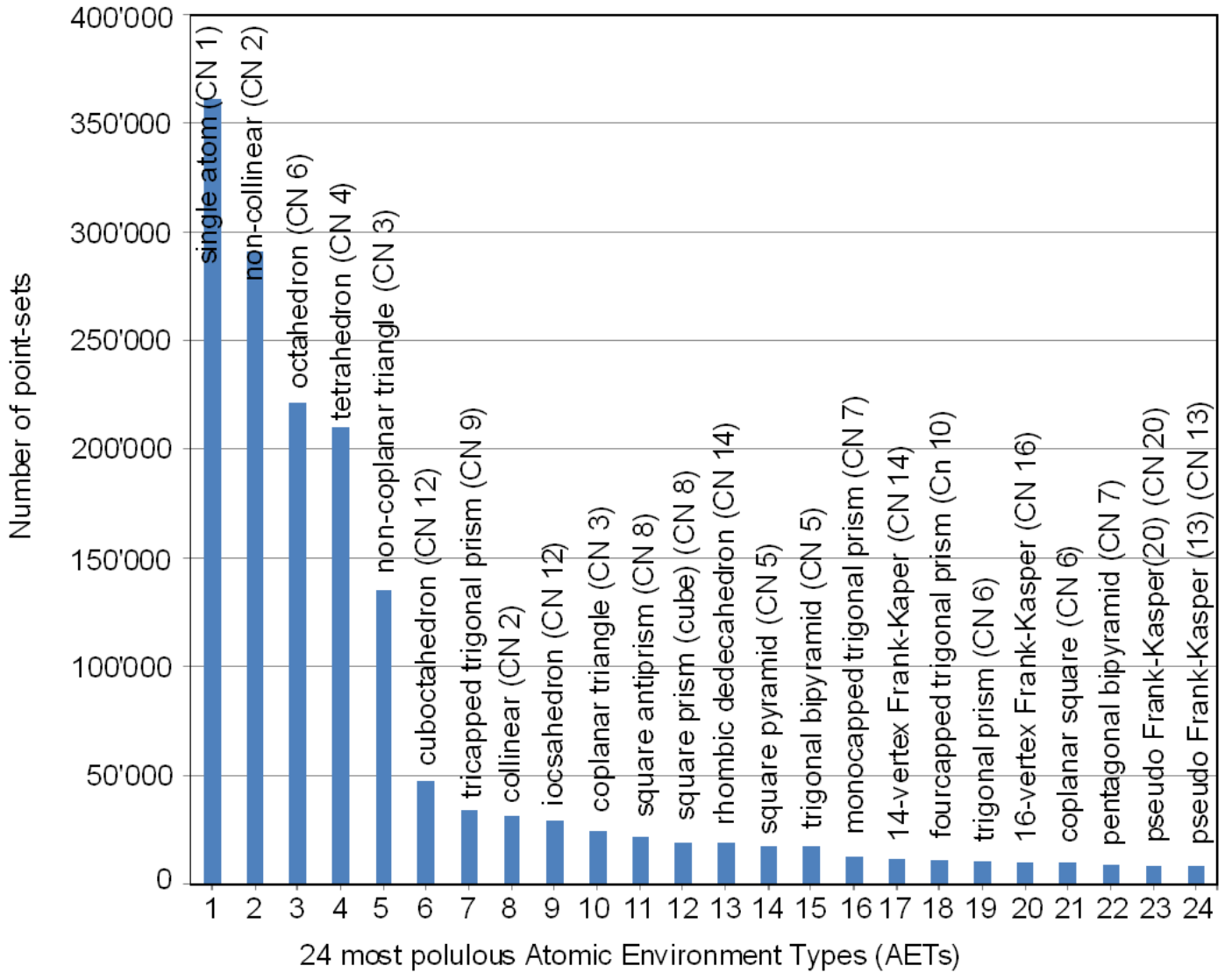

Fig. 58. Total number of point-sets for the 24 most frequent AET's considering the 1,000 most populous prototypes and their representatives.

The AET principle was formulated in 1994 **[2]** as follows: '*The vast majority of all atoms (point-sets) in intermetallic substances have as AET one of 14 polyhedra: tetrahedron, octahedron, cube, tri-capped trigonal prism, four-capped trigonal prism, icosahedron, cuboctahedron, bi-capped pentagonal pyramid, anti-cuboctahedron, pseudo Frank-Kasper (CN13), 14-vertex Frank- Kasper, rhombic dodecahedron, 15-vertex Frank-Kasper, and 16-vertex Frank-Kasper polyhedron.'* The statement made about 30 years ago that 14 AETs are highly preferred is still correct. After having added all other classes of inorganic substances to the intermetallics and alloys, the frequency order has changed (see also Table 8), and the following AETs with low coordination numbers: single atom (CN = 1), collinear (CN = 2), non-linear (CN = 2), coplanar triangle (CN = 3), non-coplanar triangle (CN = 3) and square anti-prism (CN = 8) are now among the 14 most populous AETs. In addition it can be stated that 30 out of 100 possible AETs are highly preferred, and were found for 85% of the point-sets considered here (1,371,000 of 1'532'970 point-sets). Consequently it appears that nature strongly prefers certain AETs, most of them being highly symmetrical (except the AETs with CN = 1 and CN = 2).

### *7.3.3.4. Chemical element ordering principle (eighth principle)*

Table 22 gives the number of point-sets per structure prototype (column) *versus* the number of chemical elements of its representatives (row), considering the 1,000 most populous prototypes and their representatives. Here we recall what we understand under basic structure prototype a structure prototype that has no mixed site occupancy. Looking at the numbers of Table 22 it is possible to conclude that:

- Inorganic substances with *n*-1 chemical elements cannot be representatives of a basic structure prototype containing *n* chemical elements. This is a particular feature of the structure prototype classification used here, where ordered isopointal structure prototypes are distinguished ($CaCu_5$,*hP*5,191 and $PrNi_2Al_3$, *hP*5,191 or $CeCo_3B_2$,*hP*5,191).
- Inorganic substances with *n* chemical elements that are representatives of a basic structure prototype containing *n* chemical elements have no mixed sites (and are therefore called basic representatives).
- Inorganic substances with *n* + *m* (where *m* is an integer > 0) chemical elements that are representatives of a basic structure prototype containing *n* chemical elements have mixed sites and generally belong to a solid solution of the basic structure prototype or one of the basic representatives with *n* chemical elements. The 349,481 structure entries of [62] represent only 130,823 distinct inorganic substances (chemical system + structure prototype). Among the 130,823 distinct inorganic substances, only 88,774 have no mixed sites. This means inorganic substances (basic structure prototypes or

Table 22

| Number of point-sets per prototype | | 1 | 2 | 3 | 4 | 5 | 6 | > 6 | no. distinct phases |
|---|---|---|---|---|---|---|---|---|---|
| | 1 | 434 | (1'385) | (625) | (66) | (45) | (22) | (5) | 2'582 |
| | 2 | 6 | 11'181 | (11'715) | (2'300) | (369) | (78) | (56) | 25'705 |
| | 3 | 0 | 110 | 36'745 | (11'364) | (3'665) | (814) | (0) | 52'698 |
| | 4 | 0 | 5 | 782 | 25'265 | (4'769) | (851) | (267) | 31'939 |
| | 5 | 0 | 0 | 38 | 701 | 11'161 | (995) | (311) | 13'206 |
| | 6 | 0 | 0 | 3 | 100 | 271 | 3'988 | (331) | 4'693 |

Summary: 2'016 (1.5%) 88'774 (67.9%) 40'033 (30.6%) 130'823 (100 %) 349'481 entries (2.67 entries per phase)

structures with two sites occupied by the same chemical element parent structures with no mixed site occupancies, and each site occupied by a different chemical element parent structures with mixed site occupancies, at least one site is occupied by two different chemical elements number of experimentally known distinct inorganic phases representatives) where the number of chemical elements *n* is the same as for the basic structure prototype. The remaining 40,033 inorganic substances are representatives of basic structure prototypes, but have mixed occupation sites, and are in general solid solutions of basic structure prototypes or their basic representatives. Most of them are therefore not distinct new inorganic substances. For example, replacement of a few at.% of the chemical element *A* in an *ABC* inorganic substance by a closely related chemical element *A'* will lead to a quaternary representative (*A*,*A'*)*BC*.

Table 22. Number of distinct phases comparing the number of chemical elements in the structure prototype defining database entry (rows) and the number of chemical elements in all representatives (columns). The diagonal corresponding to the same number of chemical elements as in the prototype type-defining entry is emphasized.

The four principles confirm, with the help of Figs. 55-58 and Table 22, the content of the third cornerstone: *'When chemical elements combine to form solids, their crystallographic structures are beautifully rich, yet very systematic patterns underlie this process.'* The systematic patterns lead to restraints, nicely reflected in the below listed four principles:

- Overall simplicity preferred,
- High overall symmetry preferred,
- High local symmetry reflected through the preference of regular AETs,
- High ordering tendency

The combination of the above given experimental facts reduces the number of potential structure prototypes to be considered in first priority to a few hundred, just 1-2% of the experimentally known structure prototypes. The frequency plot in Fig. 59 shows the number of representatives for the 100 most populous structure prototypes. The majority of the 39,990 experimentally known structure prototypes (about 35,000) are unique in the sense that they have no representatives. One of the reasons for the high number of unique structure prototypes is the significant number of refinements with split sites and (or) other sites with very low occupancy. Each distinct combination corresponds to a different structure prototype. For example, almost every refinement of a specific deuteride or hydride phase results in a new structure prototype, even if the phase is the same.

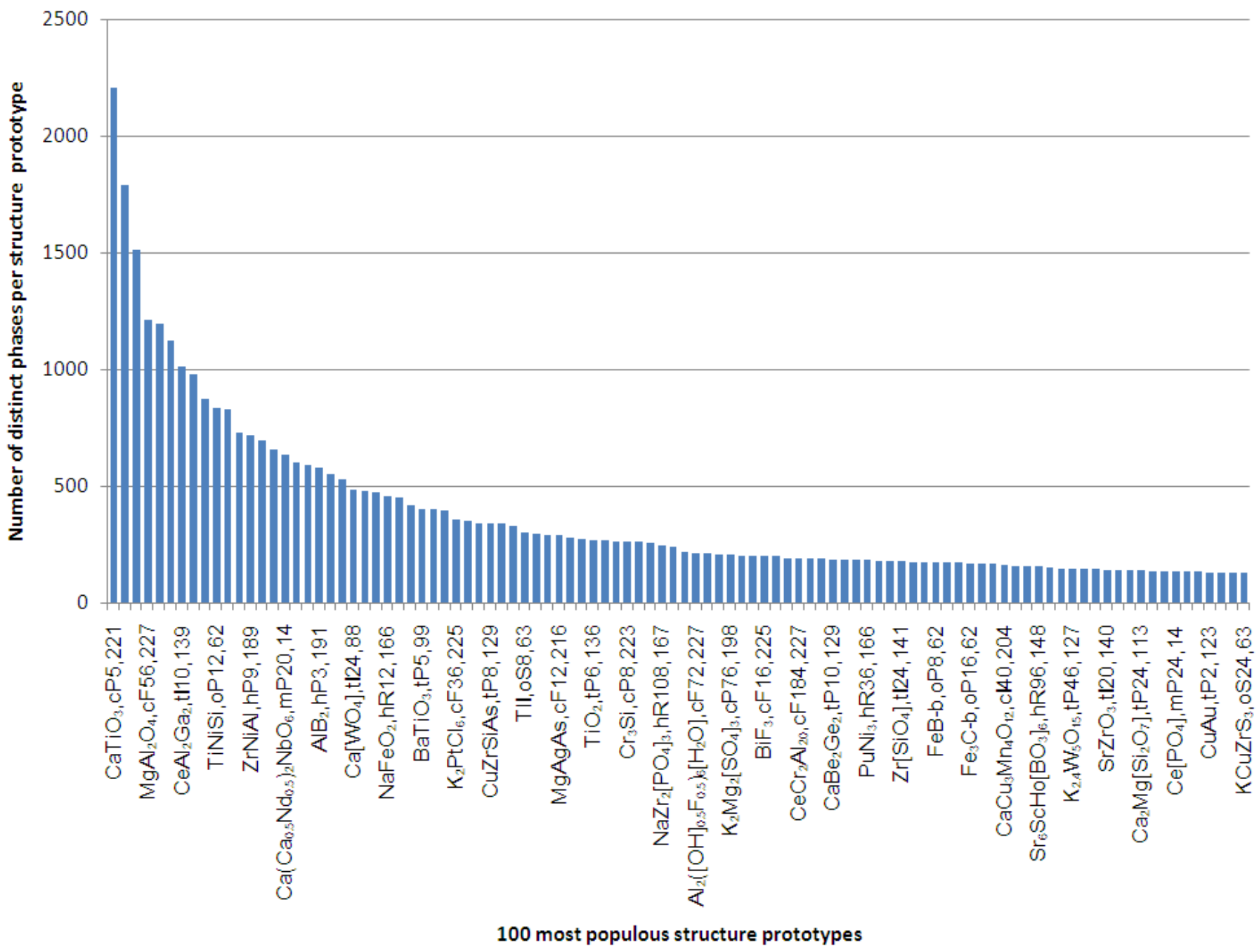

Fig. 59. Number of distinct phases of each of the 100 most populous structure prototypes.

### 7.3.4. Fourth cornerstone: Laws that link the position of chemical elements within a structure prototype and in the Periodic System

The fourth cornerstone of nature states that there exist direct links between the position of the constituent chemical element in the periodic system and its point-set occupation within the structure prototype of an inorganic substance.

By definition, structure prototypes and their representatives are geometrically very similar to each other. A closer inspection of known inorganic substances reveals that, for each of the 39,990 structure prototypes, only a relatively small subset of potential representatives is actually known. The above mentioned link may give us a possibility to predict which inorganic substances will adopt a specific structure prototype (or a limited selection of structure prototypes). This is strongly supported by the following four principles:

*7.3.4.1. Structure Prototype ↔ Periodic System correlation (chemistry) principle (ninth principle)*

This principle was formulated in 1994 **[2]** as follows: The vast majority of structure prototypes show a very strict regularity between the position of the chemical element in the periodic system (s-, p-, d-, and f-elements) and its point-set(s) occupation within a specific structure prototype. By looking at the 1,000 most populous structure prototypes we can confirm the above statement. About 50% of the structure prototypes, in general the ones with few representatives reveal a simple correlation leading to a few 100 potential combinations. The other 50%, having in general many known representatives, show a broader correlation, leading to several thousand extrapolated combinations. Fig. 60 demonstrates the chemistry principle with the help of a simple point-set – periodic system correlation for the $ErIr_3B_2$,*mS12*,12 structure prototype, where the columns of the periodic system to which the elements occupying the point-sets in the structure prototype and its representatives belong, have been highlighted. This correlation leads to 6×4×(3×2)= 144 element combinations (1:2:3), of which 23 have so far been found experimentally. Fig. 61 shows an analog correlation for a more common structure prototype, $MgCu_2$,*cF24*,227. This correlation leads to (40×85)/(1×2) = 1,700 combinations (1:2). 239 basic binary inorganic substances are already known. With the help of these two examples, it is seen that there exists a clear link between the position of the chemical element within the structure prototype and its position in the periodic system.

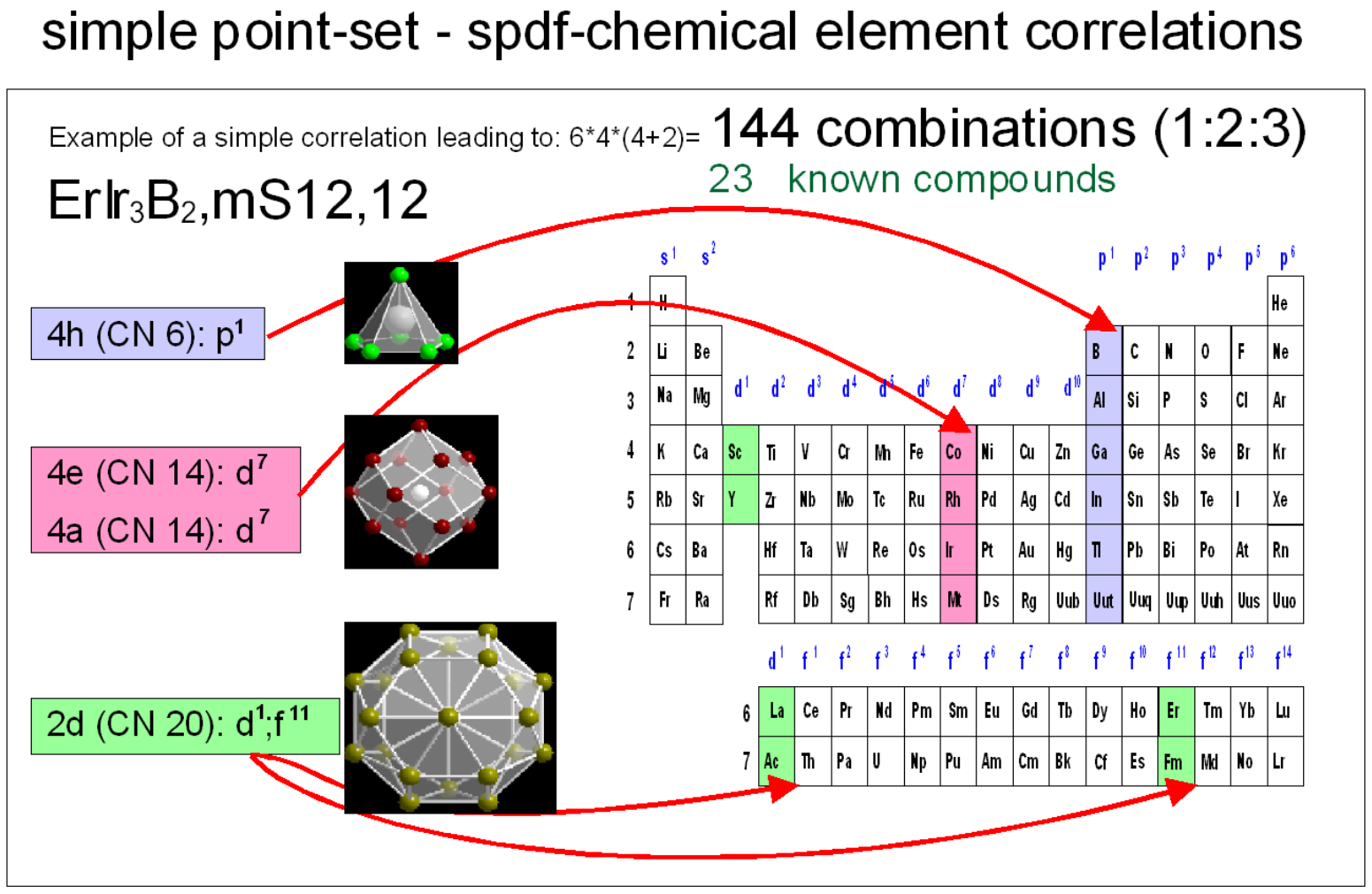

Fig. 60. Chemistry principle shown with the help of a 'simple' point-set ↔ spdf- element correlation showing the $ErIr_3B_2$,*mS12*,12 structure prototype.

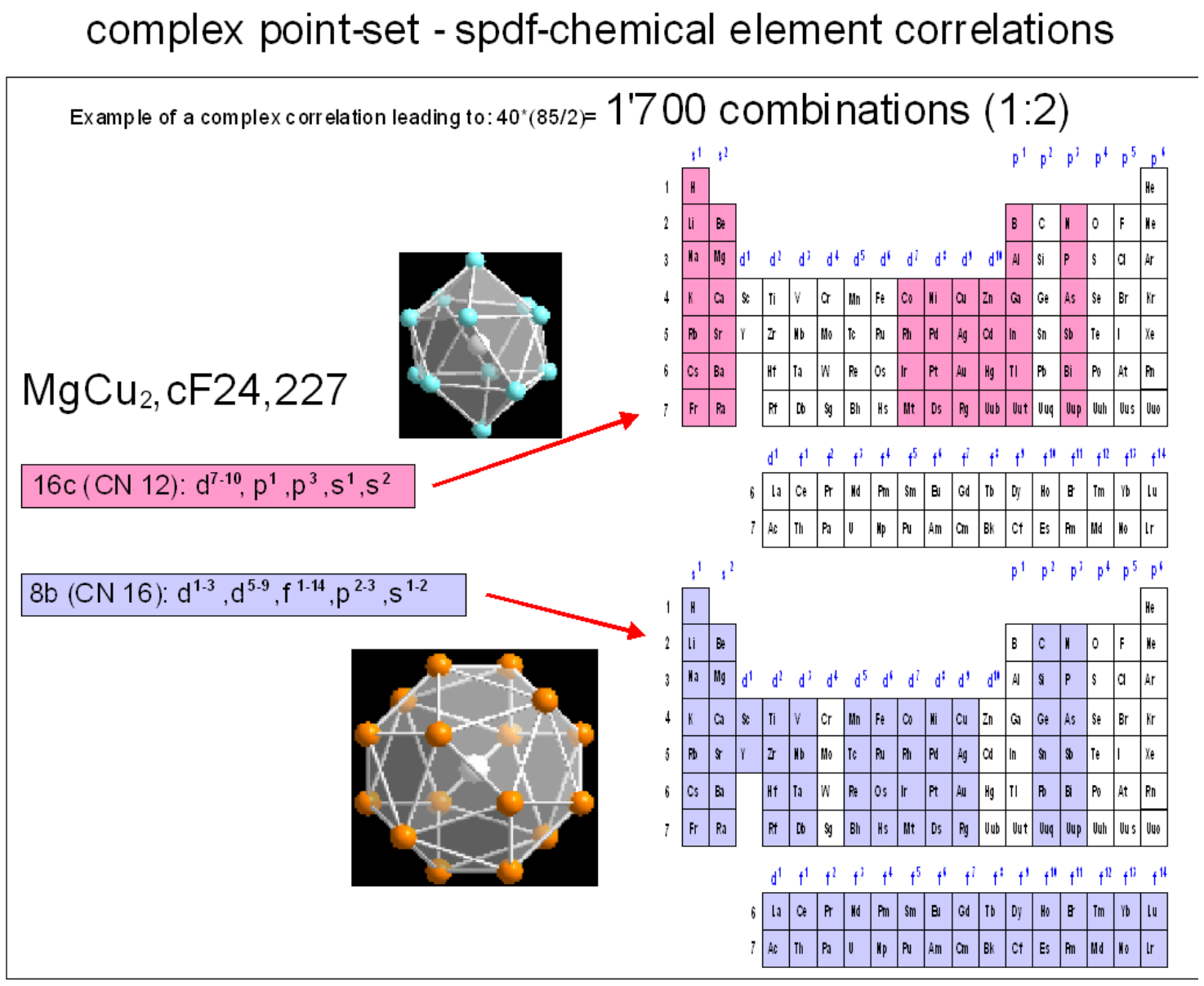

Fig. 61. Chemistry principle shown with the help of a ‘simple’ point-set ↔ spdf- element correlation showing the $MgCu_2$,*cF24*,227 structure prototype.

*7.3.4.2. Structure stability map principle (tenth principle)*

This principle was formulated in 1994 [2] as follows: *‘The atomic number, valence electron number, atomic size, and atomic reactivity are the factors governing crystallographic structures of intermetallic substances*.’ Structure stability maps separate intermetallic substances into distinct structure prototype domains. There exists a whole range of different structure stability maps for binary and ternary intermetallic substances (see Table 14). This principle has been further verified and significantly improved since the introduction of the periodic number (PN). This means using PN instead *valence electron number* as in 2008 [8]. The by far simplest and most efficient structure stability maps use the PN (or expressions of PN) as APP. The ‘best’ combination was found to be a two-dimensional $PN_{max}$ *versus* mean $PN_{min}/PN_{max}$ map (for the definition of mean $PN_{min}/PN_{max}$ see [8]), which has the following five advantages:

i) Inclusion of non-former, which nicely separates from the former systems, is possible.
ii) Applicable to binary, ternary and quaternary systems.
iii) Iso-stoichiometric structure prototypes and their representatives can be separated into distinct structure prototype stability domains in the same structure stability map, *e.g.* binary 1:3 structure stability map, ternary 1:1:4 structure stability map.
iv) 2D-structure stability maps are easy to read and provide obvious prediction ability.
v) PN is an integer number, known for all chemical elements, and no inaccuracy is coming from the choice of the actual values of the APP.

This kind of map proved to be very efficient in revealing patterns for different inorganic substances problems, *e.g.* former *versus* non-former systems; iso-stoichiometric structure prototype stability maps; complete solid solubility between binary inorganic substances having the same structure prototype. It can also separate different AETs into distinct AET stability domains in a generalized composition-independent map including binary and multinary inorganic substances [3]. Fig. 62 shows such structure prototype stability map for basic $ABC_2$ structure prototypes and their representatives, focusing on the following five most populous structure prototypes: $Cu_2MnAl$,*cF16*,225; $NaFeO_2$,*hR12*,166; $CuFeS_2$,*tI16*,122; $CuHfSi_2$,*tP8*,12; $CeNiSi_2$,*oS16*,63. From this example it is seen that there exist clearly defined domains for the representatives of the different structure prototypes within such iso-stoichiometric structure prototype stability maps. Any crystallographic structure stability maps having as axis APP or APPE, and APP= PN, provide a direct link to the periodic system.

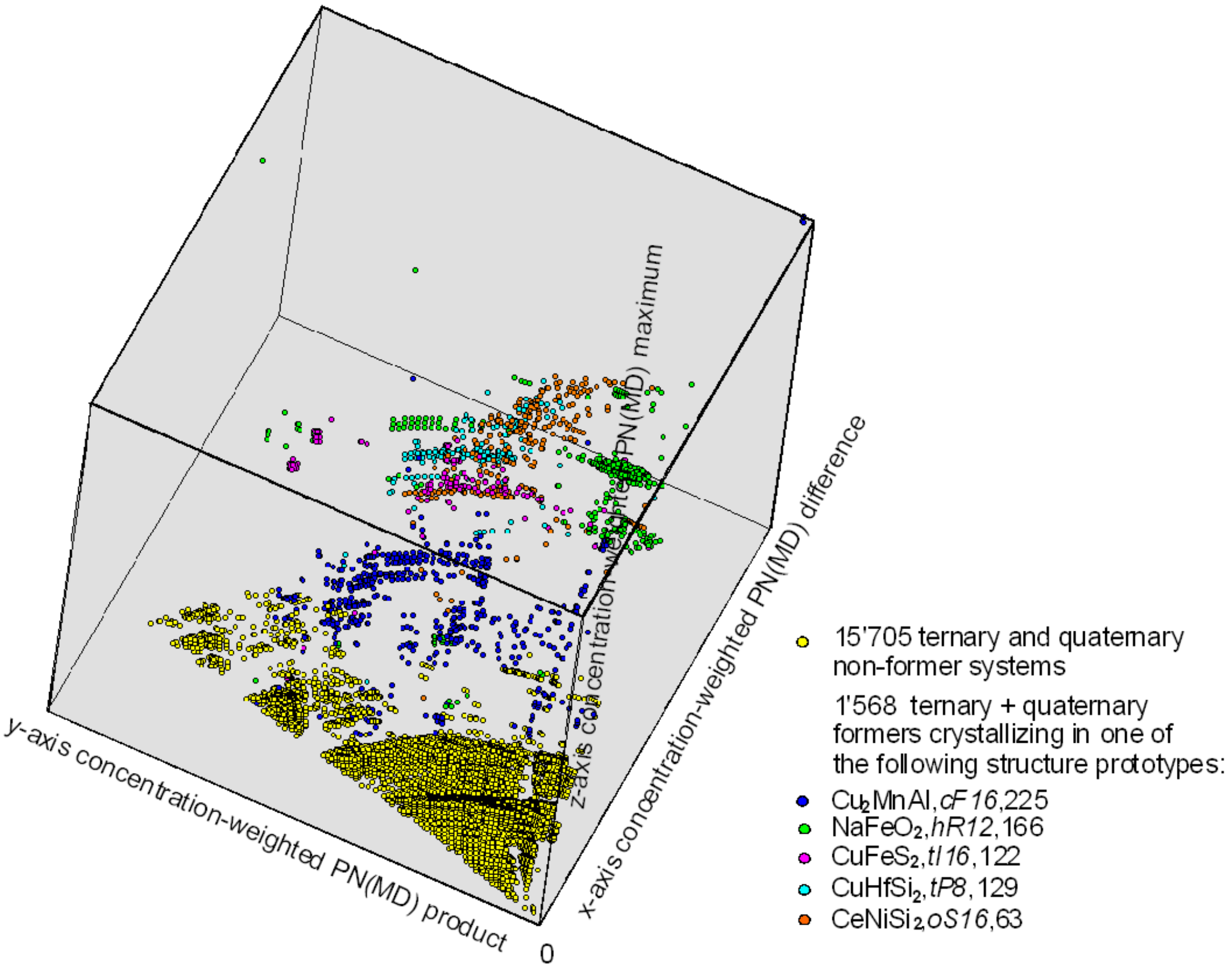

Fig. 62. Structure stability map for ternary $ABC_2$ prototypes focusing on the following three prototypes: $Cu_2MnAl$,*cF16*,225 $NaFeO_2$,*hR12*,166; $CuFeS_2$,*tI16*,122, $CuHfSi_2$,*tP8*,129, $CeNiSi_2$,*oS16*,63; including all ternary and quaternary representatives.

*7.3.4.3 Generalized Periodic System – AET graphs principle (eleventh principle)*

We formulated this principle in 2008 [3]. In a most recent and ongoing research activity we found that the big disadvantage of having to group different AETs to 'coordination number ranges', and therefore having to a certain degree limited systematization of the APP or APPE – 'coordination number ranges' can be eliminated. Therefore we developed periodic system – AET graph (Table 21a+b) by introducing a third dimension to the periodic system of Mendeleyev, this by giving to each of the 18 group numbers (GN= 1-18) a different color, which correlates the 18 lines of each group number GN, sorted according to increasing periodic number $PN_{MD}$. Further details see sub-chapter 5.3.4.

*7.3.4.4 Complete solid solution stability map principle (twelfth principle)*

This principle was formulated in 1994 [2] as follows: '*The atomic size, atomic reactivity, and valence electron number factors control solid solubility.*' Solid-solubility maps were found to separate regions of limited and extended solid solubility for a given chemical element. We have extended it to solid solution behavior from chemical elements to binary inorganic substances as solvents. Since the amount of solid solution data on binaries as solvents is very limited, we focused on a very extreme situation. We searched for ternary systems where the same structure prototype occurs in two of the binary boundary systems and either a complete solid solution or limited solid solutions are formed. This requires having access to either appropriate phase diagram data or information about cell parameter(s) *versus* concentration. We selected the structure prototype $MgCu_2$,*cF24*,227 which has 239 well established binary representatives. For a 1:2 inorganic substance the complete solid solution can be either of the type $A(B,C)_2$ or $(B,C)A_2$. From the 239 binary substances one can generate 701 $A(B,C)_2$ and 1,782 $(B,C)A_2$. Considering experimental data showing complete solid solutions, we found 61 $A(B,C)_2$ and 128 $(B,C)A_2$ cases. The prototype $MgCu_2$,*cF24*,227 has two point-sets, occupying Wyckoff positions 8*b* and 16*c* in the standardized description. The 16*c* site has as AET an icosahedron with CN = 12, and the 8*b* site a 16-vertex Frank-Kasper polyhedron with CN = 16. Since it is unlikely that the same chemical element will occupy sites with so different AETs, the following six potential solid solutions can be excluded: $A_2C\leftrightarrow AB_2$, $A_2C\leftrightarrow BC_2$, $AB_2\leftrightarrow BC_2$, $A_2B\leftrightarrow AC_2$, $AC_2\leftrightarrow B_2C$, $B_2C\leftrightarrow A_2B$ (see Fig. 63). With the help of structure stability maps and crystallographic considerations one can distinguish for two binary inorganic substances having the same structure prototype between systems having complete solid solutions or not.

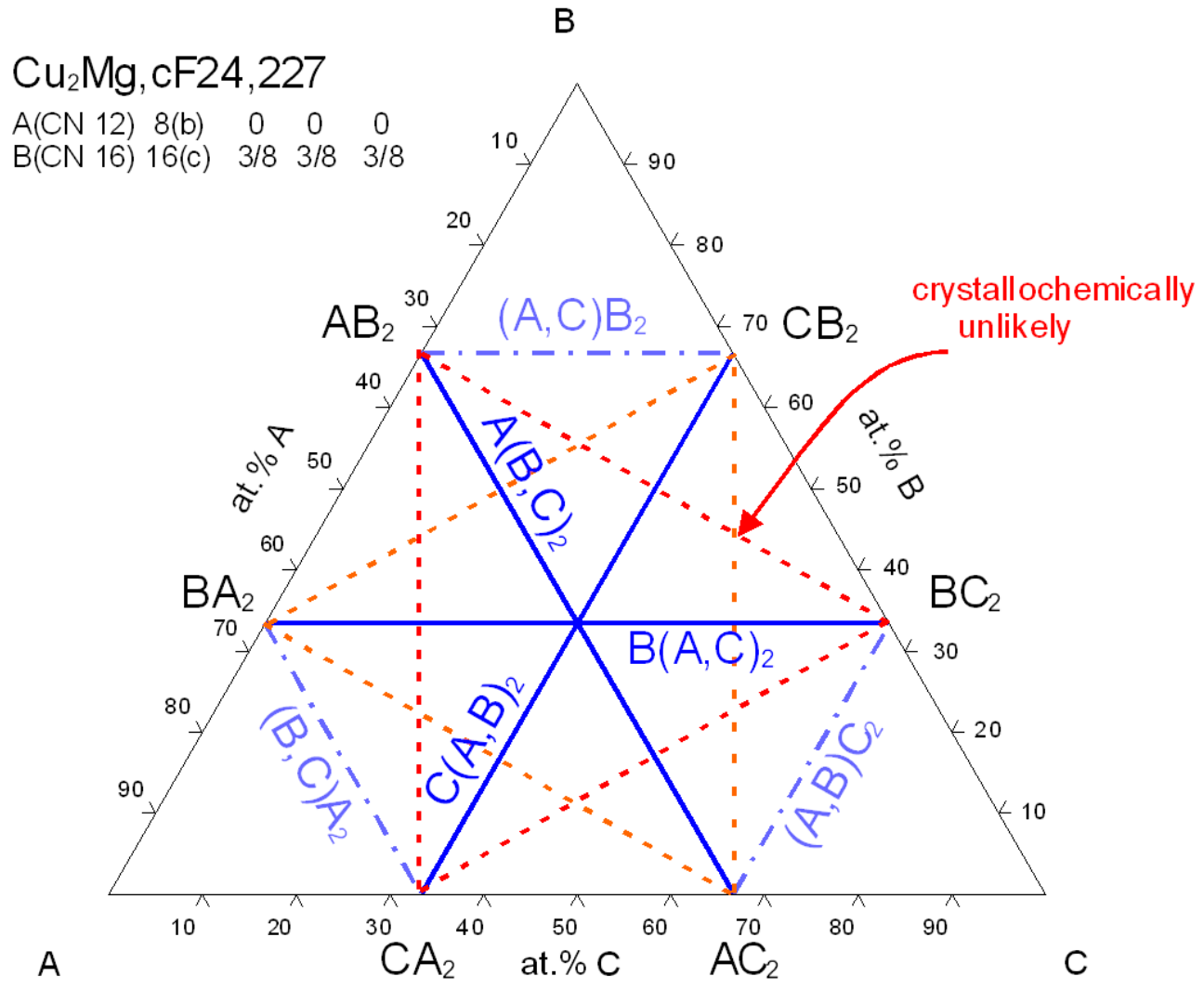

Fig. 63. Possible complete solid solution between two binaries within the same prototype, shown on the example of the prototype $MgCu_2$,*cF24*,227.

These four principles demonstrate, with the help of Figs. 60-63, the correctness of the fourth cornerstone of nature: '*There exist direct links between the position of the constituent chemical elements in the periodic system and its point-set(s) occupation within its structure prototype of an inorganic substance.'* The existence of such links is reflected in the above outlined four principles:

- Link between the position of the chemical element in the periodic system and its point-set position in the structure prototype (chemistry principle),
- Existence of stability domains for different structure prototypes and their representatives (structure stability map principle),
- Existence of stability domains for different AET (generalized AET stability map principle),
- Existence of solid solution stability domains (solid solution stability map principle).

The above given four principles reduce the number of potential inorganic substances to be considered from several thousands to a few hundred for a particular structure prototype. In addition, it efficiently helps to select from the many potential competing structure prototypes the most probable ones.

### 7.4. A realistic way to build up a trustworthy simulated part of the Materials Platform for Data Science

Based on the 12 principles enumerated above (or the four cornerstones of nature), we propose a realistic way to build up a trustworthy simulated part of the MPDS. There are potentially two major risks to make it *untrustworthy*. The first risk is by simulating potential inorganic substances belonging to the non-former systems. The second risk is by simulating potential inorganic substances with improbable structure prototypes (analog to improbable AET within the coordination type classification). One should keep in mind that it is still not feasible to simulate all possible 39,990 structure prototypes, as starting point, for each potential inorganic substance. We propose therefore the following calculation strategy, which must be dynamically linked to a reference database, such as the PAULING FILE **[8, 9,18]**:

- Focus on unaries to quaternaries,
- Exclude from all calculations chemical systems clearly belonging to non-former systems,
- Focus on experimentally established inorganic substances belonging to the most populous structure prototypes,
- For each of the most populous structure prototypes derive point-set ↔ periodic system (*s*-, *p*-, *d*-, *f*-element) correlation (based on known inorganic substances) and select, with the help of crystallographic structure stability maps (based on structure prototypes-, coordination types-, and AETs- classification), 5-10 competing structure prototypes as starting point for the simulations,
- Evaluate for each inorganic substance confidence levels by comparing experimental data with own simulated data, as well as the consistency with 'holistic views' – APP or APPE maps,
- Simulate a broad range of physical properties for already experimentally established inorganic substances, as well as for predicted phases with a high confidence level of correctness, respectively probability of existence. Evaluate the confidence level for inorganic substances where data have been published,
- Store the simulated data in a relational database consistent with the reference database, and create dynamic links so that, in case changes will be done, derived data will be recalculated automatically.

Here we have taken advantage of several statistical plots and inorganic substances-'holistic views'-APP or APPE maps. In addition, the existence of correlation between the positions of the chemical elements in the periodic system and their positions in fully standardized structure prototype. Last but not least, the correlation between the number of chemical elements of an inorganic substance and the different AET has proven to be a powerful tool. This all together makes it possible, for each of the most populous structure prototypes, to populate the atomic sites by chemical elements that are chemically meaningful, and simulate structural data and physical properties for potentially stable inorganic substances. The resulting set of simulated data will of course dynamically link with experimentally known inorganic substances contained in the reference PAULING FILE **[8,9,18]**, which thus serves as validation of the computational approach.

The reliability of the simulated data of the MPDS **[17a,b]** depends on four major factors:

1) Continuous verification of the simulated data by comparison with the reference (experimentally determined data), especially for the crystallographic structure, which constitute the starting point for all simulations. This interplay generates a 'structure reliability factor' for simulated structure prototypes of inorganic substances.
2) When the structure prototype has been confirmed, its physical properties will be simulated, and again continuously compared with the experimentally known reference data (if available). This interplay generates a 'physical property reliability factor' for the simulated physical property data. The same principle should be applied to phase diagram data.
3) Each simulated structure prototype should have an active link to the structure prototype reference data set that was used as a starting point. This with the purpose of being able to compare with the experimental values, as well as maintaining the consistency (any change of the experimental reference data will automatically trigger a new calculation procedure).
4) Thorough quality controls of the experimentally obtained data, and data simulated by different methods are required.

Thus, chapter 7 reveals the following facts:

The *ab initio* simulations require no other input than the laws of the quantum mechanics and the atomic numbers (AN) of the involved chemical elements. We trust that, through the required parameter adjustments during the simulations applying the laws of quantum mechanics, no relevant information is lost. In principal, most of the experimentally determined data contained in the PAULING FILE can be confirmed by simulations, leading to a quantum mechanical inorganic substances design approach. The cornerstones and principles presented in this work are summarized in Table 23. The first cornerstone is a direct consequence of the existence of the about 100 chemical elements given by nature and their combinations. In practice there exist an infinite number of chemical element combinations (and therefore inorganic substances). This conclusion is independent of quantum mechanics. *The second to fourth cornerstones of nature can under no way be derived from quantum mechanics, even each of the 100 chemical elements possesses all required information, given by nature, and 'knows what to do' when submitted to external conditions*. The cornerstones can only be discovered by the examination of a large amount of critically evaluated experimentally (and simulated) determined data. Since we do not have the ability to cross-link all the consequences of the interactions of the atoms under all possible conditions (*e.g.* nature of the chemical elements to be combined, stoichiometric ratios, temperature, pressure, *etc.*), we have no other choice than to focus on different consequences of the above mentioned causes. The second cornerstone focuses on the fundamental ability to form an inorganic substance. The third cornerstone focuses on the ordering of the atoms within an inorganic substance. And, finally, the fourth cornerstone reveals the link between the position of a chemical element in the periodic system and its position(s) in the crystallographic structure of a particular inorganic substance. These four cornerstones can be considered as different ways to look at the same cause. They are supported by the twelve principles outlined in Table 23.

| Selection of chemical elements to be combined | infinitely many potential inorganic substances |
|---|---|
| **First cornerstone of Nature** | |
| i) Number of chemical element combinations: | infinite |
| ii) Number of compounds per chemical system: | 0-30 |
| iii) Number of structure prototypes: | > 39,900 |
| **Principles (governing factors):** | **Restraints:** |
| **Second cornerstone of Nature** | |
| 1) Non-formers: | about 30% |
| 2) No. chemical elements per compound – AET correlation: | 4,087,975 systems |
| | 10% (2 elements); 15% (3); 38.6% (4) |
| 3) Inactive composition range: | 10 (2 elements); 769 (3) |
| 4) Most probable stoichiometry ratios: | |
| **Third cornerstone of Nature** | |
| 5) Simplicity (90% probability): | < 40 atoms/unit cell; < 10 point sets; < 5 AETs |
| 6) Symmetry (90% probability): | 23 of 230 space |
| 7) Atomic Environment Types (90% probability): | 18 of 100 AETs |
| 8) Ordering tendency: | (n-1) elements impossible |
| **Fourth cornerstone of Nature** | simple to broad |
| 9) Structure Prototype – periodic system correlation: | stability domains |
| 10) Structure stability maps: | stability domains |
| 11) Generalized AET stability map: | stability domain |
| 12) Solid solution stability map: | |

Table 23. Summary of 4 cornerstones and 12 principles proposed, and its effect on the number of potential element combinations to be considered in first priority.

These ultimately lead to restraints, which are a requirement for the development of a practicable and trustworthy inorganic substances design approach. The power of critically evaluated data is considerable, knowing that already 30 years ago, having access to only 17% of the now available data, it was possible to derive the core content for nine principles correctly. Apparently the amounts of data with information on about 30,000 inorganic substances were already sufficient to derive reliable inorganic substances knowledge. Then, in this work verified, and where necessary, extended, nine principles, as well as the newly discovered three principles are based on a much more robust database system with information on over 170,000 inorganic distinct phases. We thus conclude that the here formulated twelve principles, and from them derived restraints, MUST be considered as trustworthy.

## 8. Holistic views: Quantitative trends in physical properties via machine learning starting from the peer-reviewed PAULING FILE data

### 8.1. Introduction

Analogous to the Human Genome Project in 90s, today's efforts in materials informatics **[74-78]** tackle the 'Materials Genome' as explained in **[122,123]**, producing the vast swatches of data to be validated and interpreted. In view of this we emphasize the following critical statement. For any predictive data-intensive technique, the combination of *quality* and *quantity* of the input data must lead to the results comparable with the first-principles quantum mechanics predictions. Indeed, it is known that the input quality, as well as the input quantity, are crucial **[79,80]**.

Herewith we propose a pathway to decipher the 'Materials Genome' using a relatively cheap and unsophisticated machine learning technique – decision trees – and the large PAULING FILE materials dataset. We exemplify the predictions of 10 physical properties for the 114,872 inorganic substances from the only crystallographic classification: 'entries with its structure prototype assignment', as well as the reverse predictions of its structure prototype assignment from the values of these 10 physical properties. The considered 10 physical properties are: (*a*) isothermal bulk modulus, (*b*) enthalpy of formation, (*c*) heat capacity at constant pressure, (*d*) Seebeck coefficient, (*e*) temperature for congruent melting, (*f*) Debye temperature, (*g*) linear thermal expansion coefficient, and (*h*) energy gap (direct or indirect) for insulators, *(i)* electrical conductivity and *(k)* thermal conductivity.

Currently, the materials informatics is a collection of recipes taken from computer science and adopted for materials science. The main difficulty is purely technical from an academic point of view — how to efficiently handle inorganic substances big data. A series of modern initiatives in materials informatics is known, both academic and industrial **[74-78]**. They all have one main feature in common: they build their own software infrastructures to combat the challenge of inorganic substances big data, thus gaining the insight knowledge more efficiently. Probably the oldest effort is the PAULING FILE project **[4]**. As we learned from the previous chapters there exists presently about 170,000 *distinct phases*. Approximately one seventh of the *distinct phases* are connected between each other via the phase diagrams. About one tenth of the *distinct phases* have at least one physical property reported. However, the majority of the *distinct phases* at the PAULING FILE do not have any of the 10 mentioned physical properties reported. That is why, for their prediction, we decided to apply machine learning, namely the decision trees regression, drawing the advanced extrapolation from the well known to less known inorganic substances. Here we present a proof of concept (POC): how a relatively unsophisticated statistical model trained on the large

PAULING FILE dataset predicts a set of the 10 mentioned physical properties using as input only its structure prototype assignment and APP PN as the descriptors, on an example of the 114,872 inorganic substances, *i.e. distinct phases*. The materials predictions powered by machine learning have gained traction in the last years **[81,82]**. Among others, Isayev et al. **[82]** were training the set of regression and classification models, based on the decision trees, using the ICSD experimental database and the *ab initio* simulation repository. Our present model is similar by architecture, prepared using the bigger training set, and operates the simpler descriptors. Based on the 5 shared properties, values predicted by our model and the Isayev et al. model **[82]**, as well as their experimental values from the PAULING FILE, we made a detailed comparison. As a result, the thermal expansion coefficient is better predicted by our model, the heat capacity and the band gap is predicted by both the models similarly well, and, finally, the bulk modulus and the Debye temperature is better predicted by Isayev et al. model **[82]**.

## 8.2. Methods

### *8.2.1. Physical property taxonomy*

The physical properties of the PAULING FILE include the experimental data, and to a limited extent, the simulated data of inorganic substances in the crystalline state. The physical properties belong to 8 domains: density, phase transitions, mechanical properties, thermal and thermodynamic properties, electronic and electrical, properties, optical properties, magnetic properties, and superconductivity (see *4.5.2*). The taxonomy consists of three levels: the mentioned general domains, sub-domains, and the particular physical properties. For instance, the domain 'electronic and electrical properties' contains the subdomain 'electron energy band structure', which in turn contains the 'Fermi energy' property *etc.* Currently there are about 100 sub-domains, and nearly 500 particular numeric physical properties. The 10 particular physical properties considered in this work belong to the most common physical properties extracted by the PAULING FILE project from the world literature. The entire taxonomy was build-up by Prof. Fritz Hulliger (Swiss Federal Institute of Technology in Zurich, Switzerland), Prof. Roman Gladyshevskii (Ivan Franko National University of Lviv, Ukraine) and Dr. Karin Cenzual (University of Geneva, Switzerland). To a certain degree, it reflects the development of solid state physics during the last century. There are ongoing efforts to convert this taxonomy into an ontology, which is, in computer science, the formal naming and definition of the types, properties, and relationships of the entities for a particular knowledge domain.

### *8.2.2. MPDS access interfaces*

The online platform MPDS **[17a,b]** provides all the PAULING FILE data online via two access interfaces: browser-based graphical (GUI) and application programming (API). The GUI is intended for the traditional research, whereas the API serves various software integrations and data-mining scenarios, as described below. The MPDS API adheres to the principles of representational state transfer (REST) **[83]** and presents all the PAULING FILE data in a developer-friendly, machine-readable way, using the opened formats, such as CIF and JSON.

### *8.2.3. Decision trees regression and classification*

The decision trees regression and classification is a reliable and simple to use predictive technique. A decision tree is a statistical model, which describes the data going from the observations about some item (*e.g.* the structure prototype assignment in this work) to the conclusions about the item's target value (*e.g.* the corresponding 10 physical properties). The PAULING FILE data contain structure prototype assignments linked with the physical properties via the *distinct phases*. Therefore, it is feasible to train a model on the existing pairs 'structure prototype assignment' - 'physical property' and then to predict the absent values of physical properties, based only on the available structure prototype assignments. Multiple decision trees are built by repeatedly re-sampling training data with replacement, and voting for the trees providing the better prediction accuracy. Such an algorithm is known as a random forest **[84]**. Its presently used state-of-the-art open-source implementation scikit-learn (version 0.19.1) takes seconds to train a model from the PAULING FILE data on an average desktop PC **[85,86]**. Based on the structure prototype assignment, the random forest regressor yields all the 10 considered physical properties. Notably, the band gap predictions required additional steps. First, a subdivision of structure prototype assignments into conductors and insulators was performed via the random forest classifier. This classifier was trained on the particularly imbalanced categorical data, as the PAULING FILE contains nearly five times more conductors than insulators. The random over-sampling (*i.e.* repeating of the under-represented class samples) was therefore applied. Finally, for insulators the particular values of the band gap were obtained. The used program code is open-sourced [97].

*8.2.4. Crystallographic data with structure prototype assignment (see 4.3.4. and 4.3.6.)*

Currently the PAULING FILE contains 349,481 crystallographic structure entries [62] using the structure prototype classification. In order to give an approximate idea of the actual crystallographic structure, a complete set of atomic coordinates and site occupancies is proposed for an entry, where only a structure prototype could be assigned by the authors or editor (but no atomic coordinates have been given in the publication). In such cases the editor assigns the coordinates based on its structure prototype. The crystallographic data are stored as published, but also have been *standardized* according to the method proposed by Parthé and Gelato [34,88].

*8.2.5. Periodic numbers PN (see 2.)*

Throughout this entire work we use the periodic numbers ($PN_{ME}$) of the chemical elements. In contrast to the atomic number AN, the $PN_{ME}$ is defined using the Meyer's periodic system and represents a different enumeration of the elements, emphasizing the role of the valence electrons [8]. Along with the simplicity, this is the main reason for adopting this concept as the main APP.

*8.2.6. Structure prototype descriptors and training*

The term descriptor stands for the compact information-rich representation, allowing the convenient mathematical treatment of the encoded complex data, *i.e.* structure prototype assignment. The peer-reviewed structure prototype assignments and their corresponding 10 physical properties were obtained via the MPDS API. From each structure prototype assignment entry a certain relatively large volume comprising exactly 100 atoms was cut. Then the structure prototype descriptor was constructed as a vector, based on the following data of these atoms: (*a*) the lengths of their radius-vectors, and (*b*) their PNs [97]. To evaluate the prediction, random 33% structure prototype assignment entries together with the corresponding physical properties were isolated and excluded from training. Next, the trained machine-learning model had to predict these physical properties based on the descriptors of the isolated structure prototype assignments. The factual and predicted values were compared, and the quality metrics were calculated. The evaluation process was repeated 30 times to achieve a statistical reliability. Here a bibliographic issue arose, as the several physical property values or several structure prototype assignments could be reported for a distinct phase. In those cases the averaged values were considered. Finally, the evaluated model was trained on all the data.

*8.2.7. Structure prototype assignment entries with disorder*

About 55% of all the PAULING FILE structure prototype assignment entries are disordered. Structure ordering solved the disorder; however only a very small number of ordered structures were randomly considered. It was shown that sampling only 6 ordered structures (instead of possible millions) already leads to a relatively stable averaged prediction. That means that the machine-learning method is not very sensitive to the atomic permutations. Of course, increasing the number of the ordered samples yields more stable and consistent predictions. Here however the speed was the limiting factor, as processing each ordered sample requires computational resources.

### *8.2.8. Principal component analysis*

The principal component analysis (PCA) [89] as implemented in the *scikit-learn* program toolkit (version 0.19.1) was used with the default parameters. This technique is particularly useful for visualization of the many dimensional data, as it reduces the number of dimensions in such an optimal way to group similar values together. The PCA returns the linear combinations of the input axes, which are also known as the principal components. Using PCA, we transformed the 5-dimensional combinations of the PNs for element count $N$ = 1–5 (unaries to quinternaries) into the 2-dimensional combinations of their principal components. Thus we were able to plot all the predicted physical properties as heat-maps in two-dimensional planes.

### *8.2.9. Neighbor learning*

The radius-based neighbor learning [90] as implemented in the *scikit-learn* program toolkit (version 0.19.1) was used. The size of the leaf was increased to 300, the radius in the PN space was set to 14, and, if no neighbors detected, to 75. An exceptional quality of extrapolations was demonstrated. The values for 10 physical properties found by the neighbor learning from the neighbor chemical elements and the values predicted by decision trees regression from the structure prototype assignment entry differs on average by only 5-10%.

### *8.2.10. Inorganic substances design*

The reversed task of predicting the structure prototype assignment by the given combination of physical properties was done as follows. With the extrapolation of the 10 physical properties via the neighbor learning, the number of the new chemical element combinations was increased in about ten times. Thus, the probability of finding at least some chemical elements for an arbitrary combination of the input property ranges was also increased considerably. The obtained chemical elements in a previously unknown combination are searched in the MPDS via the fuzzy chemical matching, *i.e.* by the similar PNs. Then the elements in the found chemically similar structures are replaced with the sought elements. Under the naive assumption, that the cell and the inter-atomic distances remain the same, this technique works surprisingly well. Finally, the random forest regression is executed; the results are scored and returned.

## 8.3. Results and discussion

### *8.3.1. Overview*

The 114,872 distinct phases, missing at least one of 10 selected physical properties, were identified in the current data release of the PAULING FILE. From them, 109,446 (95%) missed all of the 10 physical properties being predicted and only 865 (0.01%) missed 4 or less physical properties being predicted. Only 14 distinct phases have 8 physical properties reported. Additionally, about 20,000 distinct phases have either incomplete or absent crystallographic structures in the PAULING FILE, and thus were also excluded. By the number of the constituent elements, there were 1,461 unary, 14,515 binary, 45,577 ternary, 34,392 quaternary, and 16,296 quinary suitable distinct phases. About 2,500 remaining phases were of higher orders. Here we report only the results for the number of constituent elements $N$= 1–5. In total, the 114,872 selected *distinct phases* belong to the 56,407 distinct chemical element systems (*e.g.* all $Ti_2O$, TiO and $TiO_2$ distinct phases belong to the Ti-O chemical element system).

### *8.3.2. Prediction quality assurance*

In general, the prediction quality of the decision trees machine-learning technique is acceptable, and, on average, may even compete with the *ab initio* simulation results. The difference is that the simulation normally requires hours or days of computation time, whereas the machine-learning model yields the results in milliseconds, even with the less powerful hardware. Another difference is that the *ab initio* simulations in practice require careful fine-tuning of the method, whereas the chosen method of machine learning is a black box, where almost no initial setup is needed. The disadvantage of the machine-learning model is that, of course, no physical meaning of predictions is implied. The underlying complex physical phenomena, as well as the lack of training data, lead to poor prediction quality. Although the size of the training dataset should not be necessarily huge, there is some minimal threshold — about several thousands of samples. In Fig. 64 the occurrences of the training and predicted values are compared. The complex distribution shape of predictions suggests that the chosen machine-learning method is very likely to be able to capture the chemical nature of solids. Table 24 presents the quality metrics of the predictions: *mean absolute error* (MAE) and *R squared coefficient*, a statistical measure of how close the data are to the fitted line. (Best possible R-squared is 1; a constant model always predicting the expected value, disregarding the input, would get an R-squared of 0.) Thus the restricted distribution range in Fig. 64 for the heat capacity at constant pressure, Seebeck coefficient, and linear thermal expansion coefficient should be taken with caution, as they have the lowest R squared coefficients. Nevertheless although all the R-squared coefficients are relatively low, the trends in predictions are visible quite well (see further). In order to estimate the prediction quality of the binary classifier model "conductor *vs.* insulator", the fraction incorrect (*i.e.* an error percentage) was controlled to be less than 1%. Based on the PAULING FILE training dataset, the decision trees regression and classification models were able to represent 10 physical properties of the wide set of various inorganic substances in an acceptable manner. Nowadays more powerful and accurate machine-learning techniques exist, so the quality of predictions may be further increased. $A_2B$ (denoted by V) and $AB_2$ (denoted by W), sorted by PNs: peer-reviewed (left) and predicted (right) heat-maps. See red bars in Figs. 65a+b for the quantitative scales.

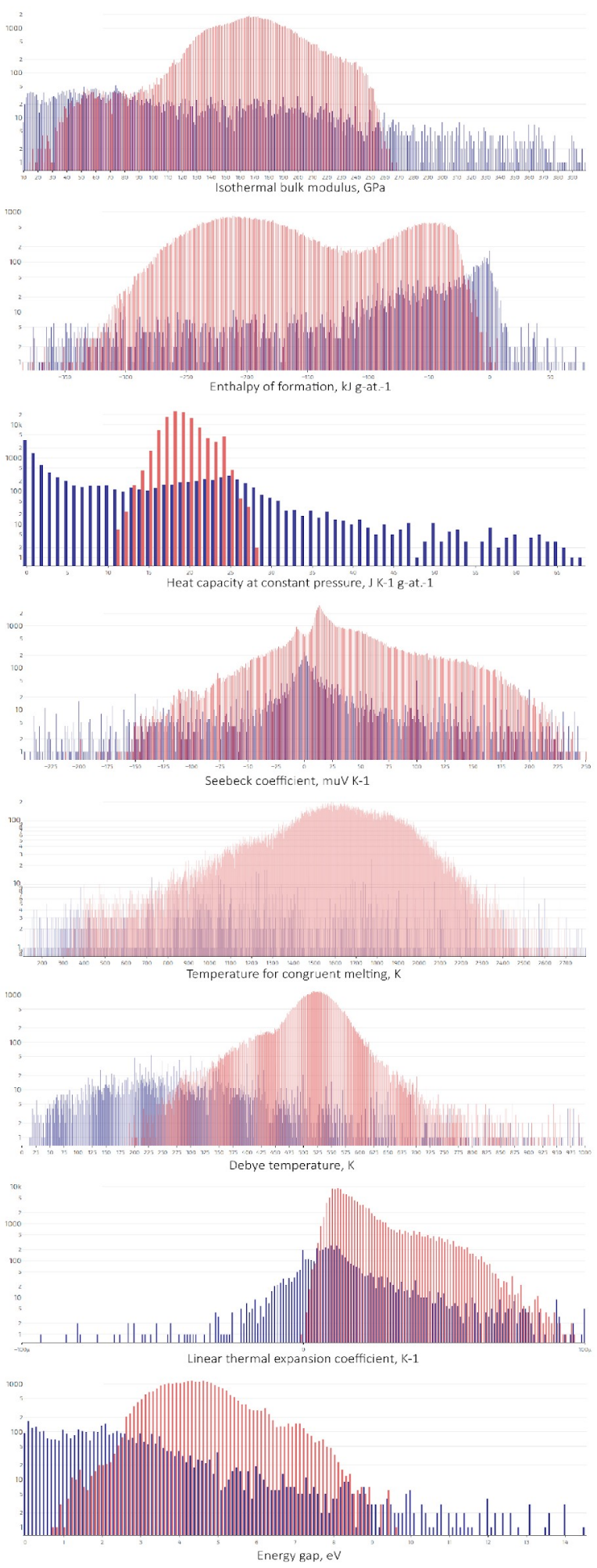

Fig. 64.
Distribution of the peer-reviewed (blue) and machine-learning (red) values. From top to bottom: isothermal bulk modulus, enthalpy of formation, heat capacity at constant pressure, Seebeck coefficient, temperature for congruent melting, Debye temperature, linear thermal expansion coefficient, and energy gap (direct or indirect) for insulators.

| Predicted physical property | Units | Mean absolute error | R-squared coeff. |
|---|---|---|---|
| isothermal bulk modulus | GPa | 45 | 0.46 |
| enthalpy of formation | kJ mol$^{-1}$ | 41 | 0.62 |
| heat capacity at constant pressure | J K$^{-1}$ mol$^{-1}$ | 4.8 | 0.08 |
| Seebeck coefficient | μV K$^{-1}$ | 87 | 0.11 |
| temperature for congruent melting | K | 280 | 0.65 |
| Debye temperature | K | 91 | 0.35 |
| linear thermal expansion coefficient | K$^{-1}$ 10$^{-5}$ | 1.2 | 0.11 |
| energy gap (if insulator predicted) | eV | 1.1 | 0.29 |
| electrical conductivity | Ω$^{-1}$m$^{-1}$ | 43 | 0.51 |
| thermal conductivity | W m$^{-1}$K$^{-1}$ | 17 | 0.10 |

Table 24. Quality estimation for our selected physical properties of our machine-learning predictions.

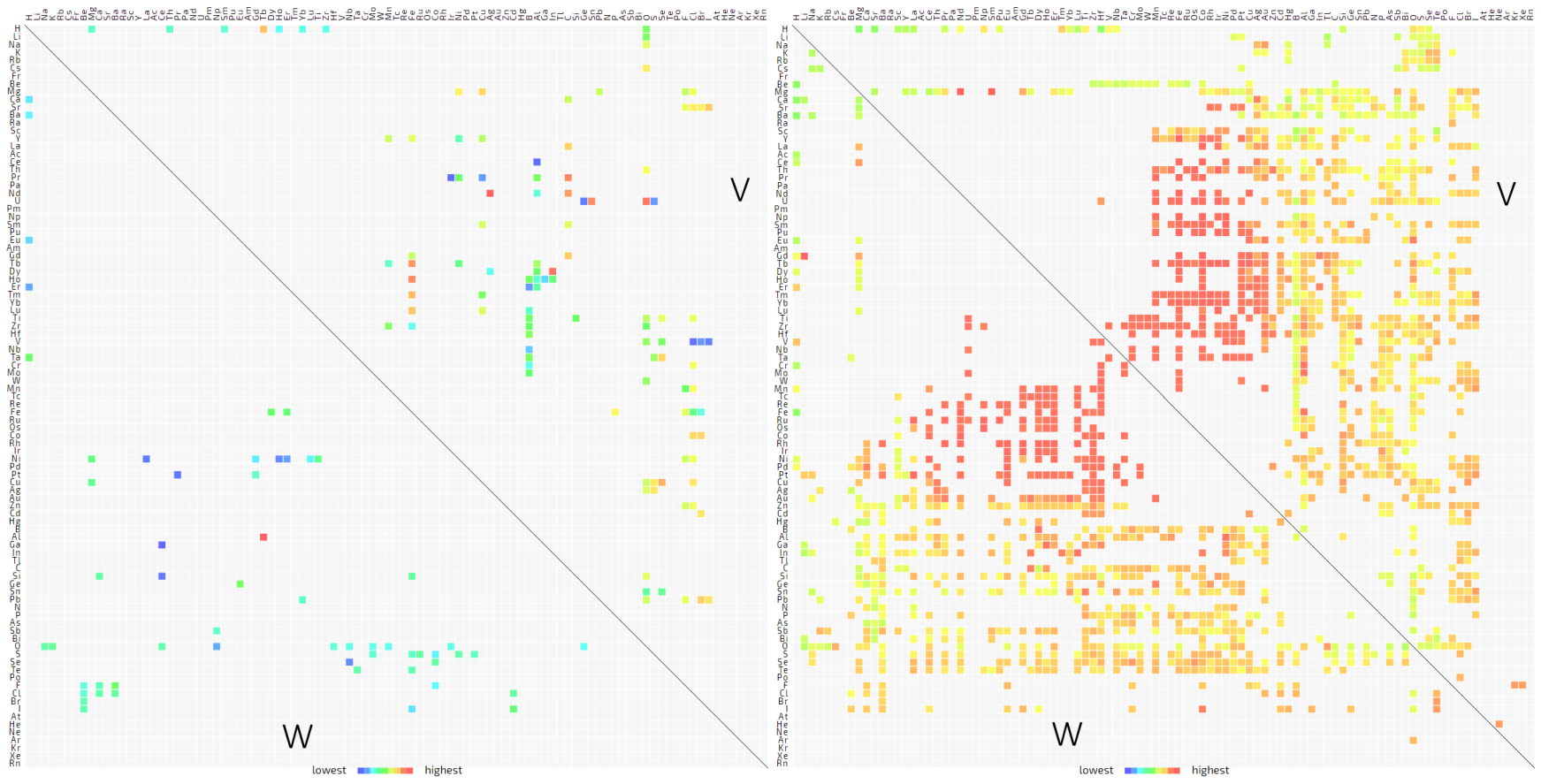

Fig. 65a-b. Periodic numbers $(PN_{ME})_A$ vs. $(PN_{ME})_B$ and the heat capacity at constant pressure (color) for the inorganic substances $A_2B$ (denoted by V) and $AB_2$ (denoted by W), sorted by PNs: peer-reviewed (left) and predicted (right) heat-maps.

*8.3.3. Binary inorganic substances (N = 2)*

In spite of Figs. 65a+ b, we have observed that in about 90% cases within the same chemical element system, the predicted values of the physical properties vary only inconsiderably, not higher than the prediction's MAE. Thus, we averaged the values of the predicted properties within the same chemical element system, which enabled us to deal with the chemical elements only, abstracting from the distinct phases. For instance, all binary Ti-O distinct phases (*e.g.* $TiO_2$ rutile, anatase, brookite, TiO, $Ti_2O$, *etc.*) were considered simply as Ti-O, or, using $PN_{ME}$, 46-100. To compare all the prediction results in terms of the APP $PN_{ME}$ for the element counts $N$= 1–5, we added trailing zeros for the systems with $N$< 5 (*e.g.* 46-100-0-0-0 in an example above). This way we represented all the predictions in 5 dimensions, with the physical property values attached. Then, using PCA for a dimensionality reduction, for each considered physical property we plotted all the prediction results at once (see *e.g.* melting temperatures in Fig. 66). Since the similar results are grouped together by PCA, a surprisingly clear subdivision by the constituent element counts was obtained for all the predicted properties. For example, in Fig. 66 it is seen that the predicted melting temperatures for inorganic substances with the element counts $N$= 1–3 demonstrate clear patterns (*cf.* Figs. 65a+b), and the vague patterns for $N$= 4. On the other hand, the quinternaries ($N$= 5) are not covered well in the literature, so there were not enough input structure prototype assignments, and no patterns in quinternaries are seen. The detailed visualizations of patterns for $N$≥ 2 (all the distinct phases within the specific chemical element systems) can be found either in the supplementary materials, or plotted online at the MPDS [17a,b].

*8.3.4. All inorganic substances (N= 1–5)*

For the binary inorganic substances AB, one can gain new, yet unknown quantitative trends with the aid of the periodic numbers $(PN_{ME})_A$ *vs.* $(PN_{ME})_B$ and the physical properties, encoded as the heat-maps, where the color stands for the particular physical property value. For example, in Figs. 65a + b the available peer-reviewed and predicted heat capacity at constant pressure for $A_2B$ and $AB_2$ substances (distinct phases, V and W, respectively) is shown. If more than one peer-reviewed value per distinct phase is available in PAULING FILE, an averaged one is taken. For the predicted values, there is always exactly one value per distinct phase. Interestingly, the maximal heat capacity values were predicted in $A_2B$ phases with the A-elements Mn to Au (PN = 58-76) and the B-elements Ce to Ta (PN = 18-52), as well as in $AB_2$ phases with the A-elements Ce to Ta (PN = 18-52) and the B-elements Cr to Pt (PN = 54-72). The predicted heat-maps for the other physical properties (here we took the average value for all binary phases (e.g. $A_2B$, AB, and $AB_2$) within the specific chemical element systems, as we have observed that in about 90% cases within the same chemical element system, the predicted values of the physical properties vary only inconsiderably, not higher than the prediction's MAE) can be found in Fig. 69. In contrast to the other APP, such as atomic numbers, atomic radii or electronegativities, using the $PN_{ME}$ gives the smooth color gradients, *i.e.* the clearest subdivisions of the predicted data.

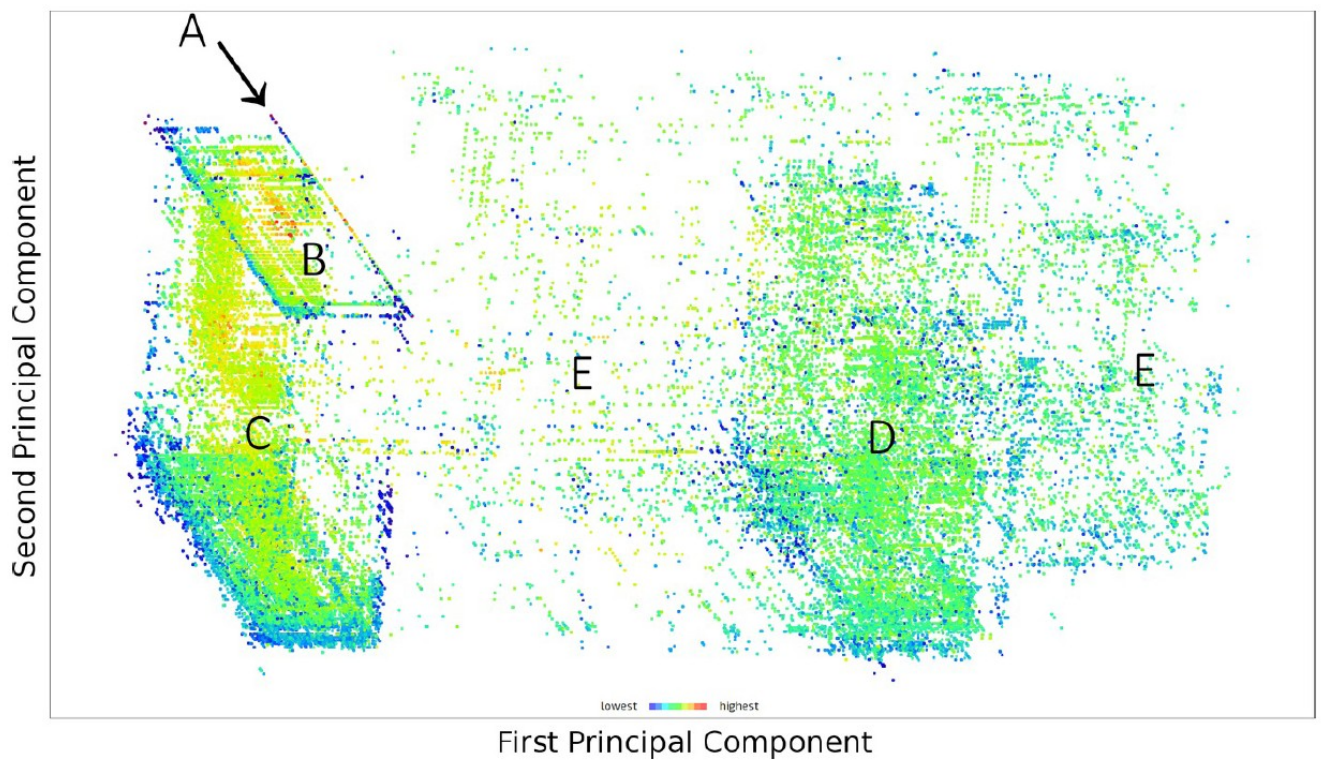

Fig. 66. Principal components of PNs and the predicted melting temperatures (color), element counts *N* = 1–5. Legend: A unaries, B binaries, C ternaries, D quaternaries, E quinternaries.

*8.3.5. Neighbor learning*

As a next step, for the 10 considered physical properties we made an attempt to fill in the numerous white spots, as seen in Fig. 67, using an advanced extrapolation technique via the radius-based neighbor learning. All the possible combinations of chemical elements (minus unrealistic ones, such as noble gases, actinides, Tc, Po) were taken to mimic the unknown inorganic substances with the element count *N*= 2–5. The element combinations of the existing PAULING FILE distinct phases were not considered, resulting in nearly seven millions of the totally unknown element combinations. For each element combination, 10 considered physical properties were predicted by the neighbor learning, based on the physical properties, obtained with the decision trees regression. Thus, each heat-map, as shown in Fig. 66 for the 10 predicted physical properties, was augmented with the nearly seven million values of the extrapolated physical property. An example of the enthalpy of formation is shown in Fig. 67. Now the patterns can be seen even for quinternaries (*N*= 5), which present the majority of the extrapolated points.

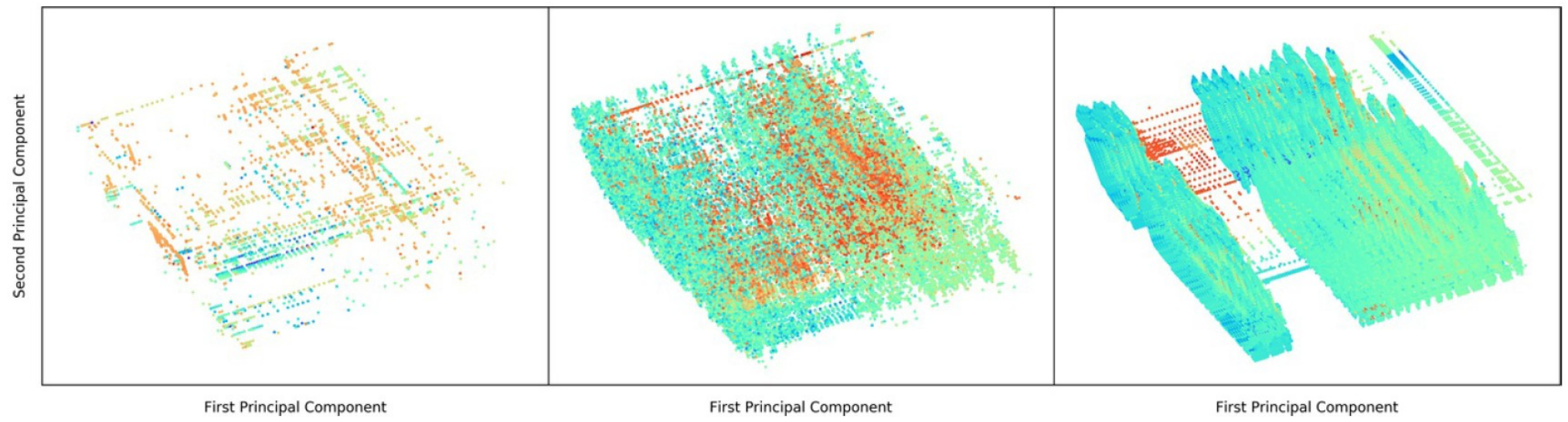

Fig. 67. From left to right: peer-reviewed, decision trees, and neighbor-learned principal components of PNs and the enthalpy of formation (color). The peer-reviewed and decision trees data stand for the existing PAULING FILE element combinations, whereas the neighbor-learned data stand for nearly seven millions of the unknown element combinations.

*8.3.6. Inorganic substances design*

The procedures discussed above were realized in an online application for the inorganic substances design **[17,87]**, which allows predictions of the structure prototype assignments, even for the exotic ranges of physical properties. This is the reversed task with respect to the previously described decision trees predictions. As the ranges of physical properties are wide, the searches are simple and return the existing machine-learning or peer-reviewed data. However, as the combination of physical property ranges becomes nontrivial, no existing data meets the search criteria. Then the design of the novel inorganic substances can be executed, based on the neighbor-learned data (*cf.* Fig. 67, right). The screenshot of the online application is shown in Fig. 68.

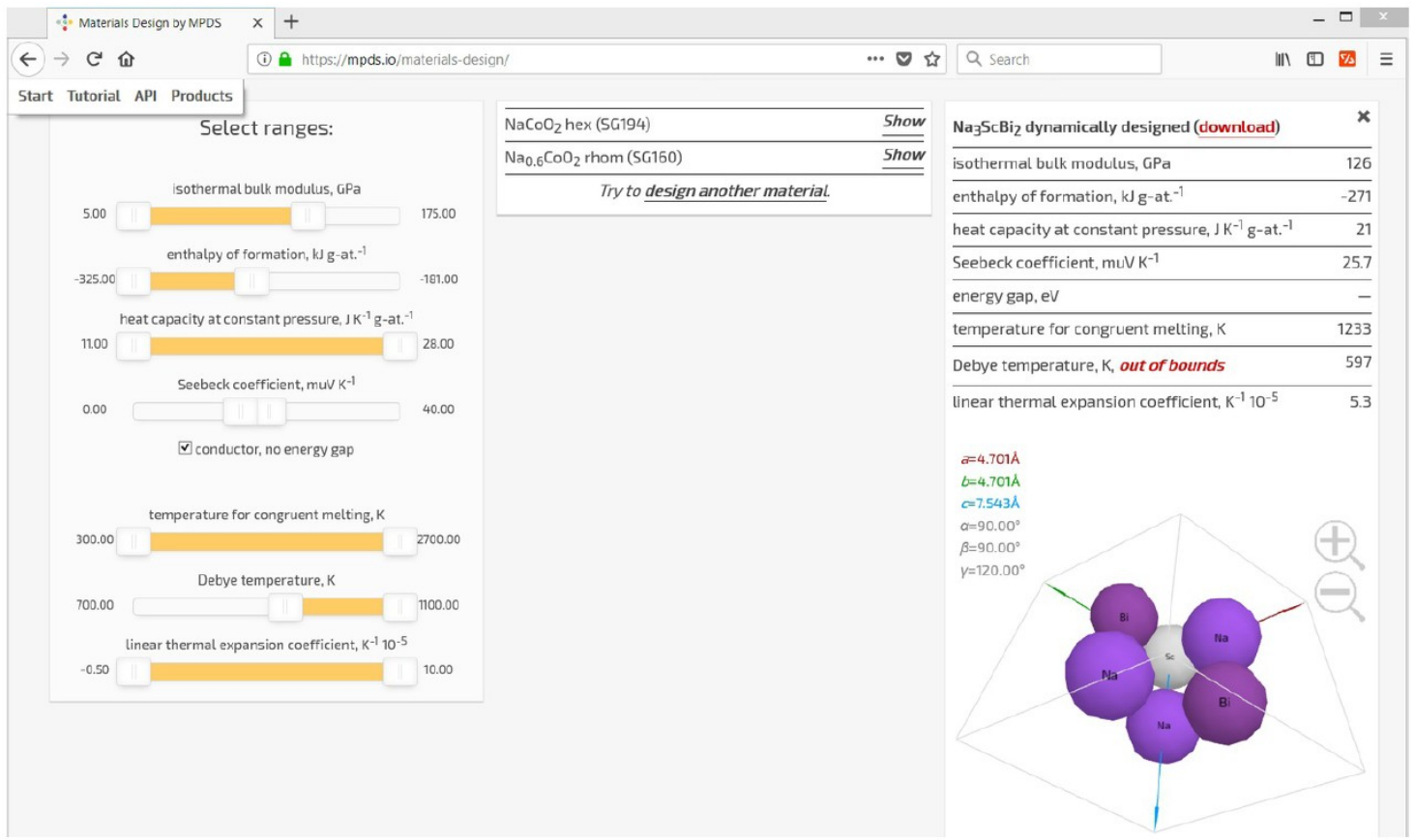

Fig. 68. An online materials design application screenshot **[17a,b]**. The structure at the right is hypothetical and was generated on the fly according to the selected ranges.

Thus, the chapter 8 reveals the following facts. Nowadays is the epoch of materials informatics and big data in materials science. Last years the US, Japanese and Chinese administrations announced the national 'Materials Genome' initiatives, and the several similar projects of large scale in EU arose. *Nonetheless we observe that the quantity of data being collected – especially simulated data – is still often prioritized in prejudice of quality.* On the other hand, PAULING FILE project, since its beginning in 1992, has been always focused on providing the data of the highest possible quality. This was possible not only considering the peer-reviewed research, but also investing tremendous manpower in the critical evaluation, such as manual structure prototype assignments, conversion of phase diagrams, systematizing the collected inorganic substances into the *distinct phases etc.* As here shown, the researchers, wishing to analyze big data in materials science, now get unprecedented power and flexibility, even with the aid of relatively simple tools, such as decision trees and neighbor learning. Using the data collected from over 309,000 publications, we trained the decision trees algorithm to predict 10 physical properties with an acceptable quality, based only on the structure prototype assignment. Such predictions were done for 114,872 distinct inorganic substances (56,407 distinct chemical element systems). Using the periodic numbers $PN_{ME}$ of the chemical elements, we have plotted the heat-maps, highlighting the patterns for all the predicted physical properties (see Figs. 69a-d; electrical conductivity and thermal conductivity are not shown). For example, based on these patterns, the unary, binary, ternary, quaternary, and quinary inorganic substances can be clearly subdivided. Also based on these patterns, one may visually estimate the particular value of the physical property of interest for any given inorganic substance. Finally, thanks to the presence of these patterns, one may compile a possible structure prototype assignment (real or hypothetical) for a given combination of values of the 10 physical properties. These results were obtained using the relatively cheap and unsophisticated machine-learning techniques. *With that we accentuate the role of the quality and quantity of the input (training) data. Moreover, except PAULING FILE, there exists no other experimental databases, able to provide the required layout of machine-readable data (i.e. structure prototype assignments and physical properties, interconnected via the Distinct Phases Concept)*.

Realizing that, we open-sourced the results of this chapter and are open to collaborations, anticipating that the other more advanced and impactful studies based on our data will follow soon.

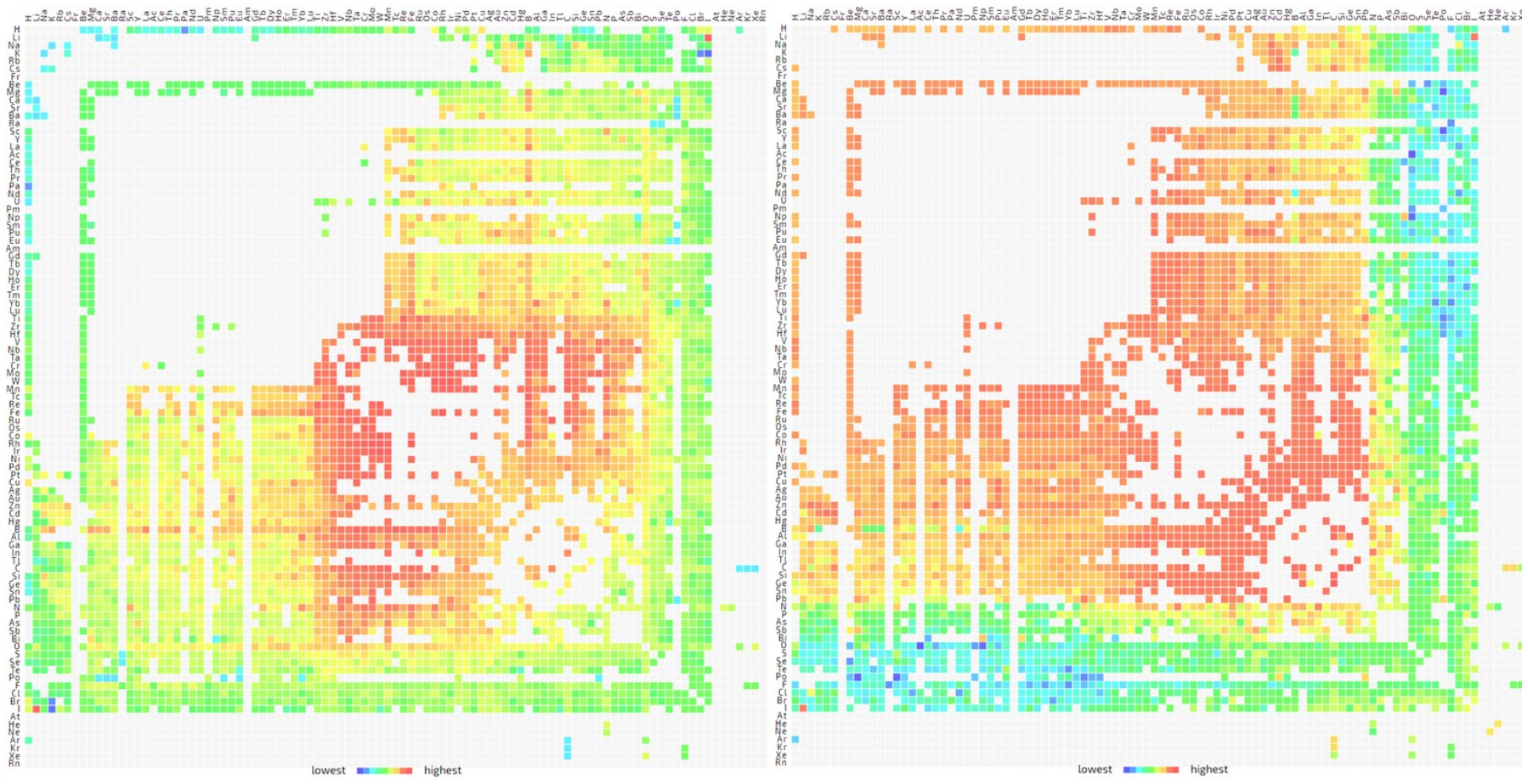

Fig. 69a. Heatmaps of $(PN_{ME})_A$ *vs.* $(PN_{ME})_B$ *vs.* predicted values (colors) for all distinct binary phases within the specific chemical element systems. Left: isothermal bulk modulus. Right: enthalpy of formation.

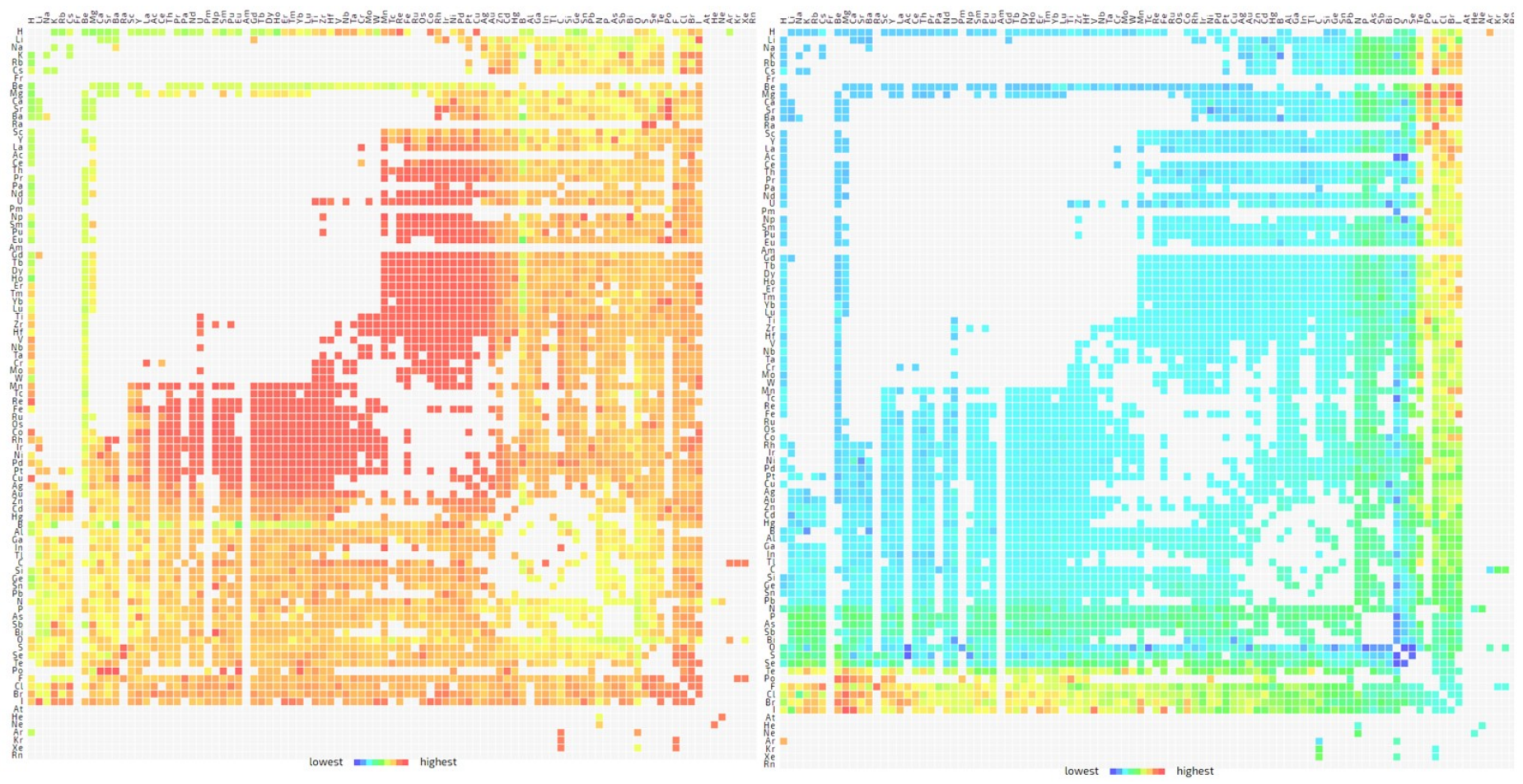

Fig. 69b. Heatmaps of $(PN_{ME})_A$ *vs.* $(PN_{ME})_B$ *vs.* predicted values (colors) for all distinct binary phases within the specific A-B systems. Left: heat capacity at constant pressure. Right: Seebeck coefficient.

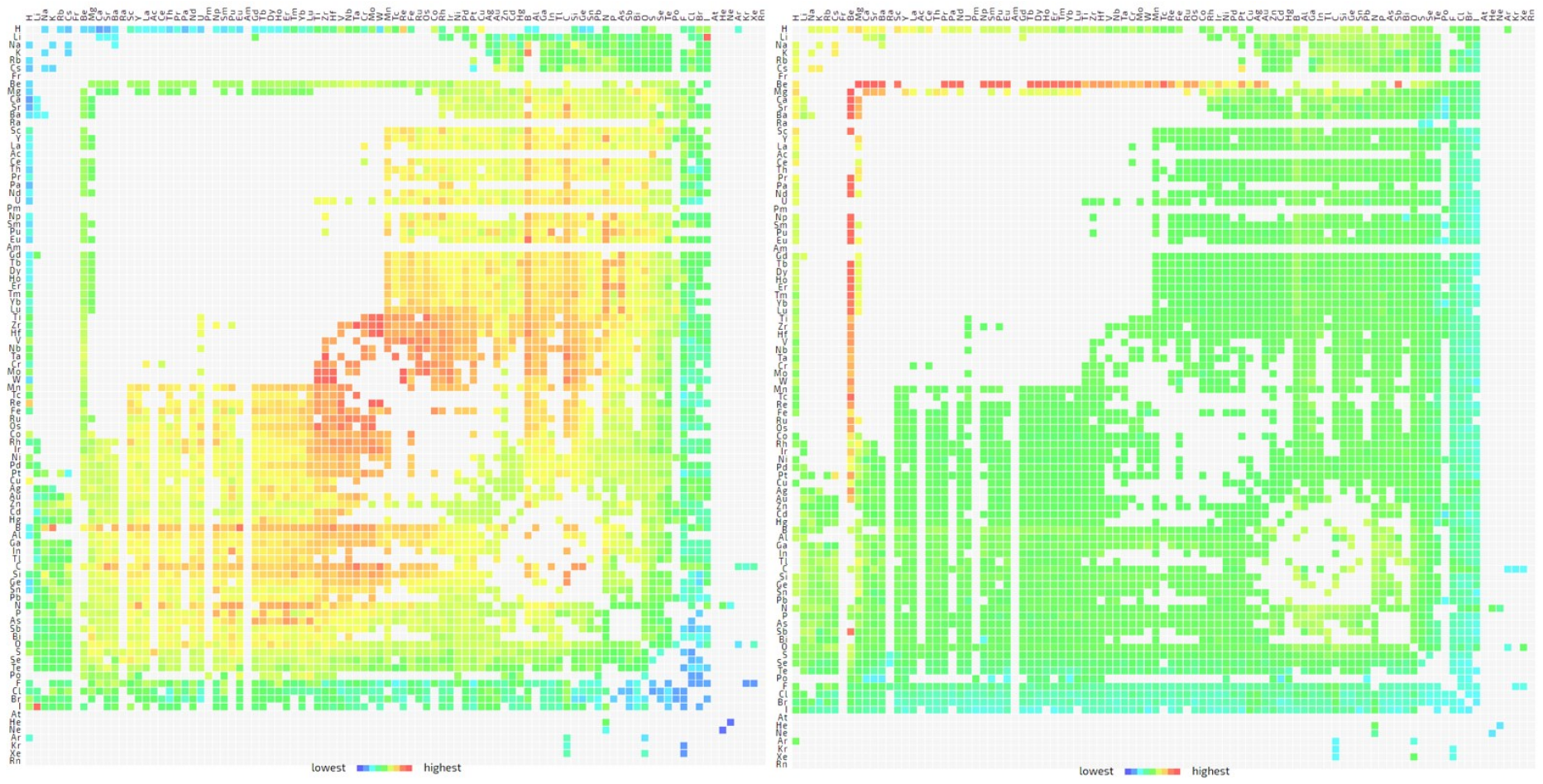

Fig. 69c. Heatmaps of $(PN_{ME})_A$ *vs.* $(PN_{ME})_B$ *vs.* predicted values (colors) for all distinct binary phases within the specific A-B systems. Left: temperature for congruent melting. Right: Debye temperature.

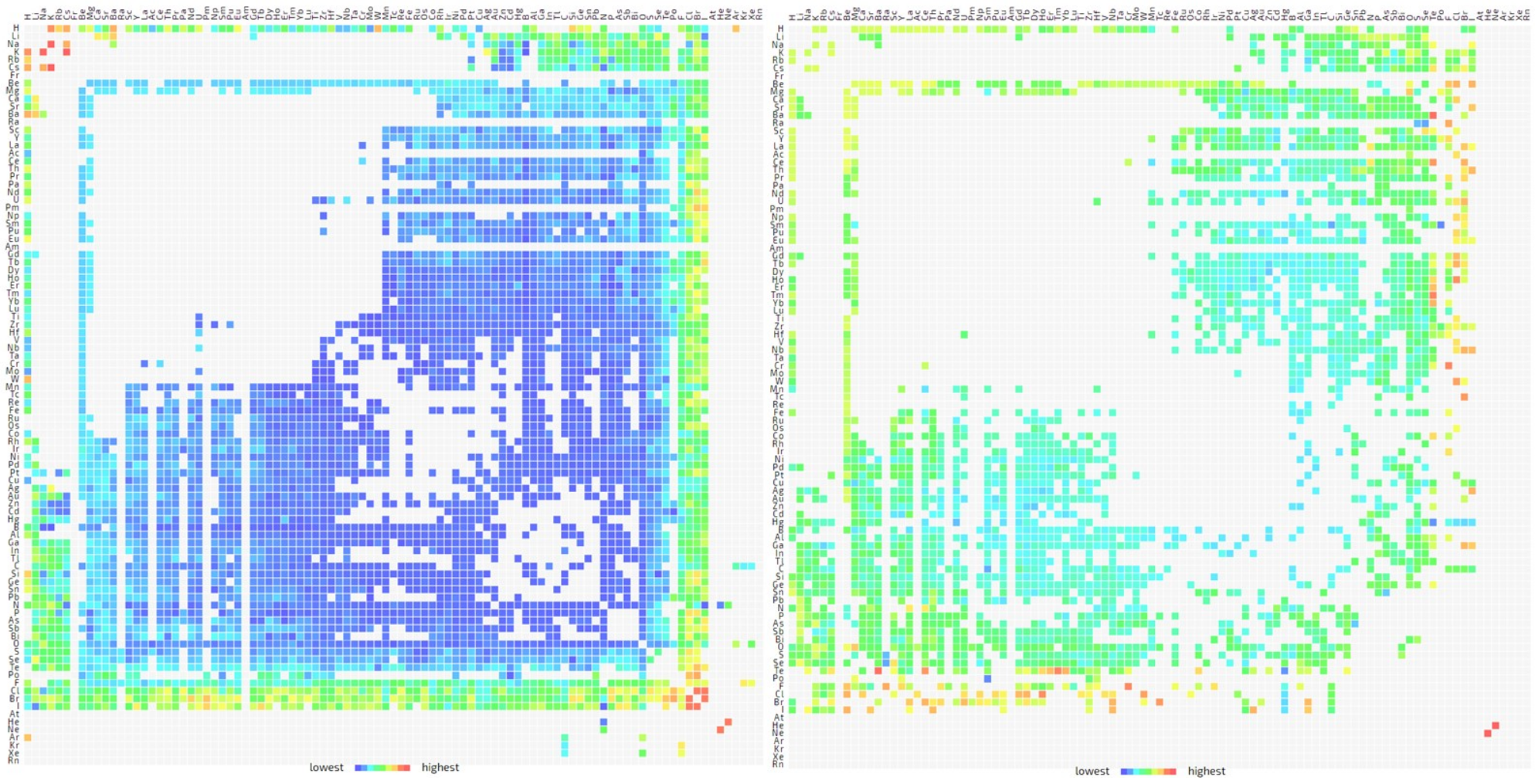

Fig. 69d. Heatmaps of $(PN_{ME})_A$ *vs.* $(PN_{ME})_B$ *vs.* predicted values (colors) for all distinct binary phases within the specific chemical element systems. Left: linear thermal expansion coefficient. Right: energy gap (direct or indirect) for insulators.

The heat plots provide 'Holistic Views' and have a strong predictive power for the systems, where no experimental data are available. Therefore they can be efficiently used for the new material design, targeting the inorganic substances with predefined property (in a well defined value range). Analogous heat plots are also available for the ternary systems (3D-cubes) and quaternary systems (2D-, 3D-projections).

## 9. Holistic views: Quantitative trends in physical properties via *ab initio* simulations starting from the peer-reviewed PAULING FILE data (work in progress)

### 9.1. Introduction

Quantum-mechanical modeling based on the density functional theory (DFT) present an efficient way for predicting many materials properties at the atomic scale. The idea to combine the large experimental databases with the high-throughput DFT simulations is not new. One of the pioneers in this direction is the Materials Project **[74]**, followed by the other research groups, e.g. **[82]**. Recently the PAULING FILE database was also used for predicting the materials properties at scale **[109]**, as well as the MPDS with its API was used for the combined ab initio and machine-learning study **[110]**. The cornerstone is the modeling automation, as the high-throughput calculations are unthinkable to perform manually. One of the well-known automation engines today is the AiiDA **[75]**, and there exist also a number of others, e.g. **[111]**. Nowadays this field is also being successfully commercialized **[76,78,112,113]**. Therefore this work logically continues the previous efforts, however employs the new modeling approximations.

Here we employ the high-throughput database-assisted DFT to get the accurate encyclopedic, reference, and benchmarking data, as well as the vast systematic training data for the machine learning. We adopt the cheap commodity cloud environment (not the expensive HPC clusters). We ensure the provenance tracking and reproducibility of our simulations using the workflow automation engine AiiDA. Finally, we open all our simulation results, including the raw simulation data, online at the MPDS. As a result, thanks to a combination of the selected software tools and the available input data, our modeling is cheap, fast, and verifiably accurate. Our simulation results and data can be unlimitedly and anonymously downloaded via the online MPDS GUI. Their advanced retrieval is also supported in the programmatic access interface (MPDS API), for that one needs to log in for free.

## 9.2. Methods

### *9.2.1. Ab initio modeling*

Several theoretical frameworks for modeling the electronic structure and the related properties from *ab initio* (i.e. from the first principles) exist. Each of them is called basis set approximation or, traditionally, electronic *ansatz*. The Gaussian and plane wave basis sets are the common choices to represent electrons in periodic crystalline systems. In this work we employ the Gaussian basis sets, as implemented in the CRYSTAL modeling engine **[114]**. Their use is better linked to a chemical experience of the molecular codes and particularly suitable to the description of crystals with the chemical bonds. On the other hand, the description of free or nearly free electrons in conductors is hard to achieve within this framework due to numerical instabilities. However, in most cases, we also successfully manage to tackle these instabilities using specifically designed numerical setups and advanced control tools. The CRYSTAL supports simulation of many physical properties, the considered ones with the respect to the PAULING FILE hierarchy are shown in the Table 25. We use the PBE0 hybrid Hartree-Fock DFT functionals for all the simulations. Our computational set-up automatically distinguishes the metallic and non-metallic systems, adopting a denser *k*-point mesh and higher tolerances for integral calculations for the metals, than for the non-metals. The tolerance of the energy convergence on the self-consistent field cycles is set to $10^{-9}$ a.u. Additionally the DFT density and grid weight tolerances are increased and an extra-large pruned DFT integration grid is adopted.

| Domain | Physical property | Data type | Calculation cost |
|---|---|---|---|
| Electron properties | direct and indirect band gaps | scalar | low |
| $T = 0$ K | electron densities of states | array | low |
| | electron band structures | array | low |
| | Mulliken effective charges | scalar | low |
| | cohesive energy | scalar | low |
| Mechanical properties | isothermal bulk modulus | scalar | moderate |
| $T = 0$ K | Young modulus | scalar | moderate |
| | shear modulus | scalar | moderate |
| | elastic tensors | array | moderate |
| | Poisson ratio | scalar | moderate |
| Optical properties | optical phonon frequencies | array | high |
| $T = 0$ K | phonon densities of states | array | very high |
| | phonon band structures | array | very high |
| | infrared intensities | array | very high |
| | Raman intensities | array | very high |
| | refractive indices | scalar | very high |
| | birefringence | scalar | very high |
| Thermodynamic properties | heat capacity at constant pressure *vs.* pVT | array | high |
| $T > 0$ K | vibrational entropy *vs.* pVT | array | high |
| | Gruneisen parameters | array | high |

Table 25. Simulated physical properties and their estimated calculation costs.

At the first step, the total energy minimization is performed in order to find the relaxed equilibrium atomic structure of the crystal. The calculated structures were used for phonon calculations via the direct frozen-phonon method in the harmonic approximation. In order to obtain the temperature dependence of the thermodynamic properties, the summation over a finite number of frequencies in the Brillouin zone center is performed. Further, the CRYSTAL engine computes the elastic tensors. All these steps are automated internally inside the CRYSTAL, and their sequence and the underlying data management are automated externally as the so called workflows (Fig. 70).

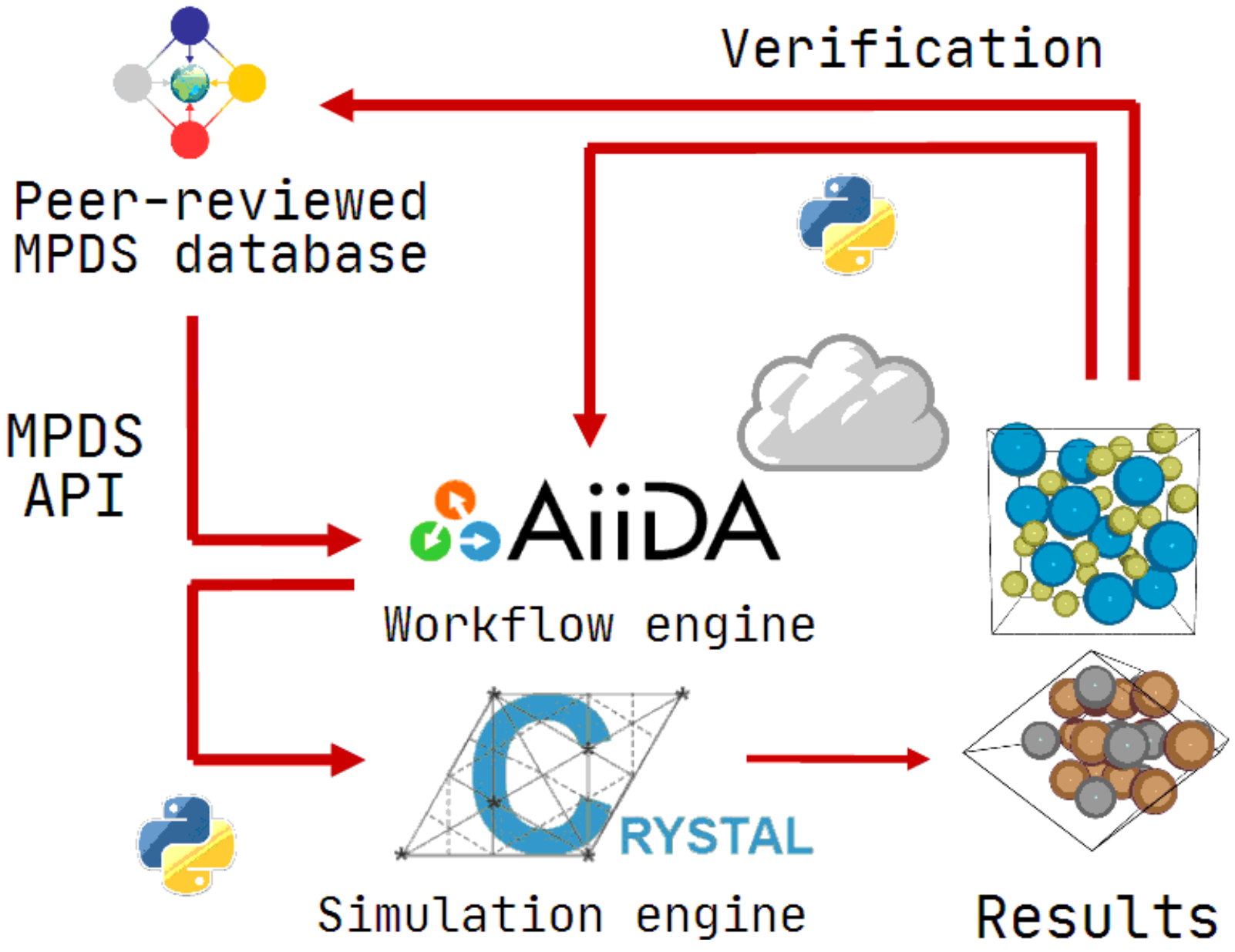

Fig. 70. A schematic computational workflow developed.

*9.2.2. Database-assisted high-throughput workflows in the cloud*

The AiiDA was chosen as the main workflow control engine. Thus, the data provenance and veracity are prioritized, and the results are accumulated in a systematic, machine-readable manner, so that every simulation step is tracked. Not only the results of simulations are represented as the data, but also the process of creation of these results is kept as the data. Currently the AiiDA does not support the commodity cloud hardware, as it was designed for the HPC clusters. Therefore a new computing control manager and scheduler was programmed specifically for inexpensive cloud environments and linked to the AiiDA engine. All the programmatic communications with the cloud hardware providers are automated. The allocation and the release of the computing resources and all the auxiliary setups are done programmatically on-demand, i.e. only while the cloud resources are actually required. Two commercial on-demand cloud providers are currently supported: Hetzner and Upcloud. In summer 2020, the chosen default Hetzner configuration (8 CPUs, 32Gb RAM) runs a test task for about 2.5 hours on average and costs €36 per month, the chosen default Upcloud configuration (8 CPUs, 4Gb RAM) runs a test task for 1.5 hours on average and costs $89 per month. Adding the other cloud providers is relatively easy and supported in a modular manner. On average, calculation of properties shown in Table 25 (several consequent tasks in total) takes about two days for each distinct phase.

*9.2.3. Automatic control and error handling*

The crystal structures to be simulated were taken from the MPDS using its API according to the following criteria: (*a*) the disordered as well as the monoclinic and triclinic systems were avoided, (*b*) the systems with the number of atoms per unit cell bigger than 40 were avoided, (*c*) the systems with the chemical elements He, Kr, Xe, Po, At, Rn, Fr, Ra, and Fm were omitted as no robust basis sets for these elements were found. Although the coverage is relatively big, it stays not fully systematic, as a lot of binary and ternary systems have to be omitted. The weak side of the selected scheme is that the full automation is implied by the AiiDA, and even the minor manual interventions in the workflows for tuning computational details are breaking and therefore not supported. On the other hand, the CRYSTAL engine requires careful fine-tuning the computational details, contradicting the high-throughput AiiDA approach. The whole idea of the provenance becomes entangled, as it needs to be re-implemented in some way outside AiiDA. Notwithstanding, all the selected parameters are kept and remain reflected in the program code and the simulation raw data, being re-usable, and tightly integrated with the AiiDA. We expect it will be helpful later, as more complex systems will be simulated.

One of the main computational difficulties while using the CRYSTAL engine is poor simulation convergence. This occurs e.g. when the studied system has different charge distributions possible, and the code needs some reasonable initial guess about the probable solutions. In other words, the assumptions have to be done by the scientist (or a guided external algorithm), and the modeling code confirms or rejects them. On the other hand, this ensures higher quality of the results, compared to the "black-box" modeling codes. Thus, the initial automatic re-distribution of electrons over the orbitals of different atoms is helpful and was implemented and applied in this work. The algorithm of electron redistribution employs the concept of known oxidation states. The idea is to redistribute the valence electrons across the atoms comprising the phase, maintaining the electro-neutrality of the phase as a whole. Also, the resultant distribution of the oxidation states should be known. Using this technique, many convergence issues were workaround.

A somewhat similar approach is used for considering the magnetic systems. The magnetism treatment in the CRYSTAL involves setting up spin-polarized DFT or unrestricted Hartree-Fock calculations and fixing the number of unpaired electrons, either across the whole cell, or on the individual atoms for the given number of steps. This is done to force the calculation converge to spin-polarized solution to a local minimum, as a non-magnetic solution always exist in CRYSTAL. We had to estimate the approximate magnetic moment across the cell before the calculation starts. We did this using Russell-Saunders coupling, a crystal field theory method. Our algorithm always prefers high-spin values. A room for improvement exists here with respect to the antiferromagnetic (AFM) solutions. They are more difficult to construct in a high-throughput way due to geometrical frustration of the lattice, i.e. several possible ground states with different relative spin configuration. The question on systematic considering the AFM phases for calculating the properties of transition metals remains open.

### 9.3. Results and discussion

#### *9.3.1. Overview*

We have simulated the vast array of systematic multi-purpose materials data, using the cheap commodity cloud environment. In total, the electron, mechanical, optical, and thermodynamic properties are computed for about 10,000 unary, binary, and ternary distinct phases (the counts are steadily growing). Herewith only the results for the binary compounds are presented, whereas the analysis of the ternary compounds is work in progress and will be reported in the future.

#### *9.3.2. Simulation quality assurance*

We automated the comparison of the simulated values and the available peer-reviewed experimental values and assured in the very high quality of simulations. The Table 26 presents some randomly taken unary and binary compounds. The absolute error does not exceed 15% in the majority of cases. In several cases, the reported experimental values are widely scattered, so the closest value is chosen. All the simulation results, including the scalar and array values, interactive plots, as well as the simulation log files, are released online at the MPDS under the permissive CC BY 4.0 license. They are accessible either by keyboarding "*ab initio simulations*", or selecting the corresponding data section at the landing screen or at the results refinement screen. The experimental values for the band gap, and isothermal bulk modulus are especially well reproduced (see further), whereas the phonon modes and effective charges are slightly worse reproduced. The values of the Poisson ratio and heat capacity at constant pressure are limited with the relatively narrow intervals (see e.g. Fig. 64, values in blue), so they are less sensitive for a direct comparison. We have also compared the other simulated properties with the experimental data available to us and will report them in our next works. Notably, these results should not be treated at the same level with the machine-learning results (chapter 8), providing only a statistical estimate. On the other hand, surprisingly, we observe that the overall results look similar.

| Compound | Calculated value (MPDS) | Experimental value (PAULING FILE) |
|---|---|---|
| | *Selected optical phonon frequencies, cm$^{-1}$* | |
| Ge | 320 | 355 |
| InSb | 194 | 181 |
| BAs | 734 | 704 |
| | *Heat capacity at constant pressure, J K$^{-1}$ mol$^{-1}$* | |
| Hf | 8 | 23 |
| FeO | 19 | 47 |
| Te | 48 | 26 |
| | *Isothermal bulk modulus, GPa* | |
| Al | 118 | 75 |
| Sb | 41 | 41 |
| TiC | 304 | 270 |
| | *Poisson ratio* | |
| Al | 0.33 | 0.35 |
| TiC | 0.22 | 0.22 |
| NiO | 0.28 | 0.31 |
| | *Effective charge* | |
| $FeBr_2$, Fe atom | 0.44 | 0.51 |
| EuTe, Eu atom | 1.92 | 3.00 |
| $YB_6$, B atom | -0.17 | -0.37 |
| | *Direct band gap, eV* | |
| $Cu_2O$ | 2.6 | 2.2 |
| AlAs | 4.1 | 3.9 |
| GeSe | 1.7 | 1.6 |

Table 26. Evaluation of the *ab initio* results for some randomly taken compounds.

*9.3.3. Binary inorganic substances (N= 2)*

Analogous to the machine-learning results in Chapter 8, we present the experimental peer-reviewed isothermal bulk modulus, direct energy gap, and heat capacity at constant pressure as the matrix heat-maps (Fig. 71-73). Again, we took the average value for all simulated binary phases (e.g. $A_2B$, AB, $AB_2$) within the specific chemical element system, since in the majority of cases the simulated values vary inconsiderably across the phases in the element system. The chemical elements are sorted according to their $PN_{ME}$s. As the disordered systems are omitted, therefore the resulted matrices look sparser than for the machine learning.

Again, one can observe clear patterns for the isothermal bulk modulus and slightly less visible patterns for the band gap. Notably, these two properties are better reproduced among others, compared to the experimental values. The machine-learning pattern for the bulk modulus is repeated (higher values are in the center, lower values are at periphery). The pattern for the direct band gap almost coincides to the indirect band gap pattern (in contrast to the machine learning, simulations clearly distinguish the band gap types). It looks inverted with respect to the bulk modulus (lower values are in the center, higher values are at periphery). Apparently it also repeats the machine-learning pattern.

Thus, the chapter 9 reveals the following facts. Counter-intuitively, the advanced high-throughput DFT simulation technique provides the results similar to the machine learning on the large scale. The simulated values distribution across the binary compounds (heat-maps) for the isothermal bulk modulus and the band gap illustrate periodic patterns, being sorted according to the constituent chemical elements $PN_{ME}$s. (The other simulation results are also publicly available online at the MPDS without restrictions.) On an example of the binary compounds, at least for the isothermal bulk modulus and the band gap prediction, our machine-learning and *ab initio* modeling approaches are complementary.

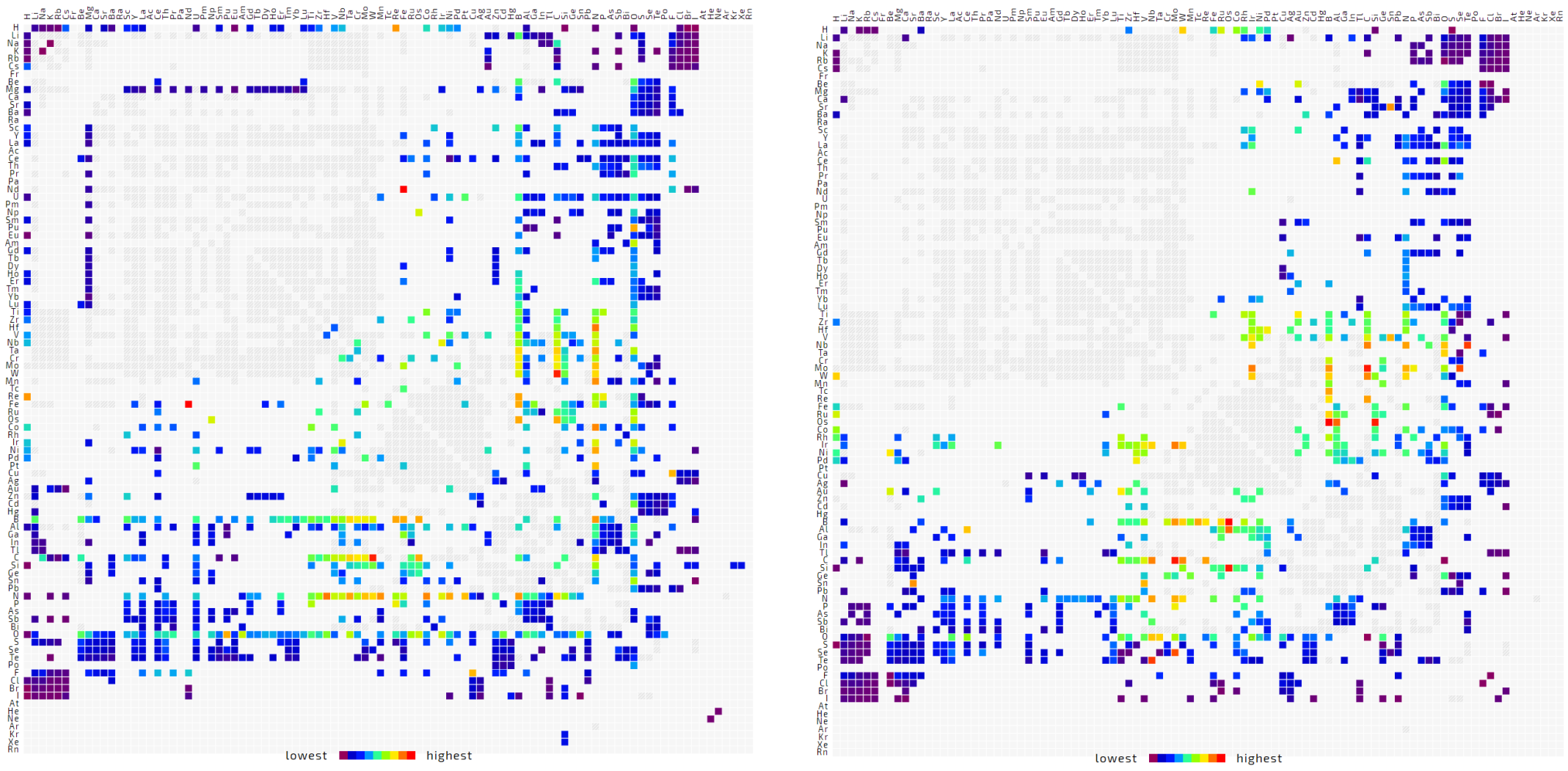

Fig. 71. Heatmaps of $(PN_{ME})_A$ *vs.* $(PN_{ME})_B$ *vs.* isothermal bulk modulus (colors) for the binary phases within the specific chemical element systems. Left: peer-reviewed experimental data. Right: simulated data.

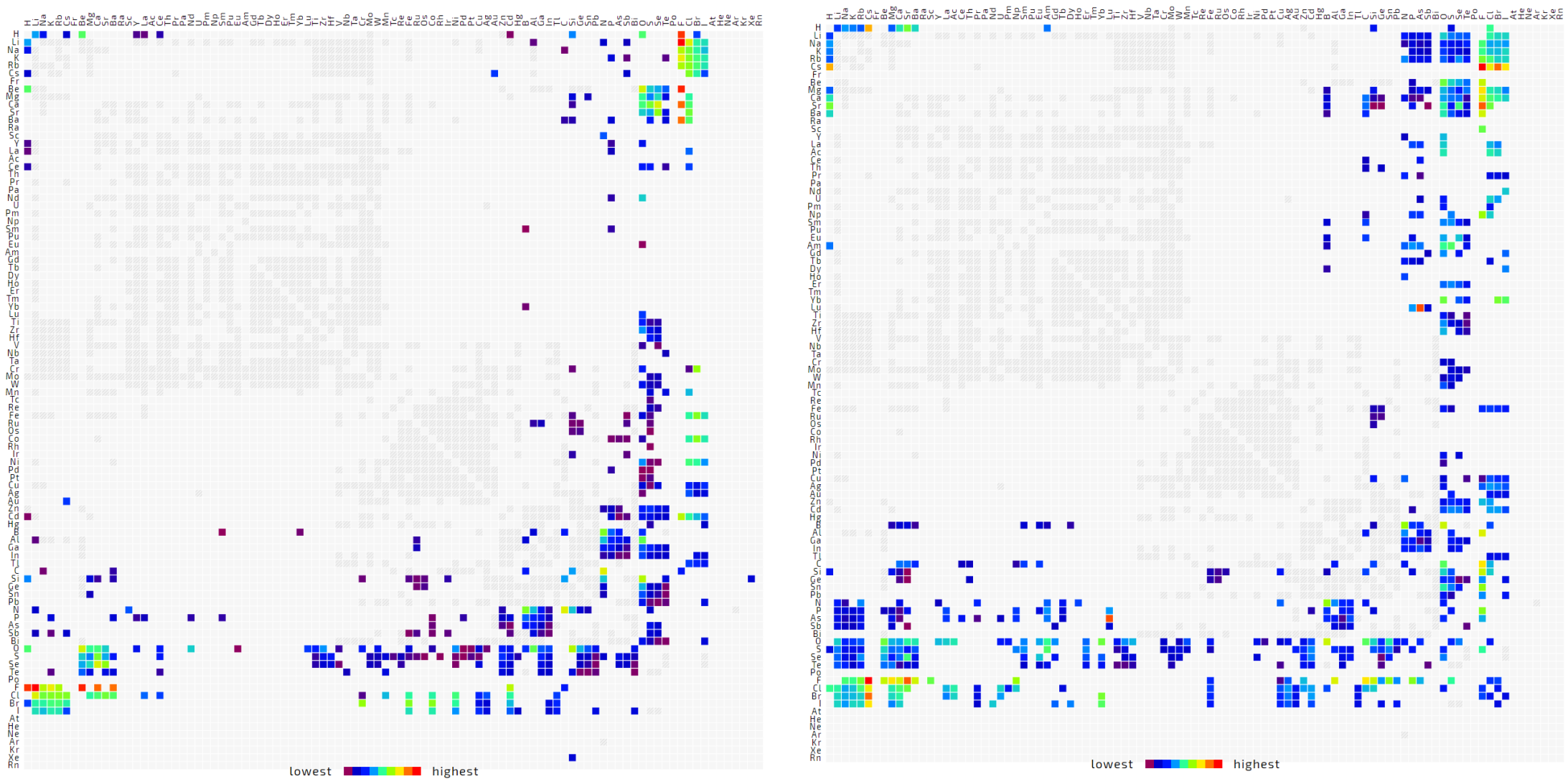

Fig. 72. Heatmaps of $(PN_{ME})_A$ *vs.* $(PN_{ME})_B$ *vs.* direct band gap (colors) for the binary phases within the specific chemical element systems. Left: peer-reviewed experimental data. Right: simulated data.

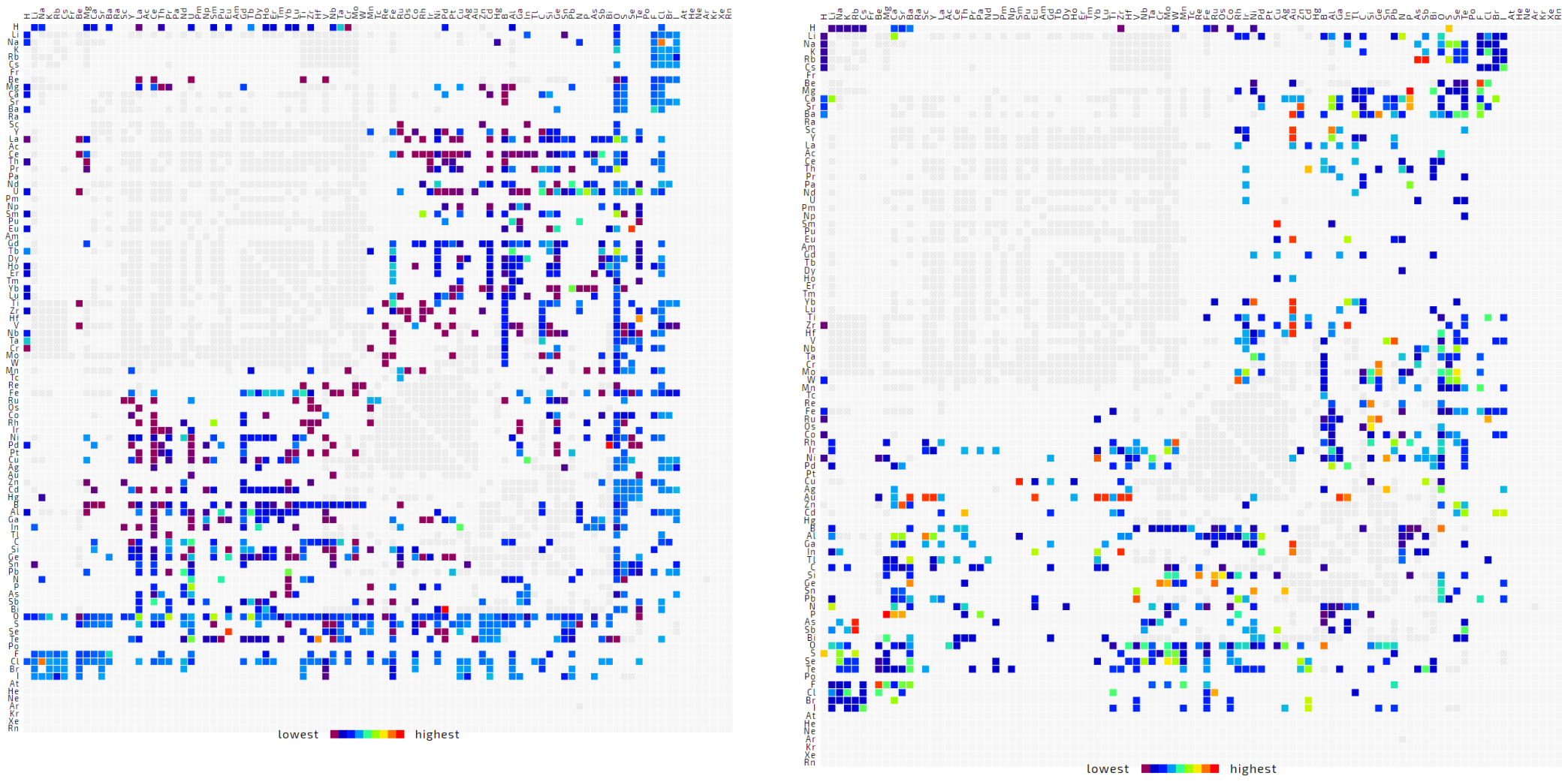

Fig. 73. Heatmaps of $(PN_{ME})_A$ *vs.* $(PN_{ME})_B$ *vs.* heat capacity for constant pressure (colors) for the binary phases within the specific chemical element systems. Left: peer-reviewed experimental data. Right: simulated data.

## 10. Materials Platform for Data Science (MPDS)

### 10.1. Introduction

As described earlier, the PAULING FILE is the relational database for materials scientists, grouping crystallographic data, phase diagrams, and physical properties of inorganic crystalline substances under the same frame. Its focus is put on the experimental observations, and the data are processed from the original publications, covering world scientific literature from 1891 to the present date. Each individual crystal structure, phase diagram, or physical property database entry originates from a particular publication. Meanwhile the PAULING FILE project is relatively well-known. In October 2019, one of the founders of the PAULING FILE project (P.V.) was acknowledged for the fundamental research supporting data-driven materials research with the NIMS Award (Japan). Chapter 4.11 lists the products based on the PAULING FILE. Among them is the MPDS which presents all the materials data, extracted by the project PAULING FILE team from the scientific publications. Even nowadays this task cannot be fully automated, so 309,460 publications in physics, chemistry, and materials science had to be manually processed and systematized. The results are now available online at the MPDS.

### 10.2. Online edition of the PAULING FILE

The MPDS is an online edition of the PAULING FILE materials database. All the data are presented in two ways (online interfaces): browser-based graphical user interface (GUI) and application programming interface (API). The MPDS data provided in API and GUI are generally the same. There are however two important points to consider. First, the GUI is intended for the human researchers, not for the automated processing. Therefore the data in GUI are less rigorously formatted, being thus closer to the original publication (which, of course, impedes machine analysis). Second, some portion of data (~15%) present in the GUI is absent in the API. This is because currently a number of MPDS entries have only textual content or even no content at all (in the process of preparation), i.e. present little value for the data mining. However, we make our best to expose the maximum MPDS data in the API, and this is the work in progress.

To summarize, the API differs from GUI in the following: (a) machine-readable (and to a certain extent machine-understandable), (b) programmer-friendly, pluggable and integration-ready, (c) massively exposing the data, i.e. all the MPDS content can be analyzed in minutes. Thus, thanks to the API, the MPDS can be used in a variety of other scientific ways, which we had never designed. So the researchers get unprecedented flexibility and power, which is unthinkable within the GUI.

### 10.2.1. GUI access

Here the browser-based user interface (GUI) is described, whereas the programmatic usage is covered in the API chapter. The GUI interface is extremely fast, in general, the retrieval results are shown within a second. Full access to all the data in all the supported formats (CIF, PDF, PNG, BIBTEX etc.) is provided by the subscription. Free access is also possible, although limited. In addition, some parts of the data are open-access. In particular, these are: (a) *cell parameters - temperature diagrams* and *cell parameters - pressure diagrams*, (b) all data for compounds containing both *Ag* and *K*, (c) all data for *binary* compounds of *oxygen*, (d) complete set of selected 'best' entries for all binaries are open-access for education, advertising and research purposes.

*10.2.1.1 Search criteria and modes*

Search of data at the MPDS is possible according to 14 criteria: 8 in physics or chemistry (materials classes, physical properties, chemical elements, chemical formulae, space groups, crystal systems, prototypes, and atomic environments) and 6 in bibliography (publication author, years, journal, geography, organization, and DOI). There are two search modes: simple and advanced. In the simple mode different search terms can be typed all in a single input field (see Fig. 74). Here the most frequently used 6 criteria are supported: materials classes, chemical elements, chemical formulae, atomic environments, crystalline classes, and physical properties. All they will be correctly recognized and attributed to the search keywords given. In the advanced mode each of the search criteria has its own input field. To use it, either click the middle search menu button, or click the criteria boxes shown at the right of the results pages. Fig. 75 shows the tab 'crystallography' with 10 different selection fields, all supported by helpful drop-downs, as well as easy to use selection of ranges. Let us get acquainted with the meaning and proper usage of each criterion of search.

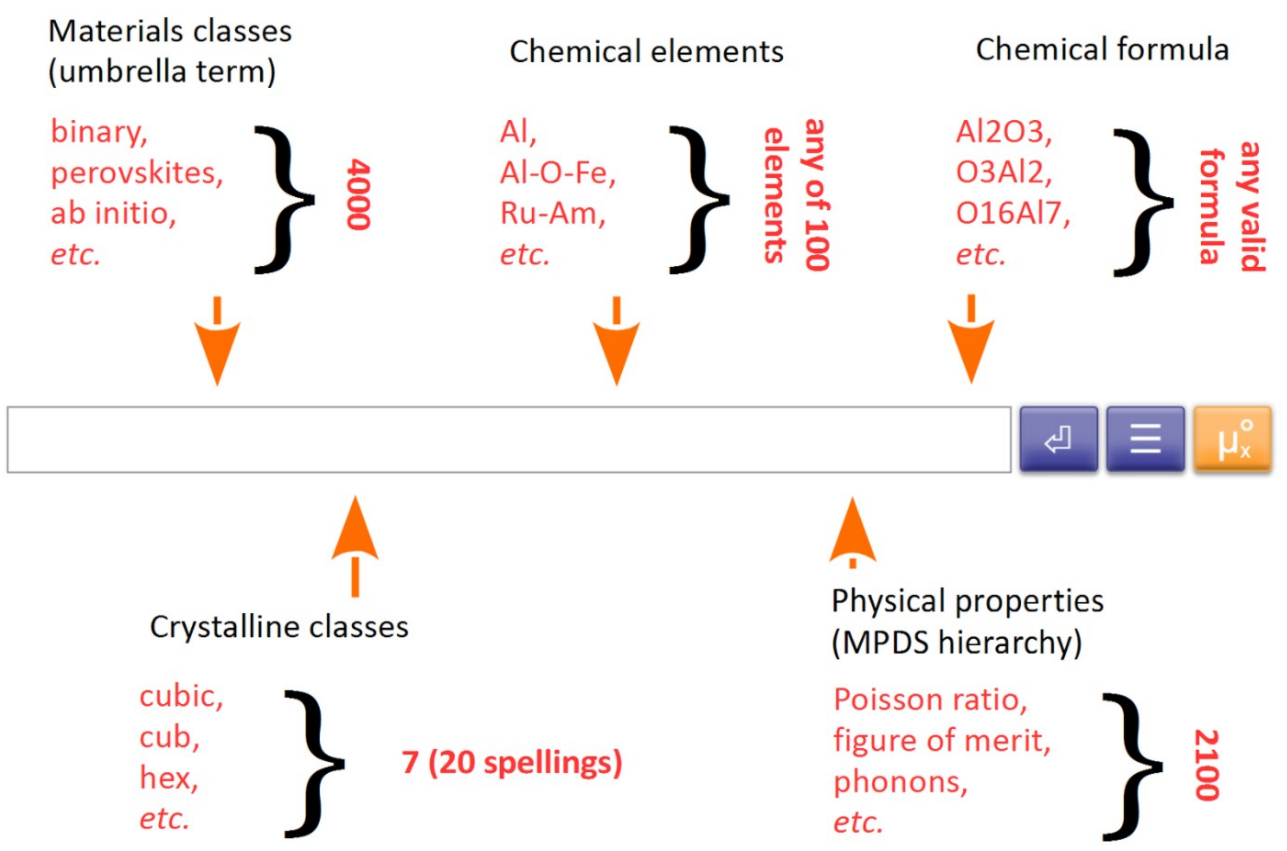

Fig. 74. Simple (one-input-field) mode of search.

*10.2.1.2. Materials classes*

In this category various materials classes are collected, ranging from technical terms to physical categories, chemical names, element counts, periodic system groups, some isotope names etc. There are lots of auxiliary terms, only applicable to the specific domains, e.g. *cell-only*, *disordered*, and *non-disordered* are valid for the crystalline structures (S-entries). Another example: the term *ab initio literature* refers to the data taken from the theoretical first-principles modeling papers. Moreover, the majority of the known mineral names are supported, e.g. *perovskite*, *baddeleyite*, *stishovite*, *yeelimite*, etc. Five special ("arity") classes *unary*, *binary*, *ternary*, *quaternary*, and *quinary* restrict the distinct element count of the results.

### *10.2.1.3. Chemical elements*

Chemical elements can be typed as names or symbols (e.g. *copper* or *Cu*). Obviously, chemical elements can be combined arbitrarily in searches, using spaces, commas, or dashes as the separators. By default, equal or greater count of elements is implied, e.g. the results for Cd-O-S may contain not only Cd, O, and S, but also Tl, H, N, K etc. To restrict the elements count, the arity materials classes *unary*, *binary*, *ternary*, *quaternary*, or *quinary* should be added, e.g. *Cd-O-S ternary*.

### *10.2.1.4. Chemical formulae*

In the chemical formulae order of elements does not matter. However the results will contain the chemical formulae with the standard order of elements (according to their electronegativity).

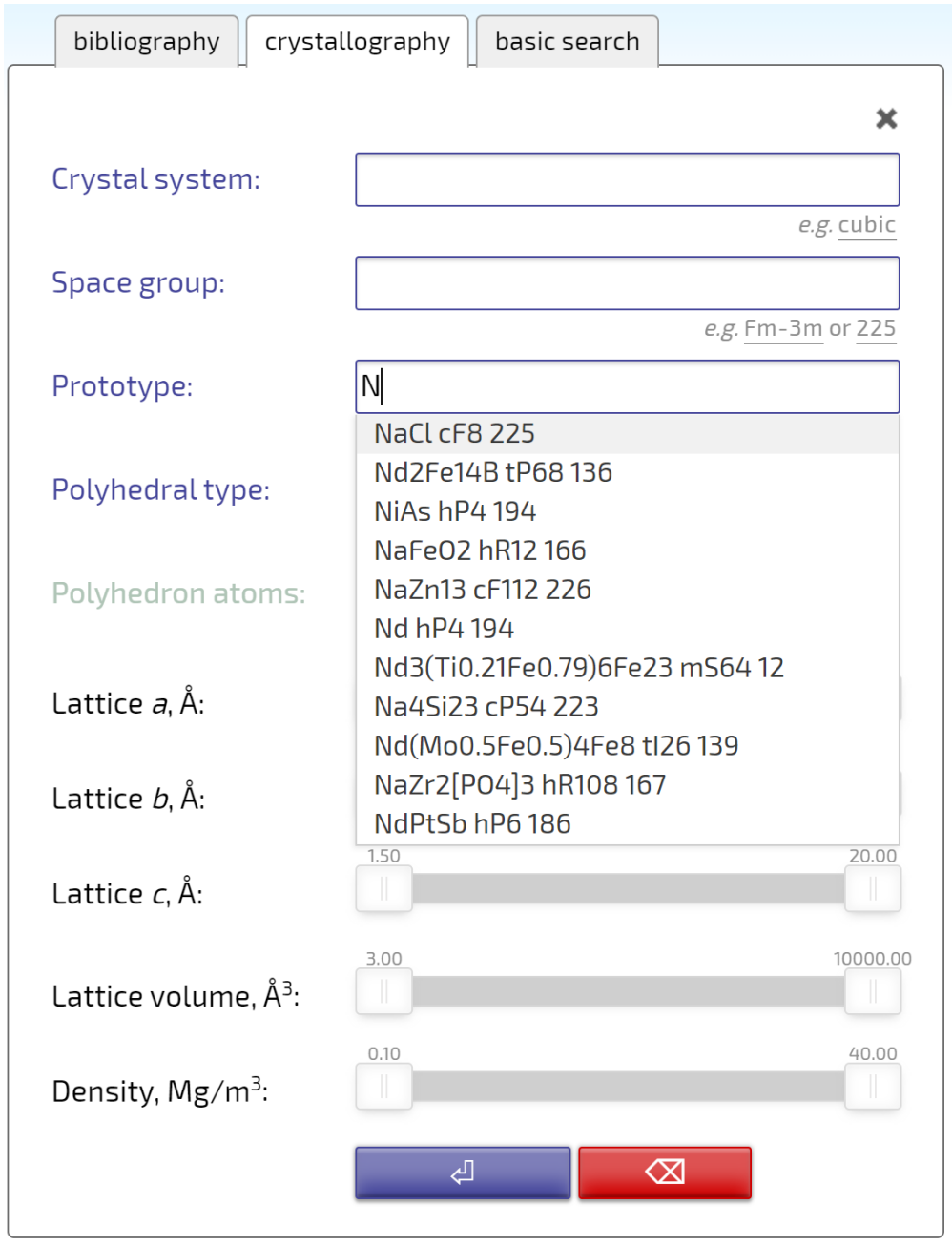

Fig. 75. Advanced search mode ("crystallography" tab).

### *10.2.1.5. Crystal systems and space groups*

Seven crystal systems and 230 space groups are fully supported. The space groups can be specified as the number or international short symbol. Full list of crystal systems and space groups can be found e.g. in Wikipedia. Note, that crystal systems, space groups, and prototype systems (see below) are mutually exclusive, i.e. not possible to combine in a search query.

### *10.2.1.6. Physical properties (also in advanced search mode)*

All the supported physical properties are given by the MPDS hierarchy. A search for a high-order property assumes all the subordinate properties included in the results. In addition, even more general terms like permittivity or pressure are supported. The physical properties containing these terms in the name will be found. A part of the physical properties in the hierarchy supports numerical searches. For that an exact name of the property should be used together with the less or more sign and the numerical value of interest (in SI units). Example: *isothermal bulk modulus > 300* (assuming SI unit: GPa).

*10.2.1.7. Structure prototype systems (only in advanced search mode)*

Structure prototype systems are supported in two notations: Strukturbericht and combined. The first notation is an old crystallographic classification system still sometimes used in the scientific literature. The second notation is given by a combination of the chemical formula, the Pearson symbol, and the space group number. For instance, the most common structure prototype in the world literature is NaCl,*cF8* 225, counting about 40,000 hits. Other important structure prototypes are *e.g.* perovskite $CaTiO_3$,*cP5*,221, zinc blende ZnS,*cF8*,216, superconducting cuprate $Ba_2Cu_3YO_{6.3}$,*tP14*,123, *etc.* There are 39,990 unique distinct structure prototypes, including about 250 Strukturbericht symbols. Strukturbericht notation is under physicists still popular in the literature, even being a very old notation.

*10.2.1.8. Atomic environment types AETs (only in advanced search mode)*

The atomic environment types in the crystalline structures are arranged in the polyhedra (e.g. $TiO_6$ or $HgX_{12}$). It is possible to search throughout the entire MPDS data by the AET and the atomic composition of these polyhedra. In this category, the first provided chemical symbol is the center of the polyhedron. It makes no sense to specify any numerical index nearby. The next provided chemical symbol is treated as a vertex. Here the numerical index is properly supported. The center and the vertices atoms can be provided together or subdivided using the space or minus sign. The X symbol stands for any chemical element. Consider the following examples: (a) U-center, any CN, any vertices (*U*), (b) any center, any CN, Se-vertices (*X-Se*), (c) U-center, CN = 6, O-vertices (*UO6*), and (d) U-center, CN = 7, any vertices (*UX7*). It is also possible to define the number of different AET's within an inorganic substance to be retrieved.

*10.2.1.9. Bibliography (only in advanced search mode)*

Since all the MPDS data were manually excerpted from the peer-reviewed publications, they are searchable by their corresponding author names, publication years, journal issues, pages, DOIs, geography etc. This information can be also used for citing. Generally citing the MPDS is desirable, but not obligatory, as all the data have already their own publishers' citing information (DOIs, etc.).

*10.2.1.10. Further examples*

We also give here 3 selected retrieval results to demonstrate the power of displaying the retrieval results, always having in mind to provide the user (without efforts of the user) with a 'Holistic View', or even 'Quality Evaluation Holistic View':

1) Fig. 76 shows the richness and the domains for paramagnetic moments of 'lowest (dark pink)' towards 'richest (red)' values in binary and ternary chemical systems (min. and max values) presented as the heat map. Each square gives for the case where several values are published for different compositions its mean value. By clicking on any square (binaries) or solid circle (ternaries) the user is connected to all available information about that chemical system. The gray squares and solid circles indicate the non-formers chemical systems.

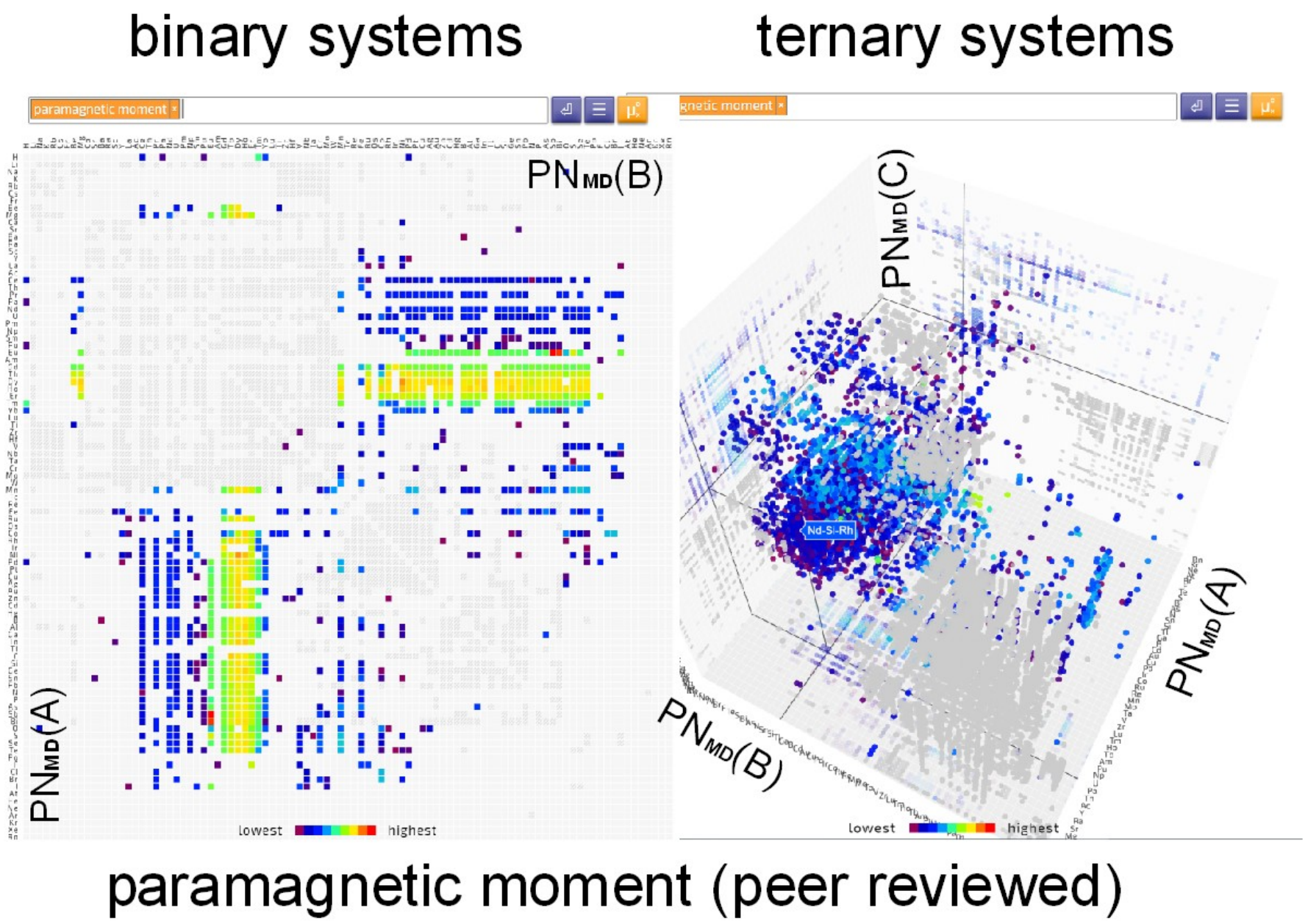

Fig. 76. Binary and ternary paramagnetic moment (peer-reviewed) as heat plots, sorted according to $PN_{MD}$.

2) Fig. 77 shows three different views of the retrieval results of the very simple search for information on 'ternary Au Ag Au'. The semantic graph gives in one glance for each distinct phase, e.g.; $Ag_{0.25}Cu_{0.25}Au_{0.50}$ ht all available information, like e.g. Debye temperature, Vickers hardness, electrical conductivity, etc. On can click on each of this property and get immediately linked to its entry(ies). At the top the Distinct Phases Table is given, in this ternary system there exists 4 different phases: $Ag_{0.25}Cu_{0.25}Au_{0.50}$ ht, $Ag_{0.04}CuAu_{0.96}$ orth rt, $Ag_{0.10}CuAu_{0.90}$ tet rt, and $Ag_{0.06}Cu_{2.94}Au$ rt. The first $Ag_{0.25}Cu_{0.25}Au_{0.50}$ ht crystallizes in space group number 225, and has in the PAULIING FILE 94 data entries from 21 different publications. And finally The preview option lists all entries 135 data entries for the ternary system Ag-Au-Cu (crystallographic structure, phase diagram, and physical properties).

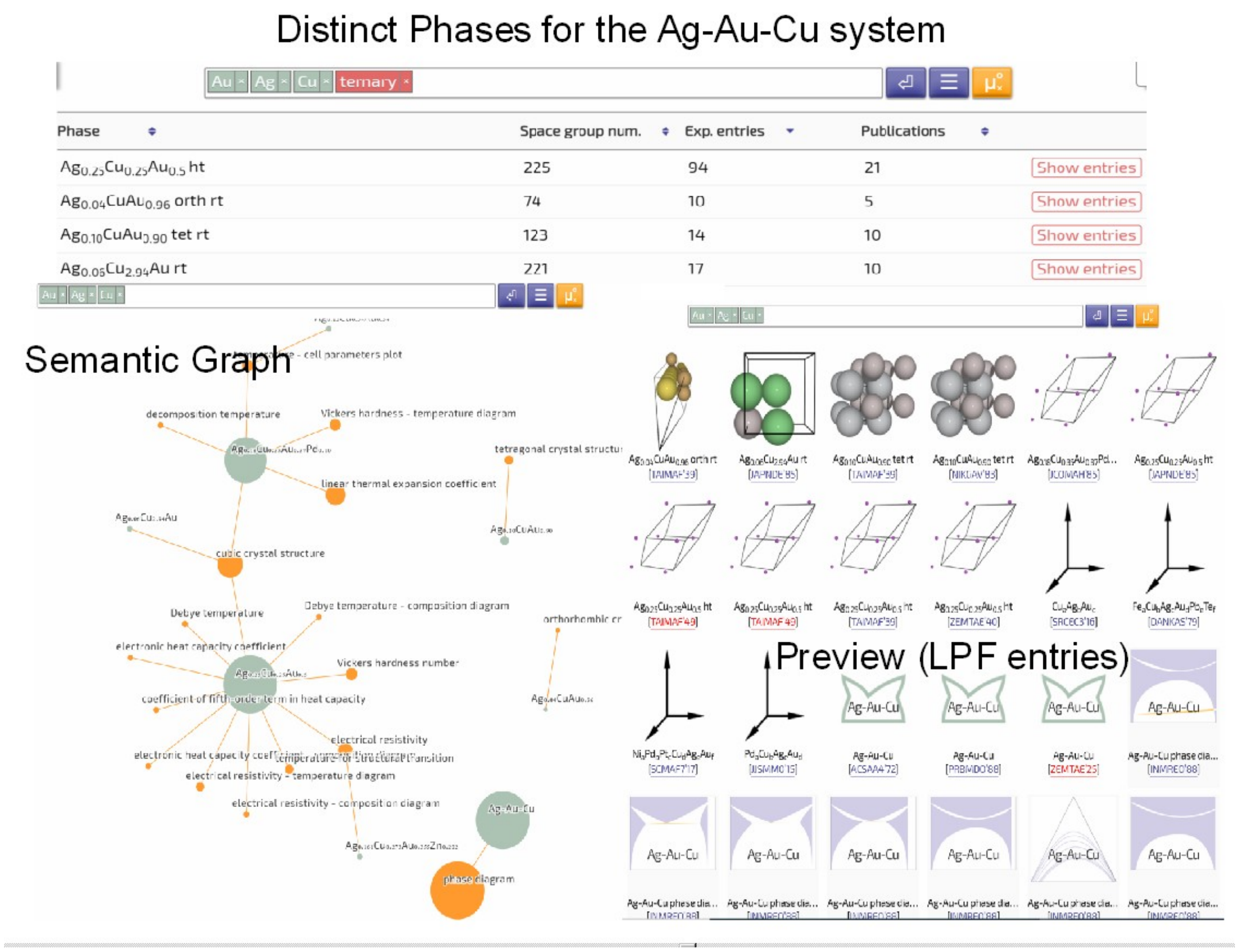

Fig. 77. Semantic graph, preview and distinct phases for Ag-Au-Cu system

3) Figure 78 displays three heat plots (y-axis: $PN_{MD(A)}$; x-axis: $PN_{MD(B)}$) for enthalpy of formation; on the left top side for ‘peer reviewed’ (from the world literature), on the right top side own created ‘machine learning’. The bottom displays the third heat plot displaying all binary systems for which both data sets exist: ‘peer-reviewed’ and ‘machine learning’. The color means now the magnetite of the difference between these two data sets. Perfect matching would result in a more or less dark pink to dark blue even distributed over the entire matrix; on the contrast poor matching would result in a more or less red to orange even distributed over the entire matrix. For our considered enthalpy of formation the overall colors are dark pink to blue telling us the matching is very good. In other words the ‘machine learning’ data are of high quality. This with the exception of oxygen-containing binaries, as well as some specific binary systems like, e.g. B-Ce, Cs-H, S-Ti, Cu-I. By clicking on any square (binaries) the user is connected to all available information about that chemical system. The gray squares and solid circles indicate the non-formers chemical systems. Therefore it can be efficiently used as ‘Quality Estimation Holistic Views’ tool, again in one glance quantitatively judging the quality of the derived data (machine learning or *ab initio*). Analogous Quality Estimation Holistic Views tools are also available for the ternary and quaternary systems.

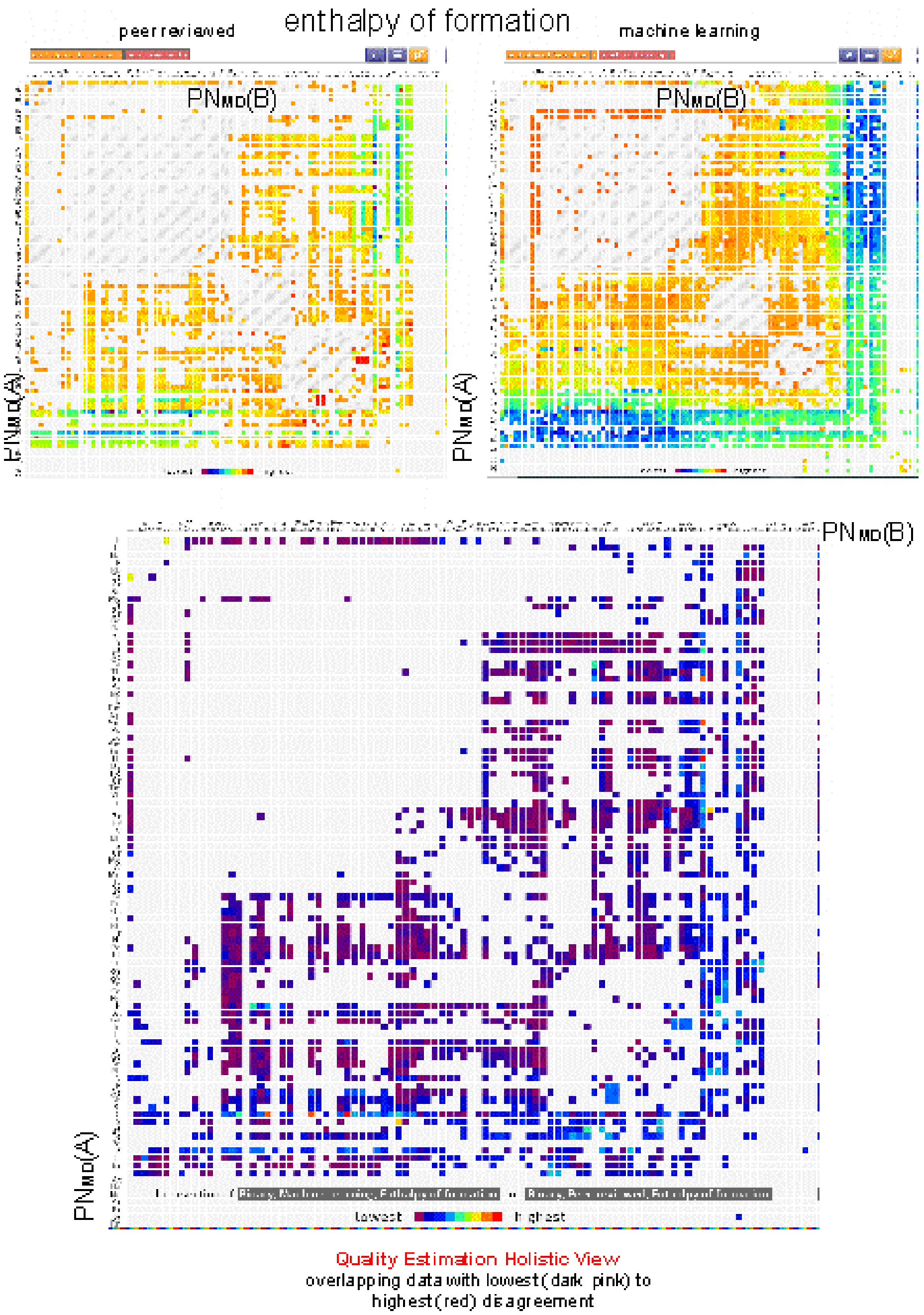

Fig. 78. Heat maps for the peer-reviewed and machine-learning enthalpies of formation and their "Quality Estimation Holistic View", respectively. Chemical elements are sorted according to $PN_{MD}$.

### 10.2.2. API access

The MPDS application programming interface (API) presents the materials data of the PAULING FILE database online in the machine-readable formats, suitable for automated processing. The intended audience is software engineers and data scientists. The full API is available by a subscription. To start using the API the reader needs a valid API key from the MPDS account.

Some parts of data are opened and freely available. In particular, these are: (a) cell parameters - temperature diagrams and cell parameters - pressure diagrams, (b) all data for compounds containing both Ag and K, (c) all data for binary compounds of oxygen, (d) complete set of selected 'best' entries for all binaries are open-access for education, advertising and research purposes. (e) all data generated by machine learning, and (f) all data generated by first-principles calculations. Please, login via GitHub to get the API access to these data. A tech-savvy reader may also explore Jupyter notebooks (also in Binder) and kickoff Python demos. Important: using the API keys within the third-party environment like Binder is potentially insecure and should be done with the greatest care.

*10.2.2.1. MPDS data structure*

The standard unit of the MPDS data is an entry. All the MPDS entries are subdivided into three kinds: crystalline structures, physical properties, and phase diagrams. They are called S-, P- or C-entries, correspondingly. Entries have persistent identifiers (similar to DOIs), e.g. S377634, P600028, C100027. Another dimension of the MPDS data is the distinct phases. The three kinds of entries are interlinked via the distinct materials phases they belong. A tremendous work was done by the PAULING FILE team in the past 30 years to manually distinguish more than 170 000 inorganic materials phases, appearing in the literature. Each phase has a unique combination of (a) chemical formula, (b) Pearson symbol, (c) space group. Each phase has an integer identifier called phase_id. Consider the following example of the entries and distinct phases. There can be the following distinct phases for the titanium dioxide: rutile with the space group 136 (let us say, phase_id 1), anatase with the space group 141 (phase_id 2), and brookite with the space group 61 (phase_id 3). Then the S- and P-entries for the titanium dioxide must refer to either 1, or 2, or 3, and the C-entries must refer to 1, 2, and 3 simultaneously.
Our new development is the machine-learning data generated from the original peer-reviewed data for the less known phases. Such data are always clearly attributed and not supplied by default. So far only the P-entries (10 physical properties) were predicted and thus can be of the machine-learning data type. See below how to enable these data in the API outputs.

*10.2.2.2. Categories of data*

The most common task is to get the MPDS data according to some criteria. We support 14 basic criteria for searching and downloading the entries, here are the categories of MPDS data listed.

| Categories of data (facet) | | *Machine-readable name* | Example |
|---|---|---|---|
| 1) | Physical properties (see MPDS physical properties hierarchy, numerical properties specifically): | *props* | conductivity |
| 2) | Chemical elements: | *elements* | Cs, Fr-O |
| 3) | Materials classes (various groups of terms: i.e. mineral names, periodic groups, physical classes, element count keywords etc.) | *classes* | perovskite, quaternary |
| 4) | Crystal system: | *lattices* | orthorhombic |
| 5) | Chemical formula: | *formulae* | SrTiO3, O3Al2 |
| 6) | Space group (number or international short symbol): | *sgs* | 221, Pm-3m |
| 7) | Prototype system (Strukturbericht notation or formula, Pearson symbol, space group number): | *protos* | D51, SiS2,oI12,72 |
| 8) | Polyhedron atoms (center, ligands): | *aeatoms* | U, X-Se, UO6, UX7 (X = any atom) |
| 9) | Polyhedral type (see all supported types): | *aetype* | icosahedrons 12-vertex |
| 10) | Publication author: | *authors* | Evarestov |
| 11) | CODEN (code of publication journal, CODEN index): | *codens* | PPCPFQ |
| 12) | Publication year: | *years* | 1960-1990 |
| 13) | Author location: | *geos* | United Arab Emirates |
| 14) | Author organization: | *orgs* | CSIRO |

Table 27. Data categories on the MPDS platform.

Thus, in an API request it is possible to combine all these criteria in a reasonable manner. Combination is always done implying conjunctive AND operator. The reasonable manner means one can combine e.g. publication authors with publication years, but cannot combine chemical formulae with chemical elements or space group with crystal system, i.e. the common sense rules are implied. We distinguish the single-term categories, such as the physical properties and crystal systems, and many-term categories, such as the materials classes and publication authors. A combination of terms from the single-term category is not supported. That is, one cannot search for both band gap and conductivity. A combination of terms from the many-term category is easily possible. That is, one can search for both binaries and perovskites.

## 11. Conclusions

*i) The PAULING FILE project is the first and only existing comprehensive purely data-driven inorganic database connecting the fundamentally different data types (crystal structures, physical properties, and phase diagrams) with the aim to discover knowledge directly from data.*

Concepts and prototypes for digital systems supporting similar interplay had also been proposed by E.J. Corey in 1969, and further developed in 1975, but the essential difference of the PAULING FILE project is to discover knowledge directly from data. This is demonstrated in this review.

*ii) To generate Holistic Views the introduction of the Distinct Phases Concept is a precondition, otherwise the different groups of inorganic substances and their properties cannot be linked.*

To create 'Holistic Views' of the database content and allow combined retrieval, it was necessary to link the different database entries from the three parts of the PAULING FILE in a more efficient way than provided through the bibliographic information and the chemical system. For this purpose the Distinct Phases concept was introduced. The linkage of the three different groups of data is achieved via a Distinct Phases relational database table, to which each individual crystallographic structure, phase diagram, and physical properties entry is linked via an identifier.

*iii) The Periodic Number ($PN_{ME}$ ,$PN_{MD}$) has the most significant role as the atomic property parameter (APP).*

All the atomic property parameters (APP) depend on the PN and AN. The atomic reactivity (many electronegativities belong to this pattern group) is simply the reciprocal of the corresponding atomic size (many radii belong to this pattern group).

*iv) The structure prototype classification and the coordination type (atomic environment AET) classifications are the direct windows of the electronic interaction within inorganic substances.*

The structure prototype classification makes principally possible the discovery of quantitative correlations between the atomic property parameters (expressions) and physical properties of the inorganic substances. In combination with the different data mining techniques, the PAULING FILE provides examples of the holistic views on inorganic substances, confirming that 'The whole is greater than the sum of its parts'. The previous conclusion suggests, that the *structure prototype*-sensitive physical properties are quantitatively described by the APPs AN and PN (or their simple mathematical functions) with respect to the constituent chemical elements. This is an important link to explore the *structure prototype*-sensitive physical properties of inorganic substances strategically.

*v) The key for machine learning is the use of efficient descriptors, not the employed machine-learning techniques, as often believed.*

The term descriptor stands for the compact information-rich representation, allowing the convenient mathematical treatment of the encoded complex data. The chemistry trends of 10 selected physical properties of inorganic substances could be 'caught' by using the structure prototype descriptor constructed as a vector, based on the following atomic data: (a) the lengths of the atomic radius-vectors, and (b) their PNs.

*vi) The Creation of the trustworthy Materials Platform for Data Science requires the knowledge of governing factors which justify general valid restraint conditions, as well as existence of a reference database (following the Distinct Phases Concept).*

Neither the Fourth Paradigm of Science, nor the Materials Genome Initiatives can be realized without the integration of restraints obtained by 'inorganic substances data exploration searching for principles (governing factors) with the aim to formulate restraints'. The infinite number of potential chemical element combinations forces us to develop approaches that are able to reduce this infinite number to a practicable number of the most probable potential inorganic substances (chemical systems), to be theoretically and experimentally investigated. Realization of the trustworthy Materials Platform for Data Science required the following two preconditions:

i) The first requirement is the introduction of the Distinct Phases Concept to link different kinds of inorganic substances data.

ii) The second requirement is the existence of a comprehensive, critically evaluated inorganic substances database system of experimentally determined single-phase inorganic substances data ('peer reviewed') from the world literature, to be used as reference.

*vii) A simple and efficient quality estimation technique for the derived (machine learning and ab initio) inorganic data is developed.*

The MPDS is a fully interlinked inorganic database, using the Distinct Phases concept to link the 'peer-reviewed' (excerpted from the world literature), 'machine learning' and '*ab initio*' inorganic substances data. With the help of '2D- (3D-)delta heat plots applied to e.g. 'peer reviewed' compared with 'machine learning' data one gets 'Quality Estimation Holistic Views' at one glance.

*viii) Overcoming the issues of "the Plato's Cave".*

Designing the new materials, it is strategic to develop the products following the conventional waterfall model and to separate the project activities into the linear sequential phases, i.e. from the requirements, design, prototyping, analysis, testing, proof of concept, verification, deployment, and scaling to maintenance, where each phase depends on the previous one and corresponds to a specialization of tasks. However, in reality, an accomplishment of this kind of project is very hard due to the limitations of resources, data, information, and knowledge. Not only a proper framing of the project design, based on the scientific rationality as well as technical feasibility, but also dynamic management of economic, social, cultural, legal, political and other incentives are required to pile up attractive results in the showcase of the materials platform. In this context our comprehensive data system is an innovation in the project management. In order to reach such a final goal, it is necessary to assume guiding principles, namely, linking data and knowledge, taking into accounts regularities, singularities and hierarchies in manifolds of the materials data sets. We believe some insights were given in this work to overcome the issues of "the Plato's Cave".

**Acknowledgements**

The authors are grateful to Dr. Karin Cenzual and Professor Roman Gladyshevskii for their contributions to the chapter 4, as well as to Dr. Karin Cenzual for her contribution to chapter 5.2. In addition the authors are grateful to Professor Andrey Sobolev for his contribution to the chapter 9. E.B. thanks Professor Bartolomeo Civalleri, Professor Lorenzo Maschio, Professor Alessandro Erba, and Professor Roberto Dovesi for their support with the CRYSTAL code.